\documentclass{elsart}
\usepackage{natbib}
\usepackage{amssymb}
\usepackage{amsmath}
\usepackage{stmaryrd}
\usepackage{marvosym}
\usepackage{graphicx}
\usepackage{url}
\usepackage{subfigure}
\usepackage{setspace}
\usepackage[footnotesize,sl,nooneline]{caption}
\usepackage{ctable}

\DeclareMathOperator*{\erf}{erf} 
 
\newcommand{\hide}[1]{\relax}
\newcommand{\unit}[1]{\ensuremath{\,\mathrm{#1}}}
\newcommand{\Og}{\ensuremath{\Omega}}

\newcommand{\meff}{m_\text{eff}}
\newcommand{\bD}{\ensuremath{\bar \Delta}}
\newcommand{\chieff}{\chi_\text{eff}}

\newcommand{\ba}{\ensuremath{\bar a}}

\newcommand{\mdagger}{\dagger}
\newcommand{\mhat}{\hat}

\newcommand{\dF}{\ensuremath{\delta\! F}}
\newcommand{\dx}{\ensuremath{\delta x}}

\newcommand{\dhqout}{\ensuremath{\delta \mhat q_\mathrm{out}}}

\newcommand{\dwdx}{\ensuremath{g_0}}

\newcommand{\etac}{\ensuremath{\eta_\mathrm{c}}}

\long
\def\symbolfootnote[#1]#2{\begingroup
\def\thefootnote{\fnsymbol{footnote}}
\footnote[#1]{#2}\endgroup}
\let\oldcite\cite
\renewcommand{\cite}[1]{(\oldcite{#1})}
\newcommand{\chapter}[1]{\relax}
\renewcommand{\intertext }[1][]{OOPS}

\begin{document}

\begin{frontmatter}

\title{Cavity optomechanics with whispering-gallery mode optical micro-resonators}

\author[mpq]{Albert Schliesser} and
\author[mpq,epfl]{Tobias J.\ Kippenberg}

\address[mpq]{Max-Planck-Institut f\"ur Quantenoptik, D-85478 Garching, Germany}
\address[epfl]{\'Ecole Polytechnique F\'ed\'erale de Lausanne (EPFL), CH-1015 Lausanne, Switzerland}

\tableofcontents

\begin{abstract}

Parametric coupling of optical and mechanical degrees of freedom forms the basis of many ultra-sensitive measurements of both force and mechanical displacement.
  An optical cavity with a mechanically compliant boundary enhances the optomechanical interaction, which gives rise to qualitatively new behavior which can modify the dynamics of the mechanical motion. 
  As early as 1967, in a pioneering work, V.~Braginsky analyzed theoretically the role of radiation pressure in the interferometric measurement process, but it has
 remained experimentally unexplored for many decades. 
  Here, we use whispering-gallery-mode optical microresonators to study these radiation pressure phenomena. Optical microresonators simultaneously host optical and mechanical modes, which are systematically analyzed and optimized to feature ultra-low mechanical dissipation, photon storage times exceeding the mechanical oscillation period (i.e. the ``resolved-sideband regime'') and large optomechanical coupling. In this manner, it is demonstrated for the first time that dynamical backaction can be employed to cool mechanical modes, i.e., to reduce their thermally excited random motion. Utilizing this novel technique together with cryogenic pre-cooling of the mechanical oscillator, the phonon occupation of mechanical radial-breathing modes could be reduced to $\langle n \rangle =63\pm20$ excitation quanta. The corresponding displacement fluctuations are monitored interferometrically with a sensitivity at the level of $1\cdot 10^{-18} \unit{m/\sqrt{Hz}}$, which is below the imprecision at the standard quantum limit (SQL). This implies that the readout is already in principle sufficient to measure the quantum mechanical zero-point position fluctuations of the mechanical mode. Moreover, it is shown that  optical measurement techniques employed here are operating in a near-ideal manner according to the principles of quantum measurement, displaying a backaction-imprecision product
close to the quantum limit.

\end{abstract}

\end{frontmatter}

\relax

\section{Introduction}

\begin{figure}[tb]
\centering
\includegraphics[width=.4\linewidth
]{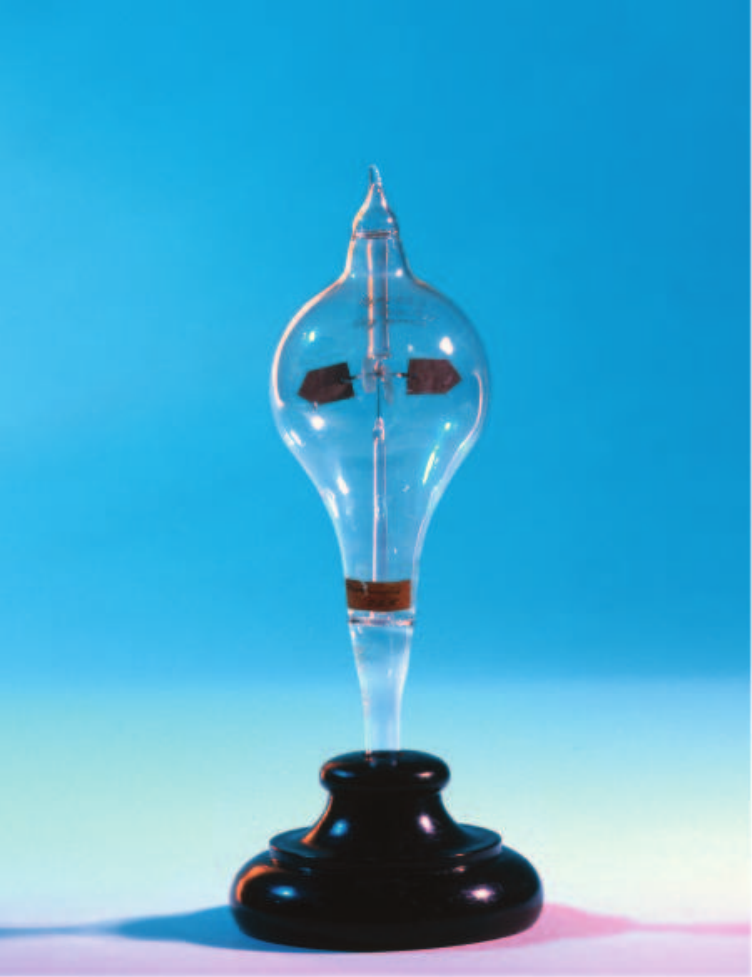}
\caption{Radiometer used by Crookes around 1877. The pivoted vanes, each
silvered and blackened on one face, are free to revolve around the central
axis. Crookes observed that the vanes started to turn when illuminated by a
strong light source. The origin of the effect was not radiation pressure, it
was mediated by the dilute gas still present in the globe. In refined
experiments by Lebedev in Russia and Hull in the US, effects of radiation
pressure on mechanical objects were observed at the beginning of the $20^{%
\mathrm{th}}$ century. Image copyright Science Museum/SSPL, London, reproduced with permission. \label{f:radiometer}}
\end{figure}

The tails of comets always point away from the sun. Kepler, in the $17^{%
\mathrm{th}}$ century, already conjectured from this observation that
sunlight exerts a force on particles in the comet tail. Some 250 years
later, Maxwell's theory of electromagnetic radiation put this conjecture on
solid theoretical grounds, yet, collecting experimental evidence for the
fact that light carries momentum has eluded even the most skilled
experimentalists of that age. As a famous example, Crookes attempted to
construct a radiometer in which the transfer of optical momentum makes a
vane spin, now famous as the \textquotedblleft light mill\textquotedblright\
(figure~\ref{f:radiometer}). After much debate, however, it was understood
that the observed rotation is mediated by the dilute gas in which the vane
is kept \cite{Woodruff1968}. True radiation-pressure effects, in agreement
with Maxwell's predictions, were not observed until the beginning of the $%
20^{\mathrm{th}}$ century, in more sophisticated experiments carried out by
Lebedew in Russia \cite{Lebedew1901} and Nichols and Hull in the United
States \cite{Nichols1901, Nichols1903a,Nichols1903b}.

The situation changed dramatically with the advent of lasers as highly
coherent light sources in the 1970s. It was soon suggested that one could \emph{utilize%
} the resulting forces to manipulate the motion of dielectric particles and
atoms or ions in a controlled manner. Pioneering work of Ashkin at Bell
Laboratories demonstrated trapping of dielectric particles using radiation
pressure, a technique still widely adopted today and known as ``optical
tweezers'' \cite{Ashkin1970}. Soon thereafter, proposals to utilize the laser field to cool the
motion of atoms emerged. H\"{a}nsch and Schawlow \cite{Hansch1975} and also
Wineland and Dehmelt \cite{Wineland1975,Wineland1979} suggested ways to use
radiation pressure to dampen the random motion of atoms or ions, now widely
known as \textquotedblleft laser cooling\textquotedblright. Ashkin and Chu
demonstrated this principle to trap and cool neutral particles using optical
forces \cite{Ashkin1978,Chu1985}. In the decades to follow, the
implementation of these ideas lead to what can be considered a true
revolution in atomic physics, pivotal for discoveries such as Bose-Einstein
condensation and the development of the most precise frequency standards
available today.

Among the most spectacular advances of this laser control is the
demonstration of quantum ground state cooling of a harmonically trapped ion 
\cite{Diedrich1989,Monroe1995, Hamann1998}, which has enabled the generation
of such exotic motional states as Fock or Schr\"{o}dinger cat states \cite%
{Leibfried2003} of a single trapped ion. However, it is a prime example of
the enigmatic quantum-classical \textquotedblleft
boundary\textquotedblright\ that such quantum effects and state preparation
have never been accomplished with mechanical oscillators. It is an
interesting, yet little known fact that the ramifications of radiation
pressure on mechanical oscillators is a subject that even predates the work
of Ashkin and co-workers.

\begin{figure}[tb]
\centering
\includegraphics[width=9.5cm]{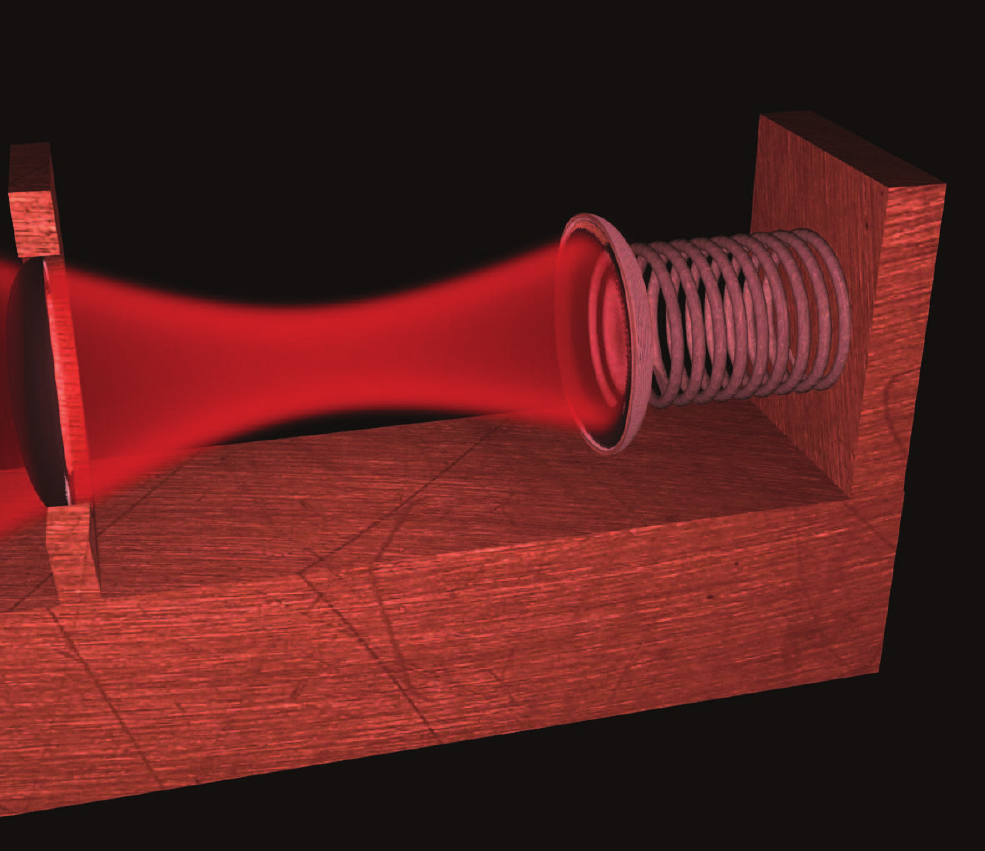} 
\caption{Artist's view of a generic optomechanical system: A Fabry-Perot\
resonator, consisting of two mirrors trapping near-resonant monochromatic
light. One of the mirrors is mechanically compliant, here it is mounted on a
spring and therefore constitutes a mechanical harmonic oscillator.}
\label{f:optomechanics}
\end{figure}

Braginsky's group at Moscow state university has played a pioneering role in
this regard. Embedded eventually in a very general theory of quantum
measurements \cite{Braginsky1980,Braginsky1992, Braginsky1996, Braginsky2003}%
, Braginsky considered already in the year 1967 the role of radiation
pressure in an optical interferometer \cite{Braginskii1967}. Since the end mirrors of a
interferometer are mechanically compliant they are subject to a radiation
pressure force. An important impetus for such \emph{theoretical} analysis
has been the use of interferometers as detectors for gravitational waves.
The exquisite sensitivity required in these experiments caused a strong
theoretical interest in understanding the limits of linear continuous
displacement measurements based on parametric transducers. Braginksy
identified two important consequences of the radiation pressure force that
photons exert on the interferometer end mirrors, and has developed a
comprehensive understanding of optomechanical interactions as they occur in
the fundamental building block of a gravitational wave observatory shown in
figure~\ref{f:optomechanics}: Monochromatic light trapped in a high-finesse
cavity (a Fabry-Perot\ resonator in this case) exerts radiation pressure on
the massive end mirrors, coupling their oscillatory motion to the light.

As early as 1967  Braginsky and coworkers \cite{Braginskii1967}
recognized that radiation pressure can change the dynamics of the mechanical
degree of freedom, effectively adding an optically induced viscous damping
to the mirror motion. It was also soon understood that this phenomenon
(termed \emph{dynamical backaction} \cite{Braginskii1967, Braginskii1970,
Braginsky1977, Braginsky2001}) could be used in principle to amplify or cool
the motion of the mirror \cite{Dykman1978, Braginsky2002} - which is in
essence very similar to the principle of laser cooling of atoms, discovered
in the decade to follow. While dynamical backaction is a classical effect,
Braginsky and co-workers also showed that the quantum nature of light within
the interferometer gives rise to an unsurmountable sensitivity limit in
displacement measurements of the mirror: For sufficiently strong input
power, quantum fluctuations of the radiation pressure force induces random
motion in the mirror, that masks the displacement to be detected \cite%
{Braginsky1992, Caves1980, Caves1981} by e.g. a gravitational wave, an
effect now referred to as the \emph{quantum backaction} of the measurement.
While the Moscow group devoted many decades in the development of parametric
transducers for gravity wave detection, it was not possible to observe the
dynamical backaction cooling or amplification, though pioneering attempts
\cite{Braginskii1970} were made by the Moscow group in 1969.

Interest in optomechanics reemerged in the 1990 in the context of quantum
optics, resulting in a variety of proposals exploring quantum effects in
these systems, ranging from proposal for quantum non-demolition measurements
of the light intensity or single quadrature measurements of the mechanical
displacement \cite{Braginsky1977a, Braginsky1980, Braginsky1996,
Heidmann1997}, as well as the generation of narrow-band squeezed light using
the mechanical oscillator as an effective third-order-nonlinearity \cite%
{Fabre1994, Mancini1994}. Later work has studied in great detail the
possibility to generate non-classical states of motion \cite{Bose1997},
including superposition states \cite{Marshall2003}, entangled states of
several oscillators \cite{Mancini2002, Zhang2003, Pinard2005,
Bhattacharya2008}, or motional states entangled with optical degrees of
freedom \cite{Vitali2007, Bhattacharya2008a}. As a special quantum state, it
has also been suggested to cool the mechanical degree of freedom to its
quantum ground state using cooling by dynamical backaction \cite%
{Wilson-Rae2007, Marquardt2007, Bhattacharya2007a, Genes2008, Dantan2008}.

Observing such phenomena \emph{experimentally} however, had been severely complicated by several
experimental challenges. First, optomechanical coupling by radiation
pressure is usually weak---the momentum transfer of a single reflected
photon changes the velocity of a (free) gram-scale mass by some $10^{-24}\,%
\mathrm{m/s}$ only. At the same time, the oscillator displacements
associated with quantum effects are typically on the scale of its zero-point
fluctuations $x_{\mathrm{ZPF}}=\sqrt{\hbar /2m\Omega _{\mathrm{m}}}$, where $%
m$ and $\Omega _{\mathrm{m}}$ are its mass and the resonance frequency,
respectively. For gram-scale oscillators, such as a small end mirror in an
interferometer, this motion is almost ten orders of magnitude smaller than
the fluctuations of a trapped atom or ion---usually at the sub-attometer
scale. Dedicated experiments \cite{Bocko1996, Tittonen1999,
Hadjar1999, Caniard2007a} have been approaching such sensitivities, but 
another challenge persists in all room-temperature
experiments: Thermal noise tends to mask quantum
signatures (caused e.\ g.\ by quantum backaction) as long as the thermal energy 
$k_{\mathrm{B}}T$ largely exceeds the energy scale $\hbar \Omega _{\mathrm{m}%
}$ of the motional quantum ($k_{\mathrm{B}}$ is the Boltzmann, and $\hbar $
the reduced Planck constant).

In recent years, the tremendous progress in micro- and nanofabrication
technologies has provided novel opportunities to engineer optomechanical devices.
For example,  toroidal whispering-gallery mode optical microresonators
were shown to exhibit optomechanical coupling to mechanical  modes of the
structure \cite{Kippenberg2005, Rokhsari2005}.  The high quality factors of their optical
and mechanical modes  allowed observing
radiation pressure amplification and coherent oscillations for the first
time in 2005 \cite{Kippenberg2005, Rokhsari2005}.
A wide variety of
other optomechanical systems has since emerged, ranging from the nano- to the
macroscale. With the
advent of these new optomechanical systems, interest in observing their
quantum mechanical properties \cite{Schwab2005} revived, an effort
pursued predominantly with electronic transducers up to that point.
The new research field, now
widely referred to as \textit{cavity (quantum) optomechanics} \cite%
{Kippenberg2007,Kippenberg2008}, uses a remarkably diverse set of
experimental systems as depicted in table \ref{t:systems}, which summarizes
the properties of optomechanical systems studied in the last ten years,
reflecting also the enormous parameter range covered.

{\tiny \ctable[ rotate, cap=tbd, caption=Key parameters of recently studied
optomechanical systems. Earlier results are summarized in ref.\
 \cite{Bocko1996}.,
label=t:systems,
]
{l|r|r|r|r|r}{
\tnote[]{
\tiny $^a$at room temperature unless noted otherwise,
\tiny $^b$experimental value, including all noise except for the mechanical mode of interest,
\tiny $^c$quoted numbers are from the references printed in bold,
\tiny $^d$at $2 \unit{K}$, at room temperature  $250{,}000$,
\tiny $^e$at room temperature, for values at cryogenic temperatures see ref.\ \cite{Arcizet2009a},
\tiny $^f$at $410 \unit{K}$, at room temperature  $32{,}000$,
\tiny $^g$at $4.2 \unit{K}$, at room temperature  $16{,}000$,
\tiny $^h$at $300 \unit{mK}$, at room temperature $1{,}100{,}000$,
\tiny $^i$the device is operated at d.c.,
\tiny $^j$superconducting microwave resonator,
\tiny $^k$the resonance is a combined mechanical and electronic resonance,
\tiny $^l$laser interferometer gravitational wave interferometer. The resonance frequency is increased by electronic feedback on actuators controlling the mirror position,
\tiny $^m$at $5.3 \unit{K}$, at room temperature  $5{,}000$%
}
}%
{
\multicolumn{1}{c}{}\\
\multicolumn{1}{c}{}\\
 System                             & \multicolumn{1}{c|}{Resonance}
                                                    &\multicolumn{1}{c|}{Quality}
                                                                        & \multicolumn{1}{c|}{Effective}
                                                                                            & \multicolumn{1}{c|}{Displacement}
                                                                                                            & References\tmark[c]\\
                                    & \multicolumn{1}{c|}{frequency}
                                                    &\multicolumn{1}{c|}{factor\tmark[a]}
                                                                        & \multicolumn{1}{c|}{mass}
                                                                                            & \multicolumn{1}{c|}{sensitivity\tmark[b]}
                                                                                                            & \\
                                    & \multicolumn{1}{c|}{(Hz)}
                                                    &\multicolumn{1}{c|}{}
                                                                        & \multicolumn{1}{c|}{ (g)}
                                                                                            & \multicolumn{1}{c|}{($\mathrm{m}/\sqrt{\mathrm{Hz}}$)}
                                                                                                            & \\
\hline
 Internal modes of a silica mirror  & $1.86\cdot10^6$   & $40{,}000$    & $230\cdot10^{-3}$ &                   & {\bf\cite{Cohadon1999}}\cite{Pinard2000,Caniard2007, Verlot2008}\\
 Silicon torsional oscillator       & $26\cdot10^3$     & $4{,}300{,}000$\tmark[d]
                                                                        & $10\cdot10^{-3}$  &$2.0\cdot10^{-16}$ & {\bf\cite{Tittonen1999}}\cite{Hahtela2004}\\
 AFM cantilevers (thermal)          & $7.28\cdot10^3$   & $2{,}000 $    &                   &                   & {\bf\cite{Hohberger2004}}\cite{Ludwig2008} \\
 Silicon micromirror                & $0.81\cdot10^6$   & $10{,}000 $    & $190\cdot10^{-6}$ &$2.0\cdot10^{-18}$& {\bf\cite{Arcizet2006}}\cite{Arcizet2006a, Arcizet2008a}\\
 Free standing Bragg mirror         & $278\cdot10^3$    & $9{,}000$    & $400\cdot10^{-9}$ & $\sim 10^{-16}$    & {\bf \cite{Bohm2006}} \cite{Gigan2006,Groblacher2008,Cole2008}\\
 Dielectric micromirror on silicon cantilever
                                    & $12.5\cdot10^3$   & $18{,}000$    & $24\cdot10^{-9}$  & $1\cdot10^{-13}$  & {\bf \cite{Kleckner2006}}\cite{Kleckner2006a}\\
 Silica microtoroids                & $40.6\cdot10^6$   & $31{,}000$\tmark[e]    & $10\cdot10^{-9}$  & $1.0\cdot10^{-18}$& {\bf\cite{Schliesser2008}}\cite{Schliesser2006,Schliesser2008b,Schliesser2009a}\\
 Spoky silica microtoroids          & $38.0\cdot10^6$   & $80{,}000$\tmark[f]    &                   &                   & {\bf \cite{Anetsberger2008}}\\
 Silica microspheres                & $87.2\cdot10^6$   & $15{,}600$    &                   &                   & \cite{Ma2007}{\bf\cite{Park2007a}}\\
 Chip-based silica microsphere      & $1.08\cdot10^9$   &               &                   &                   & {\bf \cite{Carmon2007}}\\
 Big suspended mirrors              & $173$             & $3{,}200$     & $1$               &                   & {\bf \cite{Corbitt2007}},\cite{Corbitt2007a}\\
 Ultrasoft silicon cantilever       & $3.3\cdot10^3$    & $44{,}200$\tmark[g]
                                                                        & $140\cdot10^{-12}$& $1\cdot10^{-12}$  & {\bf \cite{Poggio2007}}\\
 Silicon cantilever                 & $7\cdot10^3$      & $20{,}000$    & $2.4\cdot10^{-6}$ &                   & {\bf \cite{Brown2007}}\\
 Silicon nitride membrane           & $135\cdot10^3$    & $12{,}000{,}000$\tmark[h]&$40\cdot10^{-9}$  & $4.5\cdot10^{-16}$     & {\bf \cite{Thomson2007}}\cite{Zwickl2008}\\
 Gold-coated silicon micropaddle    & $547\cdot10^3$    & $1{,}060$     & $11\cdot10^{-12}$ &                   & {\bf \cite{Favero2007}}\\
 Silica nanowire                    & $193$\tmark[i]    &               & $26\cdot10^{-9}$ &                    & {\bf \cite{Eichenfield2007}}\\
 Mirror on a flexure mount          & $85$              & $44{,}500$    & $0.69$            & $3\cdot10^{-15}$   & {\bf \cite{Mow-Lowry2008}}\\
 Nanomechanical resonator coupled to SCMWR\tmark[j]
                                    & $1.53\cdot10^6$   & $300{,}000$   & $6.2\cdot10^{-12}$ & $45\cdot10^{-15}$ & {\bf \cite{Teufel2008}}\cite{Regal2008, Hertzberg2009, Rocheleau2009}\\
 Resonant-bar gravitational wave detector\tmark[k]
                                    & $865$             & $1{,}200{,}000$& $1.1\cdot10^{+6}$ &                  & {\bf \cite{Vinante2008}}\\
 Silicon photonic circuit           & $8.87\cdot10^6$   & $1{,}850$    &                   & $18\cdot10^{-15}$  & {\bf \cite{Li2008}}\\
 Silicon nitride ``zipper'' cavity  & $8.2\cdot10^6$    & $\sim100$     & $43\cdot10^{-12}$ & $2\cdot10^{-15}$  & {\bf \cite{Eichenfield2009a}}\\
 Suspended mirrors in LIGO\tmark[l] & $140$             &               & $2.7\cdot10^{+3}$ & $2\cdot10^{-19}$   & {\bf \cite{Abbott2009b}}\\
 Ta$_2$O$_5$/SiO$_2$ Bragg mirror on Si$_3$N$_4$ beam
                                    & $945\cdot10^3$    & $30{,}000$\tmark[m]
                                                                        & $43\cdot10^{-9}$  & $1\cdot10^{-16}$  & {\bf \cite{Groblacher2009}}\\
$e^-$-beam deposited nanorod    & $1.9\cdot10^6$    & $6{,}500$     &                   & $1\cdot10^{-13}$  & {\bf \cite{Favero2009}}\\
 Silica double-disk resonator       & $8.5\cdot10^6$    & $4{,}070$     &                   & $2\cdot10^{-15}$  & {\bf \cite{Lin2009}}\\%
 Silicon optomechanical crystal     & $2.3\cdot10^9$    & $1{,}300$     & $330\cdot10^{-15}$&                   & {\bf \cite{Eichenfield2009}}\\%            \\
 Silicon nitride nanoresonator      & $10.7\cdot10^6$   & $53{,}000$    & $3.6\cdot10^{-12}$& $6.4\cdot10^{-16}$& {\bf \cite{Anetsberger2009}}\\%
 Trapped ion                        & $71\cdot10^3$     &               & $4.0\cdot10^{-23}$&                   & {\bf \cite{Vahala2009}}\\%
CaF$_2$ WGM resonator     & $1.06\cdot10^6$   & $136{,}000$   & $0.5$    &                   & {\bf \cite{Hofer2009}}\\%
 \hline
} }

Within this work, we have developed optomechanical systems based on
silica whispering-gallery mode (WGM) microresonators. 
Their compactly co-located high-quality optical and
mechanical modes render microresonators a very favorable system for the
study of radiation-pressure induced optomechanical coupling. Indeed, we show
here that basically all experimental challenges for the observation of
quantum effects can be successfully addressed with this system, providing a
route to ground state cooling of a mechanical oscillator or measurement of
its zero point motion. As a key enabling step towards these studies, it has
been possible for the first time to demonstrate radiation-pressure cooling
of a mesoscopic mechanical oscillator based on optically induced dynamical
backaction \cite{Schliesser2006}. Adapting, in addition, advanced quantum
optical and cryogenic techniques to this setting, we show experimentally
that these systems are capable of closely approaching fundamental quantum
limits---both in terms of the quality of the displacement sensitivity, reaching an imprecision below that at the 
SQL---and the occupation of the mechanical oscillator.

\clearpage

\section{Theory of optomechanical interactions}

\label{s:omtheory}

\subsection{Classical description and elementary phenomenology}

\label{ss:classical}

\begin{figure}[!b]
\centering
\includegraphics[width=\linewidth]{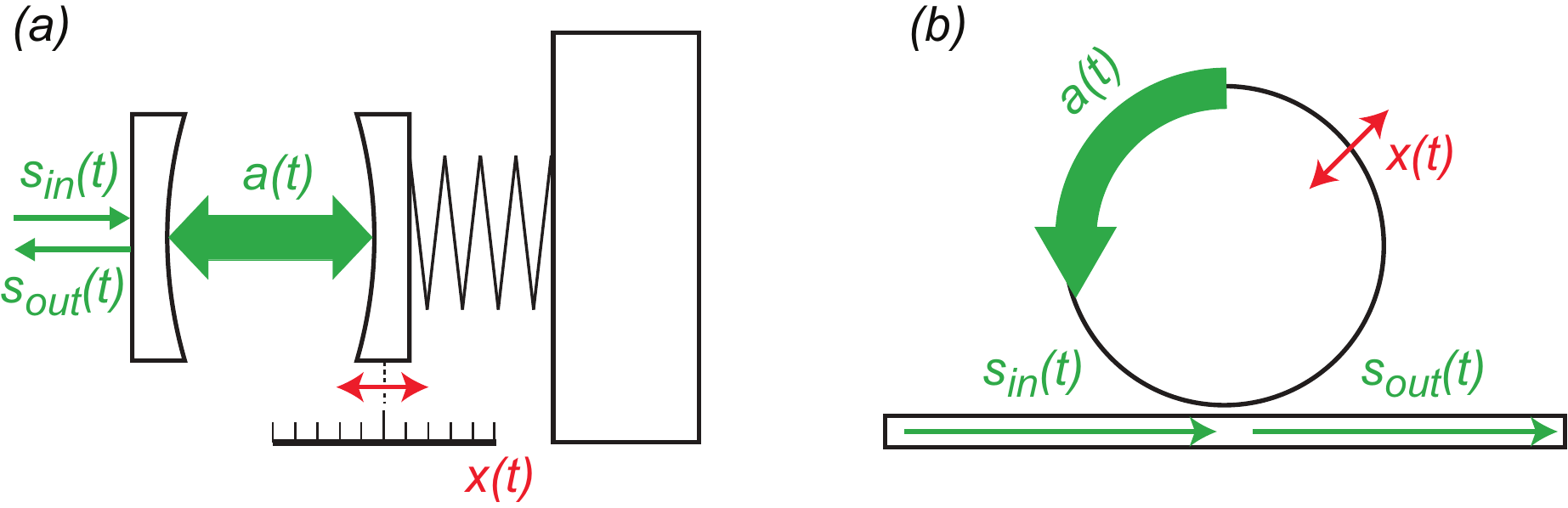}  
\caption{Schematic of two generic geometries of optomechanical systems. (a)
Linear Fabry-Perot-type cavity with a movable mirror, pumped through a
partially transparent mirror. (b) Whispering-gallery mode resonator pumped
by evanescent coupling to a waveguide. In both cases, an incoming field $s_%
\mathrm{in}(t)$ drives the intracavity field $a(t)$. The cavity resonance
frequency depends on the displacement $x(t)$ of a cavity boundary from its
equilibrium position.}
\label{f:genericom}
\end{figure}

To begin the discussion of cavity optomechanics, it is useful to review some
basic underlying physical concepts and simple limiting cases. Consider the
generic optomechanical system depicted in figure~\ref{f:genericom}(a). The
impinging field $s_\mathrm{in}(t)$ drives the cavity mode amplitude $a(t)$.
Part of the boundary of this mode---in the simplest case, one of the end
mirrors of a Fabry-Perot\ cavity---is free to move, and its displacement is
described by the one-dimensional variable $x(t)$. Irrespective of the
spatial structure of both the optical mode and the mechanical displacement
pattern, we assume that the displacement $x(t)$ shifts the resonance
frequency of the optical mode in a linear fashion, 
\begin{equation}  \label{e:dwdxdef}
\omega_\mathrm{c} ^{\prime }(t)=\omega_\mathrm{c} +g_0  x(t),
\end{equation}
where $\omega_\mathrm{c} $ is the cavity resonance frequency for $x=0$ and 
\begin{equation}
g_0 \equiv\frac{\partial \omega_\mathrm{c} ^{\prime }}{\partial x}
\label{e:dwdx}
\end{equation}
is the optomechanical coupling constant. For the cases depicted in figure~%
\ref{f:genericom}, $g_0 =-\omega_\mathrm{c} /L$ for a Fabry-Perot\ cavity of
length $L$, and $g_0 =-\omega_\mathrm{c} /R$ for a WGM resonator of radius $R
$.

\subsubsection{A moving cavity boundary}

We will first discuss the effect of a moving cavity boundary on the optical
mode, and neglect the backaction (radiation pressure) of the light. For a
monochromatic pump wave $\bar s_\mathrm{in}  e^{-i\omega_\mathrm{l}  t}$,
the resulting equation of motion for the intracavity field amplitude reads
(following the notation of \cite{Haus1984}) 
\begin{align}
\dot a(t) &= \left(-i \left(\omega_\mathrm{c} +g_0  x(t)\right) -\frac{\kappa%
}{2}\right) a(t) + \sqrt{\eta_\mathrm{c}  \kappa} \,\bar s_\mathrm{in} 
e^{-i\omega_\mathrm{l}  t}
\end{align}
Here, $a(t)$ and $\bar s_\mathrm{in} (t)$ are the amplitudes of the field in
the cavity, and the driving field, respectively, normalized such that $%
|a(t)|^2$ is the intracavity photon number and $|\bar s_\mathrm{in} (t)|^2$
the photon flux impinging on the cavity. Furthermore, $\omega_\mathrm{l}
$ is the laser field angular frequency, $\kappa$ the cavity energy decay
rate, and $\eta_\mathrm{c} \equiv\tau_0 /(\tau_0 +\tau_\mathrm{ex} )\in[%
0\ldots1]$ describes the degree of overcoupling, $\tau_0 $ and $\tau_\mathrm{%
ex} $ being the cavity photon lifetimes due to intrinsic losses of the
cavity and coupling to the taper waveguide, respectively. As one of the
simplest cases, we analyze the response of the driven cavity to a sinusoidal
oscillation in the mechanical degree of freedom. For $x(t)=x_0 \sin(\Omega_%
\mathrm{m}  t)$, the solution for the mode amplitude reads %
\begin{equation}
a(t)=\sqrt{\eta_\mathrm{c}  \kappa} \,\bar s_\mathrm{in} 
\sum_{n=-\infty}^{+\infty} \frac{i^n J_n(\beta)}{-i(\omega_\mathrm{l} +n
\Omega_\mathrm{m} -\omega_\mathrm{c} )+\kappa/2}e^{-i(\omega_\mathrm{l}
+n\Omega_\mathrm{m} )t-i \beta \cos(\Omega_\mathrm{m}  t)}
\end{equation}
after all transients have decayed within a timescale of $\kappa^{-1}$ \cite%
{Schliesser2008}. Here, the $J_n$ are the Bessel functions of the first kind
and $\beta=g_0  x_0/\Omega_\mathrm{m} $ is the modulation index. For small
amplitudes $x_0$, so that $|\beta|\ll 1$, the intracavity field can be
approximated to 
\begin{align}
a(t)&\approx a_0(t)+a_1(t)+\mathcal{O}(\beta^2) \\
a_0(t)&= \frac{\sqrt{\eta_\mathrm{c}  \kappa} \,\bar s_\mathrm{in}  }{%
-i\Delta+\kappa/2} e^{-i \omega_\mathrm{l}  t} \\
a_1(t)&= \frac{g_0  x_0}{2} \frac{ \sqrt{\eta_\mathrm{c}  \kappa} \,\bar s_%
\mathrm{in}  }{-i\Delta+\kappa/2}\bigg( \underbrace{\frac{e^{- i (\omega_%
\mathrm{l} +\Omega_\mathrm{m} ) t}}{-i(\Delta+\Omega_\mathrm{m} )+\kappa/2}}_%
\text{anti-Stokes}- \underbrace{\frac{e^{- i (\omega_\mathrm{l} -\Omega_%
\mathrm{m} ) t}}{-i(\Delta-\Omega_\mathrm{m} )+\kappa/2}}_\text{Stokes} %
\bigg),  \label{e:sidebands}
\end{align}
where $\Delta=\omega_\mathrm{l} -\omega_\mathrm{c} $ is the laser detuning.
Evidently, the moving boundary acts as a modulator, building up a pair of
sidebands $a_1$ in the cavity, with weights proportional to the cavity
Lorentzian evaluated at the frequencies $\omega_\mathrm{l} +\Omega_\mathrm{m}
$ and $\omega_\mathrm{l} -\Omega_\mathrm{m} $ for the upper and lower
sideband, respectively. These sidebands, schematically shown in figure~\ref%
{f:sidebands}, are also often referred to as anti-Stokes and Stokes
sidebands \cite{Kippenberg2005}. 
\begin{figure}[tb]
\centering
\includegraphics[width=.8\linewidth]{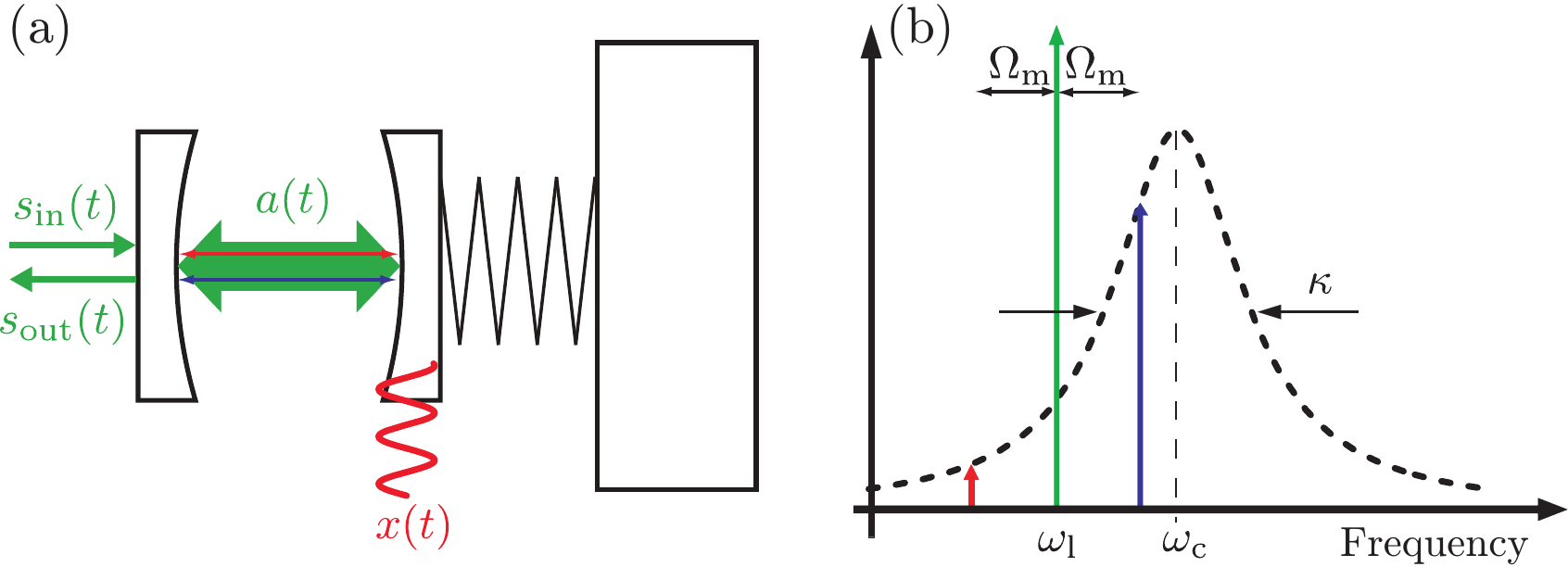}  
\caption{Response of a driven cavity to an oscillating end mirror (a), which
leads ti the generation of anti-Stokes and Stokes sidebands at frequencies $%
\protect\omega_\mathrm{l} +\Omega_\mathrm{m} $ and $\protect\omega_\mathrm{l}
-\Omega_\mathrm{m} $, weighted by the cavity Lorentzian (b). }
\label{f:sidebands}
\end{figure}

\begin{figure}[tb]
\centering
\includegraphics[width=.8\linewidth]{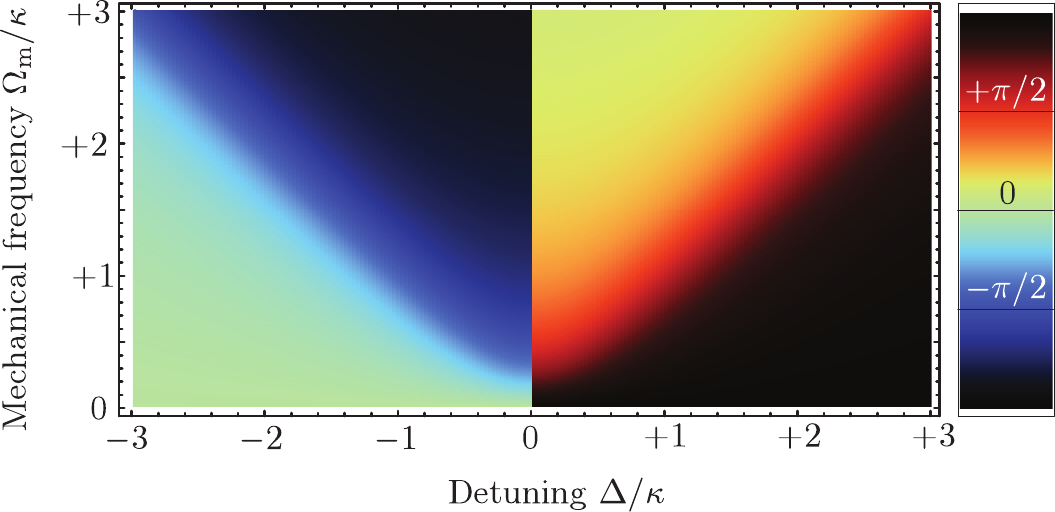}  
\caption{Phase lag $\protect\phi_\mathrm{lag}$ between the oscillation of
the mechanical degree of freedom and the stored optical energy (or
circulating power) in the cavity, assuming $g_0 <0$. For $\pm\Delta=\Omega_%
\mathrm{m} \gg\protect\kappa$ the phase lag is exactly $\mp \protect\pi/2$.
The phase jump at $\Delta=0$ is due to a zero crossing of the oscillation
amplitude of the stored optical energy.}
\label{f:phaselag}
\end{figure}

\begin{figure}[tb]
\centering
\includegraphics[width=.5\linewidth]{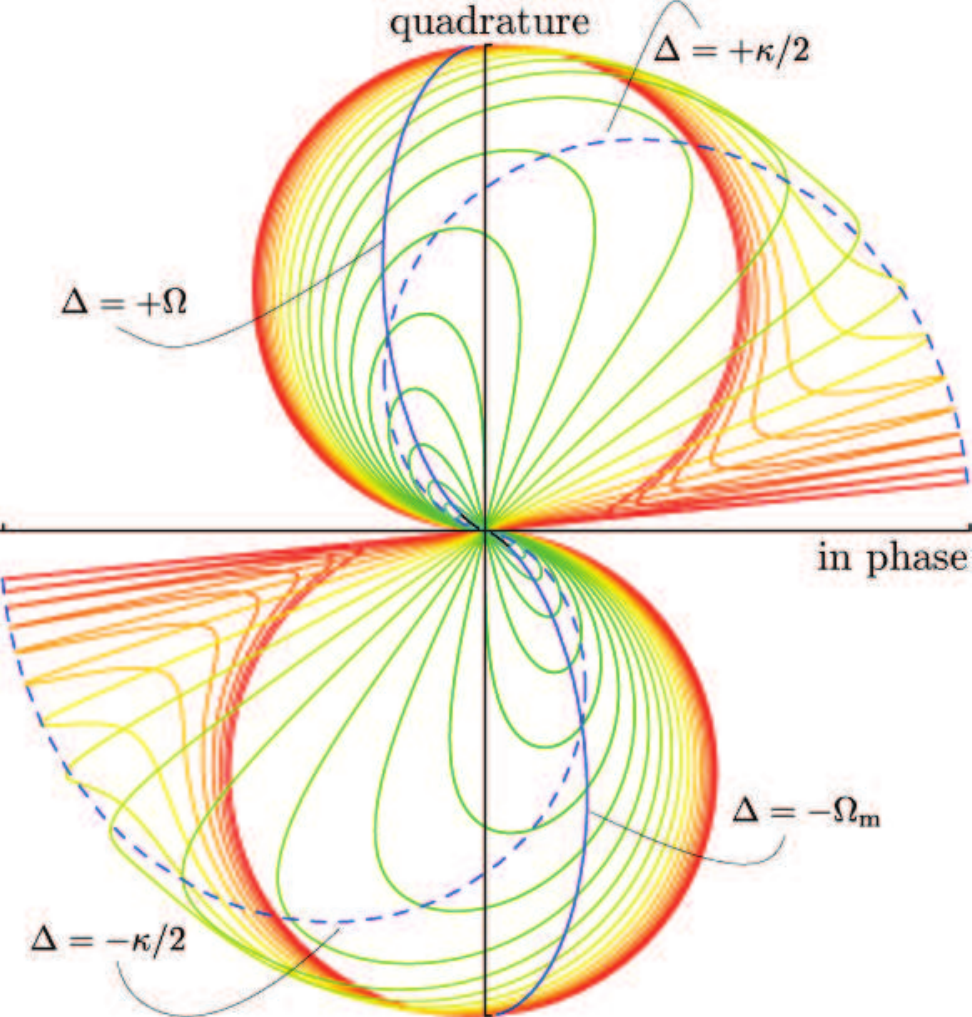}  
\caption{In-phase and quadrature component of the intracavity energy (or
circulating power) modulation when the detuning is varied (colored curves).
Different colors represent different choices of the cavity linewidth $%
\protect\kappa$, which was changed from $\Omega_\mathrm{m} /10$ (green) to $%
10\,\Omega_\mathrm{m} $ (red) in logarithmic steps. In the blue curves, a
fixed detuning was chosen (as indicated by the labels), and the cavity
linewidth was varied from $\Omega_\mathrm{m} /10$ to $10\,\Omega_\mathrm{m} $%
.}
\label{f:iq}
\end{figure}

The presence of these sidebands corresponds to a modulation of the
intracavity photon number $|a(t)|^2$, and it is easy to show that 
\begin{align}  \label{e:icenergy}
|a(t)|^2&\approx |a_0(t)|^2+ a_0(t) a_1^*(t) + a_0^*(t) a_1(t)= \\
&=\frac{\eta_\mathrm{c}  \kappa |\bar s_\mathrm{in} |^2}{\Delta^2+(%
\kappa/2)^2}\left(1+\right.  \notag \\
&\qquad g_0  x_0 \left(\frac{\Delta+\Omega_\mathrm{m} }{(\Delta+\Omega_%
\mathrm{m} )^2+(\kappa/2)^2}+ \frac{\Delta-\Omega_\mathrm{m} }{%
(\Delta-\Omega_\mathrm{m} )^2+(\kappa/2)^2}\right)\sin(\Omega_\mathrm{m}  t)+
\notag \\
&\qquad g_0  x_0 \left.\left(\frac{\kappa/2}{(\Delta+\Omega_\mathrm{m}
)^2+(\kappa/2)^2}- \frac{\kappa/2}{(\Delta-\Omega_\mathrm{m} )^2+(\kappa/2)^2%
}\right)\cos(\Omega_\mathrm{m}  t)\right).
\end{align}
The intracavity optical photon number is modulated at the oscillation
frequency $\Omega_\mathrm{m} $, however, the modulation does not necessarily
occur in phase with the mechanical oscillation as is evident from the
cosine-term. In fact, the phase lag depends in a non-trivial manner on the
laser detuning $\Delta$, oscillation frequency $\Omega_\mathrm{m} $ and the
cavity buildup time $\kappa^{-1}$, a simple calculation yields for the phase
lag 
\begin{equation}
\phi_\mathrm{lag}=\arg\left(g_0  \Delta(\Delta^2+(\kappa/2)^2-\Omega_\mathrm{%
m} ^2- i \Omega_\mathrm{m}  \kappa)\right),
\end{equation}
which is shown in figure~\ref{f:phaselag}. %
The quadrature component ($\propto \cos(\Omega_\mathrm{m}  t)$) 
can become significant if the cavity buildup time $\kappa^{-1}$ is
comparable or larger than the oscillation period $\Omega_\mathrm{m} ^{-1}$,
leading to a viscous radiation pressure force as detailed below. Both
in-phase and quadrature components are shown in the parametric plot in figure~%
\ref{f:iq}. 

\subsubsection{Radiation-pressure backaction: static phenomena}

\label{sss:static}

So far, only the effect of the mechanical on the optical degree of freedom
has been considered. However, the richness of optomechanical phenomena
arises only when the \emph{mutual} coupling of optical and mechanical
degrees of freedom is taken into account. The physical origin of the
``back-action'' of light on the movable cavity boundary is due to radiation
pressure. In the simple case of a Fabry-Perot\ cavity, the force arises from
the momentum flips of the photons reflected from the movable mirror, and
thus 
\begin{equation}  \label{e:simplefrp}
F_\mathrm{rp}(t)={|a(t)|^2}\frac{c}{2 L}{2\hbar k}=\frac{|a(t)|^2 \hbar
\omega_\mathrm{l} }{L}=-\hbar g_0  {|a(t)|^2},
\end{equation}
where $\hbar k$ with $k=\omega_\mathrm{l} /c$ is the momentum of the
photons.

\begin{figure}[bt]
\centering
\includegraphics[width=\linewidth]{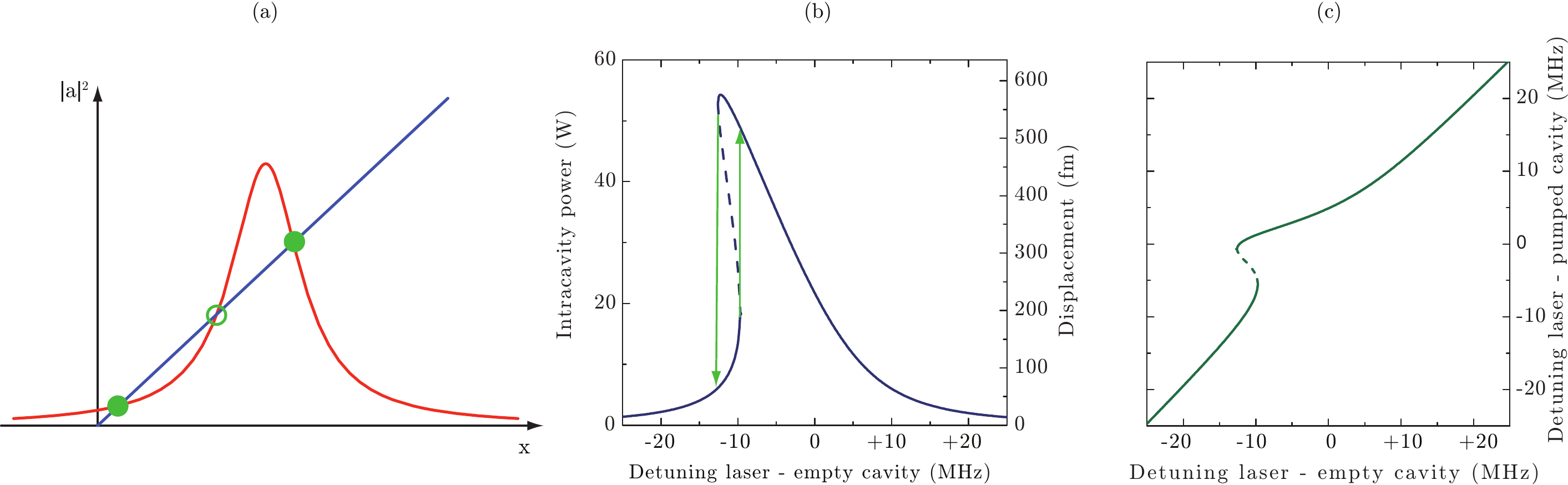} 
\caption{(a) Graphical representation of the self-consistent solutions of
the coupled equations \eqref{e:aux1} (red) and \eqref{e:aux2} (blue) for the
intracavity photon number $|\bar a|^2$ and the radiation-pressure induced
displacement $\bar x$ of the cavity radius. Intersections of the curves
indicate possible stable (full circles) and unstable (empty circles)
solutions. (b) Plot of numerical solutions for the power $|\bar a|^2/\protect%
\tau_\mathrm{rt} $ circulating in the cavity and the radiation-pressure
induced displacement for typical parameters of a silica microtoroidal
optomechanical oscillator ($R=25\,\mathrm{\protect\mu m} $, $\protect\kappa=2%
\protect\pi\, 8\,\mathrm{MHz} $, $\Omega_\mathrm{m} =2\protect\pi\, 50\,%
\mathrm{MHz} $, $m_\text{eff}=20\,\mathrm{ng} $, $\protect\omega_\mathrm{c}
=2\protect\pi\, 380 \,\mathrm{THz} $, $|\bar s_\mathrm{in} |^2=1\,\mathrm{mW}
$). The dashed line indicates the unstable solutions. (c) The actual
detuning $\Delta-g_0  \bar x$ as a function of the detuning $\Delta$ of the
laser from the undriven cavity resonance. }
\label{f:rpbistability}
\end{figure}

The relation $F_\mathrm{rp}(t)=-\hbar g_0  {|a(t)|^2}$ is of general
validity and also applies to the WGM resonators, as will be derived in
section \ref{ss:omcoupling}. If the cavity boundary is free to move, the
coupled equations describing the optomechanical system will read (in a frame
rotating at the laser frequency) 
\begin{align}
\dot a(t)=(i ( \Delta -g_0  x(t)) - \kappa/2)\,a(t)+\sqrt{\eta_\mathrm{c} 
\kappa} s_\mathrm{in}(t)  \label{e:omsa} \\
\ddot x(t)+\Gamma_\mathrm{m}  \, \dot{x}(t)+\Omega_\mathrm{m} ^2\, x(t) =
-\hbar g_0  \frac{|a(t)|^2}{m_\text{eff}},  \label{e:omsx}
\end{align}
where, for the mechanical oscillation, a resonance frequency $\Omega_\mathrm{%
m} $, viscous damping at a rate $\Gamma_\mathrm{m} $, and an effective mass $%
m_\text{eff}$ (cf.\ section \ref{ss:omcoupling}) are assumed.  For a
constant drive amplitude $\bar s_\mathrm{in} $, these coupled nonlinear
equations can be analyzed in a first step by finding stable solutions $%
a(t)=\bar a $, $x(t)=\bar x$ in which all time derivatives ($\dot a(t), \dot
x(t), \ddot x(t)$) vanish, requiring simultaneously 
\begin{align}
\bar a &= \frac{1}{-i (\Delta-g_0  \bar x )+\kappa/2}\,\sqrt{\eta_\mathrm{c}
\kappa} \bar s_\mathrm{in}  \qquad \text{and}  \label{e:aux1} \\
m_\text{eff} \Omega_\mathrm{m} ^2 \bar x &=-\hbar g_0  {|\bar a |^2}.
\label{e:aux2}
\end{align}
Equations \eqref{e:aux1} and \eqref{e:aux2} can be both understood as
functions mapping the displacement $\bar x$ to an intracavity photon number $%
|\bar a|^2$, as shown in figure~\ref{f:rpbistability}. The self-consistent,
physically possible solutions are given by the intersections of the two
curves.\footnote{%
Note that the phase of the complex entity $\bar a$ can always be adjusted to
fulfill \eqref{e:aux1}, as it does not affect \eqref{e:aux2}.} Evidently,
the system has at least one solution for arbitrary parameters. For
sufficiently high power or finesse, and/or floppy enough mechanical
oscillators, two additional solutions are physically possible. The condition
for their appearance is given by 
\begin{equation}
|\bar s_\mathrm{in} |^2\geq\frac{\sqrt{3}}{9}\frac{\Omega_\mathrm{m} ^2 m_%
\text{eff} \kappa^2 }{\eta_\mathrm{c}  \hbar g_0 ^2 },
\end{equation}
which is derived in a straightforward manner from the requirement that the
maximum slope $3 \sqrt{3} |\bar s_\mathrm{in} |^2 \eta_\mathrm{c}  |g_0 |/
\kappa^2 $ of the Lorentzian square modulus of \eqref{e:aux1} must exceed
the slope $m_\text{eff} \Omega_\mathrm{m} ^2 /\hbar |g_0 |$ corresponding to %
\eqref{e:aux2}. Above this threshold, the optomechanical system displays a
well-known bistable behavior, resulting, for example, in a hysteretic
transmission behavior upon the variation of the input power. This effect was
observed in a pioneering experiment \cite{Dorsel1983} at the
Max-Planck-Institute of Quantum Optics (MPQ) as early as 1983 and is
referred to as optical bistability. Reports in the microwave domain followed
soon thereafter \cite{Gozzini1985}.

\subsubsection{Radiation-pressure backaction: dynamical effects}

\label{sss:dynbatheo}

Quantitatively novel behavior occurs when analyzing the  dynamical response
of  the system around an equilibrium $(\bar a ,\bar x )$. As first pointed
out by Braginsky and co-workers in 1967 \cite{Braginskii1967, Braginsky2001,
Braginsky2002}, the dynamics of fluctuations around the equilibrium not
only display new physical effects, 
but are also of experimental relevance, in particular in the most sensitive
gravitational wave interferometers. Keeping with the simple illustrative
approach of this section, let us assume the system is in a stable
equilibrium if $a=\bar a $ and $x=\bar x $ and analyze the dynamics of small
excursions $\delta a (t)$ and $\delta x (t)$ if a small external force $%
\delta F(t)$ is applied to the mechanical oscillator. Substituting $a(t)=\bar a
+\delta a (t)$, $x(t)=\bar x +\delta x (t)$ into \eqref{e:omsa}--%
\eqref{e:omsx} and introducing the equilibrium detuning 
\begin{equation}  \label{e:bDdef}
\bar \Delta \equiv \omega_\mathrm{l} -(\omega_\mathrm{c} +g_0  \bar x ),
\end{equation}
one finds the linearized equations 
\begin{align}
\dot \delta a (t)=(+i \bar \Delta  -\kappa/2) \delta a (t)- i g_0  \bar a
\,\delta x (t)  \label{e:omsda} \\
m_\text{eff}\left(\ddot \delta x (t)+\Gamma_\mathrm{m}  \, \dot{\delta x }%
(t)+\Omega_\mathrm{m} ^2\, \delta x (t) \right)= -\hbar g_0  {\bar a (\delta
a (t)+\delta a ^*(t)) }+\delta F(t),  \label{e:omsdx}
\end{align}
where \eqref{e:aux1} and \eqref{e:aux2} were used, and second-order terms $%
\propto \delta a (t)\delta x (t)$ or $\propto |\delta a (t)|^2$ were
dropped, as we assume $|\delta a |\ll|\bar a |$. Furthermore, without loss
of generality, we have assumed real $\bar a =\bar a ^*$, which can always be
attained by adjusting the (physically irrelevant) phase of the incoming
driving wave $\bar s_\mathrm{in} $.

This equation system is most easily solved in the frequency domain, by
applying a Fourier transformation to all involved time-dependent variables.%
\footnote{%
We choose the convention $f(\Omega )=\int_{-\infty}^{+\infty} f(t) e^{+i
\Omega  t} dt$.} One then obtains 
\begin{align}
-i \Omega  \delta a (\Omega ) &=(+i \bar \Delta  -\kappa/2) \delta a (\Omega
)- i g_0  \bar a \,\delta x (\Omega ) \\
-i \Omega  {\delta a ^*}(\Omega ) &=(-i \bar \Delta  -\kappa/2) {\delta a ^*}%
(\Omega )+ i g_0  \bar a \,\delta x (\Omega ) \\
m_\text{eff}\left(-\Omega ^2-i \Gamma_\mathrm{m}  \Omega +\Omega_\mathrm{m}
^2\right) \delta x (\Omega ) &=-\hbar g_0  {\bar a  \left(\delta a (\Omega )+%
{\delta a ^*}(\Omega )\right)}+ \delta\! F (\Omega ),  \label{e:mo}
\end{align}
where ${\delta a ^*}(\Omega )=\left(\delta a (-\Omega )\right)^*$ was used.
Analogous to the previous section which considered a moving boundary, we now
find that a non-zero displacement amplitude $\delta x (\Omega )$ at Fourier
frequency $\Omega $ induces anti-Stokes and Stokes sidebands of amplitudes 
\begin{align}
\delta a (\Omega ) &=\frac{- i g_0  \bar a }{-i (\bar \Delta +\Omega )
+\kappa/2}\,\delta x (\Omega ) \\
{\delta a ^*}(\Omega ) &=\frac{+ i g_0  \bar a }{+i (\bar \Delta -\Omega )
+\kappa/2}\,\delta x (\Omega ),
\end{align}
respectively. As a result, the intracavity energy is modulated, giving rise
to an oscillating force of magnitude 
\begin{align}  \label{e:frpx}
\delta\! F _\mathrm{rp}(\Omega ) &=-\hbar g_0  {\bar a  \left(\delta a
(\Omega )+{\delta a ^*}(\Omega )\right)}= \\
&=-\hbar {g_0 ^2 \bar a ^2} \left(\frac{\bar \Delta +\Omega }{(\bar \Delta
+\Omega )^2+(\kappa/2)^2}+\frac{\bar \Delta -\Omega }{(\bar \Delta -\Omega
)^2+(\kappa/2)^2}\right)\delta x (\Omega )  \notag \\
&{}\qquad+i\hbar {g_0 ^2 \bar a ^2} \left(\frac{\kappa/2}{(\bar \Delta
+\Omega )^2+(\kappa/2)^2}-\frac{\kappa/2}{(\bar \Delta -\Omega
)^2+(\kappa/2)^2}\right)\delta x (\Omega ).
\end{align}
The real and imaginary parts of the radiation pressure force in this
representation are identified as being due to the in-phase and quadrature
modulation of the circulating power in the cavity. The additional force
acting on the mechanical oscillator changes its dynamical behavior, in
particular its response to the external perturbation. This effect is known
as \emph{dynamical backaction} \cite{Braginskii1967}.

Specifically, substituting \eqref{e:frpx} back into \eqref{e:mo}, a modified
response of the oscillator to an external force is found, 
\begin{align}  \label{e:chieff}
\dx(\Og) &= \chieff(\Og) \dF(\Og)
\intertext{with the effective susceptibility $\chieff(\Og)$,}
\chi_\text{eff}(\Omega )^{-1} &=m_\text{eff}\left(- \Omega ^2-i (\Gamma_%
\mathrm{m} +\Gamma_\mathrm{dba}(\Omega )) \Omega +\left(\Omega_\mathrm{m} ^2+%
\frac{k_\mathrm{dba}(\Omega )}{m_\text{eff}}\right)\right).
\end{align}
The damping and spring constant induced by dynamical backaction are given by 
\begin{align}
\Gamma_\mathrm{dba}&= \frac{\hbar g_0 ^2 \bar a ^2}{ m_\text{eff} \Omega }
\left(\frac{\kappa/2}{(\bar \Delta +\Omega )^2+(\kappa/2)^2}-\frac{\kappa/2}{%
(\bar \Delta -\Omega )^2+(\kappa/2)^2}\right)  \label{e:Gdba} \\
k_\mathrm{dba}&= \frac{\hbar g_0 ^2 \bar a ^2}{ m_\text{eff}} \left(\frac{%
\bar \Delta +\Omega }{(\bar \Delta +\Omega )^2+(\kappa/2)^2}+\frac{\bar
\Delta -\Omega }{(\bar \Delta -\Omega )^2+(\kappa/2)^2}\right).
\label{e:Odba}
\end{align}
If the induced changes of the mechanical oscillator's dynamics are small,
the oscillator still behaves as a damped harmonic oscillator  with effective
damping and resonance frequency \cite{Kippenberg2005,Schliesser2006} 
\begin{align}  
\label{e:Geff}
\Gamma_\mathrm{eff} &\approx\Gamma_\mathrm{m} +\frac{\hbar g_0 ^2 \bar a ^2}{
m_\text{eff} \Omega_\mathrm{m} } \left(\frac{\kappa/2}{(\bar \Delta +\Omega_%
\mathrm{m} )^2+(\kappa/2)^2}-\frac{\kappa/2}{(\bar \Delta -\Omega_\mathrm{m}
)^2+(\kappa/2)^2}\right) \\
\label{e:Oeff}
\Omega_\mathrm{eff} &\approx\Omega_\mathrm{m} +\frac{\hbar g_0 ^2 \bar a ^2}{%
2 m_\text{eff} \Omega_\mathrm{m} } \left(\frac{\bar \Delta +\Omega_\mathrm{m}
}{(\bar \Delta +\Omega_\mathrm{m} )^2+(\kappa/2)^2}+\frac{\bar \Delta
-\Omega_\mathrm{m} }{(\bar \Delta -\Omega_\mathrm{m} )^2+(\kappa/2)^2}%
\right).
\end{align}

\subsection{Formal framework: quantum Langevin equations}

A more general formulation of optomechanical interactions than the simple
but illustrative considerations in the previous section is possible within
the framework of a quantum Langevin approach. This enables the full
description of the quantum dynamics of an optomechanical system. In
particular, effects related to the quantum nature of light can be treated in
an adequate manner. %

\subsubsection{Hamiltonian of cavity optomechanics}

Starting point of the analysis is a Hamiltonian formulation of a generic
optomechanical system \cite{Law1995}. If the mechanical oscillation
frequency is much smaller than the free spectral range of the cavity, such
that a only a single optical mode has to be considered, the system
Hamiltonian can be written as 
\begin{align}
\hat H&=\hat H_\mathrm{mech}+\hat H_\mathrm{opt}+\hat H_\mathrm{int}+\hat H_%
\mathrm{drive}  \label{e:h} \\
\hat H_\mathrm{mech} &= \frac{\hat p ^2}{2 m_\text{eff}}+ \frac{1}{2} m_%
\text{eff} \Omega_\mathrm{m} ^2 \hat x ^2 \\
\hat H_\mathrm{opt}&=\hbar \omega_\mathrm{c}  \left(\hat a^\dagger  \hat a +%
\frac{1}{2}\right) \\
\hat H_\mathrm{int}&=\hbar g_0  \hat x \,\hat a^\dagger  \hat a  \\
\hat H_\mathrm{drive}&= {i \hbar} {\ \sqrt{\eta_\mathrm{c}  \kappa}}
\left(\bar s_\mathrm{in}  \hat a^\dagger  e^{-i \omega_\mathrm{l}  t}-\bar s_%
\mathrm{in} ^* \hat a  e^{+i\omega_\mathrm{l}  t}\right)
\end{align}
where $\hat x $ and $\hat p $ are the mechanical displacement and momentum
operators, and $\hat a^\dagger $ and $\hat a $ are the creation and
annihilation operators of the considered optical mode, i.\ e.\ $\hat n=\hat
a^\dagger  \hat a $ is the intracavity photon operator, and correspondingly,
the drive amplitude $\bar s_\mathrm{in} $ is now normalized to photon flux
at the input of the cavity %
$|\bar s_\mathrm{in} |^2={P_\mathrm{in}}/{\hbar\omega_\mathrm{l} }$.
Evidently, this Hamiltonian reproduces the optical resonance frequency shift
upon mechanical displacement, as 
\begin{equation}
\hat H_\mathrm{opt}+\hat H_\mathrm{int}=\hbar (\omega_\mathrm{c} +g_0  \hat
x )\,\hat a^\dagger  \hat a ,
\end{equation}
and simultaneously describes the radiation pressure force, with 
\begin{equation}  \label{e:frp}
\hat F_\mathrm{rp}=-\frac{\partial \hat H_\mathrm{int}}{\partial \hat x }%
=-\hbar g_0  \hat a^\dagger  \hat a .
\end{equation}

\subsubsection{Quantum Langevin equations}

From the Hamiltonian, the time evolution of the operators of interest can be
derived. In addition to the conservative dynamics described by \eqref{e:h},
dissipation of both the optical and mechanical modes, and the corresponding
fluctuations are taken into account by the following set of quantum Langevin
equations (QLEs) \cite{Giovannetti2001} (in a frame rotating at $\omega_%
\mathrm{l} $) 
\begin{align}
\frac{d}{dt}\hat a (t)&=\left(+i \Delta-\frac{\kappa}{2}\right) \hat a (t)-
i g_0  \hat x (t) \hat a (t) + \frac{ (\bar s_\mathrm{in} +\delta \hat s_%
\mathrm{in} (t))}{\sqrt{\tau_\mathrm{ex} }} + \frac{\delta \hat s_\mathrm{vac%
} (t)}{\sqrt{\tau_0 }} \\
\frac{d}{dt}\hat x (t)&=\frac{\hat p (t)}{m_\text{eff}}  \label{e:pQLE} \\
\frac{d}{dt}\hat p (t)&=-m_\text{eff} \Omega_\mathrm{m} ^2 \hat x (t)- \hbar
g_0  \hat a^\dagger (t) \hat a (t)-{\Gamma_\mathrm{m} } \hat p (t) +
\delta\! \hat F_\mathrm{th} (t)  \label{e:xQLE}
\end{align}
where the noise terms $\delta \hat s_\mathrm{in} $, $\delta \hat s_\mathrm{%
vac} $, and $\delta\! \hat F_\mathrm{th} $ were introduced. They fulfill the
commutation relations \cite{Gardiner2004} 
\begin{equation}
[\delta \hat s_\mathrm{in} (t),\delta \hat s_\mathrm{in}^{\dagger}
(t^{\prime })]=[\delta \hat s_\mathrm{vac} (t),\delta \hat s_\mathrm{vac}%
^{\dagger} (t^{\prime })]=\delta(t-t^{\prime })
\end{equation}
and the quantities
\begin{align}
\langle \delta \hat s_\mathrm{in} (t) \delta \hat s_\mathrm{in}^{\dagger}
(t^{\prime })\rangle=\langle \delta \hat s_\mathrm{vac} (t) \delta \hat s_%
\mathrm{vac}^{\dagger} (t^{\prime })\rangle=\delta(t-t^{\prime })
\end{align}
are the only non-zero correlation functions for the quantum vacuum entering the cavity
from its two ports: the one through which it is pumped ($\delta \hat s_%
\mathrm{in} $), and the second port ($\delta \hat s_\mathrm{vac} $)
representing all other loss channels. %
Here, zero thermal excitation of the optical mode has been assumed which is
valid in the optical domain. In order to adequately describe the mirror
undergoing Brownian motion the correlation function of the mechanical driving term can
be shown \cite{Giovannetti2001} to have the form 
\begin{equation}
\langle \delta\! \hat F_\mathrm{th} (t) \delta\! \hat F_\mathrm{th}
(t^{\prime })\rangle=\hbar \meff \Gamma_\mathrm{m}  \int e^{-i\Omega
(t-t^{\prime })} \Omega  \left(\coth\left(\frac{\hbar \Omega }{2 k_\mathrm{B}
T}\right)+1\right) \, \frac{d\Omega }{2 \pi}  \label{e:dhfth}
\end{equation}

As in the previous subsection, the QLEs are simplified in the first place by
considering static and the dynamical effects separately. To this end, the
unitary transformations $\hat a (t)=\bar a +\delta \hat a (t)$ and $\hat x
(t)=\bar x +\delta \hat x (t)$ with $\langle\delta \hat a
(t)\rangle=\langle\delta \hat x (t)\rangle=0$ yields again the requirements %
\eqref{e:aux1}--(\ref{e:aux2}) for the steady state values of intracavity
field amplitude $\bar a $ and displacement $\bar x $. %

\subsubsection{Dynamics of the fluctuations}

Assuming $(\bar a ,\bar x )$ to be known as a stable solution of the system
(which can, for example, be tested for using the Routh-Hurwitz criterion,
see \cite{Fabre1994}), the Heisenberg equation of motion for the
fluctuations $\delta \hat a $, $\delta \hat a^{\dagger} $ and $\delta \hat x 
$ can be derived. Choosing again the phase of the input field $\bar s_%
\mathrm{in} $ such that $\bar a $ is real and positive, and assuming again a
strong coherent drive 
\begin{equation}
\bar a \gg 1
\end{equation}
it is possible to derive linearized quantum Langevin equations for the
fluctuations by dropping terms $\propto \delta \hat a  \delta \hat x $, $%
\delta \hat a^{\dagger}  \delta \hat x $ or $\delta \hat a^{\dagger}  \delta
\hat a $, yielding 
\begin{align}
\frac{d}{dt}\, \delta \hat a (t) &=\left(+i \bar \Delta  - \frac{\kappa}{2}%
\right)\delta \hat a (t)-i g_0 \bar a  \delta \hat x (t)+\frac{\delta \hat s_%
\mathrm{in} (t)}{\sqrt{\tau_\mathrm{ex} }}+\frac{\delta \hat s_\mathrm{vac}
(t)}{\sqrt{\tau_0 }} \\
\frac{d}{dt}\, \delta \hat a^{\dagger} (t) &= \left(-i \bar \Delta  - \frac{%
\kappa}{2}\right)\delta \hat a^{\dagger} (t)+i g_0 \bar a  \delta \hat x (t)+%
\frac{\delta \hat s_\mathrm{in}^{\dagger} (t)}{\sqrt{\tau_\mathrm{ex} }}+%
\frac{\delta \hat s_\mathrm{vac}^{\dagger} (t)}{\sqrt{\tau_0 }} \\
\frac{d^2}{dt^2}\, \delta \hat x (t)&+ \Gamma_\mathrm{m}  \frac{d}{dt} \,
\delta \hat x (t)+\Omega_\mathrm{m} ^2 \delta \hat x (t)=-\frac{\hbar g_0}{m_%
\text{eff}}\bar a \left( \delta \hat a (t)+\delta \hat a^{\dagger}
(t)\right)+ \frac{\delta\! \hat F_\mathrm{th} (t)}{m_\text{eff}}
\end{align}
where the Hermitian property $\delta \hat x (t)=\delta \hat x^{\dagger} (t)$
was used. This set of equations is most easily solved in the Fourier domain: 
\begin{align}  \label{e:eomx}
\left(-i (\bar \Delta +\Omega ) +\kappa/2\right)\delta \hat a (\Omega )
&=-ig_0 \bar a  \delta \hat x (\Omega )+\frac{\delta \hat s_\mathrm{in}
(\Omega )}{\sqrt{\tau_\mathrm{ex} }}+\frac{\delta \hat s_\mathrm{vac}
(\Omega )}{\sqrt{\tau_0 }} \\
\left(+i (\bar \Delta -\Omega ) +\kappa/2\right)\delta \hat a^{\dagger}
(\Omega ) &=+ig_0 \bar a  \delta \hat x (\Omega )+\frac{\delta \hat s_%
\mathrm{in}^{\dagger} (\Omega )}{\sqrt{\tau_\mathrm{ex} }}+\frac{\delta \hat
s_\mathrm{vac}^{\dagger} (\Omega )}{\sqrt{\tau_0 }} \\
\label{e:aux00}
m_\text{eff}\left(\Omega_\mathrm{m} ^2-\Omega ^2-i \Gamma_\mathrm{m}  \Omega
\right) \delta \hat x (\Omega ) &= -\hbar g_0  \bar a  \left(\delta \hat a
(\Omega )+\delta \hat a^{\dagger} (\Omega )\right)+\delta\! \hat F_\mathrm{th%
} (\Omega ).
\end{align}
In the frequency domain, 
\begin{align}  
\label{e:firstcorrelator}
\langle \delta \hat s_\mathrm{in} (\Omega ) \delta \hat s_\mathrm{in}%
^{\dagger} [\Omega ^{\prime }]\rangle&=2\pi \delta(\Omega +\Omega ^{\prime })
\\
\label{e:lastcorrelator}
\langle \delta \hat s_\mathrm{vac} (\Omega ) \delta \hat s_\mathrm{vac}%
^{\dagger} [\Omega ^{\prime }]\rangle&=2\pi \delta(\Omega +\Omega ^{\prime })
\end{align}
and 
\begin{equation}  \label{e:thermalforcefourier}
\langle \delta\! \hat F_\mathrm{th} (\Omega ) \delta\! \hat F_\mathrm{th}
[\Omega ^{\prime }] \rangle = 2 \pi \delta(\Omega +\Omega ^{\prime }) \hbar\meff \Gamma_\mathrm{m}  \Omega  \left(\coth\left(\frac{\hbar \Omega }{%
2 k_\mathrm{B}  T}\right)+1\right)
\end{equation}
are the only non-zero correlation functions. Together with the input-output relations
for the fluctuations 
\begin{align}
\delta \hat s_\mathrm{out} (\Omega )&=\delta \hat s_\mathrm{in} (\Omega )-%
\sqrt{\eta_\mathrm{c}  \kappa}\,{\delta \hat a (\Omega )} \\
\delta \hat s_\mathrm{out}^{\dagger} (\Omega )&=\delta \hat s_\mathrm{in}%
^{\dagger} (\Omega )-\sqrt{\eta_\mathrm{c}  \kappa}\,{\delta \hat
a^{\dagger} (\Omega )}  \label{e:last}
\end{align}
these equations constitute the theoretical description of the most important
effects in cavity optomechanics.

\section[WGM resonators as optomechanical systems]{Whispering gallery-mode
microresonators as optomechanical systems}

\label{s:WGMom}

\begin{figure}[tbp]
\centering
\includegraphics[width=\linewidth]{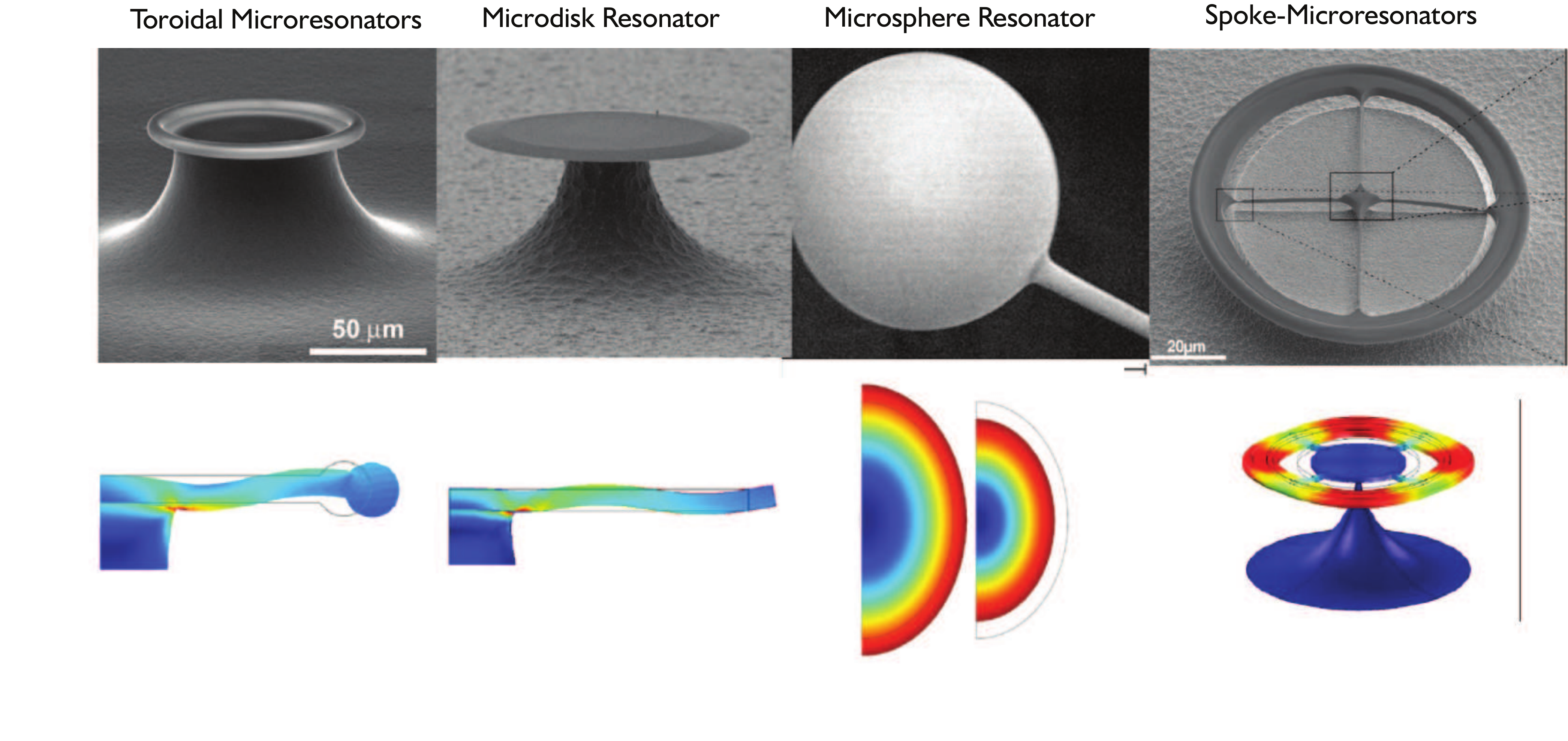}  
\caption{Optomechanical systems based on WGM optical microresonators. Top row shows
scanning electron micrographs of four different systems studied experimentally. The corresponding
panels in the bottom row depict the displacement patterns of mechanical modes which couple 
to the optical modes particularly strongly. The mechanical modes are simulated using the finite-element
method (first three panels: 2-dimensional cross-section, last panel: full 3-dimensional simulation).
}
\label{f:tjk}
\end{figure}

Optical microcavities are dielectric structures that confine light to small
volumes for extended amounts of time. In 1989 Braginsky and co-workers
in Moscow discovered that small dielectric microspheres from glass exhibit
optical whispering gallery modes that exhibit giant photon lifetimes, corresponding to ultra high
Q-factors exceeding $10^{9}$ \cite{Braginsky1989, Collot1993, Vernooy1998}.
The ultra high Q of microspheres were subsequently investigated
for a variety of applications, including laser stabilization \cite{Vassiliev1998}, cavity Quantum
Electrodynamics (cQED) \cite{Vernooy1998a, Aoki2006}, narrow linewidth lasers \cite{Sandoghdar1996},
nonlinear
optics at ultra low light powers \cite{Chang1996, Spillane2002} and biophysical
sensing \cite{Vollmer2002, Vollmer2009}. That optical microresonators
also exhibit mechanical modes that are co-located in the same structure and
are coupled via radiation pressure, was only realized in 2005, in toroidal
microresonators \cite{Kippenberg2005, Rokhsari2005}. It became
immediately evident that optomechanical coupling in this manner pertains to
virtually any optical microresonator \cite{Kippenberg2007} and has
been demonstrated in a variety of microresonator geometries, including
microspheres \cite{Ma2007} and microdisks or optimized optomechanical
resonators \cite{Schliesser2008, Anetsberger2008} as shown in figure \ref{f:tjk}. In this section, prior to
discussing the optomechanical phenomena that can occur in optical whispering
gallery mode microresonators, we analyze the mechanical modes of WGM microresonators,
discuss their mechanical
eigenfrequencies and address the question of the sources of mechanical
dissipation and other relevant optomechanical properties.

\subsection{Optical properties of WGM silica microresonators}

An essential precondition to render the weak effects of radiation pressure
experimentally accessible is a high optical finesse. In a cavity of finesse $%
\mathcal{F}$, the circulating power---and thus also the radiation pressure
force---is enhanced by a factor of $\sim\mathcal{F}/\pi$ compared to the
launched power. Silica microresonators can achieve finesses on the order of $%
10^{6}$, exceeding even the best
results achieved with Fabry-Perot\ cavities for cavity QED \cite{Vernooy1998}%
.

At the same time, in order to reveal dynamical effects of radiation-pressure
coupling, the photon storage time should be on the order of---or ideally
exceed---the period of the mechanical oscillator coupled to the resonator.
In spite of the very short round-trip time of about $1 \,\mathrm{ps} $, the
storage time can amount to several hundreds of nanoseconds owing to the high
finesse. This is about 10 times longer than the oscillation period of the
mechanical modes in these structures. To ensure well-controlled
optomechanical interaction, the optical mode spectrum must be well
understood.

\subsection{Optomechanical coupling in silica WGM resonators}

\subsubsection{Silica microspheres}

The mechanical degree(s) of freedom coupled parametrically to the WGMs in
silica microresonators are given simply by the intrinsic acoustic modes of
the structure. Each of the eigenmodes can be viewed, to a good
approximation, as a damped harmonic oscillator, driven by thermal forces. In
the following, the nature of these modes, in particular, their
eigenfrequency, the damping mechanisms, (effective) mass, mode shapes, and
coupling to the optical degrees of freedom will be discussed for two
canonical systems; silica microspheres and silica microtoroids.

\subsubsection{Acoustic modes in silica microresonators}

\label{sss:acousticModes}

The deformation induced by mechanical modes is described by a vector field $%
\vec u(\vec r,t)$, which denotes the displacement of an (infinitesimally
small) cubic volume element at position $\vec r$ and time $t$ from its
initial position. In an isotropic homogenous medium, to which no external
forces are applied, the equation of motion for the displacement field reads 
\cite{Landau1970} 
\begin{equation}
\rho \ddot{\vec{ u}}(\vec r,t)=(\lambda+\mu)\vec\nabla(\vec\nabla\cdot\vec
u(\vec r,t))+\mu {\vec\nabla}^2 \vec u(\vec r,t)  \label{e:elasticeom}
\end{equation}
where the density $\rho$ and the Lam\'e constants 
\begin{align}
\lambda &= \frac{\sigma E}{(1+\sigma)(1-2\sigma)} \\
\mu &= \frac{E}{2(1+\sigma)},
\end{align}
with $\sigma$ Poisson's ratio and $E$ Young's modulus, characterize the
elastic properties of the material. %

While for an infinitely extended medium, a continuum of solutions for
equation (\ref{e:elasticeom}) are obtained, for a finite-size body such as a
silica sphere or toroid, the boundary conditions lead to a discrete spectrum
of solutions, such that the total displacement $\vec u(\vec r,t)$ can be
decomposed into modes oscillating harmonically at a set of frequencies $%
\Omega_n$, 
\begin{equation}  \label{e:decomposition}
\vec u(\vec r,t)=\sum_n \vec u_n(\vec r,t)= \sum_n c_n(t) \vec u_n^0(\vec r)
= \sum_n \bar c_n \vec u_n^0(\vec r) e^{- i \Omega_{n} t},
\end{equation}
where $c_n(t)$ is the displacement amplitude of a mode with index $n$, $%
\Omega_n$ its eigenfrequency and $\vec u_n^0(\vec r)$ is the spatial
displacement pattern of the mode, normalized such that 
\begin{equation}
\frac{\int_V \vec u_n^0(\vec r) \vec u_{n^{\prime }}^0(\vec r) d^3r}{\int_V
\,d^3r}=\delta_{n n^{\prime }}. 
\end{equation}

For more sophisticated geometries (e.\ g.\ a silica toroid supported by a
silicon pillar), it is difficult to obtain analytical solutions for the mode
shapes and frequencies.\footnote{%
For simple cylinders, approximate solutions have been developed \cite%
{Hutchinson1979, Hutchinson1980, Tamura2009}.} In this case, equation (\ref%
{e:elasticeom}) with suitable boundary conditions is solved using the
finite-element method (FEM). As an illustrative example, and to validate the
accuracy of the FEM results, an analytical solution available for spheres is
developed in the following \cite{Love1906, Ma2007}. In spherical
coordinates, equation (\ref{e:elasticeom}) is solved in a homogeneous medium
by functions of the form 
\begin{equation}
\vec u(\vec r,t)=\vec \nabla \phi_0(\vec r,t)+\vec \nabla\times
\vec\Phi_1(\vec r,t)+\vec \nabla\times \vec \nabla\times \vec\Phi_2(\vec r,t)
\end{equation}
derived from a scalar potential $\phi_{0}$ and two vector potentials $\vec
\Phi_{1}=(r\phi_{1},0,0)$ and $\vec \Phi _{2}=(r\phi_{2},0,0)$ with 
\begin{equation}
\phi_{q}(\vec r,t)=\sum_{l,m}A_{qnlm}\,j_{l}\!\left(\frac{\Omega_{nlm}}{v_q}
r\right)Y_{l}^{m}(\theta,\varphi)e^{- i \Omega_{nlm}t}
\end{equation}
where $q=0,1,2$; $j_{l}$ denotes the spherical Bessel function, $Y_{\ell}^{m}
$ is the spherical harmonic function, $v_{0}=\sqrt{(\lambda+2\mu)/\rho}$ is
the longitudinal sound velocity, and $v_{1}=v_{2}=\sqrt{\mu/\rho}$ is the
transverse sound velocity. %
The acoustic modes are characterized by an angular momentum mode number $l$ (%
$l=0,1,2,\ldots$), an azimuthal mode number $m$ ($-l\leq m\leq l$) and a
radial mode number $n$ $(n=1,2,\ldots)$. Here, $n=1$ corresponds to the
surface mode, $n\geq2$ to inner modes and $\Omega_{nlm} $ denotes the
frequency of the vibration characterized by the mode numbers $(n,l,m)$.

We focus now on the fundamental spheroidal mode $(n,l,m)=(1,0,0)$, for
higher-order modes, cf.\ \cite{Ma2007} and references therein. In
particular, the displacement vector field is purely radial, 
\begin{equation}
\vec u^0_{1,0,0}(\vec r)= A_{0,1,0,0} \frac{\sin(k_{1,0,0} r)-k_{1,0,0} r
\cos(k_{1,0,0} r)}{r^2} \vec e_r,
\end{equation}
where $k_{1,0,0}=\Omega_{1,0,0}/v_0$, $A_{0,1,0,0}$ is a normalization
constant, and $\vec e_r$ the radial unit vector. In the following, we drop
the mode index $(1,0,0)$ for better legibility, and an index to a vector
field now denotes one of its components in a given coordinate system.

To determine the allowed values of $k$ (and therefore $\Omega $), the strain
tensor 
\begin{align}
\varepsilon_{ij}=\frac{1}{2}\left(\frac{\partial u_i}{\partial x_j}+\frac{%
\partial u_j}{\partial x_i}\right)  \label{e:straindef}
\end{align}
is evaluated in a spherical coordinate system. For the considered mode, all
non-diagonal elements vanish, and the diagonal elements read\footnote{%
For the definitions of the strain tensor in spherical coordinates see \cite%
{Landau1970}.} 
\begin{align}  
\label{e:spherestrainr}
\varepsilon_{rr}=\frac{\partial u_{r}}{\partial r}&=c(t) \, A\frac{(k^2
r^2-2)\sin(k r)+2 k r \cos(k r)}{r^3} \\
\varepsilon_{\theta\theta}=\frac{u_{r}}{r}&= c(t) \, A \frac{\sin(k r)-k r
\cos(k r)}{r^3} \\
\label{e:spherestrainphi}
\varepsilon_{\varphi\varphi}=\frac{u_{r}}{r}&= c(t) \,A \frac{\sin(k r)-k r
\cos(k r)}{r^3},
\end{align}
where $c(t)$ is the excitation amplitude of the mode. The stress is related
to the strain by Hooke's law, which in a lossless isotropic medium is given
by 
\begin{equation}
\sigma_{ij}=2 \mu \varepsilon_{ij}+\lambda \delta_{ij} \sum_k
\varepsilon_{kk}  \label{e:hookeslaw}
\end{equation}
rendering in this case the stress tensor $\sigma_{ij}$ diagonal as well, 
\begin{align}
\sigma_{rr}&=c(t)\, A\frac{((\lambda+2\mu)k^2 r^2-4\mu)\sin(k r)+4\mu k r
\cos(k r)}{r^3} \\
\sigma_{\theta\theta}&=c(t)\, A\frac{(\lambda k^2 r^2+2 \mu) \sin( k r)-2
\mu k r \cos( k r)}{r^3} \\
\sigma_{\varphi\varphi}&=c(t)\, A\frac{(\lambda k^2 r^2+2 \mu) \sin( k r)-2
\mu k r \cos( k r)}{r^3}.
\end{align}
Applying the boundary conditions for a freely oscillating sphere of radius $R
$ leads to the requirement that the stress is zero at its boundary ($r=R$),
leading to the characteristic equation 
\begin{equation}
\left( 1-\frac{1}{4} \frac{v_0^2}{v_1^2} k^2 R^2\right) \frac{\tan(k R)}{k R}%
-1=0.
\end{equation}
For the parameters of fused silica, this equation is solved for $k
R\approx2.4005\ldots$ or 
\begin{equation}  \label{e:Omspherenum}
\frac{\Omega_\mathrm{m} }{2\pi}\approx\frac{2280 \,\mathrm{m/s} }{R}, 
\end{equation}
corresponding to a resonance frequency of $91.2\,\mathrm{MHz} $ for a 50-$\,%
\mathrm{\mu m} $\, diameter sphere. Finite element modeling %
yields the same result within less than 1\%, testifying to the integrity of our
employed numerical simulation.

\begin{figure}[tbp]
\centering
\includegraphics[width=.9\linewidth]{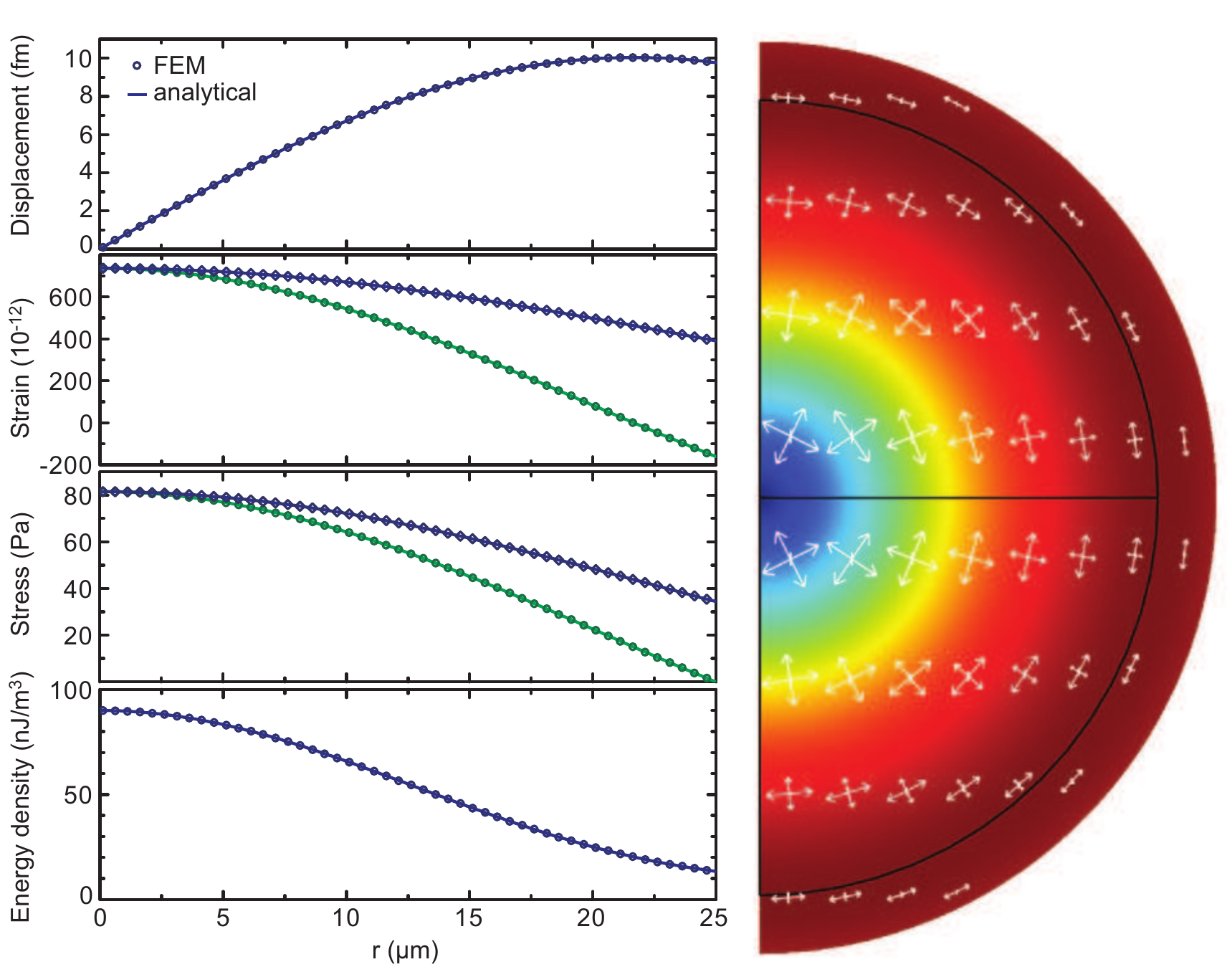}  
\caption{The $(n,l,m)=(1,0,0)$ mode of a silica sphere. Left panels show
(from top to bottom) the radial displacement $\vec u(r,\protect\theta,%
\protect\varphi)\cdot \vec e_r$, the strains $\protect\varepsilon_{rr}$
(green) and $\protect\varepsilon_{\protect\theta\protect\theta}=\protect%
\varepsilon_{\protect\varphi\protect\varphi}$ (blue), stresses $\protect%
\sigma_{rr}$ (green) and $\protect\sigma_{\protect\theta\protect\theta}=%
\protect\sigma_{\protect\varphi\protect\varphi}$ (blue), and the strain
energy density $\frac{1}{2}\sum_{ij}\protect\sigma_{ij}\protect\varepsilon%
_{ij}$. Symbols are results of finite-element modeling and lines are derived
from the analytical calculations, showing excellent agreement. The given
magnitudes correspond to a $50\,\mathrm{\protect\mu m} $-diameter sphere
containing a total strain energy of $k_\mathrm{B} (300\,\mathrm{K} )/2$ .
The right panel shows the exaggerated displacement profile (original sphere
outlined in black), and magnitude (color coded) as well as the principal
stresses (indicated by arrow lengths). }
\label{f:sphere}
\end{figure}

Another interesting physical entity is the potential energy stored in the
deformation. It given by 
\begin{align}
U&=\sum_{i,j} \int_V \frac{1}{2} \sigma_{ij} \varepsilon_{ij} \mathrm{d}^3r.
\label{e:stressstrainenergy}
\end{align}
The diagonal form of strain and stress tensors for a sphere again
facilitates the analytical evaluation of this integral simple yielding 
\begin{equation}  \label{e:Uspherenum}
U\approx 8.69\cdot10^{11} \,\mathrm{\frac{J}{m^3}} \cdot R \cdot x^2
\end{equation}
where $x=\vec u(R,\theta,\varphi)\cdot \vec e_r=c(t)\,A(\sin(k R)-k R\cos(k
R))/R^2$ is the radial displacement of the boundary, with $A\approx 0.427 R^2
$ for silica. Figure~\ref{f:sphere} shows displacement, strain, stress and
strain energy density of a silica sphere when the fundamental mode is
excited. Results from analytic calculations and the FEM agree very well.

Figure~\ref{f:spherefrequencies} compares experimentally measured resonance
frequencies of several silica microspheres with diameters between 30 and 100$%
\,\mathrm{\mu m} $\, with analytical calculations. Again, very good
agreement is obtained. 

\begin{figure}[tbp]
\centering
\includegraphics[width=.75\linewidth]{Figures/sphere_frequencies_v01r}  
\caption{Resonance frequencies of silica microspheres for the $(n,l)=(1,0)$
(upper branch) and $(n,l)=(1,2)$ (lower branch) spheroidal resonance modes 
\protect\cite{Ma2007}. Curves are expected resonance locations for free
boundary conditions, following an inverse dependence on the sphere radius
with $\Omega_\mathrm{m} /2\protect\pi\approx 2280\,\mathrm{m/s} /R$ for the $%
l=0$ and $\Omega_\mathrm{m} /2\protect\pi\approx 1580\,\mathrm{m/s} /R$ for
the $l=2$ mode. Symbols represent measured spheres. The panels on the right
show the corresponding displacement patterns as obtained from the FEM. The
left and right halves of the sphere show the displacement of the volume
elements at two different times separated by half the oscillation period. }
\label{f:spherefrequencies}
\end{figure}

\subsubsection{Silica microtoroids}

\begin{figure}[bp]
\centering
\includegraphics[width=.7\linewidth]{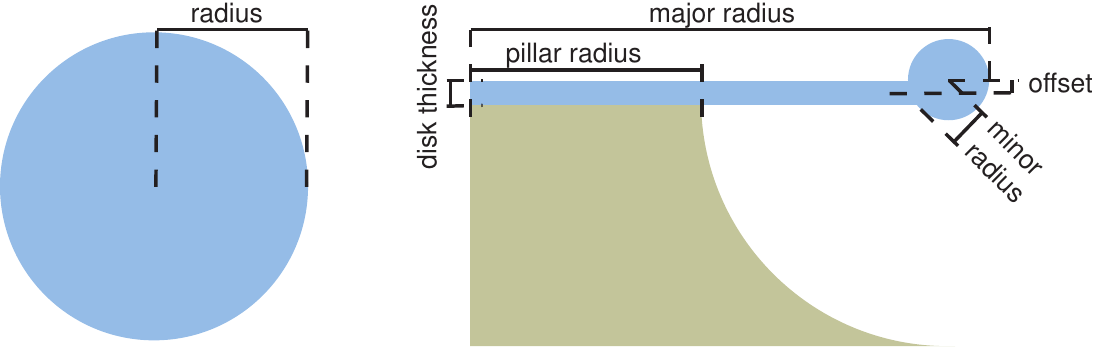}  
\caption{Geometric parameters required to describe a sphere (right) and a
toroidal WGM resonator, where silica is shown in blue and silicon in grey. }
\label{f:geometryParameters}
\end{figure}

\emph{Toroidal} silica resonators, attached a silicon pillar, have fewer
symmetries than spheres. This structure must be parameterized by several
parameters, at least by the major and minor radii of the silica torus, the
radius of the silicon pillar, the thickness of the silica disk, and the
offset of the symmetry planes of the disk and torus along the $z$-axis \cite%
{Kippenberg2005}. The shape of the silicon pillar is assumed to be
rotationally symmetric, and to constitute a quarter circle in the $r$-$z$%
-plane. The radius of this circular arch is another degree of freedom, but is
usually assumed to be similar to the difference of toroid and pillar radii
(figure~\ref{f:geometryParameters}) due to the isotropic nature of the
silicon etch.

\begin{figure}[tbp]
\centering
\includegraphics[width=\linewidth]{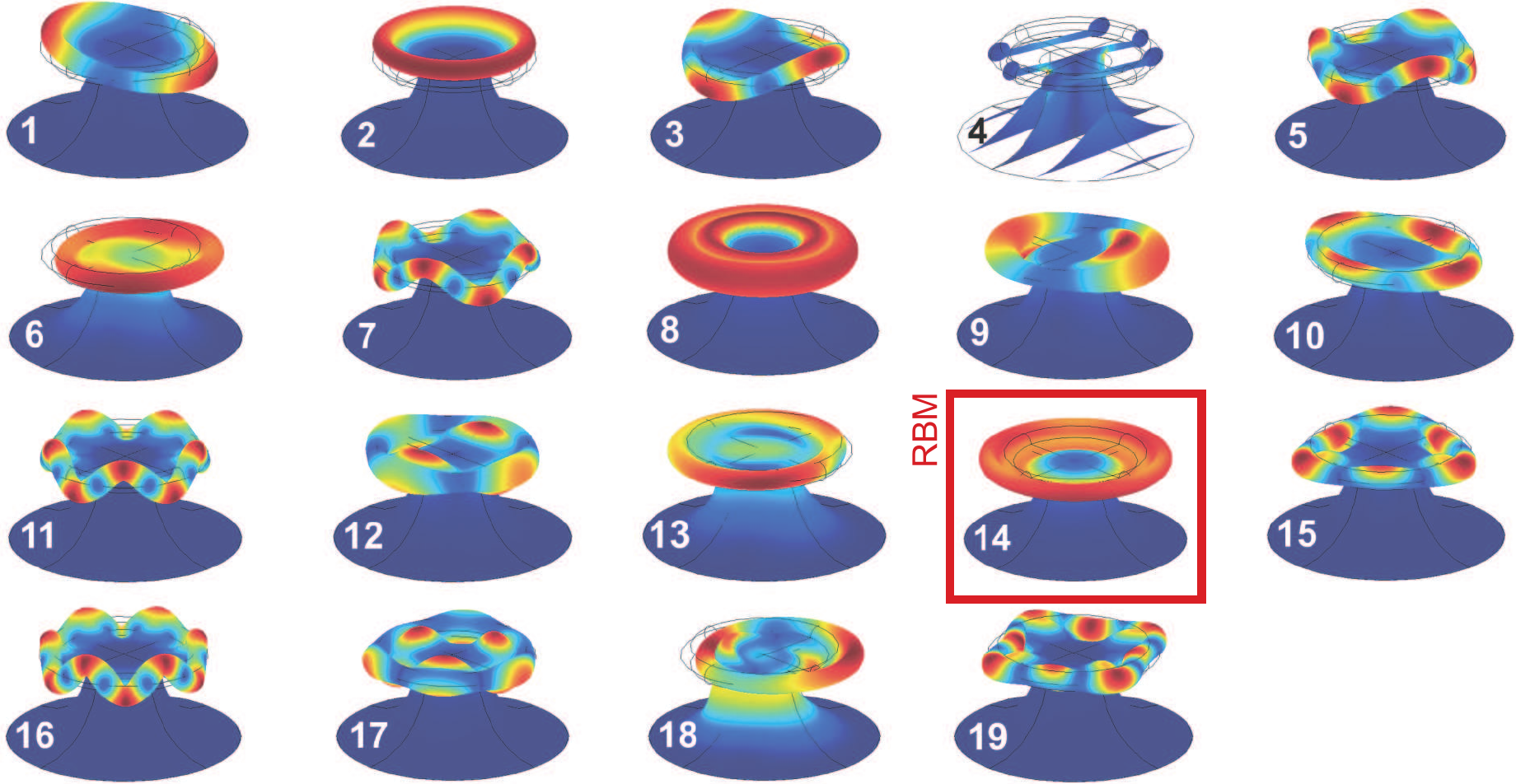}  
\caption{Displacement patterns of the 19 lowest-frequency modes (see text)
of a toroidal microresonator as calculated using the FEM, indicated both as
the deformed shape and in the color code (increasing displacement from dark
blue to dark red). Mode number 4 involves mainly torsional motion, which is
illustrated by plotting the displacement of originally parallel slices
through the structure. Figure from ref.\ \protect\cite{Schliesser2008b}.}
\label{f:toroidModes}
\end{figure}

The complex boundary conditions (all surfaces free, except for the bottom of
the silicon pillar) render an analytical solution prohibitively difficult.
We therefore use a commercial finite-element software to solve equation (\ref%
{e:elasticeom}) on a discrete mesh consisting typically of more than $10{,}%
000$ nodes which are automatically distributed in the simulation volume.
Simulations can be run both assuming rotational symmetry for the modes, and
in full three dimensions. Figure~\ref{f:toroidModes} shows the obtained
displacement patterns for the 19 lowest-frequency modes of a toroid of major
radius $23.0\,\mathrm{\mu m}  $, pillar radius $13.23\,\mathrm{\mu m}  $,
minor radius $2.63\,\mathrm{\mu m}  $, disk thickness $2\,\mathrm{\mu m}  $
and no offset, in a three-dimensional simulation. As shown in figure~\ref%
{f:comparison}, the simulation can reproduce the frequencies measured on a
real toroid to a very high accuracy \cite{Schliesser2008b}. Indeed, the
average deviation between measured and simulated frequency is, on average,
below 2\%. This emphasizes the reliability of the FEM. Note also that
probing of the modal displacement patterns using a scanning probe technique
has confirmed the shape expected from simulations in an earlier experiment 
\cite{Kippenberg2006}.

Various mode families with qualitatively different displacement patterns are
recognized. Some modes involve mainly motion of the silicon pillar (number
6, 13 and 18 in figure~\ref{f:toroidModes}) and therefore couple only
negligibly to the optical mode, rendering it  irrelevant for the purpose of
cavity optomechanics. A torsional mode (number 4) is also observed, but it
couples only weakly to the optical modes (cf.\ section \ref{ss:omcoupling}).
Modes number 1, 3, 5, 7, 11 and 16 are characterized by an $n$-fold symmetry
under rotation by an angle $\pi/n$, where $n=1\ldots 6$ in this case. These
modes, which we refer to as ``crown'' modes, involve mainly sinusoidal
oscillation of the toroid in the $z$-direction. They typically follow a
quadratic dispersion relation, that is, their eigenfrequency is proportional
to the square of the number of nodes along the circumference \cite%
{Schliesser2008b}. Modes of this family are doubly frequency degenerate as
for each mode in this family there exists a second mode in which the
positions of nodes and antinodes are swapped. In measurements on real
toroids, this degeneracy is often lifted due to small deviations from
perfect rotational symmetry of the structure, leading to a splitting into a
resonance doublet.

In the context of cavity optomechanics, the most interesting modes coupling
strongly to the optical field are the ones with (nearly) radially symmetric
displacement patterns (number 2, 8, and 14). One usually distinguishes
flexural modes, in which the displacement is mainly along the $z$-axis
(number 2 and 8), and the radial modes involving mainly radial displacement.
Strongest coupling is caused by the fundamental radial mode, which we refer
to as the radial breathing mode (RBM) with number 14 in figure~\ref%
{f:toroidModes}. The RBM in the measured toroid has a frequency of $75.1\,%
\mathrm{MHz} $, and usually lies between $30$ and $120\,\mathrm{MHz} $ for
typical toroidal geometries.

\begin{figure}[tbp]
\centering
\includegraphics[width=.5\linewidth]{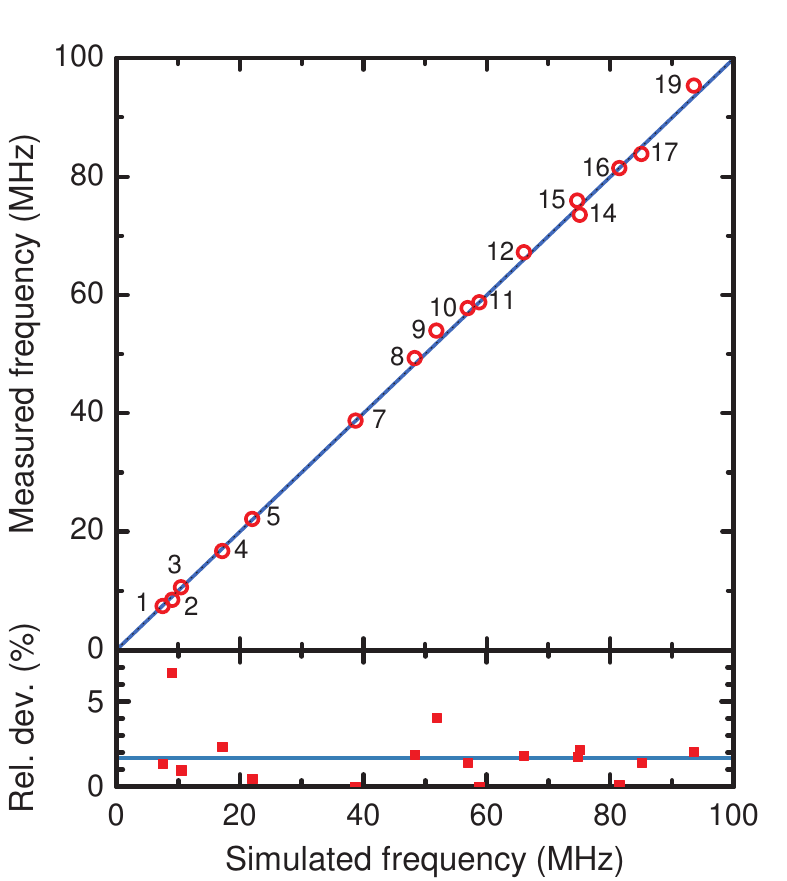}  
\caption{Comparison of simulated and measured frequencies of a toroidal
silica microresonator. The relative deviation is below 2\% on average.
Figure from ref.\ \protect\cite{Schliesser2008b}.}
\label{f:comparison}
\end{figure}

\subsubsection{Mechanical dissipation}

\label{sss:dissipation}

Another very important property of the RBM is its damping rate. As in the
case of the optical modes, different dissipation mechanisms lead to a
release of mechanical energy stored in the RBM to other degrees of freedom.
The mechanical damping rate $\Gamma_\mathrm{m} $ quantifies the rate at
which this takes place, and is again expressed in terms of a quality factor 
\begin{equation}  \label{e:Qmdef}
Q_\mathrm{m}=\Omega_\mathrm{m} /\Gamma_\mathrm{m} .
\end{equation}
The different damping mechanisms all contribute to the total damping, and
the resulting quality factor can be written 
\begin{equation}
Q^{-1}_\mathrm{m}=Q^{-1}_\mathrm{gas}+Q^{-1}_\mathrm{clamp}+Q^{-1}_\mathrm{%
TLS}+Q^{-1}_\mathrm{other}
\end{equation}
with contributions by the surrounding gas ($Q^{-1}_\mathrm{gas}$), by
clamping losses ($Q^{-1}_\mathrm{clamp}$), by two-level systems ($Q^{-1}_%
\mathrm{TLS}$) and by other damping mechanisms ($Q^{-1}_\mathrm{other}$).

In the initial optomechanical experiments with silica microtoroids the
mechanical quality factors reached values of 3{,}000 \cite{Kippenberg2005,
Schliesser2006}. By replacing the surrounding nitrogen gas with helium
(which has lower viscosity $\eta$ and molecular mass $M$), the quality
factors rose beyond 5{,}000, clearly indicating that the losses were
dominated by gas damping. From the known drag on a sphere oscillating in a
viscous fluid \cite{Landau1987} one can expect a scaling $Q_\mathrm{gas}%
\propto (\eta p M)^{-1/2}$ in this regime. Reducing the pressure $p$ of the
ambient gas should therefore raise $Q_\mathrm{gas}$. Indeed, at pressures
above $10\,\mathrm{mbar} $ the quality factor was observed to increase as $Q_%
\mathrm{gas}\propto p^{-1/2}$, indicative of the viscous regime, while for
lower pressures, in the molecular regime, $Q_\mathrm{gas}\propto p^{-1}$ was
observed \cite{Anetsberger2008}. At pressures below $1 \,\mathrm{mbar} $,
gas damping becomes irrelevant ($Q_\mathrm{gas}\gg Q_\mathrm{m}$) and the
quality factor is observed to converge towards a saturation value at lower
pressures. Among different toroids, there are significant variations in the
range between $1{,}000$ and $30{,}000$ for this value. This is due to
clamping losses, which strongly depend on the geometry of the sample as is
detailed below.

To study this effect of clamping losses more systematically, a toroidal
resonator was underetched in several steps, to reduce the radius of the
silicon pillar, and thereby increase the relative undercut, which we define
as 
\begin{equation}
\text{relative undercut}=1-\frac{\text{silicon pillar radius}}{\text{silica
toroid major radius}}.
\end{equation}
The quality factors and resonance frequencies of the RBM of six toroidal
resonators on the chip were measured for each etching step (figure~\ref%
{f:undercutSeries}). Strikingly, while the resonance frequency reduces
monotonically, the quality factors vary in a non-monotonic manner, with a
distinct minimum for an undercut of approximately 0.7. This behavior was highly
reproducible and observed for different independent samples.

\begin{figure}[htb]
\centering
\includegraphics[width=\linewidth]{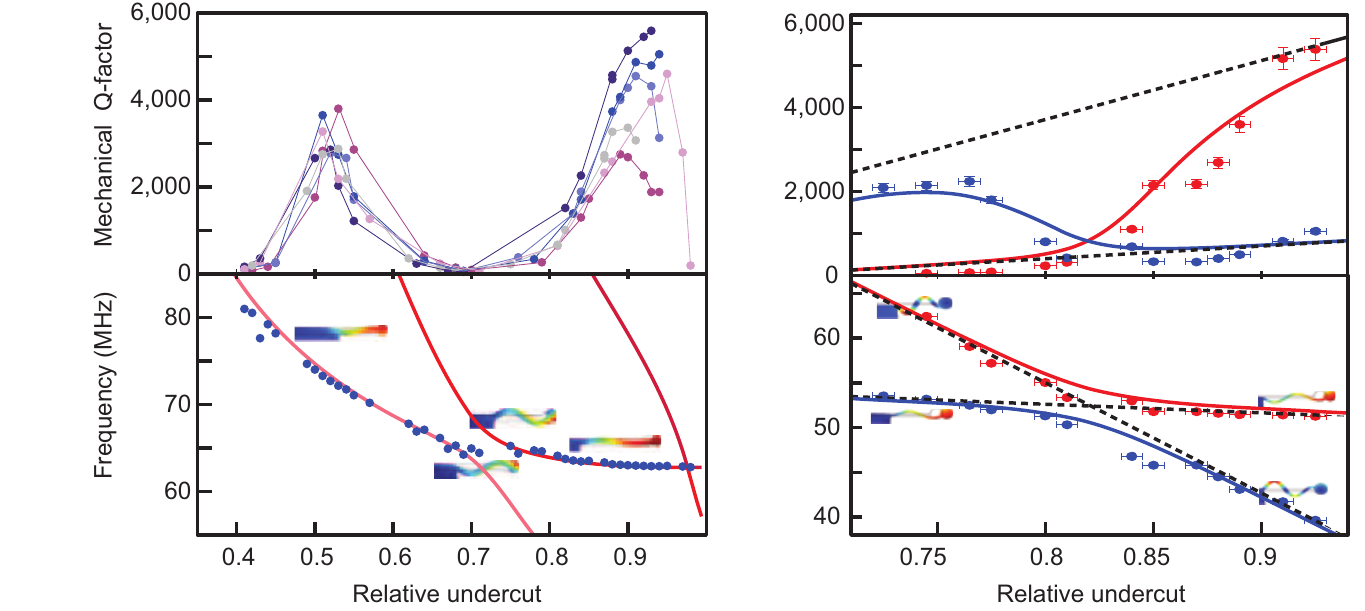}  
\caption{Normal mode coupling between the RBM and a flexural mode. Left
panels: Measured quality factors of the RBM of six samples on the same chip
as a function of relative undercut, and the resonance frequencies of one of
the samples (dots). Lines in the lower panel are results of simulations, and
show not only the frequency of the RBM, but also of radially symmetric
flexural modes. In the region where the frequencies are similar, for a
relative undercut of $\sim 0{.}7$, the quality factors are strongly reduced.
Right panels: Evidence of normal mode coupling between the RBM and a
flexural mode in another sample. The frequencies $\Omega _\pm$ and quality
factors $Q_\pm$ of the experimentally observed modes (dots) can be
reproduced using a model of two coupled harmonic oscillators (red and blue
lines). The frequencies and quality factors of the uncoupled modes,
corresponding to a pure RBM and flexural mode are assumed to depend only
linearly on the relative undercut (dashed lines). Adapted from \protect\cite%
{Anetsberger2008}.}
\label{f:undercutSeries}
\end{figure}

A series of simulations with increasing undercut reveals that the resonance
frequencies of the RBM and a radially symmetric flexural mode are crossing
each other for this undercut, as they have a different undercut dependence.
However, the actual resonance frequencies of the structure exhibit an
avoided crossing. At the same time, the modal shapes of the RBM and flexural
mode hybridize in the crossing region. These two facts imply a normal mode
coupling between the RBM and the flexural mode.

To corroborate this conjecture experimentally, the undercut in a different
sample was again systematically increased. For each etching step, a highly
sensitive measurement technique (polarization spectroscopy, cf.\ section \ref%
{s:UHS}) allowed the determination of the frequencies of both the RBM ($%
\Omega _\mathrm{R}$) and also the adjacent flexural ($\Omega _\mathrm{F}$)
mode. Indeed, the measured frequencies and quality factors can be
simultaneously reproduced using a simple model of two coupled harmonic
oscillators, the intrinsic frequencies $\Omega _\mathrm{R}$ and $\Omega _%
\mathrm{F}$ and quality factors $Q_\mathrm{R}$ and $Q_\mathrm{F}$ of which
are linearly dependent on the undercut. The frequencies $\Omega _\pm$ and
quality factors $Q_\pm$ of the new eigenmodes are given by \cite%
{Anetsberger2008} 
\begin{align}
\Omega _\pm+\frac{i}{2}\frac{\Omega _\pm}{Q_\pm}=&\frac{1}{2}\left(\Omega _%
\mathrm{R}+\Omega _\mathrm{F}\right)+ \frac{i}{4}\left(\frac{\Omega _\mathrm{%
R}}{Q_\mathrm{R}}+\frac{\Omega _\mathrm{F}}{Q_\mathrm{F}}\right)  \notag \\
&\quad \pm \sqrt{\left(\frac{1}{2}\left(\Omega _\mathrm{R}+\Omega _\mathrm{F}%
\right)+ \frac{i}{4}\left(\frac{\Omega _\mathrm{R}}{Q_\mathrm{R}}-\frac{%
\Omega _\mathrm{F}}{Q_\mathrm{F}}\right)\right)^2+\frac{g_\mathrm{im}^4}{%
4\Omega _\mathrm{R}\Omega _\mathrm{F}}}
\end{align}
where the intermode coupling $g_\mathrm{im}$ is an adjustable parameter
(figure~\ref{f:undercutSeries}). The data in figure~\ref{f:undercutSeries}
can be fit using $g_\mathrm{im}/2 \pi=14\,\mathrm{MHz} $. This rather strong
coupling is attributed to the asymmetry of the structure in the axial
direction due to both the offset of the toroid from the disk \cite%
{Kippenberg2005}, and the fact that the silicon pillar supports the disk
only from below.

Summarizing the previous observations, we conclude that the admixture of a
flexural displacement pattern to the RBM reduces the quality factor of the
latter. This can be explained by noting that the flexural modes induce axial
displacement also in the region where the silica disk is supported by the
silicon pillar. The periodic oscillation in axial direction launches
acoustic waves into the pillar, through which the energy of the mode is
dissipated, thereby deteriorating the quality factor.

\begin{figure}[ptb]
\centering
\includegraphics[width=\linewidth]{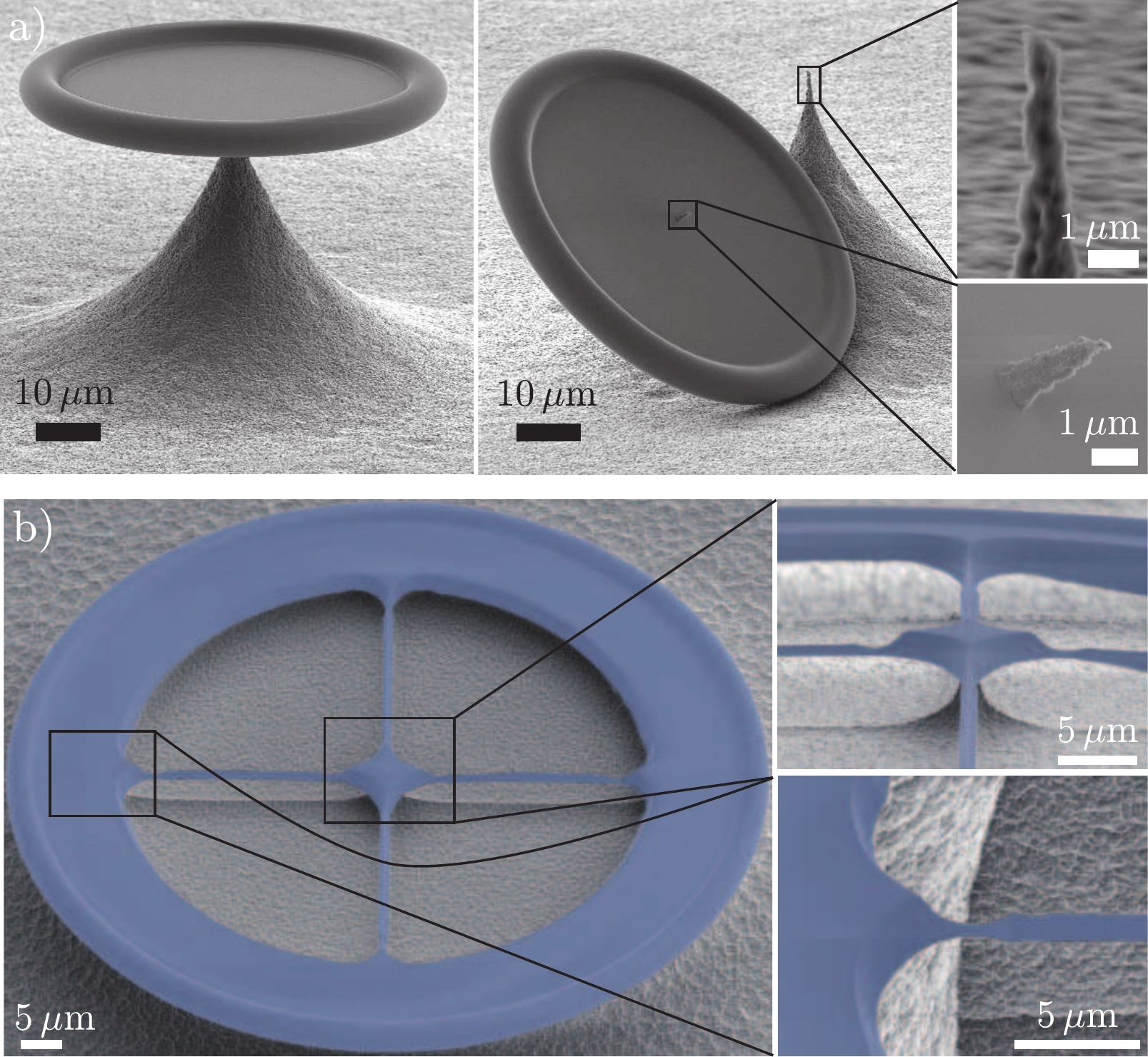}  
\caption{Silica toroidal resonators with ultralow mechanical damping. a)
Reduced clamping loss by supporting the silica disk with a ``needle''
pillar. The central and right panels show the tip of an intentionally broken
pillar, which has a sub-micron diameter. A quality factor of $30{,}000$ is
reached with such structures for a $40\,\mathrm{MHz} $-RBM \protect\cite%
{Schliesser2008}. b) Reducing clamping loss by engineering the axial
displacement in the clamping region using a ``spokes'' design. Such
resonators achieved mechanical quality factors up to $32{,}000$ at $38\,%
\mathrm{MHz} $, and $50{,}000$ at $24 \,\mathrm{MHz} $ at room temperature 
\protect\cite{Anetsberger2008} limited by TLS losses. }
\label{f:highmechQ}
\end{figure}

This qualitative understanding is quantitatively supported by an analytical
model \cite{Anetsberger2008}, in which the acoustic energy loss is estimated
as the power radiated by a membrane of area $A_\mathrm{p}$ (area of the
silicon pillar) oscillating with an axial displacement $\Delta z(\vec r)$ at
frequency $\Omega_\mathrm{m} $ 
\begin{equation}
P_\mathrm{mech}=v \rho \Omega_\mathrm{m} ^2 \int_{A_\mathrm{p}} \left|
\Delta z(\vec r) \right|^2 d^2 r,  \label{e:dparameter}
\end{equation}
where $v$ is the sound velocity and $\rho$ the density. For geometry parameters
close to the modal crossing, it was indeed found experimentally that 
\begin{equation}
Q_\mathrm{clamp}\propto\frac{\Omega_\mathrm{m}  E_\mathrm{mech}}{P_\mathrm{%
mech}},  \label{e:qcla}
\end{equation}
where $E_\mathrm{mech}$ is the total mechanical energy of the mode.
Advantageously, both $E_\mathrm{mech}$ and $P_\mathrm{mech}$ can be
simulated using FEM \cite{Anetsberger2008}. As an aside we note that an
expression similar to (\ref{e:dparameter}) and (\ref{e:qcla}) was obtained
in a rigorous theoretical analysis based on a phonon tunneling approach \cite%
{Wilson-Rae2008}. To reduce $P_\mathrm{mech}$, two strategies are
immediately evident: Either the clamping area $A_\mathrm{p}$ is minimized,
or the axial displacement $\Delta z(\vec r)$ in this region is reduced.
Minimizing the clamping area is possible, for example, by fabricating
toroids with a very strong undercut, as shown in figure~\ref{f:highmechQ},
in which the silica disk is supported by a ``needle pillar'' of
sub-micrometric diameter \cite{Schliesser2008}.

Alternatively, by introducing spokes into the silica disk, it was possible
to strongly reduce the coupling of the radial motion to an axial
displacement in the clamping region \cite{Anetsberger2008}. Both presented
approaches yield good results with quality factors exceeding $30{,}000$ at
frequencies around $40 \,\mathrm{MHz} $ \cite{Schliesser2008,
Anetsberger2008}.

\begin{figure}[ptb]
\centering
\includegraphics[width=\linewidth]{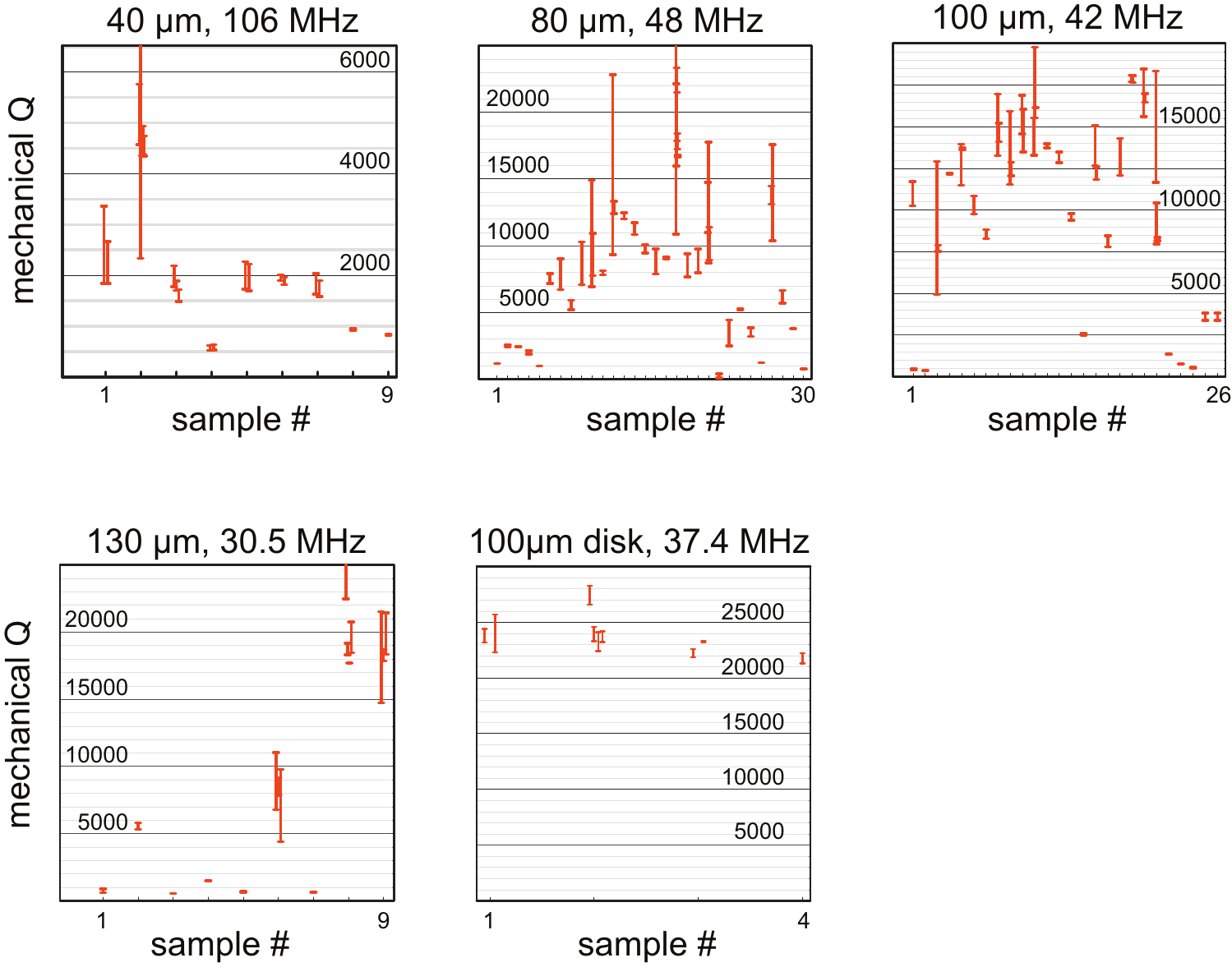}  
\caption{Overview of mechanical quality factors of the RBM measured in
typical toroids. Major diameters of the pre-reflow disk and the approximate
resonance frequency are indicated in the figure captions. Each bracket
indicates the span within which the mechanical quality factor was found in a
pair of measurements, in which the probing laser was red and blue detuned
(dynamical backaction modifies the measured effective mechanical quality
factor in opposite directions in the two cases). Groups of brackets belong
to the same toroid. The last panel shows measurements taken on silica disks
prior to the reflow. }
\label{f:qs}
\end{figure}

In figure \ref{f:qs} we show an overview of mechanical quality factors
achieved in typical samples with different major diameters, all strongly
undercut ($>90\%$) and measured in vacuum. A clear trend to higher quality
factors for larger cavity sizes (and therefore lower frequencies) is
observed. The scatter in the data of neighboring toroids (with very similar
reflow preform and final pillar shapes) indicates that clamping losses
depend sensitively on geometry parameters. Finally, we note that we have
consistently observed higher quality factors in disks than in toroids. This
is attributed to a reduced offset (figure \ref{f:geometryParameters}) of the
oscillating mass, which, in toroids, mediates the coupling of the RBM to
radially symmetric flexural modes with strong dissipation to the pillar.

The highest quality factors which were experimentally achieved at room
temperature are $Q_\mathrm{m}\sim 50{,}000$ above $20\,\mathrm{MHz} $. For
these structures however, simulations clearly indicate $Q_\mathrm{clamp}\gg
Q_\mathrm{m}$ according to equation (\ref{e:qcla}), indicating that a
different dissipation mechanism must now be dominant. A strong temperature
dependence of the quality factor (allowing for values up to $80{,}000$ at $%
110^\circ\mathrm{C}$) suggests a temperature-dependent dissipation mechanism 
\cite{Anetsberger2008}.

Indeed, the intrinsic damping of acoustic excitations in glass is known to
follow a universal behavior observed in many amorphous solids \cite{Pohl2002}%
. This effect is attributed to the coupling of strain fields to structural
defects in the material. While the microscopic nature of these defects is
not precisely known, it can be successfully modeled by a distribution of
effective two-level systems (TLS) with two stable equilibria, represented by
a particle in an asymmetric double-well potential \cite{Jackle1972}. These
potentials are characterized by the energy asymmetry $\Delta$ of the ground
states in both potentials (the two levels involved), and the height $V$ of
the energy barrier separating the two wells as shown in figure~\ref%
{f:doublewell}. Oscillating strain fields associated with an acoustic
excitation modulate the energy asymmetry $\Delta$ between the two potential
minima, and thereby couple to the TLS.

\begin{figure}[bth]
\centering
\includegraphics[width=.33\linewidth]{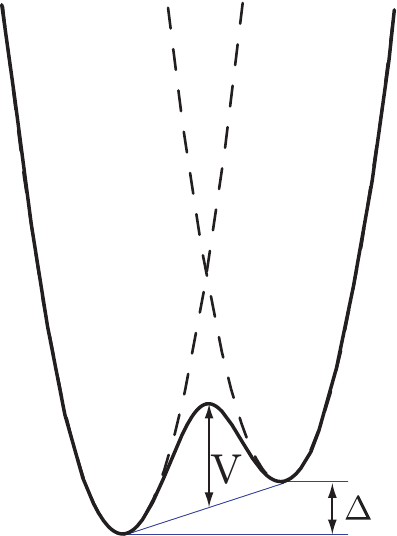}  
\caption{Double-well potential used to model the structural defects in
glass. The two individual wells (dashed lines) are usually assumed to be
identical, but to have a ground-state energy that differs by an energy
asymmetry $\Delta$. The two stable equilibria are separated by a barrier of
height $V$. }
\label{f:doublewell}
\end{figure}

To a very good approximation, the resulting Debye relaxation of acoustic
excitations leads to a quality factor given by \cite{Vacher2005} 
\begin{equation}
Q^{-1}_\mathrm{TLS}= \frac{\gamma^2}{\rho v^2 k_\mathrm{B} T}
\int_{-\infty}^{+\infty} \int_0^{+\infty} P(\Delta, V)\, \mathrm{sech}%
^2\left(\frac{\Delta}{2 k_\mathrm{B} T}\right) \frac{\Omega  \tau}{1+\Omega
^2 \tau^2} dV d\Delta,
\end{equation}
where $\gamma=\frac{1}{2} \partial \Delta /\partial \varepsilon$ is the
change of the potential asymmetry as a function of strain $\varepsilon$, $%
\rho$ is the density, $v$ sound velocity, and $P(\Delta,V)$ is the
distribution of TLS in the energy parameters $\Delta$ and $V$ in the sense
that $P(\Delta,V) dV d\Delta$ is a volume density of defects. 

At room temperature, the relaxation between the two stable states is
dominated by thermally activated processes. The relaxation time is thus
given by an Arrhenius-type law 
\begin{equation}
\tau^{-1}=\tau_0^{-1} e^{-V/k_\mathrm{B} T} \cosh\left(\frac{\Delta}{2k_%
\mathrm{B} T} \right).
\end{equation}
Following the arguments given in reference \cite{Vacher2005}, a sensible
choice of the distribution function $P(\Delta, V)$ yields eventually a
quality factor of 
\begin{equation}
Q_\mathrm{TLS}^{-1}= C \cdot\erf\left(\frac{\sqrt{2} k_\mathrm{B} T}{\Delta_%
\mathrm{c}}\right) \frac{1}{k_\mathrm{B} T}\int_0^{\infty} \left(\frac{V}{V_0%
}\right)^{-\zeta} e^{-{V^2}/2{V_0^2}} \frac{\Omega  \tau_0 e^{V/k_\mathrm{B}
T}}{1+\Omega ^2 \tau_0^2 e^{2 V/k_\mathrm{B} T}} dV  \label{e:qta}
\end{equation}
reproducing the experimental data over four orders of magnitude in frequency
($11\,\mathrm{kHz} \ldots 200\,\mathrm{MHz} $) and two orders of magnitude
in temperature (from a few Kelvin to above room temperature) for the
parameters $V_0=(667\pm 21)\,\mathrm{K} \cdot k_\mathrm{B}$, $\zeta=0{.}%
28\pm0{.}03$, $\log_{10}(\tau_0/\mathrm{s})=-12{.}2\pm 0{.}18$ and $%
V_0/\Delta_\mathrm{c}=7{.}7\pm 0{.}7$.

\begin{figure}[htb]
\centering
\includegraphics[width=.8\linewidth]{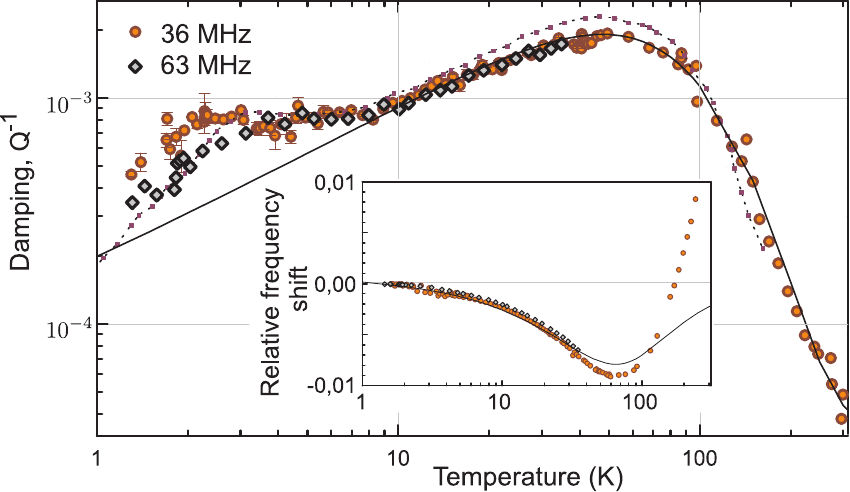}  
\caption{Quality factor of the RBMs of two samples which are dominated by
the damping due to coupling to structural defects in glass. Solid line is a
fit using equation (\protect\ref{e:qta}), and the dashed lines are
experimental data from an acoustic wave attenuation experiment at $40 \,%
\mathrm{MHz} $ \protect\cite{Bartell1982}. The inset shows the relative
frequency shift of the RBM induced by the relaxation of TLS. From ref.\ 
\protect\cite{Arcizet2009a}.}
\label{f:QvsT}
\end{figure}

While usually measured as the attenuation of large-amplitude acoustic waves
in bulk material, the very same temperature dependence of the quality factor
was found for the RBM of spokes toroids with sufficiently low clamping
losses \cite{Arcizet2009a}. Figure~\ref{f:QvsT} shows the measured quality
factor of the RBMs of two samples at frequencies of $36$ and $63 \,\mathrm{%
MHz} $. Simultaneously with the damping, relaxation of the TLS also leads to
a change in the sound velocity, giving rise to a frequency shift of the
mechanical modes \cite{Vacher2005}, 
\begin{equation}
\left(\frac{\delta\Omega_\mathrm{m} }{\Omega_\mathrm{m} }\right)=-\frac{C}{2}
\cdot\erf\left(\frac{\sqrt{2} k_\mathrm{B} T}{\Delta_\mathrm{c}}\right) 
\frac{1}{k_\mathrm{B} T}\int_0^{\infty} \left(\frac{V}{V_0}\right)^{-\zeta}
e^{-{V^2}/2{V_0^2}} \frac{1}{1+\Omega ^2 \tau_0^2 e^{2 V/k_\mathrm{B} T}} dV,
\end{equation}
which is also shown in figure~\ref{f:QvsT}. At temperatures above $10 \,%
\mathrm{K} $, the damping can be accurately described by 
equation (\ref{e:qta}), with a peak damping at about $50\,\mathrm{K} $
leading to a minimum quality factor of $Q_\mathrm{TLS}\approx 500$.

Below $10\,\mathrm{K} $, the relaxation is dominated by tunneling processes
between the two equilibria \cite{Jackle1972, Tielburger1992}, instead of the
thermally-activated relaxation. The tunneling relaxation is responsible both
for the low-temperature plateau ($Q_\mathrm{m} \approx 1200$ at $5\,\mathrm{K%
} $) and the roll-off at very low temperatures with $Q_\mathrm{TLS}%
^{-1}\propto T^3/\Omega $. For completeness, we note that apart from damping
via relaxation, at sufficiently low temperatures, direct absorption of
acoustic waves by the TLS also leads to damping, which saturates at high enough amplitudes \cite{Jackle1972, Hunklinger1973}.

Other damping mechanisms, as described by $Q_\mathrm{other}$, include
thermoelastic damping \cite{Zener1937, Zener1938}, damping by anharmonicity 
\cite{Vacher2005}, and surface effects \cite{Ekinci2005}. For silica
microtoroids or spheres, these effects are individually estimated to lead to
limiting quality factors on the order of $10^5$ or more. In particular when
operating the resonators at cryogenic temperatures---as required for
advanced experiments in cavity optomechanics---these damping mechanisms can
be safely neglected compared with the damping due to TLS.

\clearpage

\subsection{Optomechanical coupling}

\label{ss:omcoupling}

\begin{figure}[hbt]
\centering
\includegraphics[width= \linewidth]{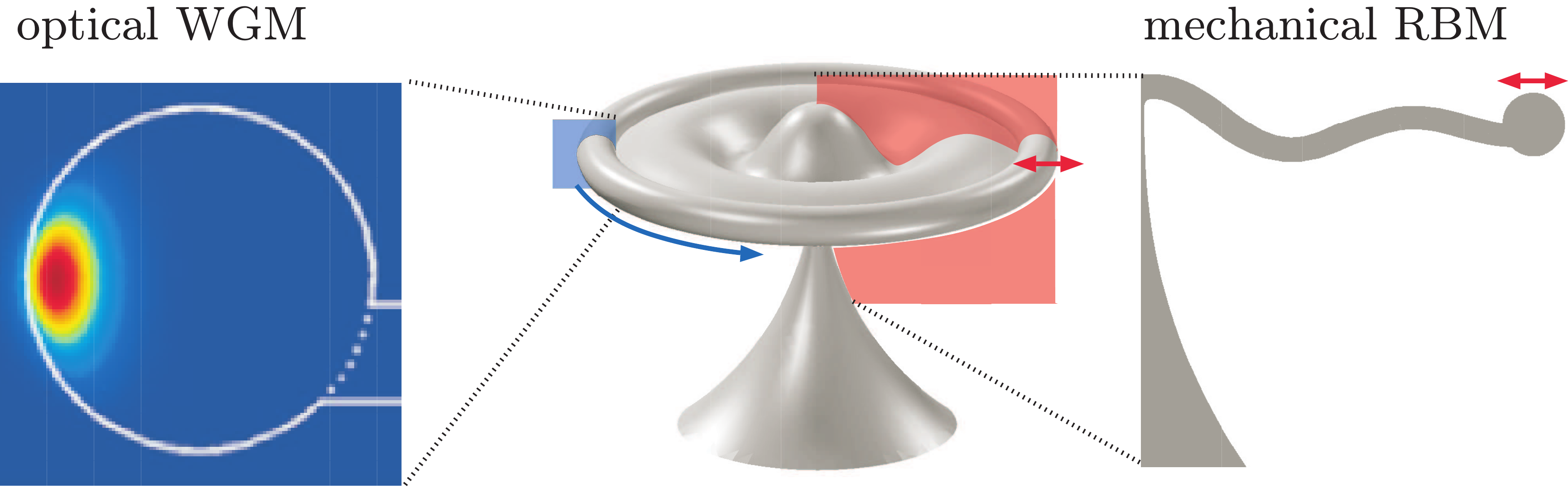}
\caption{Mode shapes of optical and mechanical modes in a silica
microtoroidal resonator (FEM simulations), which have to be taken into
account to quantify the strength of optomechanical interaction. }
\label{f:om}
\end{figure}

In order to describe optomechanical coupling in microspheres and
microtoroids introduced in the previous section, it is advantageous to map
the mechanical modes of interest to an effective one-dimensional mechanical
oscillator, described by a displacement $x$, which parametrically modulates
the optical resonance frequency through a non-zero $g_0 =d\omega_\mathrm{c}
/dx$. For optomechanical devices which host optical and mechanical modes
with complex three-dimensional mode distributions such as silica
microtoroidal resonators (figure~\ref{f:om}), this mapping can be
non-trivial.

\subsubsection{Mapping to a scalar displacement}

To formally describe this mapping, a weighting function $\vec w(\vec r)$ is
introduced, mapping the displacement field $u(\vec r,t)$ to a scalar
displacement $x$ according to 
\begin{align}  \label{e:xdef}
x(t)&=\int_V \vec w(\vec r) \cdot \vec u(\vec r,t) d^3r=\sum_n c_n(t) \int_V
\vec w(\vec r) \cdot \vec u_n^0(\vec r) d^3r \\
& \equiv \sum_n c_n(t)\left\langle \vec w, \vec u_n^0 \right\rangle,
\end{align}
where the decomposition of the displacement pattern $\vec u$ as presented in
eq.\ (\ref{e:decomposition}) is used. For each mechanical mode, the overlap
integral $\langle \vec w , \vec u_n^0\rangle $ determines the relative
strength of the transduction of the $n$-th mode amplitude $c_n(t)$ into the
optically sampled displacement $x(t)$. The \emph{global} normalization of $%
\vec w(\vec r)$ is, in principle, arbitrary, as it can be compensated by an
adequate choice of $g_0 =d\omega_\mathrm{c} /dx$. For example, it can be
chosen such that the displacement of a particular part of one mode's
displacement pattern (such as the antinode of a beam's fundamental mode)
directly corresponds to $x$ \cite{Eichenfield2009}.

For the most generic optomechanical systems, there are other obvious
choices. For example, for a Fabry-Perot\ cavity, the normalization is chosen
such that $x$ corresponds to the center-of-mass mirror movement if it was
displaced as a whole. Thus, if the laser spot on a mirror at $z=z_0$, which
contains the mechanical modes, is given by a rotationally symmetric Gaussian
with a waist of $w_0$, the weighting function reads \cite%
{Pinard1999,Braginsky2001} 
\begin{equation}
\vec w(\vec r)=\frac{2}{\pi w_0^2} e^{-2 r^2/w_0^2} \delta(z-z_0) \vec e_z.
\end{equation}
With this weighting function, the physically correct frequency shift is
obtained using the coupling constant $g_0 =-\omega_\mathrm{c} /L$, where $L$
is the total length of the cavity.

In the case of silica WGM resonators, with their three-dimensional
distributions of optical and mechanical fields, the calculation of the
effective displacement is more difficult. A sensible approximation may be
obtained by considering the mechanical displacement as a perturbation, which
does not modify the optical fields, but only displaces polarizable matter
within the optical field distribution. The resulting relative frequency
shift equals the relative change in the electromagnetic energy stored in the
mode, as the number of stored photons, each of energy $\hbar \omega_\mathrm{c%
} $, is conserved in the cavity \cite{Arnold2003}. The resulting frequency
shift is 
\begin{equation}
\frac{\delta \omega_\mathrm{c} }{\omega_\mathrm{c} }=\frac{\delta E_\mathrm{%
em}}{E_\mathrm{em}}=\frac{\int \frac{1}{2} E^2(\vec r)\vec \nabla
\varepsilon(\vec r)\cdot \vec u(\vec r) d^3 r} {2\int \frac{1}{2} E^2(\vec
r) \varepsilon(\vec r) d^3 r},
\end{equation}
where $E^2(\vec r)$ is the squared magnitude of the electric field, and the
factor $2$ in the denominator is due to the fact that equal amounts of
energy are stored in both the electric and magnetic fields. Note however
that the magnetic energy does not shift upon a displacement of the boundary,
as the magnetic permeability of glass is very close to unity. For a
resonator made out of a homogeneous dielectric material, the integral in the
nominator is essentially a surface integral over the boundary, as $\vec
\nabla \varepsilon(\vec r)$ is zero everywhere except for the surface. We
use the coupling constant $g_0 =-\omega_\mathrm{c} /R$, expressing the
frequency shift as a consequence of an effective radius change. One then
finally obtains 
\begin{equation}
\vec w(\vec r)=-\frac{R}{E_\mathrm{em}}\cdot\frac{1}{2} {\ E^2(\vec r) \vec
\nabla \varepsilon(\vec r)}.   \label{e:wwgm}
\end{equation}
As equation (\ref{e:wwgm}) is difficult to analytically evaluate in complex
geometries, a useful approximation is given by 
\begin{equation}
\vec w(\vec r)\approx\frac{1}{2\pi R}\delta(z-z_0) \delta(r-R) \vec e_r. 
\label{e:wwgmsimple}
\end{equation}
where $R$ is the major radius and $z=z_0$ the plane of the equator of the
toroid. This weighting function essentially considers the transverse size of
the optical mode as negligibly small compared to the scales of the
displacement patterns, and the resulting displacement $x$ corresponds to the
change of the cavity radius, which is sampled by the optical mode. In this
work, equation (\ref{e:wwgmsimple}) is applied to derive the displacement
induced by the excitation of a particular WGM from the results of FEM
simulations.

We finally note that in a dielectric resonator, it is important to also
consider strain-optical effects, that is, a strain-dependent refractive
index leading to additional resonance frequency shifts for a given
excitation of the mechanical mode. This effect was found to dominate
the optomechanical coupling in a cryogenic sapphire microwave WGM resonator 
\cite{Locke2004}. To assess the relative contribution in silica WGM
microresonators, we may use the analytic expressions for the strain field (%
\ref{e:spherestrainr})--(\ref{e:spherestrainphi}) in a micro\emph{sphere} to
calculate the corresponding change in the refractive index as experienced by
the optical mode. Due to the homogeneity of the strain fields on the scale
of the optical mode cross section, it can be well approximated by just
evaluating the strain fields at the edge of the sphere using \cite%
{Ilchenko1998} 
\begin{align}
\delta \left(n^{-2}\right)_\mathrm{TE}&= p_2 \varepsilon_{rr}+ p_1
\varepsilon_{\theta\theta}+ p_2 \varepsilon_{\phi\phi} \\
\delta \left(n^{-2}\right)_\mathrm{TM}&= p_1 \varepsilon_{rr}+ p_2
\varepsilon_{\theta\theta}+ p_2 \varepsilon_{\varphi\varphi},
\end{align}
where the required coefficients of the photo-elastic tensor are given by $%
p_1=0.121$ and $p_2=0.270$ \cite{Dixon1967}. This leads to an extra
frequency shift of about 30\% (TE modes) and 50\% (TM modes), as compared to
the shift induced by the displacement of the boundary alone. Evaluation of
the strain-optical coupling in a toroid is not possible analytically. For
typical torus geometries as used in this work, however, we can extract an
extra frequency shift of less than 20\% from FEM simulations, in agreement
with an earlier estimate \cite{Kippenberg2005}.

\subsubsection{Effective mass}

\label{sss:effective mass}

As a global coupling coefficient $g_0 $ is used to quantify the coupling
strength of an effective displacement $x$ to the resonance frequency $\omega_%
\mathrm{c} $ of the cavity, it is necessary to absorb the different coupling
strengths of different mechanical modes into another parameter. This
parameter is referred to as the \emph{effective} mass of the individual
mechanical modes \cite{Gillespie1995, Pinard1999}.

An operational definition of the effective mass of one particular mode
(labeled in the following with an index $n$) can be derived from its
potential energy $U_n$ which can be recast from equation (\ref%
{e:stressstrainenergy}) to 
\begin{equation}
U_n=\frac{1}{2}M_n \,\Omega_{n}\, (c_n(t))^2  \label{e:potentailergy}
\end{equation}
using the free-boundary conditions \cite{Pinard1999} and the definition of
the moving mass 
\begin{equation}
M_n=\int_V \rho |\vec u_n^0(\vec r)|^2d^3r.  \label{e:Mndef}
\end{equation}
As $c_n(t)$ is experimentally not accessible, we want to express the
potential energy in terms of the measured displacement of the mode, 
\begin{align}  \label{e:xndef}
x_n(t)&=\int_V \vec w(\vec r) \cdot \vec u_n(\vec r,t) d^3r=
c_n(t)\left\langle \vec w, \vec u_n^0 \right\rangle,
\end{align}
and therefore require 
\begin{equation}
U_n=\frac{1}{2} {m_\text{eff}}_{,n} \,\Omega_{n}^2\, (x_n(t))^2.
\label{e:effectivemassoperational}
\end{equation}
This immediately leads to the formal definition 
\begin{equation}
{m_\text{eff}}_{,n}\equiv\frac{M_n}{\left\langle \vec w, \vec u_n^0
\right\rangle^2}.  \label{e:meffdef}
\end{equation}
In practice, equation (\ref{e:effectivemassoperational}) is used to
calculate the effective masses from experimental data (where, for a given $%
g_0 $, $x$ is directly measured, and $U_n\approx k_\mathrm{B} T/2$) or FEM
simulations, from which $x_n$ and $U_n$ can be simultaneously extracted.

The effective masses of the fundamental sphere modes can be calculated
analytically, as both the potential energy (\ref{e:Uspherenum}) and the
resonance frequency (\ref{e:Omspherenum}) as a function of the radius $R$
are known, yielding the numeric relation $m_\text{eff} = 8470\,\mathrm{kg/m^3%
} \cdot R^3$ for silica, i.\ e.\ approximately $30\,\mathrm{ng} $ for a $30\,\mathrm{%
\mu m}  $-diameter sphere. The numerical values for the RBMs in silica
microtoroids are lower, in the range of $3$ to $20\,\mathrm{ng} $ for the
typical dimensions used.

\subsubsection{Forces acting on the mechanical modes}

To calculate the radiation pressure force acting on the mechanical modes, it
is necessary to assess the momentum transfer from the optical mode to the
dielectric medium. %
The fundamental starting point for such an analysis in a complex geometry
such as a silica WGM resonator is the flux of momentum density of the
electromagnetic field in the medium \cite{Landau1984, Pfeifer2007} 
\begin{equation}
T_{ij}=-E_i D_j-H_i B_j+\frac{1}{2} \delta_{ij} \left( \sum_k E_k D_k +
\sum_k H_k B_k\right),
\end{equation}
where $E$, $D$, $H$ and $B$ denote the usual electric and magnetic fields,
indices $i$ and $j$ denote the cartesian components of vectors and tensors,
and $\delta_{ij}$ is the Kronecker-delta. In vacuum, the entity $T_{ij}$ is
usually referred to as Maxwell's stress tensor. The body force $\vec f$,
that is, the force density acting on the medium, is given by the divergence
of this flux, plus a contribution from a temporal change of the flux density 
\begin{align}
f_i=-\sum_j \frac{\partial T_{ij}}{\partial x_j}-\frac{\partial}{\partial t} 
\frac{1}{c^2}S_i
\end{align}
where 
\begin{equation}
\vec S=\vec E \times \vec H
\end{equation}
is the real-valued Poynting vector. With $\vec D=\varepsilon \vec E$ and $%
\vec B=\mu \vec H$ we can write this as 
\begin{align}
\vec f &= -\frac{1}{2}\vec E \vec E \,\vec \nabla\varepsilon-\frac{1}{2}
\vec H \vec H \,\vec \nabla \mu+\left(\frac{\varepsilon\mu}{\varepsilon_%
\mathrm{0}\mu_\mathrm{0}}-1\right)\frac{1}{c^2} \frac{\partial}{\partial t}%
(\vec E\times \vec H)  \label{e:fullbodyforce}
\end{align}
where Maxwell's equations $\vec \nabla\cdot \vec D=0$, $\vec \nabla\cdot
\vec B=0$, $\vec \nabla\times \vec E=-\mu\partial\! \vec H/\partial t $ and $%
\vec \nabla\times \vec H=+\varepsilon \partial\! \vec E/\partial t$ were
used \cite{Landau1984}. We will disregard in the following the last term,
the so-called Abraham force\footnote{%
The nature of this force has also remained a contentious issue for decades 
\cite{Pfeifer2007}.} as it is usually small, and oscillates at the optical
carrier frequency, so that it averages out on the timescale of a mechanical
oscillation. Furthermore, we neglect the second term due to the
close-to-unity magnetic permeability of normal glass and obtain finally 
\begin{equation}
\vec f(\vec r)\approx -\frac{1}{2} E^2(\vec r)^2 \,\vec
\nabla\varepsilon(\vec r).
\end{equation}
This body force can be recast to 
\begin{equation}
\vec f(\vec r)=\vec w(\vec r) \frac{E_\mathrm{em}}{R}\equiv \vec w(\vec r) F_%
\mathrm{rp}   \label{e:bodyforce}
\end{equation}
using equation (\ref{e:wwgm}) and the scalar radiation pressure force from
equation (\ref{e:frp}) 
\begin{equation}
F_\mathrm{rp} = -g_0  \frac{E_\mathrm{em}}{\omega}.
\end{equation}

Note that equation (\ref{e:bodyforce}) also holds for the simplified
assumptions leading to the simplified weighting function (\ref{e:wwgmsimple}%
): If all optical power is concentrated to the rim of the toroid, an
estimate for the force density can be derived from a simple consideration: A
line element $R \, d\varphi$ contains the fraction $d\varphi/2\pi$ of the
total number of $E_\mathrm{em}/\hbar \omega$ intracavity photons, where $E_%
\mathrm{em}$ is the total electromagnetic energy stored in the resonator.
Within the time $n R\, d\varphi/c$, each photon transfers a fraction of $%
\sin d\varphi\approx d\varphi$ of its momentum $\hbar k$ to the wall,
yielding eventually 
\begin{equation}
\vec f(\vec r)=\frac{1}{R d\varphi}\frac{d\varphi}{2\pi}\frac{E_\mathrm{em}}{%
\hbar \omega}\frac{\hbar k \,d\varphi}{n R \,d\varphi/c}\delta(z-z_0)
\delta(r-R)\vec e_r= \vec w(\vec r)\, \frac{E_\mathrm{em}}{R} .
\end{equation}

Irrespective of the detailed form of the weighting function, the  energy of
the mechanical system is \cite{Pinard1999} 
\begin{equation}
H=\sum_n \frac{1}{2} M_n \left(\dot c_n(t)\right)^2+\frac{1}{2}M_n
\Omega_{n}^2 \left(c_n(t)\right)^2-\langle \vec f,\vec u\rangle
\end{equation}
where strain-optical effects are neglected for simplicity. This leads to the
following equations of motion for the mechanical mode amplitudes 
\begin{equation}
\ddot c_n + \Omega_{n}^2 c_n=\frac{1}{M_n} \langle \vec f, \vec u_n^0\rangle
\end{equation}
or, in the Fourier domain, 
\begin{equation}
c_n(\Omega )=\frac{1}{M_n(\Omega_{n}^2-\Omega ^2- i \Omega  \Gamma_n)} \cdot
\left(\langle \vec f(\Omega ), \vec u_n^0 \rangle+\delta F_{\mathrm{T}%
,n}(\Omega )\right)
\end{equation}
where the damping $\Gamma_n$ and the corresponding fluctuational force $%
\delta F_T$ have been introduced as well. As only the projection $x_n$ of
the excitation of the mechanical mode is measured, one obtains 
\begin{align}
x_n(\Omega )&=\langle \vec w, \vec u_n^0\rangle c_n(\Omega ) =\frac{\langle
\vec w, \vec u_n^0\rangle}{M_n(\Omega_n-\Omega ^2- i \Omega  \Gamma_n)}
\cdot \left(\langle \vec f(\Omega ), \vec u_n^0 \rangle+\delta F_{\mathrm{T}%
,n}(\Omega )\right)  \notag \\
&=\frac{\langle \vec w, \vec u_n^0\rangle^2}{M_n (\Omega_n-\Omega ^2- i
\Omega  \Gamma_n)}\left(F_\mathrm{rp} (\Omega )+\frac{\delta F_{\mathrm{T}%
,n}(\Omega )}{\langle \vec w, \vec u_n^0\rangle}\right)  \notag \\
&=\frac{1}{m_{\mathrm{eff},n} (\Omega_n-\Omega ^2- i \Omega  \Gamma_n)}%
\left(F_\mathrm{rp} (\Omega )+\frac{\delta F_{\mathrm{T},n}(\Omega )}{%
\langle \vec w, \vec u_n^0\rangle}\right).
\end{align}
The fluctuational thermal force $\delta F_{\mathrm{T},n}$ used here obeys $%
\langle\delta F_{\mathrm{T},n}(\Omega ) \,\delta F_{\mathrm{T},n}(\Omega
^{\prime })\rangle\propto M_n$, but the commonly employed approach is to
introduce an effective thermal force $\delta F_{\mathrm{th},n}$ with $M_n$
replaced by $m_{\mathrm{eff},n}$ leading to the form (\ref%
{e:thermalforcefourier}), and yielding finally 
\begin{align}  \label{e:xn}
x_n(\Omega )&=\frac{1}{m_{\mathrm{eff},n} (\Omega_n-\Omega ^2- i \Omega 
\Gamma_n)}\left(F_\mathrm{rp} (\Omega )+\delta F_{\mathrm{th},n}(\Omega
)\right) 
\end{align}
the one-dimensional description used in section \ref{s:omtheory}.

As an illustration of the influence of the effective mass, figure~\ref%
{f:dispersion} shows the spectrum $\bar S_{xx}(\Omega )$ of the fluctuations
of $x=\sum_n x_n$ for a toroid driven by thermal noise according to equation
(\ref{e:xn}). Data were extracted from FEM simulations for toroids the major
radius of which has been continuously swept from $35$ to $100\,\mathrm{\mu m}
$. Frequencies are a direct simulation result, and the effective masses were
extracted using equation (\ref{e:wwgmsimple}). All quality factors were, for
simplicity, assumed to equal 5000. Clearly, the strong signature of the RBM
can be discerned for its low effective mass, on top of the background of
weaker flexural modes. At a major diameter of $\sim75\,\mathrm{\mu m}  $,
the avoided crossing discussed in section \ref{sss:dissipation} is also
apparent.

\begin{figure}[htb]
\centering
\includegraphics[width=.55\linewidth]{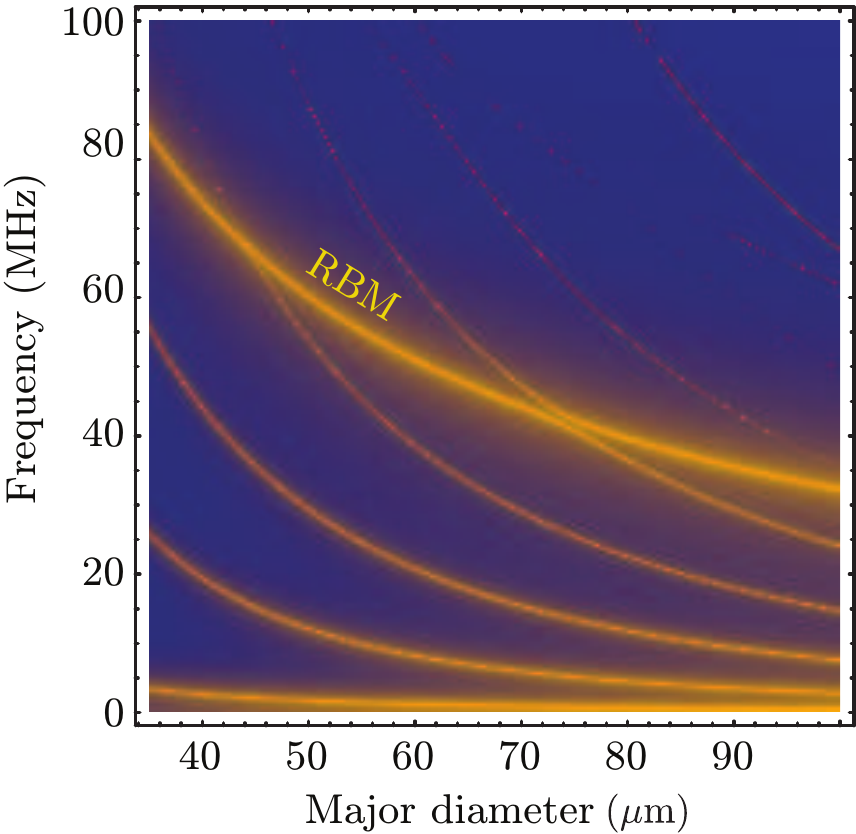}  
\caption{Color-coded displacement noise spectrum of a toroid with a $1 \,%
\mathrm{\protect\mu m}  $ thick silica disk, a $4\,\mathrm{\protect\mu m}  $
minor diameter, 90\% undercut and a varying major diameter, as simulated
using FEM. The strong signature is from the RBM, the other traces are from
flexural modes.}
\label{f:dispersion}
\end{figure}

\clearpage

\section{Ultrahigh-sensitivity interferometric motion transduction}

\label{s:UHS}

\begin{figure}[htb]
\centering
\includegraphics[width=.9\linewidth]{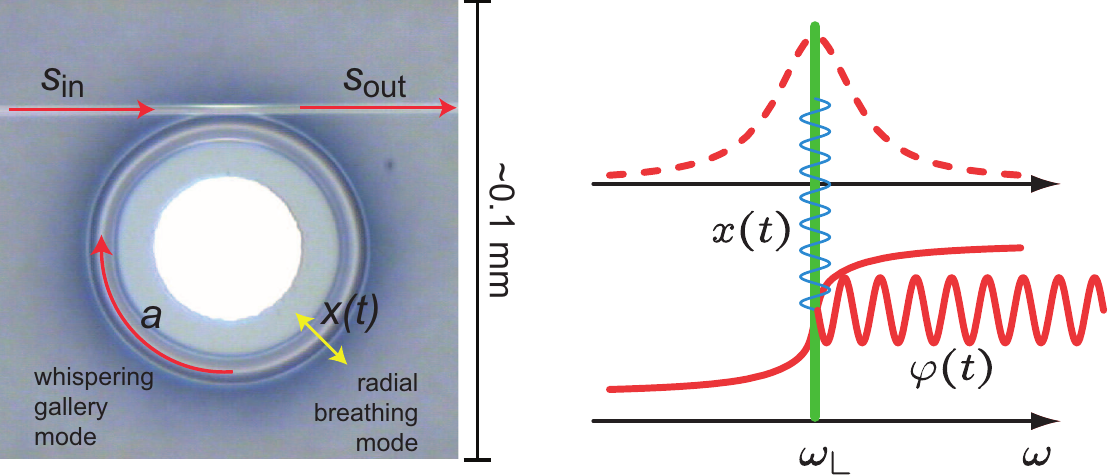}  
\caption{High-sensitivity displacement sensing using optomechanical
coupling. Left panel: an input field $s_\mathrm{in}$ is launched into the
taper and resonantly coupled to the WGM. The properties of the intracavity
field $a$ and the field $s_\mathrm{out}$ coupled back to the fiber taper are
modified by the displacement $x(t)$ of the RBM. Right panel: the launched
field at frequency $\protect\omega_\mathrm{l} $ (green line) acquires a
phase shift that depends on the mutual detuning of laser and cavity
resonance frequency (lower red curve, dashed red curve indicates the WGM
Lorentzian as a reference). If the displacement $x(t)$ modulates the WGM
resonance frequency, the phase $\protect\varphi(t)$ of the emerging field $s_%
\mathrm{out}$ is also modulated. This phase modulation can be detected with
quantum-limited sensitivity by comparison with an optical phase reference. }
\label{f:uhs}
\end{figure}

As a first application of optomechanical coupling present in WGM resonators,
we consider monitoring of  mechanical displacements using the optical degree of
freedom. While in early work the separation of two resonators was measured
in this manner \cite{Ilchenko1994}, we focus here on the measurement of
displacements related to the internal mechanical modes of WGM resonators 
\cite{Schliesser2008b}. The principal idea of such a measurement is
illustrated in figure~\ref{f:uhs}. In the following, the theoretical limits,
possible experimental implementations, and experimental results will be
presented.

\subsection{Theoretical limits of displacement sensing}

\label{ss:sensitivetheory}

We first explore the theoretical limits in the sensitivity of the
measurement, and restrict ourselves to the simple case of resonant probing $%
\bar \Delta =0$ (for a more general discussion, see section \ref{s:cryo}).
In this case, the \emph{dynamical properties} of the mechanical oscillator
are not affected by the presence of the light in the cavity, that is $\Gamma_%
\mathrm{dba}=\Omega_\mathrm{dba}=0$ in equations (\ref{e:Gdba}) and (\ref%
{e:Odba}). Still, the light can be used as a sensitive probe for the
mechanical mode; in the following we discuss the performance and theoretical
limitations of this method.

\subsubsection{Quantum noise}

To analyze the fundamental quantum limits, we directly calculate from
equations (\ref{e:eomx})--(\ref{e:last}) the noise in the light field at
the output 
\begin{equation}
\delta \hat s_\mathrm{out} (\Omega )=\delta \hat s_\mathrm{in} (\Omega )-%
\frac{\sqrt{\eta_\mathrm{c}  \kappa}}{-i\Omega +\kappa/2}\left(-i g_0  \bar
a  \delta \hat x (\Omega )+\frac{\delta \hat s_\mathrm{in} (\Omega )}{\sqrt{%
\tau_\mathrm{ex} }}+\frac{\delta \hat s_\mathrm{vac} (\Omega )}{\sqrt{\tau_0 
}} \right)
\end{equation}
with an analogous equation for $\delta \hat s_\mathrm{out}^{\dagger} (\Omega )$. For
resonant probing, the mean field at the cavity output is real, and therefore
the phase quadrature is directly given by 
\begin{equation}
\delta \hat q_\mathrm{out} (\Omega )=i\left(-\delta \hat s_\mathrm{out}
(\Omega )+\delta \hat s_\mathrm{out}^{\dagger} (\Omega )\right).
\end{equation}
For the symmetrized noise spectral density\footnote{%
Note that double-sided spectral densities are calculated in this chapter.}
of the phase quadrature at the output, defined by 
\begin{align}
  \bar S_{qq}^{\mathrm{out}}(\Og)&\equiv\frac{1}{2}\left(S_{qq}^{\mathrm{out}}(+\Og)+S_{qq}^{\mathrm{out}}(-\Og)\right)
  \intertext{with}
  2 \pi \delta(\Og+\Og')S_{qq}^{\mathrm{out}}(\Og)&=\langle
  \dhqout(\Og) \dhqout^\dagger(\Og')\rangle
  \intertext{we obtain}
  \label{e:sqqresonant}
  \bar S_{qq}^{\mathrm{out}}(\Og)&=1+\frac{4 \ba^2 \dwdx^2 \etac
  \kappa}{\Og^2+(\kappa/2)^2} \bar S_{xx}(\Og)
\end{align}
with the correlation functions from equations (\ref{e:firstcorrelator})-(\ref%
{e:lastcorrelator}). Evidently, the noise spectrum of the phase quadrature
contains information on the mechanical displacement spectrum $\bar S_{xx}$,
but also a background term (in this normalization equal to 1) which is due
to the quantum noise. This background constitutes the fundamental
imprecision of the measurement, and is given by 
\begin{equation}  \label{e:sximp}
S_{xx}^\mathrm{im,qn}(\Omega )=\frac{\Omega ^2+(\kappa/2)^2}{4 \bar a ^2 g_0
^2 \eta_\mathrm{c}  \kappa}
\end{equation}
if expressed as an equivalent displacement noise. In other words, the
imprecision is the \emph{apparent} displacement noise measured in such an
experiment, due to the inevitable quantum noise in the measurement of the
probing light's phase quadrature.

Recast to experimentally more accessible parameters, equation (\ref{e:sximp}%
) determines the smallest possible displacement $\delta x_\mathrm{min}%
(\Omega )$ which can be measured using a WGM resonator \cite{Schliesser2008b},
\begin{equation}  \label{e:dxmin}
\frac{\delta x_\mathrm{min}(\Omega )}{\sqrt {\Delta f}}=\sqrt {\bar S_{xx}^%
\mathrm{im,qn}(\Omega )}=\frac{\lambda}{16 \pi \mathcal{F}  \eta_\mathrm{c}  
\sqrt{P_\mathrm{in}/\hbar \omega_\mathrm{l} }} \sqrt{1+\left(\frac{\Omega }{%
\kappa/2}\right)^2},
\end{equation}
where $\lambda$ is the wavelength in the medium and $\Delta f$ the
measurement bandwidth. Written this way, the importance of high-finesse
cavities is directly evident: The smallest displacement that can be measured
is roughly given by the wavelength, divided by the cavity finesse and the
square root of the number of photons accumulated in the measurement time
(inverse bandwidth). The term $\sqrt{ 1+ 4\Omega ^2/\kappa^2}$ is due to a
less efficient transduction of the motion into phase shift for Fourier
frequencies beyond the cavity cutoff (which can, in principle, be avoided
using multiple cavity modes \cite{Dobrindt2009}).

As an aside, we remark that the same result can be obtained by considering
the classical transduction of a displacement into the phase of the light
exiting the cavity and comparing the result with the shot noise in the
measurement process of the light phase \cite{Schliesser2008}. Importantly,
equations (\ref{e:sximp}) and (\ref{e:dxmin}) are independent of the particular
strategy used to detect the light's phase, as long as it can be achieved in
a quantum-limited manner. Two strategies to accomplish this, namely homodyne
and polarization spectroscopy, are described in section \ref{s:sensitive}.

As first discussed by Braginsky \cite{Braginsky1992}, it is inevitable that
the measurement of the oscillator's position disturbs it (``measurement
backaction''). In the case of an optomechanical system, this is due to the
fluctuations of intracavity radiation pressure \cite{Caves1980}, which can
be written as 
\begin{equation}
\delta\! \hat F_\mathrm{rp} (\Omega )=-\hbar g_0  \bar a  \left(\delta \hat
a (\Omega )+\delta \hat a^{\dagger} (\Omega )\right)
\end{equation}
in equation (\ref{e:eomx}). Again from the known correlation functions, we obtain here 
\begin{equation}  \label{e:sfba}
\bar S_{FF}^\mathrm{ba,qn}(\Omega )=\frac{\bar a ^2 g_0 ^2 \kappa \hbar^2}{%
\Omega ^2+(\kappa/2)^2},
\end{equation}
if the input noise is again only quantum noise. In this case, the force
noise (\ref{e:sfba}) is referred to as \emph{quantum backaction}. Evidently,
equations (\ref{e:sximp}) and (\ref{e:sfba}) fulfill the
quantum-mechanically required inequality of the imprecision-back\-action
product \cite{Braginsky1992} 
\begin{equation}
\bar S_{xx}^\mathrm{im,qn}(\Omega )\cdot \bar S_{FF}^{\mathrm{ba}}(\Omega )=%
\frac{\hbar^2}{4 \eta_\mathrm{c} } \geq \frac{\hbar^2}{4}.
\end{equation}

\begin{figure}[tb]
\centering
\includegraphics[width=.6 \linewidth]{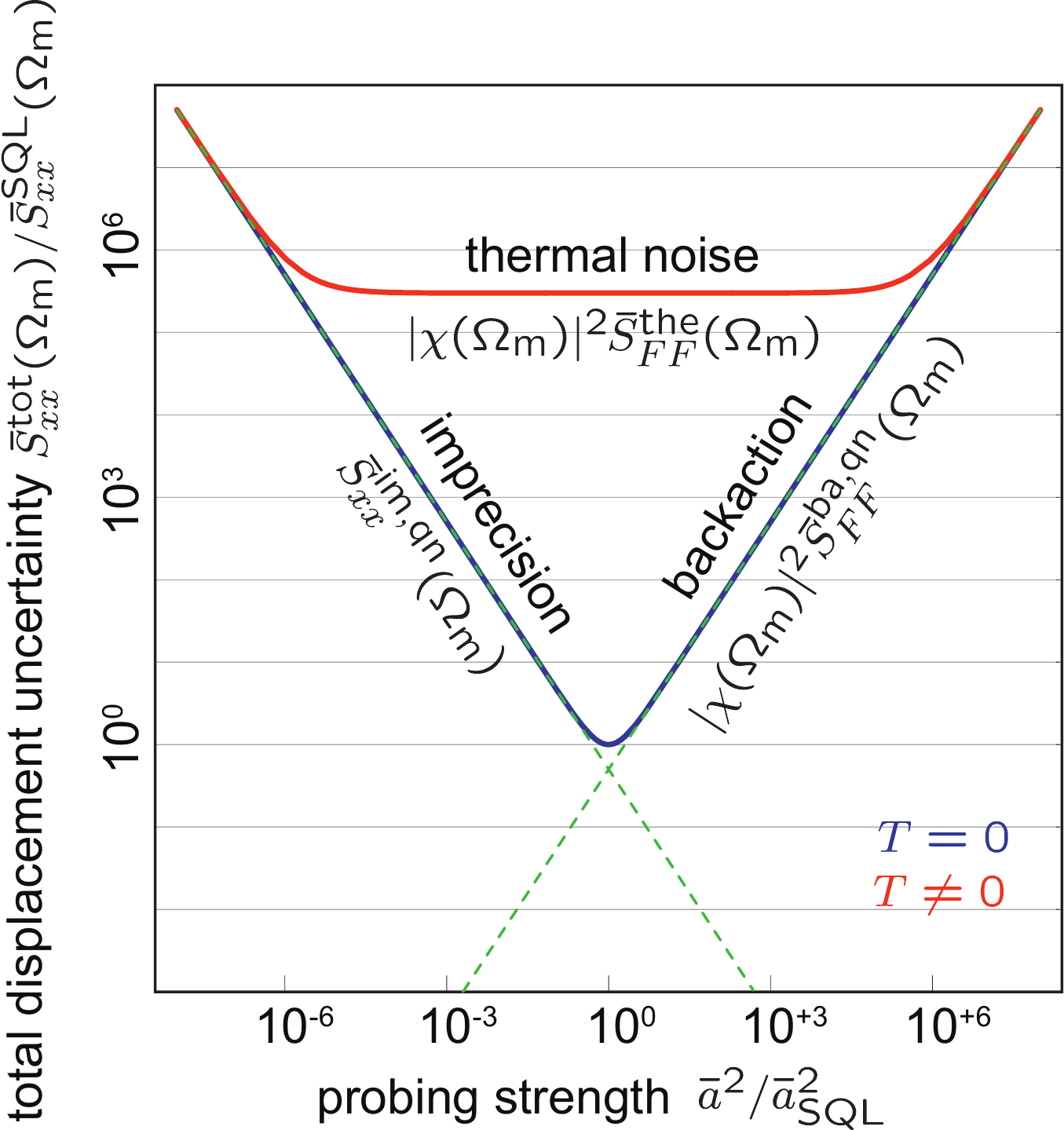} 
\caption{ Quantum limits in the measurement of mechanical displacements
(blue line). For weak probing $\bar a ^2<\bar a ^2_\mathrm{SQL}$,
measurement imprecision dominates the total uncertainty, while for stronger
probing $\bar a ^2>\bar a ^2_\mathrm{SQL}$, the noise in the mechanical
oscillator induced by quantum backaction dominates the uncertainty. For
optimum measurements with $\bar a ^2=\bar a ^2_\mathrm{SQL}$, the
uncertainty is at the standard quantum limit $\bar S_{xx}^\mathrm{SQL}%
(\Omega_\mathrm{m} )=\hbar/m_\text{eff} \Gamma_\mathrm{m}  \Omega_\mathrm{m} 
$. Under laboratory conditions, thermal noise is additionally present (red
line). }
\label{f:sql}
\end{figure}

By causing \emph{additional} displacement fluctuations in the mechanical
oscillator, backaction noise also impedes the determination of the
oscillator's displacement. The total uncertainty in the measurement is
therefore given by 
\begin{equation}
\bar S_{xx}^{\mathrm{tot}}(\Omega )= \bar S_{xx}^\mathrm{im,qn}(\Omega )+ |
\chi(\Omega )|^2 \bar S_{FF}^\mathrm{ba,qn}(\Omega ),
\end{equation}
where 
\begin{equation}  \label{e:chidef}
\chi(\Omega )=\frac{1}{m_\text{eff}(\Omega_\mathrm{m} ^2-\Omega ^2-i \Omega
\Gamma_\mathrm{m} )}
\end{equation}
is the susceptibility of the mechanical oscillator. Obviously, a tradeoff in
terms of the ``strength'' $\propto g_0 ^2 \bar a ^2$ of the measurement has
to be made, as imprecision reduces, but backaction increases for
``stronger'' measurements \cite{Caves1980,Caves1981}, as illustrated in
figure~\ref{f:sql}. Optimum measurement conditions are reached for 
\begin{equation}
\bar a ^2=\bar a ^2_\mathrm{SQL}=\frac{m_\text{eff} \Gamma_\mathrm{m} 
\Omega_\mathrm{m} }{2 g_0 ^2 \hbar \kappa \sqrt{\eta_\mathrm{c} }} (\Omega_%
\mathrm{m} ^2+(\kappa/2)^2),
\end{equation}
or, equivalently an input power of $P_\mathrm{SQL}=\hbar \omega  \kappa \bar
a ^2_\mathrm{SQL}/4 \eta_\mathrm{c} $. In this case, one obtains a total
uncertainty of \cite{Schliesser2008b} 
\begin{equation}
\bar S_{xx}^\mathrm{SQL}(\Omega ) = \frac{ \hbar\left|\chi(\Omega )\right|}{%
\sqrt{\eta_\mathrm{c}}}= \frac{\hbar}{m_\text{eff} \sqrt{\eta_\mathrm{c}
((\Omega_\mathrm{m} ^2-\Omega ^2)^2+\Gamma_\mathrm{m} ^2\Omega ^2)}},
\label{e:SQL}
\end{equation}
called the \textit{standard quantum limit} \cite{Braginsky1992, Caves1980}
in the case $\eta_\mathrm{c}=1$. Its peak value is calculated at $\Omega_%
\mathrm{m} $, 
\begin{equation}
\bar S_{xx}^\mathrm{SQL}(\Omega_\mathrm{m} )=\frac{1}{\sqrt{\eta_\mathrm{c}}}%
\frac{\hbar}{ m_\text{eff} \Gamma_\mathrm{m}  \Omega_\mathrm{m} }.
\end{equation}

In this calculation we have explicitly considered the effect of the coupling
conditions to the cavity, which can---as a unique feature---be varied
continuously in the experiment by adjusting the gap between the coupling
waveguide and the WGM resonator. The SQL is approached most closely in the
overcoupled limit $\tau_\mathrm{ex} \ll\tau_0 $. It is noteworthy that the
fibre-taper coupling technique to microtoroids can deeply enter this regime,
and $100\cdot \tau_\mathrm{ex} <\tau_0 $ ($\eta_\mathrm{c}=99\%$) has been
demonstrated \cite{Spillane2003}.

\subsubsection{Laser technical noise}

The previous derivation deals with the fundamental sensitivity limits. A
frequent \emph{technical} limitation is due to excess (beyond the
fundamental) noise of the laser used for probing. %
Frequency noise in the laser, characterized by a power spectral density $%
\bar S_{\omega \omega }(\Omega )$, corresponds to a higher level of
fluctuations in the input phase quadrature, 
\begin{equation}
\bar S_{qq}^\mathrm{in}(\Omega )=1+\frac{4 |\bar s_\mathrm{in} |^2}{\Omega ^2%
} \bar S_{\omega \omega }(\Omega ),
\end{equation}
raising the background on top of which the displacement spectrum has to be
observed. Note also that most schemes to measure the phase of the light rely
on a phase reference. If this reference is noisy, because it is derived from
the same noisy laser, the imprecision in the displacement measurement is
given by 
\begin{align}
\bar S_{xx}^{\mathrm{im,fn}}(\Omega )&=\frac{\bar S_{\omega \omega }(\Omega )%
}{g_0 ^2}  \label{e:fn}
\end{align}
if the frequency noise overwhelms quantum noise in the measurement.

\subsubsection{Thermorefractive noise}

Another important source of noise potentially preventing the measurement of
mechanical displacements with quantum-limited sensitivity is  fluctuation
of the resonance frequency of the WGM \cite{Gorodetsky2004, Matsko2007}
which are not related to the mechanical oscillators. For a dielectric
resonator as silica microspheres or -toroids, the dominant effect to be
considered here are fluctuations of the refractive index due to temperature
fluctuations. At any finite mean temperature $\bar T$, the actual average
temperature $T_V$ in a volume $V$ fluctuates according to \cite{Landau1980} 
\begin{equation}
\left\langle( T_V-\bar T)^2 \right\rangle=\frac{k_\mathrm{B} \bar T^2}{c_p V
\rho},
\end{equation}
where $\rho$ is the material density, and $c_p$ the specific heat capacity.
This applies in particular also to the mode volume, within which the WGM
samples the temperature-dependent refractive index.

To calculate the frequency spectrum of the resulting fluctuations,
Gorodetsky and Grudinin \cite{Gorodetsky2004} have used a Langevin approach,
introducing fluctuational driving terms into the dynamic equations of
temperature diffusion \cite{Braginsky1999,Braginsky2000}. As a result, the
imprecision due to thermorefractive noise in a silica WGM resonator can be
estimated to amount to 
\begin{align}
\bar S_{xx}^\mathrm{im,tr}(\Omega )&=R^2 \bar S_{\delta n/n}(\Omega)\approx 
\notag \\
&\approx\frac{k_\mathrm{B} T^2 k R}{\pi^{5/2} n^2 \rho^2 c_\mathrm{p}^2 } 
\frac{1}{\sqrt{d^2-b^2}}\left(\frac{d n}{d T}\right)^2 \int_0^{+\infty} 
\frac{q^2 e^{-\frac{q^2 b^2}{2}}}{D^2 q^4+\Omega ^2} \frac{dq}{2\pi},
\label{e:trn}
\end{align}
where $k$ is heat conductivity, $R$ the cavity radius, and $d$ and $b$ the
transverse mode dimensions. At low frequency ($\lesssim 10 \,\mathrm{MHz} $%
), this noise indeed can dominate the measurement imprecision (cf.\ section %
\ref{ss:noises}).

\subsection{Experimental techniques}

\label{s:sensitive}

\subsubsection{Homodyne spectroscopy}

\label{ss:homodyne}

\begin{figure}[tbp]
\centering
\includegraphics[width=\linewidth]{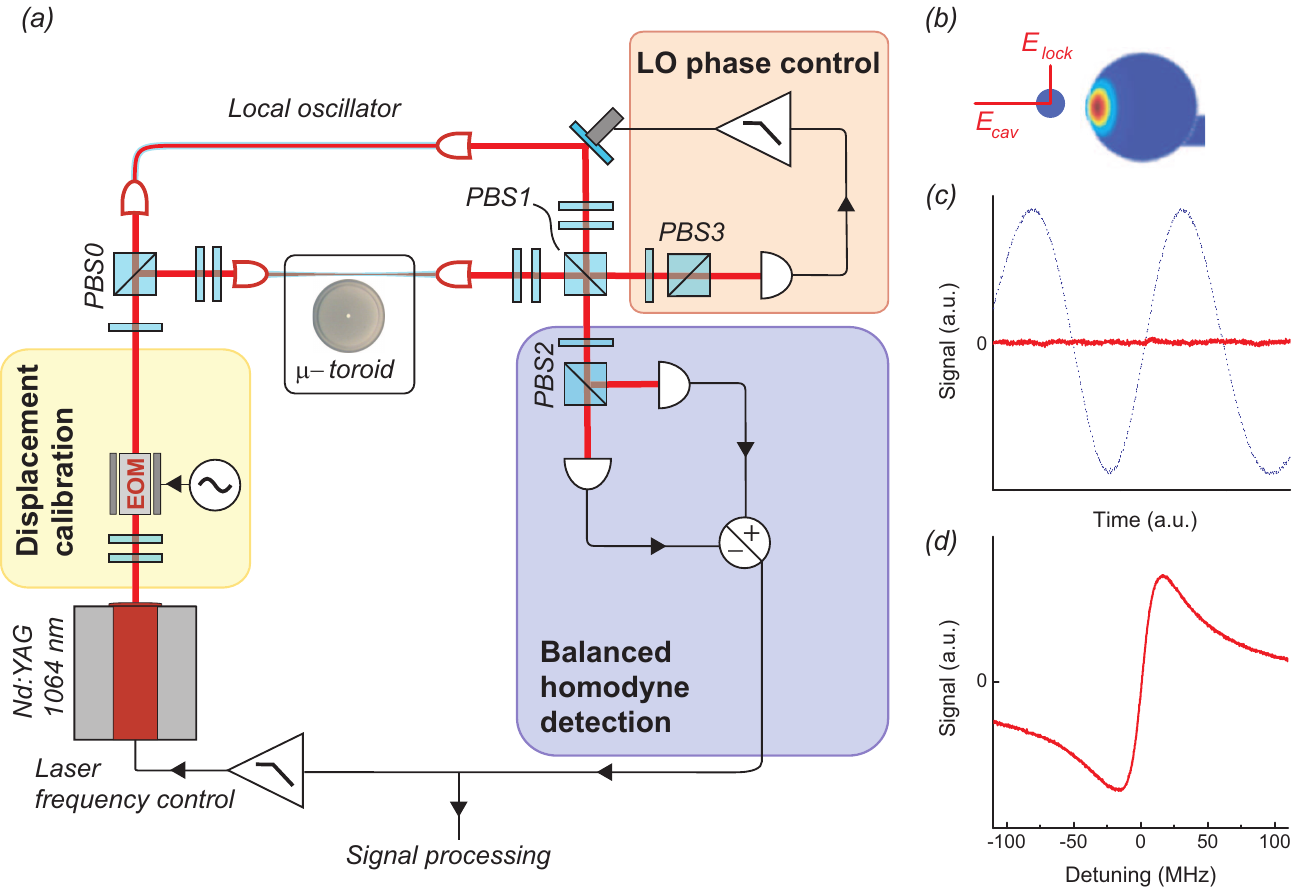} 
\caption{(a) Optical interferometric displacement transducer based on
homodyne spectroscopy of light transmitted past the cavity (``$\protect\mu$%
-toroid''). The phase of the local oscillator is actively stabilized (``LO
phase control''). Details are given in the text. PBS0-PBS3, polarizing beam
splitters. (b) Cross section through the fiber taper and the toroidal rim in
the coupling region. The polarization in the taper is slightly mismatched
with the polarization of the cavity mode. Thus only the part $E_\mathrm{cav}$
of the total field couples to the WGM, the other component $E_\mathrm{lock}$
can be used for the stabilization of the local oscillator phase. The
components $E_\mathrm{cav}$ and $E_\mathrm{lock}$ are separated in PBS1. (c)
Signal in the balanced receiver for a scanning local oscillator (dotted,
blue) at low power, and for the locked LO (red). The shown locked trace was
recorded for about 5 seconds. (d) Typical experimental error signal in the
balanced receiver when the laser is scanned over a cavity resonance with the
local oscillator locked to the appropriate phase. Figure from ref. 
\protect\cite{Schliesser2008b}. }
\label{f:homodynesetup}
\end{figure}

A commonly employed technique for quantum-limited phase measurement is a
balanced homodyne receiver \cite{Yuen1983}, which has been used in earlier
optomechanical experiments \cite{Hadjar1999, Briant2003b, Caniard2007a}. One
possible adaptation of this technique to the ring topology of a WGM
resonator is shown in figure~\ref{f:homodynesetup}. The probing (or signal)
beam and a phase reference beam, referred to as the local oscillator (LO)
are derived from the same laser, in this case a monolithic Nd:YAG laser
operating at $\lambda=1064\,\mathrm{nm} $. This source exhibits
quantum-limited amplitude and phase noise at Fourier frequencies $%
\Omega/2\pi \gtrsim 5 \,\mathrm{MHz} $ and power levels $P_\mathrm{LO}+P
\lesssim 5 \,\mathrm{mW} $ of interest. As tuning speed and range of this
laser are limited, it was found advantageous to use a home-built
external-cavity diode laser for pre-characterization of several samples
until a suitable toroid was found. The probing beam is sent through the
coupling taper and interacts with the WGM of the studied toroid. The LO
travels in the reference arm of a Mach-Zehnder interferometer over the same
distance. It is finally recombined with the signal beam at a polarizing beam
splitter (PBS1). Spatial mode matching of the incident beams is enhanced by
using single-mode fiber as mode filter on the local oscillator. After
spatial recombination, interference is enforced using a retarder plate and
another polarizing beam splitter (PBS2).

As the relative phase of the two interfering beams is also subject to drifts
and fluctuations, due to, for example, temperature drift of the fiber in
which the reference beam propagates, active stabilization is necessary here.
In one possible implementation, this is accomplished by purposely
introducing a small polarization mismatch between the light in the taper
region and the either predominantly TE- or TM-like WGM modes of the
microcavity. The polarization component of the probing beam which does not
interact with the WGM resonance can then be used to stabilize the phase of
the LO (see figure~\ref{f:homodynesetup}). In this case, a dispersive signal, 
\begin{equation}
h(\Delta)=\frac{2 \eta_\mathrm{c} \kappa \, \Delta}{\Delta^2+(\kappa/2)^2} 
\sqrt{P_\mathrm{cav} P_\mathrm{LO}},  \label{e:homodyne}
\end{equation}
is obtained at the output of the balanced receiver comparing the other
polarization component (which is coupled to the WGM) with the LO. Here $P_%
\mathrm{cav}$ and $P_\mathrm{LO}$ are the powers of the probing and local
oscillator beams and $h(\Delta)$ is the power difference measured between
the two employed receivers. Figure~\ref{f:homodynesetup} shows an example of
an experimental trace obtained when scanning the laser detuning. Evidently,
due to its dispersive shape, the d.c.-component of this signal can be used
to lock the laser frequency to the WGM resonance frequency using electronic
feedback.

In this way, the mean detuning can be stabilized to $\bar \Delta =0$ and
with $\Delta=\bar \Delta -g_0  \delta \hat x (\Omega )$ the signal is
directly given by 
\begin{equation}
h_{\bar \Delta =0}(\delta x,\Omega )\approx -\frac{8 \eta_\mathrm{c}  g_0 
\delta x(\Omega )}{\kappa}\sqrt{\frac{P_\mathrm{cav} P_\mathrm{LO}}{%
1+(\Omega /(\kappa/2))^2}},  \label{e:slope}
\end{equation}
where the reduced signal strength for Fourier frequencies $\Omega $ beyond
the cavity cutoff can be viewed as a consequence of the reduced buildup of
intracavity sidebands (\ref{e:sidebands})\footnote{%
A more detailed calculation is presented in the supplementary information of
ref.\ \cite{Schliesser2008}.}. For a strong local oscillator $P_\mathrm{LO}%
\gg P_\mathrm{cav}$, the detection noise is given by shot noise caused by
the local oscillator beam, and the fluctuations in the detected differential
power are simply $\delta h\approx\sqrt{P_\mathrm{LO}\, \hbar \omega_\mathrm{l%
} }$. Comparison with the signal (\ref{e:slope}) induced by displacements $%
\delta x$ then directly give the sensitivity derived in equation (\ref%
{e:sximp}). %

Furthermore, as equation (\ref{e:homodyne}) evidently only depends on the
mutual detuning of laser and WGM resonance, frequency fluctuations of the
laser are indistinguishable from fluctuations due to mechanical
displacement. In the case of frequency noise of the laser, this leads
directly to the imprecision described by equation (\ref{e:fn}). On the other
hand, an intentional frequency modulation of controlled modulation depth of
the laser can be utilized to calibrate the measured signals: a frequency
modulation of $\delta \omega $ corresponds to a displacement of amplitude $%
g_0  \delta x$, independent of the detuning and coupling conditions\footnote{%
We emphasize that for this relation to be valid for arbitrary modulation
frequencies, it is necessary that the lengths of the two arms of the
Mach-Zehnder interferometer are equal.} \cite{Tittonen1999, Hadjar1999,
Gorodetsky2004, Schliesser2008}. With the calibration at one particular
modulation frequency $\Omega _\mathrm{mod}$, the measured spectra can be
absolutely calibrated at all Fourier frequencies, taking into account the
reduced sensitivity beyond the cavity cutoff at $\kappa/2$.

\subsubsection{Polarization spectroscopy (H\"ansch-Couillaud method)}

\label{ss:HC}

A simplified setup may be obtained by copropagating the local oscillator field in
the same spatial, but orthogonal polarization mode as compared to the signal
beam \cite{Schliesser2008}. Since the WGM modes have predominantly TE or TM
character and are not degenerate, this guarantees that the local oscillator
is not affected by the cavity. Due to common-mode rejection of many sources
of noise in the relative phase between signal and LO (for example, frequency
noise in the optical fiber), the passive stability is sufficiently enhanced
to enable operation without active stabilization (figure~\ref%
{f:haenschcouillaud}).

\begin{figure}[tbp]
\centering
\includegraphics[width=\linewidth]{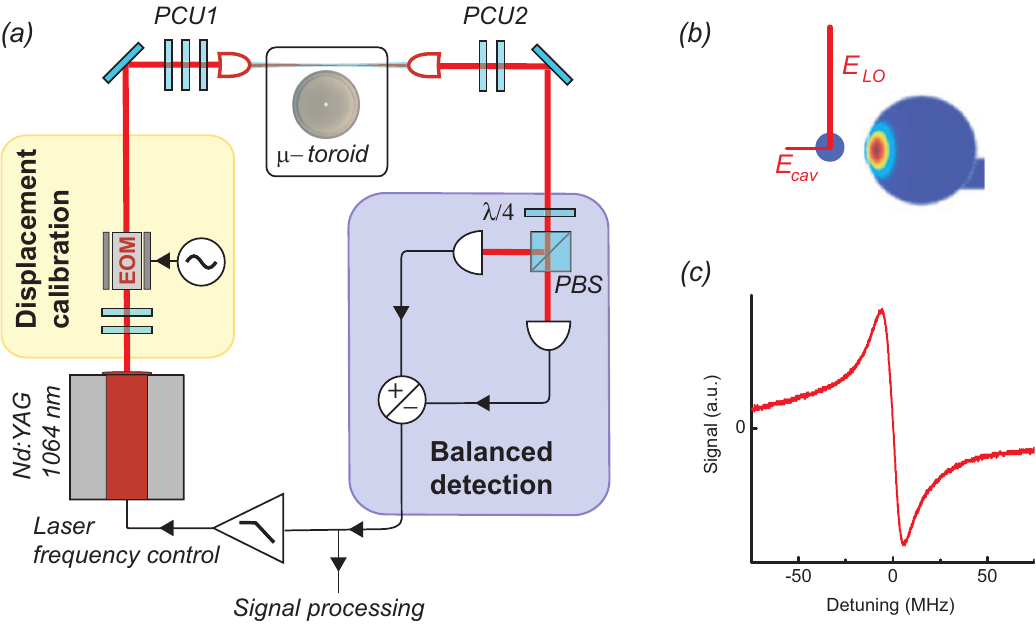}  
\caption{Optical interferometric displacement transducer based on
polarization spectroscopy of light transmitted in the taper past the cavity
(``$\protect\mu$-toroid''). (a) After phase modulation with an electro-optic
modulator, the polarization is prepared with a first polarization control
unit (PCU1). The cavity WGM defines signal and LO polarizations. A second
polarization control unit (PCU2) compensates for fibre birefringence.
Polarization analysis using a $\protect\lambda/4$ plate and a polarizing
beam splitter enforces interference of the signal and LO fields. (b) Due to
the polarization non-degeneracy of the WGM in the cavity, only one
polarization component of the light interacts with the mode. (c) Typical
error signal obtained when the laser is scanned over a cavity resonance.
Figure from ref.\ \protect\cite{Schliesser2008b}. }
\label{f:haenschcouillaud}
\end{figure}

Enforcing interference between local oscillator and signal beams then
corresponds to polarization analysis of the light (comprising both signal
and LO) emerging from the cavity. While novel in the present context of a
tapered fibre coupled microcavity, this is a well established technique to
derive a dispersive error signal from the light reflected from a Fabry-Perot
type reference cavity, named after their inventors H{\"a}nsch and Couillaud 
\cite{Hansch1980}.

If fiber birefringence is adequately compensated, the error signal is 
\begin{equation}
h(\Delta)=\frac{2 \eta_\mathrm{c} \kappa \, \Delta}{\Delta^2+(\kappa/2)^2} 
\sqrt{P_\mathrm{cav} P_\mathrm{LO}},
\end{equation}
identical to (\ref{e:homodyne}), and a typical trace is shown in figure~\ref%
{f:haenschcouillaud}(c). This is used to lock the laser at resonance $\bar
\Delta = 0$ with a bandwidth of about $10 \,\mathrm{kHz} $. Calibration of
the spectra may be performed as described in the previous section.

While this approach obviously allows one to reduce the complexity of the
experiment, this arrangement proved more sensitive to slow temperature
drifts in the polarization mode dispersion of the fibers employed, due to
the large ratio of signal and LO powers, the magnitudes of which are only
defined by the polarization state of the light in the fiber taper region.
Improved stability may be obtained by reducing the fiber length to its minimum
of value of aproximately $0.5 \,\mathrm{m} $. For reasons of flexibility and convenience, the
actual fiber length totaled to several meters in our experiment.
Nonetheless, sensitivities of $10^{-18}\,\mathrm{m} /\sqrt{\mathrm{Hz}}$ are
achieved in toroids using this method \cite{Schliesser2008}. The intrinsic
polarization selectivity of WGM renders the introduction of an additional
polarizer, mandatory in the original implementation \cite{Hansch1980},
obsolete. In an earlier experiment with a Fabry-Perot\ cavity \cite%
{Hahtela2004}, the losses associated with an intracavity polarization
element limited the finesse, and therefore the attained sensitivity to $\sim
10^{-14}\,\mathrm{m} /\sqrt{\mathrm{Hz}}$.

\subsubsection{Frequency modulation spectroscopy (Pound-Drever-Hall method)}

\label{sss:PDH}

\begin{figure}[btp]
\centering
\includegraphics[width=.7\linewidth]{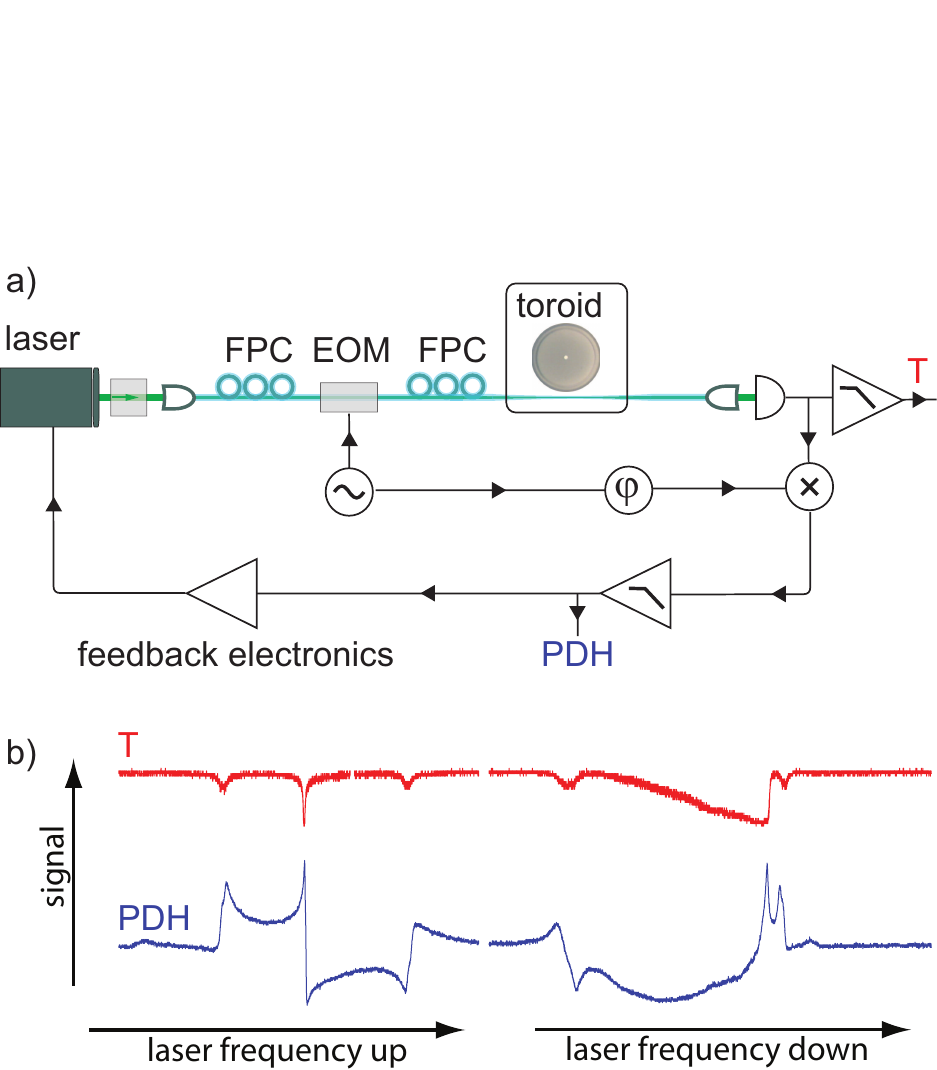}  
\caption{Displacement measurement using the Pound-Drever-Hall method 
\protect\cite{Drever1983}. a) The phase of the probing laser is
phase-modulated at a radio-frequency of typically 50--100 MHz using a
fiber-coupled electro-optic modulator (EOM). After interaction with the WGM,
the detected signal is demodulated at the same frequency. The demodulation
quadrature is chosen by adjusting the phase $\protect\varphi$ of the
radio-frequency wave. In the simplest case, this can be accomplished by
adjusting the length of the cable carrying the signal. The demodulated
signal is low-pass filtered at a bandwidth well below the modulation signal.
The resulting signal ``PDH'' can be used to monitor the detuning of laser
and WGM resonance, and is also suited to electronically stabilize the laser
frequency to the WGM. In addition, the transmission ``T'' of the WGM can be
directly monitored. FPC, fiber polarization controller. b) Typical traces of
transmission and PDH signals obtained with a silica toroidal WGM upon a
laser frequency scan in the presence of thermal bistability. The satellite
dips in the transmission signal appear when the modulation sidebands are
scanned over the WGM frequency.}
\label{f:pdh}
\end{figure}

Another possible method to experimentally determine the detuning of laser
and WGM resonance is frequency modulation spectroscopy as introduced by
Pound, Drever and Hall \cite{Drever1983} and discussed in great detail by
Black \cite{Black2001}. Figure~\ref{f:pdh} shows a possible application of
this scheme to WGM resonators. The dispersive shape of the signal can be
used to lock the laser to the WGM resonance using electronic feedback.
Fluctuations of the Pound-Drever-Hall (PDH) signal beyond the feedback
bandwidth then indicate fluctuations of the WGM resonance frequency with
respect to the laser frequency, and can therefore be used to monitor
displacements.

While both homodyne and polarization spectroscopy can attain the fundamental
quantum-limited displacement sensitivity (\ref{e:sximp}), the sensitivity of
the PDH method is reduced by a factor $1+\eta_\mathrm{c} +(1-2\eta_\mathrm{c}
)^2/2 J_1^2(\beta)$, where $\beta$ is the phase modulation depth and $J_1$ a
Bessel function of the first kind \cite{Arcizet2007}. Note that even for a
maximally overcoupled cavity with $\eta_\mathrm{c} \rightarrow 1$ this
yields $2+1/2 J_1^2(\beta)$. In practice, the displacement sensitivity using
this method is often limited by electronic noise in the detector. As
essentially all light used in this scheme interacts with the WGM (there is
no phase reference beam in a different spatial or polarization mode), the
total power levels must be kept low in order to prevent strong thermal
nonlinearities. These light levels (typically a few microwatts) are not
sufficient to overwhelm the electronic noise of broadband light receivers.
If available at the particular wavelength of interest, a low-noise optical
amplifier such as an erbium-doped fiber amplifier (EDFA) can however
ameliorate this drawback \cite{Arcizet2009a} at the expense of a higher
noise figure.

\subsection{Observation and analysis of quantum and thermal noises}

\label{ss:noises}

\begin{figure}[tbp]
\centering
\includegraphics[width=.8\linewidth]{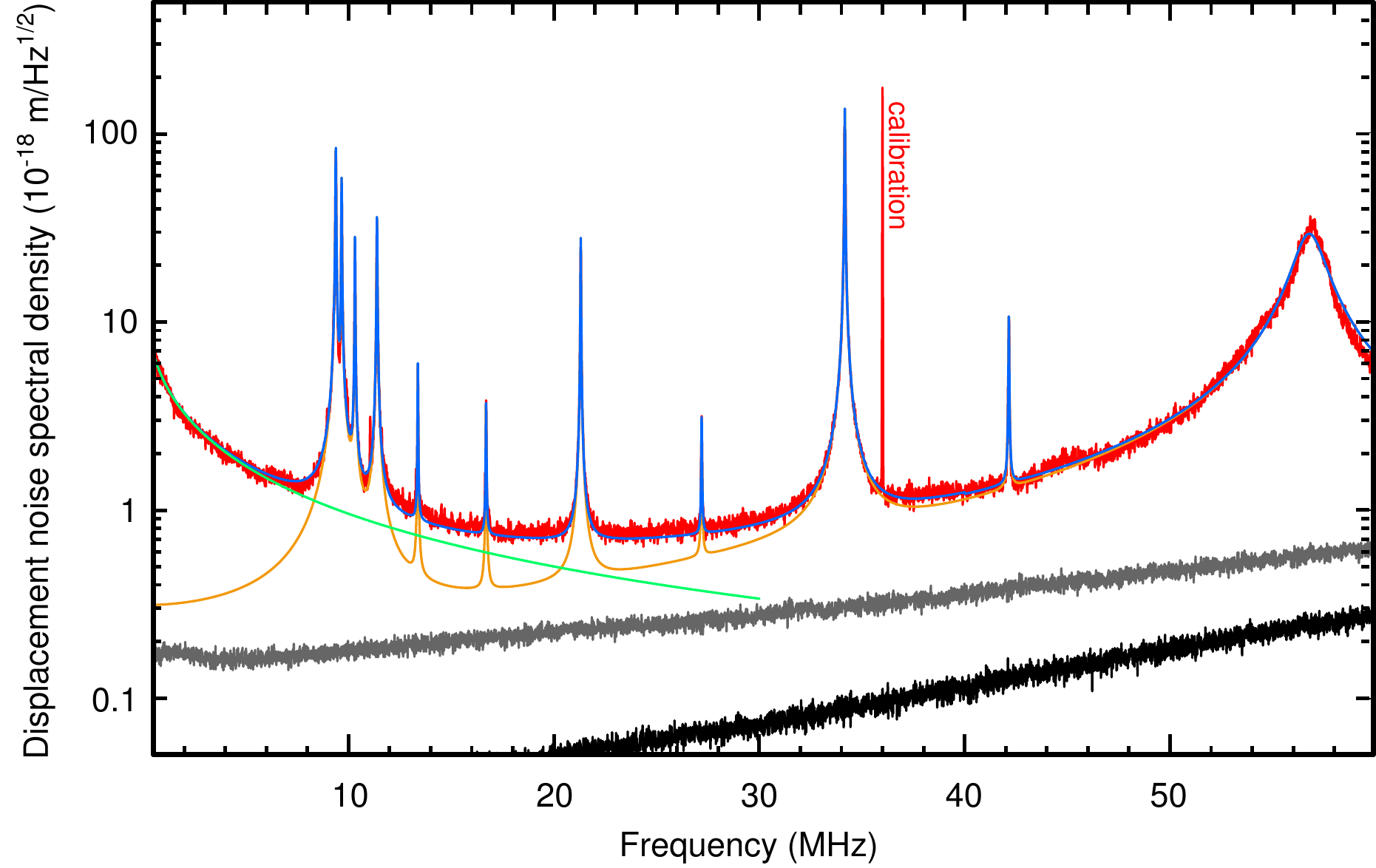}  
\caption{Equivalent displacement noise measured in a $\sim 90\,\mathrm{%
\protect\mu m}  $-diameter silica toroidal cavity. Red, measured trace with
laser coupled to a cavity resonance, including a peak at 36 $\,\mathrm{MHz}$
due to phase modulation for calibration purposes. Gray, measured shot noise
with taper retracted from the cavity and black, detector noise. Models for
mechanical noise (orange line) and thermorefractive noise (green line), and
sum of the two models plus the shot noise background (blue line) are also
shown. Figure from ref.\ \protect\cite{Schliesser2008b}.}
\label{f:broadbandcutoff}
\end{figure}

Figure~\ref{f:broadbandcutoff} shows data obtained using homodyne
spectroscopy on a toroid of about $45 \,\mathrm{\mu m}  $ radius. As long as
the taper is retracted from the WGM evanescent field, quantum shot noise is
observed to exceed the electronic detector noise. Note that while the
detected shot noise (due to the local oscillator) is spectrally flat (white
noise) to a good approximation, the equivalent displacement noise plotted in
figure~\ref{f:broadbandcutoff} exhibits a calculated $\sqrt{%
1+\Omega^2/(\kappa/2)^2}$ frequency dependence beyond the cavity cutoff at $%
\kappa/2\approx2\pi \cdot 17 \,\mathrm{MHz} $ due to the Fourier
frequency-dependent transduction (\ref{e:slope}).

Approaching the taper to the resonator, and locking the laser to the WGM
resonance, a substantially different spectrum is observed (figure~\ref%
{f:broadbandcutoff}). The equivalent displacement noise is calibrated in
absolute terms using an \textit{a priori} known phase modulation at $36 \,%
\mathrm{MHz} $, as explained in subsection \ref{ss:homodyne}, taking also
the cavity cutoff into account. While the background due to quantum
measurement imprecision is at a level of $10^{-19}\,\mathrm{m} /\sqrt{%
\mathrm{Hz}}$ at low Fourier frequency, a significantly higher equivalent
displacement noise level is observed when coupling the laser to the WGM.

The broad background particularly strong at low frequency can be
quantitatively reproduced by the model for thermorefractive noise (equation (%
\ref{e:trn})), when no parameters except $b$ and the absolute magnitude are
adjusted by factors of order 2. This is justified considering the
approximations made in the derivation, and the incomplete knowledge on the
transverse mode shape of the WGM probed in this experiment. We note here as
an aside that thermorefractive noise, measured here for the first time in a
toroidal microresonator, is an important limitation for the generation of
Kerr squeezing in these devices \cite{Rehbein2005}.

On top of these backgrounds, a sparse spectrum of peaks is observed, which
are due to different mechanical modes in the spectrum. In this measurement
at room temperature, the thermal Langevin force largely dominates over
radiation pressure force fluctuations. Therefore, each individual mode is
driven by a random thermal force according to equation (\ref{e:xn}), and the
thermal displacement noise spectra of the individual modes add up to the
total measured equivalent displacement noise 
\begin{equation}
\bar S_{xx}^\mathrm{tot}(\Omega )\approx\bar S_{xx}^\mathrm{im,qn}(\Omega
)+\bar S_{xx}^\mathrm{im,tr}(\Omega )+\sum_n |\chi_n(\Omega )|^2 \bar
S_{FF}^{\mathrm{th},n}(\Omega ),
\end{equation}
where the symmetrized spectrum of the Langevin force is given, from equation
(\ref{e:thermalforcefourier}), by 
\begin{equation}  \label{e:sff}
\bar S_{FF}^{\mathrm{th},n}(\Omega ) = \hbar\, m_{\mathrm{eff},n} \,\Gamma_n
\Omega \,\coth\left(\frac{\hbar \Omega }{2 k_\mathrm{B}  T}\right) \approx 2
m_{\mathrm{eff},n}\, \Gamma_n \,k_\mathrm{B}  T,
\end{equation}
where the second relation is valid as long as $k_\mathrm{B} T \gg \hbar
\Omega $ for the frequencies of interest.

\begin{figure}[btp]
\centering
\includegraphics[width=\linewidth]{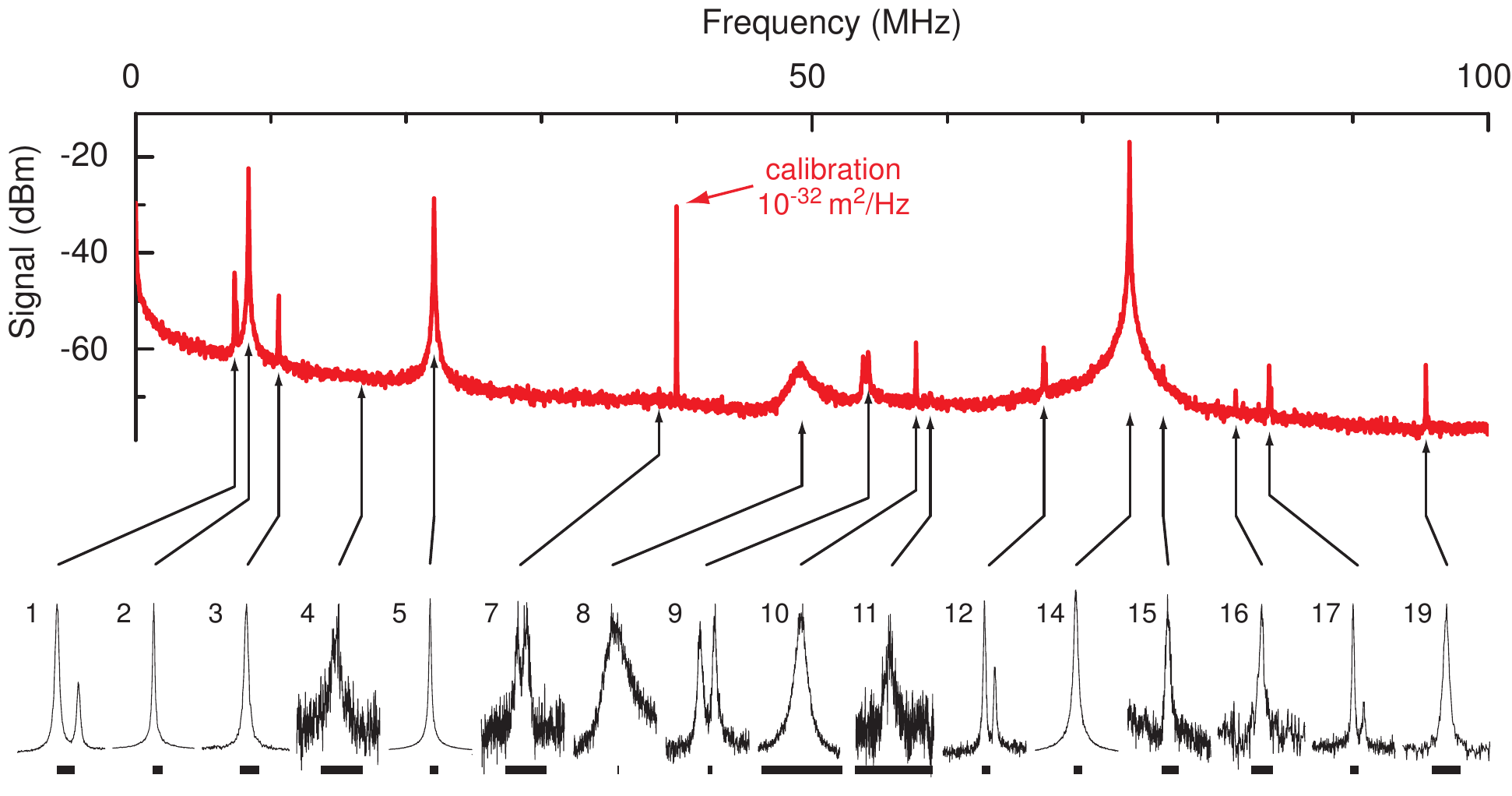} 
\caption{Broadband displacement noise spectrum recorded using the
H\"ansch-Couillaud technique (top panel). Zooming in on the individual peaks
(lower panels) reveals the precise frequency and linewidth of the modes
(scale bar corresponds to a $100\,\mathrm{kHz} $ frequency span), some of
which are split due to a lifted degeneracy. All observed peaks could be
attributed to mechanical modes using finite element modeling (section 
\protect\ref{sss:acousticModes}). Figure from ref.\ \protect\cite%
{Schliesser2008b}. }
\label{f:modes}
\end{figure}

Figure~\ref{f:modes} shows another example of a highly sensitive
measurement, using the H\"ansch-Couillaud technique in this case. Beyond the
clear signatures of the RBM at around $73\,\mathrm{MHz} $, fifteen other
peaks related to mechanical modes are observed. Zooming in on the individual
peaks reveals that some are split, typically in modes whose
degeneracy is lifted by residual asymmetry of the sample. The mode
frequencies can be reproduced very accurately using finite element modeling;
the peaks shown in this figure correspond to the modes discussed in section %
\ref{sss:acousticModes} (cf.\ also figures \ref{f:toroidModes} and \ref%
{f:comparison}).

\begin{figure}[tbh]
\centering
\includegraphics[width=.7\linewidth]{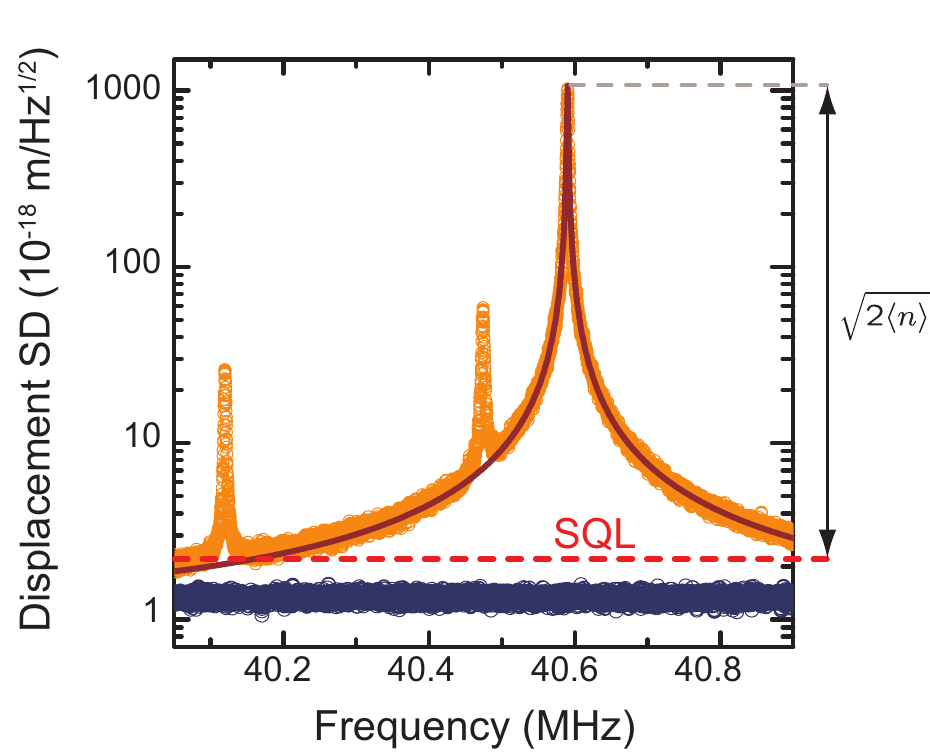} 
\caption{High sensitivity-measurement of the RBM of a larger sample using
the H{\"a}nsch-Couillaud technique. Orange circles represent measured data,
revealing also the signatures of two neighboring modes. The red line is a
Lorentzian fit. Blue circles are the recorded measurement imprecision due to
detection shot noise. An excellent signal-to-background ratio of nearly $60\,%
\mathrm{dB} $ is attained, corresponding to a measurement imprecision of $1{.%
}1 \,\mathrm{am/\protect\sqrt{Hz}} $, which is well below the standard
quantum limit at $2{.}2 \,\mathrm{am/\protect\sqrt{Hz}} $ . }
\label{f:highsens}
\end{figure}

Finally, in figure~\ref{f:highsens}, we zoom in on a frequency interval that
has the signature of a radial breathing mode (RBM) of a
larger sample ($R=38\,\mathrm{\mu m}  $). The measurement achieves
a signal-to-background ratio of nearly $60\,\mathrm{dB} $ determined by
measurement imprecision due to detection shot noise. This dynamic range
exceeds the ratio 
\begin{equation}
\frac{|\chi(\Omega_\mathrm{m} )|^2 \bar S_{FF}^\mathrm{the}(\Omega_\mathrm{m}
)}{\bar S_{xx}^\mathrm{SQL}(\Omega_\mathrm{m} )}\approx2 \langle n \rangle.
\end{equation}
We can therefore conclude that the imprecision background, at a level of $1{.%
}1 \,\mathrm{am/\sqrt{Hz}} $, is \emph{below} the standard quantum limit,
calculated to $\sqrt{\bar S_\mathrm{xx}^\mathrm{SQL}(\Omega_\mathrm{m} ) }=2{%
.}2 \,\mathrm{am/\sqrt{Hz}} $ for this sample with $\Omega_\mathrm{m}
/2\pi=40{.}6\,\mathrm{MHz} $, $\Gamma_\mathrm{m} /2\pi=1{.}3\,\mathrm{kHz} $
and $m_\text{eff}=10\,\mathrm{ng} $. 
We emphasize however, that this does not imply that measurements with a
better \emph{total} uncertainty than the standard quantum limit are
possible. Quantum backaction-induced fluctuations in the mechanical
displacement increase the position uncertainty, but are masked by thermal
noise in this measurement.

\clearpage

\section{Observation of dynamical backaction}

\label{s:dynba}

In contrast to the resonant probing scenario discussed in the previous
section, the dynamics of the mechanical oscillator is modified by
radiation-pressure backaction if the optical mode is pumped in a detuned
manner. This so-called ``dynamical backaction'' not only modifies the
effective damping and spring constant of the mechanical oscillators, but, as
we will show in the following, also leads to an energy exchange between
optical and mechanical modes.

Predicted as early as the 1960s by Braginsky \cite{Braginskii1967},
dynamical backaction has been observed early on in mechanical devices
coupled to microwave resonators \cite{Braginskii1970,
Braginsky1977,Blair1995, Cuthbertson1996, Locke1998}. In the optical domain,
dynamical backaction induced by radiation pressure has been first observed
by the Vahala group in 2005 in the form of an oscillatory instability and
studied in great detail \cite{Kippenberg2005, Carmon2005, Rokhsari2005,
Rokhsari2006a,Hossein-Zadeh2006}.

Here, we present a systematic study of dynamical backaction as observed in
silica microresonators, in particular its dependence on the relevant
frequencies $\kappa$, $\Omega_\mathrm{m} $ and $\bar \Delta $, and briefly
introduce the oscillatory instability described above. In the second part,
we focus on the case of negative detuning $\bar \Delta <0$. In this case,
the light field extracts energy from the mechanical mode, leading to the
cooling of the latter. This effect was first reported by our group at the
MPQ in Garching \cite{Schliesser2006} and groups in Paris \cite{Arcizet2006a}
and Vienna \cite{Gigan2006}. Finally, we rule out thermal nonlinearities as
the origin of optomechanical interactions in silica microtoroids.

\subsection{Optical spring and optical damping}

For a detuned optical pump, we have found in section \ref{sss:dynbatheo}
that the presence of light modifies the dynamics of the mechanical degree of
freedom when it responds to an external force. In an intuitive picture, this
can be understood as the consequence of the in-phase and quadrature response
of the radiation-pressure force, when the mechanical oscillator is driven by
the external force. The same result is formally attained using the quantum
Langevin approach. Disregarding, in a first step, the quantum fluctuations
of the light ($\delta \hat s_\mathrm{in} =\delta \hat s_\mathrm{vac}
\rightarrow 0$)\footnote{%
This simplification is justified as long as the thermal Langevin force
largely exceeds force fluctuations due to the quantum nature of the light.},
the radiation pressure force fluctuations in Fourier space are 
\begin{align}
\delta\! \hat F_\mathrm{rp} (\Omega )&=i \hbar g_0 ^2 \bar a ^2\, \delta
\hat x (\Omega ) \left(\frac{1}{-i(\bar \Delta +\Omega )+\kappa/2} -\frac{1}{%
+i(\bar \Delta -\Omega )+\kappa/2}\right),
\end{align}
equivalent to equation (\ref{e:frpx}). As a consequence, the mechanical
oscillator reacts to the thermal Langevin force with the effective
susceptibility already derived in equation (\ref{e:chieff}).

\begin{figure}[bht]
\centering
\includegraphics[width= \linewidth]{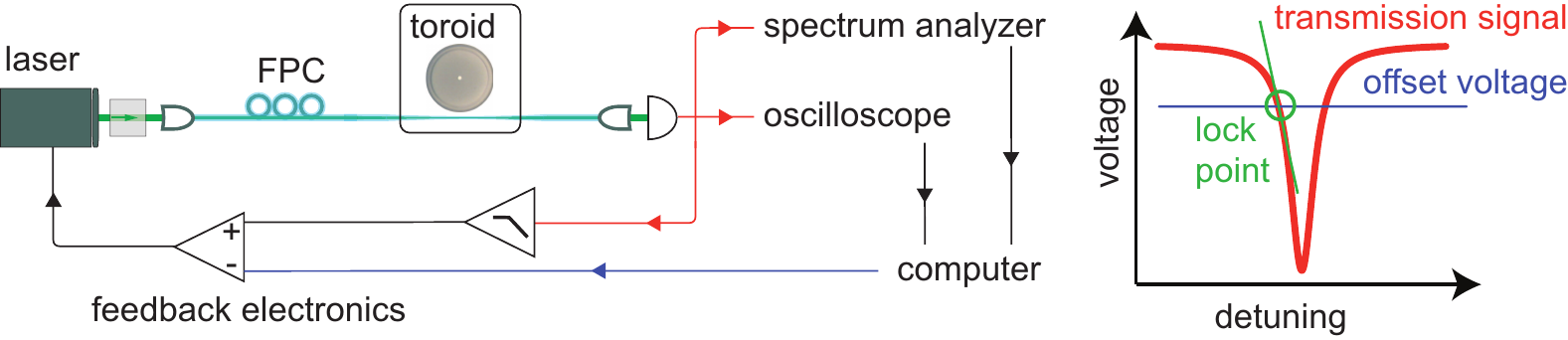}
\caption{Schematic illustration of the setup used for the measurement of
dynamical backaction. A diode laser is locked to the side of the optical
fringe by applying an electronic, computer-controlled offset to the
transmission signal. This differential signal is used as an error signal in
a feedback loop actuating the laser frequency by regulating both the
position of the grating in the laser cavity, and the current pumping the
laser diode. Once the laser is locked, the detuning is varied in small steps
by adjusting the electronic offset. For each detuning, a trace from the
oscilloscope (to determine the actual residual transmission) and a noise
spectrum from the electronic spectrum analyzer are taken. FPC stands for fiber
polarization controller.}
\label{f:directDetection}
\end{figure}

To confirm these predictions, a series of measurements was taken using the
setup described in figure~\ref{f:directDetection}. A $980\,\mathrm{nm} $%
-wavelength diode laser was locked to a resonance of the silica microtoroid,
simply using the transmission signal as an error signal, from which an
offset can be subtracted to control the detuning. Moderate optical quality
factors ($Q<10^7$) and low optical powers ($P_\mathrm{in}\sim 200 \,\mathrm{%
\mu W} $) ensure that thermal nonlinearities are weak enough to still allow
stable locking.

Applying this procedure to both the red ($\bar \Delta <0$) and blue ($\bar
\Delta >0$) wing of the optical resonance by changing the sign of the error
signal, a detuning series can be recorded. From the transmission signal
level with the laser locked to the side of the fringe, the relative detuning 
$\bar \Delta /\kappa$ can be determined. At the same time, the fluctuations
of the transmitted power, as recorded by the spectrum analyzer, reflect the
position fluctuations of the mechanical modes.

\begin{figure}[bt]
\centering
\includegraphics[width= \linewidth]{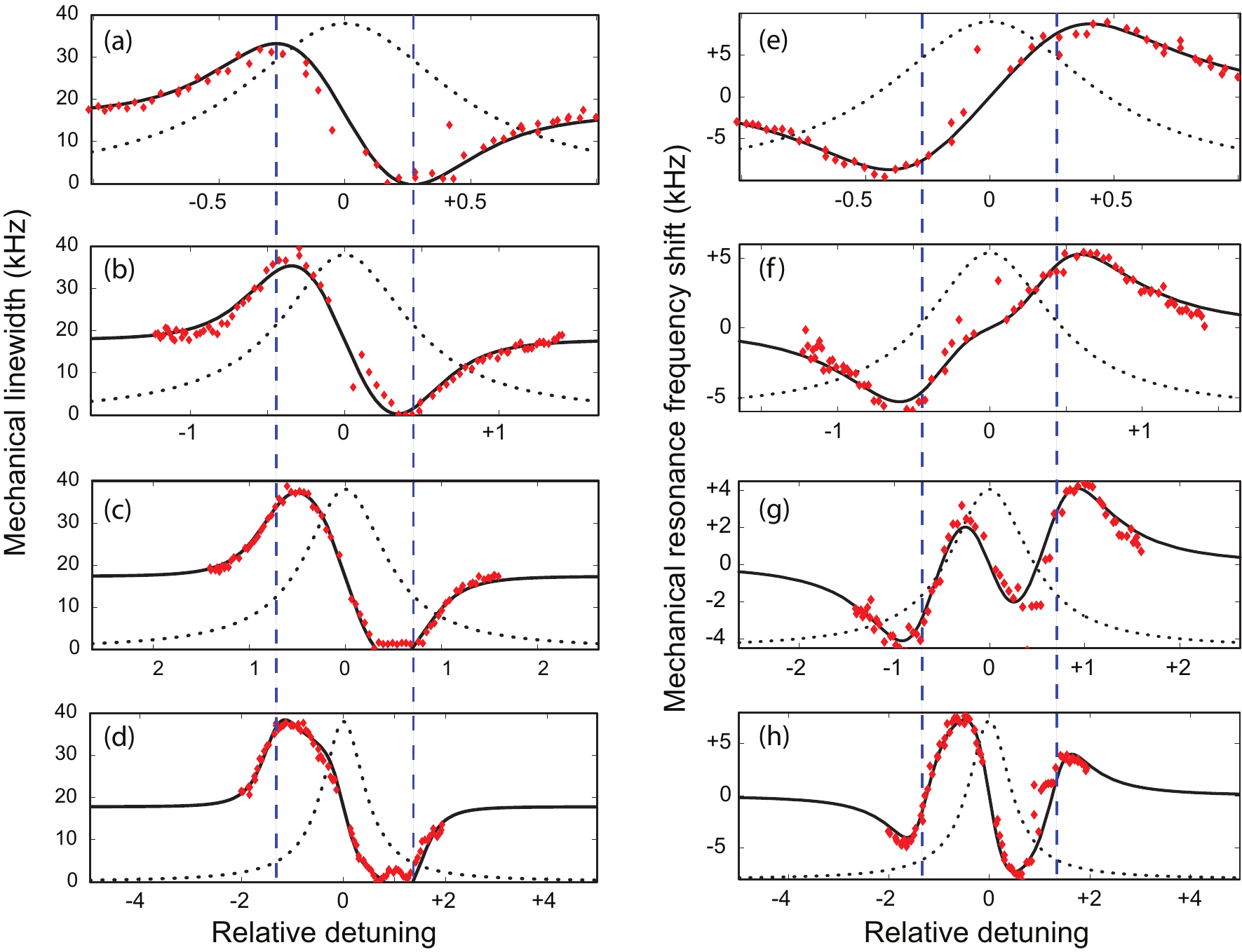}
\caption{Linewidth (a--d) and mechanical resonance frequency shift (e--h) of
the mechanical mode as a function of relative laser detuning $\bar \Delta /%
\protect\kappa$ for optical resonance linewidths $\protect\kappa/2\protect\pi
$ of about $207\,\mathrm{MHz} $ (a,e), $127\,\mathrm{MHz} $ (b,f), $79\,%
\mathrm{MHz} $ (c,g) and $42\,\mathrm{MHz} $ (d,h). Dotted lines indicate
the corresponding optical resonance Lorentzian, over which the laser was
tuned. Dashed blue lines marks a detuning equal to the mechanical
oscillator's frequency of $56.5 \,\mathrm{MHz} $ and full lines are fits
from the models for dynamical backaction.}
\label{f:lws+fs}
\end{figure}

Driven predominantly by the thermal Langevin force with its essentially
frequency-independent spectrum (\ref{e:sff}), the measured displacement
spectrum directly reveals the effective susceptibility of the mechanical
mode. It is therefore possible to extract the effective damping and
resonance frequency of the mode using the fit model (\ref{e:chieff}). Figure~%
\ref{f:lws+fs} shows the data obtained from the $56.5\,\mathrm{MHz} $-RBM of
a silica microtoroid together with fits by the models (\ref{e:Oeff}) and (%
\ref{e:Geff}). The measured changes in both damping and resonance frequency
agree well with expectation.

The resonance frequency shift is often referred to as ``optical spring''
effect \cite{Sheard2004}, as it originates from an optical restoring force
proportional to the displacement of the resonator. It is interesting to note
that this optical force can even exceed the natural restoring force of the
mechanical oscillator, and thereby totally dominate the mechanical resonance
frequency \cite{Corbitt2007}. For silica microresonators, due to the
stiffness of the structure, this is typically not the case. In these devices
however, another interesting effect occurs for narrow optical resonances ($%
\kappa<\Omega_\mathrm{m} $): in this case, the optical spring force changes
its sign for small detunings, turning a restoring into an anti-restoring
force and vice versa \cite{Schliesser2006}, an effect not observed in other
systematic studies of dynamical backaction \cite%
{Arcizet2006a,Hossein-Zadeh2007}.

The optically induced damping can provide both positive and negative
damping. For positive detuning $\bar \Delta >0$, the total damping can reach
zero. In this case, the mechanical mode, initially driven by thermal noise,
starts to oscillate regeneratively. Specifically, solving $\Gamma_\mathrm{eff%
} =0$ for a threshold power, one obtains 
\begin{align}
P_\mathrm{thresh}&=\Gamma_\mathrm{m} \frac{\bar \Delta ^2+(\kappa/2)^2}{\eta_%
\mathrm{c}  \kappa} \frac{\omega_\mathrm{c}  m_\text{eff} \Omega_\mathrm{m} 
}{g_0 ^2}  \notag \\
&\qquad\times\left(\frac{\kappa/2}{(\bar \Delta -\Omega_\mathrm{m}
)^2+(\kappa/2)^2}-\frac{\kappa/2}{(\bar \Delta +\Omega_\mathrm{m}
)^2+(\kappa/2)^2}\right)^{-1}
\end{align}
for this optically driven mechanical oscillation to occur. This effect,
often referred to as parametric oscillatory instability (POI), has been
reported for various systems, including silica microspheres \cite{Ma2007}
with mechanical modes at up to GHz-frequencies \cite{Carmon2007}. For light
powers largely exceeding the threshold, nonlinearities neglected in the
linearized models presented in this work lead to complex behavior such as
multistability \cite{Marquardt2006} and chaos \cite{Carmon2007a}. For an
in-depth theoretical discussion of the oscillatory instability, including
also quantum effects, we refer the reader to references \cite{Vahala2008,
Ludwig2008a}.

As an aside we note that in the regime $\kappa\ll\Omega_\mathrm{m} =\bar
\Delta $ discussed in greater detail in section \ref{s:rsc}, one finds the
interesting relation 
\begin{equation}
P_\mathrm{thresh}=4 \sqrt{\eta_\mathrm{c} } P_\mathrm{SQL}.
\end{equation}
This universally relevant power scale (for both dynamical and
quantum backaction) is at the level of $30\,\mathrm{\mu W} $ for typical
parameters of silica microtoroids.

\subsection{Radiation pressure cooling by dynamical backaction}

\label{ss:coolingByDBA}

In the preceding sections, we have only discussed the damping and resonance
frequency of the mechanical mode and its modification by dynamical
backaction. However, a major feature of light-induced damping is that it
also changes the temperature of the mechanical mode. To introduce the
concept of a ``mode temperature'', let us first evaluate the amplitude of
the displacement of a specific mode, which is driven by the thermal Langevin
force, by integrating its noise spectrum over all Fourier frequencies 
\begin{align}
\langle \delta x ^2\rangle &= \int_{-\infty}^{+\infty} \bar S_{xx}(\Omega )%
\frac{d\Omega }{2\pi} = \int_{-\infty}^{+\infty} |\chi(\Omega )|^2 \bar
S_{FF}^\mathrm{th}(\Omega ) \frac{d\Omega }{2\pi}\approx  \notag \\
&\approx \int_{-\infty}^{+\infty} \frac{2 m_\text{eff} \Gamma_\mathrm{m}  k_%
\mathrm{B} T} {m_\text{eff}^2\left((\Omega ^2-\Omega_\mathrm{m} ^2)^2+\Omega
^2 \Gamma_\mathrm{m} ^2\right)} \frac{d\Omega }{2\pi}.
\end{align}
This integral can be evaluated using the residue theorem, one obtains 
\begin{equation}
\frac{1}{2} m_\text{eff} \Omega_\mathrm{m} ^2 \langle \delta x ^2 \rangle = 
\frac{1}{2} k_\mathrm{B}T.
\end{equation}
We may turn this result around and use it to introduce the mechanical mode
temperature 
\begin{equation}  \label{e:Tmdef}
T_\mathrm{m} = m_\text{eff} \Omega_\mathrm{m} ^2 \langle \delta x ^2 \rangle
/ k_\mathrm{B}.
\end{equation}
With this definition, if the mechanical mode is only driven by the Langevin
force, it ends up in thermal equilibrium with its environment, and $T_%
\mathrm{m}=T$. As an example, the root-mean-square (RMS) displacement $%
\langle \delta x ^2 \rangle^{1/2}$ of the RBM of a silica microtoroid is
typically a few tens of femtometers at room temperature (figure~\ref%
{f:timeDomain}). 
\begin{figure}[hbt]
\centering
\includegraphics[width=\linewidth]{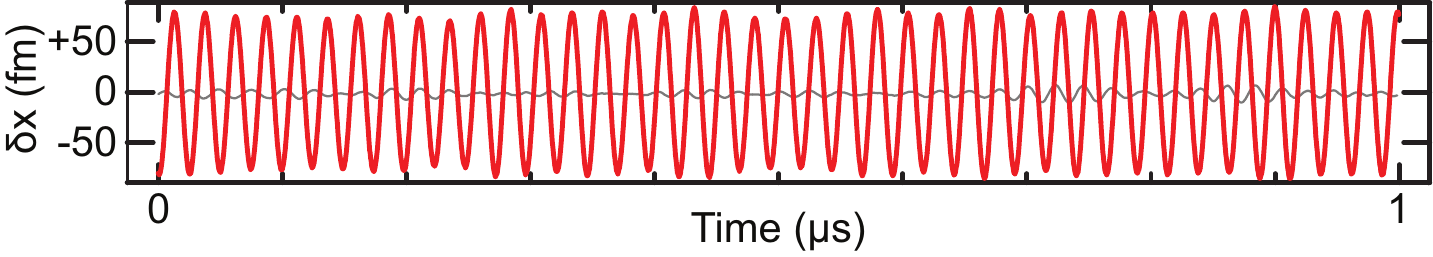}
\caption{Real-time recording of the displacement of a 40.6-MHz RBM of a
silica microtoroid measured using the H\"ansch-Couillaud technique. A $2 \,%
\mathrm{MHz} $-wide band was filtered out of the displacement signal and
selectively amplified to record the mechanical trace (red line). The gray
line is a background trace recorded with the taper retracted from the
cavity. Figure from ref.\ \protect\cite{Schliesser2008}.}
\label{f:timeDomain}
\end{figure}

In the presence of detuned pumping, the mechanical susceptibility is
modified due to dynamical backaction, \emph{the thermal Langevin force,
however, is not.} If, therefore, the RMS displacement is calculated from the
modified spectrum 
\begin{equation}
\bar S_{xx}(\Omega )=|\chi_\text{eff}(\Omega )|^2 \bar S_{FF}^\mathrm{th}%
(\Omega )
\end{equation}
one obtains 
\begin{equation}
\frac{1}{2} m_\text{eff} \Omega_\mathrm{m} ^2 \langle \delta x ^2 \rangle
\approx \frac{1}{2}\frac{\Gamma_\mathrm{m} }{\Gamma_\mathrm{eff} } k_\mathrm{%
B} T
\end{equation}
as long as the mechanical oscillator can be described with its
frequency-independent effective damping $\Gamma_\mathrm{eff} $ and resonance
frequency $\Omega_\mathrm{eff} $ for not too strong backaction effects%
\footnote{%
See e.\ g.\ \cite{Genes2008} for more general calculations}. The mode
temperature of the mechanical oscillator therefore is changed to 
\begin{equation}  \label{e:simplecooling}
T_\mathrm{m}=\frac{\Gamma_\mathrm{m} }{\Gamma_\mathrm{eff} } T=\frac{\Gamma_%
\mathrm{m} }{\Gamma_\mathrm{m} +\Gamma_\mathrm{dba}} T.
\end{equation}
As for $\Gamma_\mathrm{dba}>0$ one has $T_\mathrm{m}<T$ the damping rate $%
\Gamma_\mathrm{dba}$ induced by dynamical backaction is often referred to as
the laser \emph{cooling rate}.

Indeed it can be shown that $\Gamma_\mathrm{dba}$ is the rate with which
energy is transferred from the mechanical resonator to the optical field.
Returning, for simplicity, to the example of a sinusoidally oscillating
boundary considered in section \ref{ss:classical} ($x(t)=x_0 \sin(\Omega_%
\mathrm{m}  t)$), the cycle-averaged work done by the mechanical oscillator
on the optical field can be calculated to 
\begin{align}
P_\mathrm{dba}&=-\frac{2 \pi}{\Omega_\mathrm{m} }\int_0^{2\pi/\Omega_\mathrm{%
m} } F_\mathrm{rp}(t) \dot x(t) dt \approx\Gamma_\mathrm{dba}\,\left(\frac{1%
}{2} m_\text{eff} \Omega_\mathrm{m} ^2 x_0^2\right)
\end{align}
using only the elementary relations (\ref{e:icenergy}) and (\ref{e:simplefrp}%
).

In very general terms, this cooling effect arises by coupling the mechanical
oscillator not only to the reservoir---consisting of all other mechanical
modes present in the device, the gas and thermal radiation field surrounding
it etc.---at room temperature, but also to the cooling laser field. Laser
cooling therefore disequilibrates the oscillator with the reservoir, and
brings it into a new equilibrium in which it is coupled both to the
reservoir and the laser field. This field possesses an effective temperature
very close to zero (see section \ref{s:rsc} for the limitations), and acts
as a ``cold damper'', by introducing dissipation, but only very little
fluctuations to the mechanical mode.

We note here that the application of cold damping schemes has a long history
in physics, and has been successfully applied in systems as diverse as
electrometers and particle storage rings \cite{Milatz1953, Milatz1953a,
Mohl1980} (here often referred to as ``stochastic cooling''). Interestingly,
cold damping has also been used to cool a mechanical mode of a mirror in a
pioneering 1999 experiment at Laboratoire Kastler Brossel in Paris \cite%
{Cohadon1999}, and subsequently in many other experiments \cite%
{Arcizet2006,Kleckner2006,Weld2006,Poggio2007,Corbitt2007a}. However, these
experiments all involve a complex hybrid electronic/optical feedback loop,
whereas the method presented here relies solely on the intrinsic dynamics of
radiation pressure.

\begin{figure}[bth]
\centering
\includegraphics[width= \linewidth]{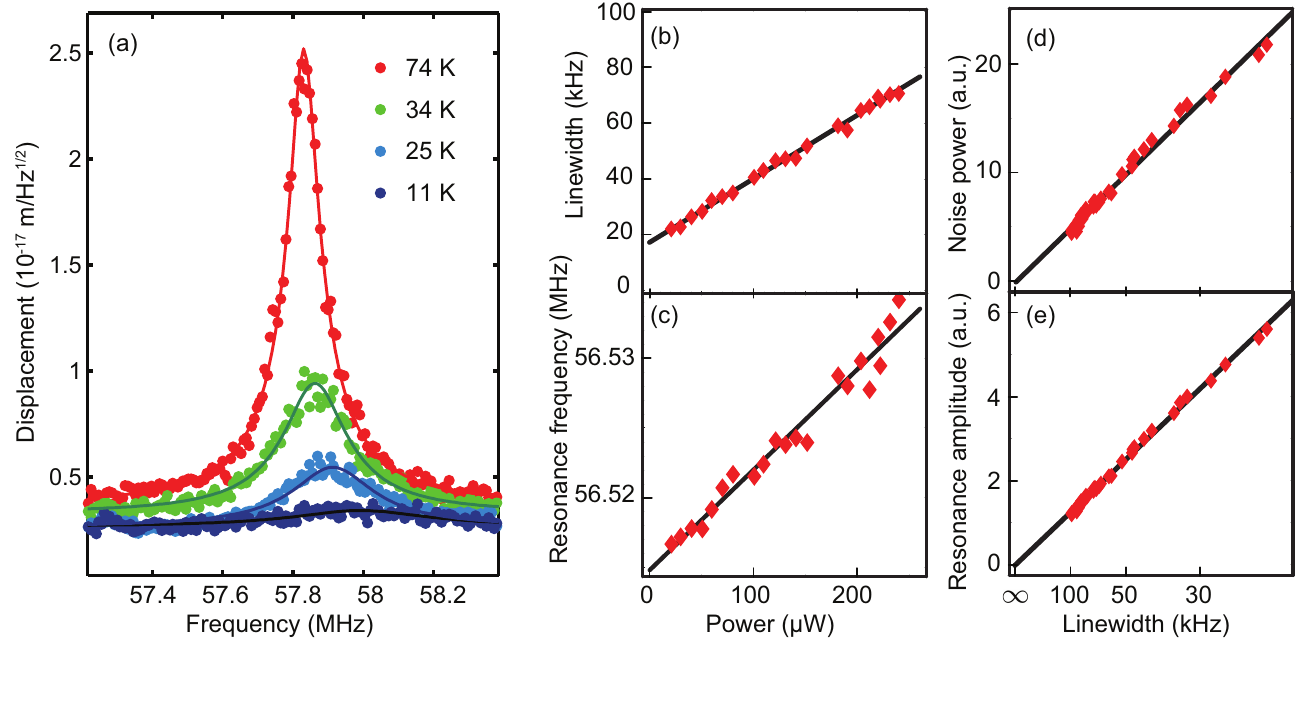}
\caption{Cooling of the RBM of silica microtoroids. (a) Noise spectra of the
RBM when a $50\,\mathrm{MHz} $-wide WGM resonance is pumped at $\bar \Delta
\approx-\protect\kappa/2$ with increasing power ($0{.}25\,\mathrm{mW} $, $0{.%
}75$, $1{.}25 \,\mathrm{mW} $ and $1{.}75 \,\mathrm{mW} $). The extracted
increased damping rates $\Gamma_\mathrm{eff} $ correspond to mode
temperatures $T_\mathrm{m}$ given in the legend. (b)--(e) Dependence of
mechanical mode properties on the launched laser power (measured on the RBM
of a different toroid). The linewidth $\Gamma_\mathrm{eff} $ and resonance
frequency $\Omega_\mathrm{eff} $ vary linearly with input power in this
range, and both the noise power $\protect\int_{-\infty}^{+\infty} \bar
S_{xx}(\Omega )d\Omega /2\protect\pi\propto T_\mathrm{m}$ and the resonance
amplitude $\protect\sqrt{\bar S_{xx}(\Omega_\mathrm{m} )}$ vary linearly
with the inverse linewidth $\Gamma_\mathrm{eff} ^{-1}$ as expected. }
\label{f:cooling}
\end{figure}

In figure~\ref{f:cooling} we show cooling results obtained on the RBM of
silica microresonators at frequencies around $57 \,\mathrm{MHz} $. In an
experimental setup essentially identical to the one described in figure~\ref%
{f:directDetection}, a 980-nm wavelength diode laser was locked to the red
wing of an optical resonance. Note that as the thermal bistability renders
the red wing dynamically unstable under laser or cavity frequency
fluctuations \cite{Carmon2004a}, special care has to be taken in the
implementation of the feedback loop stabilizing the laser frequency to a
given detuning.

As the launched laser power is increased, the total damping increases, and
correspondingly so does the width of the Lorentzian resonance. At the same time, the
temperature $T_\mathrm{m}$ of the mode is reduced by the optical pumping.
Panel (a) shows four traces for detuned pumping of a WGM mode with $\bar
\Delta \approx-\kappa/2\approx-25\,\mathrm{MHz} $. For the highest pump
powers, the damping is increased beyond $450\,\mathrm{kHz} $, and the
corresponding reduced mode temperature is $11\,\mathrm{K} $. The other
panels show a systematic power series, in which a different torus was pumped
at $\bar \Delta \approx -0.7 \kappa$ with powers between $20$ and $200 \,%
\mathrm{\mu W} $.

\subsection{Radiation pressure versus thermal effects}

\label{ss:rpvsthe}

In many early experiments, the optomechanical interaction has been mediated
by thermal effects, sometimes referred to as ``photothermal pressure''
instead of radiation pressure \cite{Hohberger2004, Gigan2006, Harris2007,
Favero2007, Ludwig2008}. In the following we present evidence for the fact
that the optomechanical interaction in silica microtoroids is strongly
dominated by radiation pressure due to the high ($\Omega_\mathrm{m} /2\pi
\gtrsim 50 \,\mathrm{MHz} $) frequencies of the RBM.

Prior work on the radiation-pressure-induced parametric oscillatory
instability has provided independent evidence that radiation-pressure
dominates the interaction between optical and mechanical modes \cite%
{Kippenberg2005, Rokhsari2006a} in this system. For instance, it has been
shown that the mechanical gain depends on the cavity finesse. This
observation demonstrates that radiation pressure is responsible, since
thermal forces only depend on the absorbed power and not on the cavity
enhanced circulating power.

\subsubsection{Theoretical estimate}

To theoretically estimate the forces due to light absorption it is necessary to
study the coupling of mechanical and thermal waves (or modes) of the
structure. Thermodynamical considerations allow us to derive the mutually
coupled differential equations for temperature and displacement
distributions \cite{Nowacki1975}, adding a coupling  term proportional to
the linear expansion coefficient $\alpha$ to the heat diffusion equation and
the equations of motion of the displacement field (\ref{e:elasticeom}).%

For small perturbations, we may however assume that the solutions of the
uncoupled system ($\alpha\rightarrow 0$) are still approximately valid, and
heat transport is still dominated by diffusion. In this case, the resulting
temperature gradients give rise to a thermoelastic body force \cite%
{Nowacki1975} 
\begin{equation}
\vec f_\mathrm{te}(\vec r)=-(3 \lambda +2 \mu) \alpha\, \vec \nabla \delta
T(\vec r)
\end{equation}
driving the mechanical modes ($\lambda$ and $\mu$ are the Lam\'e constants).
Importantly, this body force depends on the gradient of the temperature
distribution.

The effective scalar thermal force on a mechanical mode with a displacement
pattern $\vec u_n^0$ is determined by an overlap integral $\langle \vec f_%
\mathrm{te}\, \vec u_n^0 \rangle $. Due to the very high mechanical
resonance frequencies, the diffusion length for the temperature distribution 
$\lambda_\mathrm{D}=\sqrt{{2 k}/{c_\mathrm{p} \rho \Omega }}$ becomes very
short, for example about $50\,\mathrm{nm} $ for $\Omega /2 \pi=50\,\mathrm{%
MHz} $. If absorption takes place in the silica, we can therefore conclude
that the temperature distribution is essentially given by the energy
distribution of the optical mode, and the resulting temperature modulation
in the volume can be estimated at $\delta T(\Omega )\approx 2P_\mathrm{abs}%
/\Omega  c_\mathrm{p} \rho V_\mathrm{mode}$, typically $\lesssim 10 \,%
\mathrm{K/W} $ for a typical $30\,\mathrm{\mu m}  $-radius toroid---a value
confirmed by finite element modeling of heat diffusion in such a geometry.
Due to the nearly \emph{symmetric} temperature distribution in the radial
direction with respect to the center of the optical mode, the overlap
integral over the gradient of the temperature distribution can be expected
to yield only a small total contribution. For a rough estimate, we may use 
\begin{align}
\langle \vec f_\mathrm{te}\, \vec u_n^0 \rangle &\propto \int_V \vec \nabla
\delta T(\vec r)\, \vec u_n^0(\vec r) d^3r\approx |\vec u_n^0(R)| 2\pi R d_%
\mathrm{m}\int_{R-d_\mathrm{m}}^{R} \partial \delta T /\partial r dr  \notag
\\
&\approx |\vec u_n^0(R)| 2\pi R d_\mathrm{m} \delta T(R),
\end{align}
where $d_\mathrm{m}\ll R$ is the transverse diameter of the optical mode.
Using $\delta T(R)/P_\mathrm{abs}\approx 2\,\mathrm{K/W} $ for the
parameters described above (again confirmed by FEM), we obtain $ {\langle
\vec f_\mathrm{rp} \,\vec u_n^0 \rangle}/  {\langle \vec f_\mathrm{te}
\,\vec u_n^0  \rangle}\approx\mathcal{O}(10^{2})$. We note however that this
result depends on the exact location of the heat source (absorption may also
take place in a water or helium surface layer), and the cooling mechanisms
provided by a surrounding medium, effects presently investigated in our
group \cite{Zhou2009}. We also note that at cryogenic temperatures,
thermoelastic coupling is weaker as the expansion coefficient drops below $%
2\cdot10^{-9}\,\mathrm{K^{-1}} $ at $1.6\,\mathrm{K} $ \cite{White1975,
Arcizet2009a}.

\subsubsection{Response measurements}

Experimentally, we have made response measurements \cite%
{Rokhsari2005} to quantify the different nonlinearities---due to the
thermal, Kerr and radiation-pressure effects---encountered in silica
microtoroids. To that end, two lasers at different wavelengths ($980$ and $%
1550 \,\mathrm{nm} $ in this case) are coupled to two WGM resonances of a
single toroid. One laser, referred to as the ``pump'', is amplitude
modulated at a variable frequency $\Omega $, while the other ``probe'' laser
is used to measured the response of the WGM frequency to the pump laser
power modulation. In the simplest case, this is accomplished by tuning the
probe laser to the wing of a WGM resonance, and measuring the variation of
its transmission at the same frequency $\Omega $, most conveniently
implemented using a network analyzer. Care is taken to suppress direct
optical or electronic cross-talk of the pump modulation into this signal.

\begin{figure}[htbp]
\centering
\includegraphics[width=\linewidth]{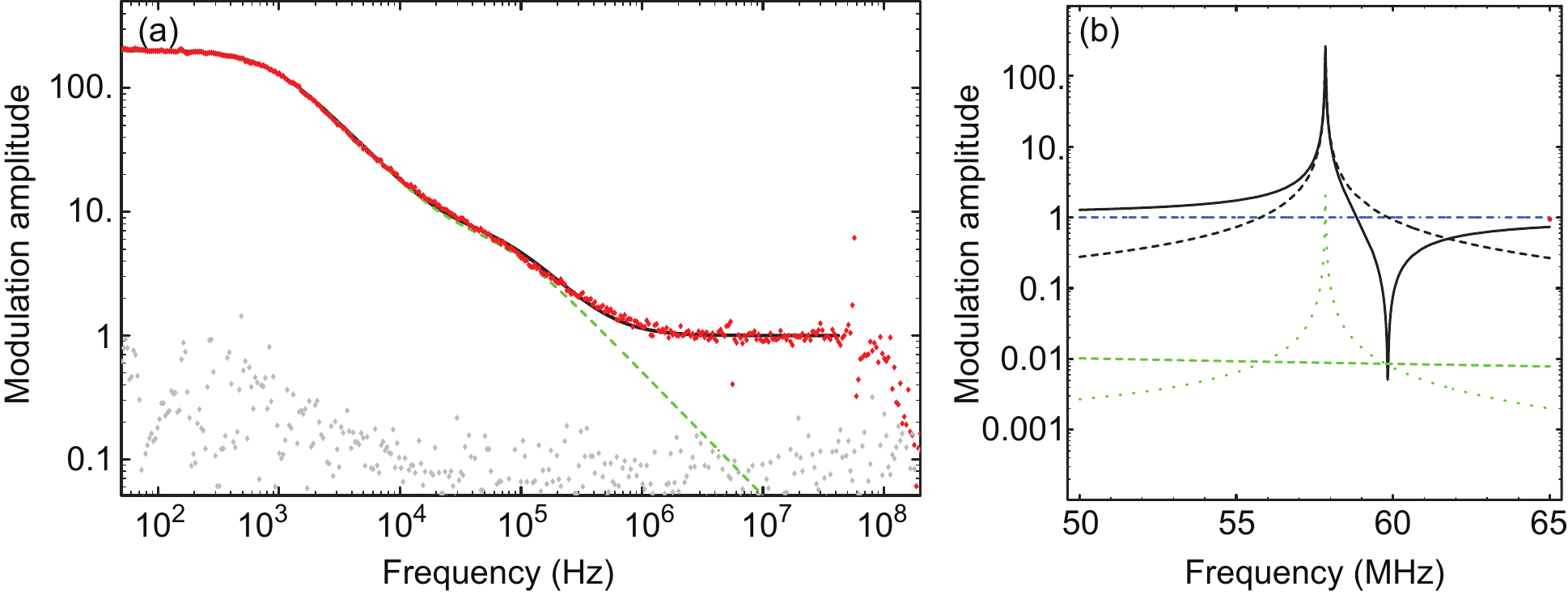}
\caption{Pump-probe type response measurements of the nonlinearities present
in toroidal silica microcavities. (a) Measured response (red dots) and fit
(black line) modeling a two-pole low-pass thermal nonlinearity and an
instantaneous Kerr effect, which becomes significant above 1~MHz. The data
furthermore show a detector and cavity-induced cutoff above 100~MHz as well
as an indication of mechanical resonances at 5{.}6 and 58 MHz, though not
resolved. The green dashed line shows the extrapolation of the thermal
effects to higher frequencies and below the measurement background at about
0.1 (gray dots). All data are normalized to the fitted Kerr response. (b)
Higher-resolution data (red points) as recorded in the vicinity of the
radial-breathing mechanical resonance at $\sim 58\,\mathrm{MHz} $. Full
black line, model for the harmonic response plus a constant background due
to the Kerr effect. Black and blue dashed lines illustrate the individual
contributions from the harmonic response and the Kerr effect, respectively.
Green dashed line indicates the extrapolated thermal response, and the green
dotted line the expected response if the thermally induced displacement was
resonantly enhanced. All data are normalized to Kerr response.}
\label{f:response}
\end{figure}

Figure~\ref{f:response} shows the result of such a measurement on a $29\,%
\mathrm{\mu m}  $-major radius toroid. Clearly, at low frequencies ($\Omega
/2\pi<1\,\mathrm{MHz} $), a strong modulation of the probe WGM frequency is
apparent. This is due to absorption of pump light and the consequent
modulation of the temperature-dependent expansion and refractive index of
the toroid material. Above this frequency, a plateau is observed in the
response, due to the modulation of the refractive index seen by the probe
WGM, which varies with pump power due to the non-linear refractive index of
silica (Kerr effect). Finally, around a mechanical resonance, a dispersive
peak is observed. This is due to the excitation of mechanical modes by the
modulated pump power.

To fit the response, we use the model 
\begin{equation}
\delta \omega _\mathrm{probe}(\Omega )= \delta \omega _\mathrm{th}(\Omega
)+\delta \omega _\mathrm{K}(\Omega )+\delta\omega _\mathrm{rp}(\Omega )
\end{equation}
with 
\begin{align}
\delta\omega _\mathrm{th}(\Omega )&=-\omega_\mathrm{c}  \left(\alpha+\frac{1%
}{n}\frac{dn}{dT}\right) \underbrace{\left(\frac{\beta_1}{1+i\Omega /\Omega
_1}+\frac{\beta_2}{1+i\Omega /\Omega _2}\right) \frac{2\pi n R}{c}\frac{%
\delta P_\mathrm{IC}(\Omega )}{\tau_\mathrm{abs}}}_{\delta T_\mathrm{eff}%
(\Omega )} \\
\delta \omega _\mathrm{K}(\Omega )&=-\omega_\mathrm{c} \frac{n_2}{n}\frac{%
\delta P_\mathrm{IC}(\Omega )}{A_\mathrm{eff}} \\
\delta\omega _\mathrm{rp}(\Omega )&=g_0 \chi(\Omega ) \delta F_\mathrm{rp}
(\Omega )=  \notag \\
&=-\frac{\omega_\mathrm{c} }{R} \frac{1}{m_\mathrm{eff} (\Omega_\mathrm{m}
^2-\Omega^2+i\Gamma_\mathrm{m} \Omega)} \frac{2 \pi n}{c} \delta P_\mathrm{IC%
}(\Omega ).
\end{align}
For the thermal effect, we extract cutoff frequencies $\Omega _1\approx
2\pi\cdot 900 \,\mathrm{Hz} $ and $\Omega _2\approx 2 \pi\cdot 69\,\mathrm{%
kHz} $. Furthermore, using the material parameters of silica and $R=29 \,%
\mathrm{\mu m} $ as well as $A_\mathrm{eff}\approx2{.}5 \,\mathrm{\,\mathrm{%
\mu m}  ^2} $, we can normalize all results to the measured Kerr response.
In this manner, we extract $\beta_1/\tau_\mathrm{abs}\approx1{.}8\cdot 10^4\,%
\mathrm{K} /\mathrm{W} / 100\,\mathrm{ns} $ and $\beta_2/\tau_\mathrm{abs}%
\approx 570\,\mathrm{K} /\mathrm{W}/100\,\mathrm{ns} $. All these parameters
are very well in the range expected for the thermal effects.

The combination of the radiation-pressure and Kerr responses gives rise to
the dispersive signature around the mechanical resonance frequency: At
modulation frequencies slightly above the mechanical resonance, the
mechanical degree of freedom oscillates out of phase and therefore
counteracts the Kerr effect, which always reacts instantaneously to power
changes. Furthermore, at the resonance frequency $\Omega_\mathrm{m} $ of the
RBM we expect 
\begin{equation}
\left|\frac{\delta \omega _\mathrm{rp}(\Omega_\mathrm{m} )} {\delta \omega _%
\mathrm{K}(\Omega_\mathrm{m} )}\right|= \frac{2 \pi n^2 A_\mathrm{eff} }{R
n_2 \Gamma_\mathrm{m}  \Omega_\mathrm{m}  c m_\mathrm{eff} }\approx 240
\end{equation}
with $\Gamma_\mathrm{m} =15.7\,\mathrm{kHz} $ derived from the fit and the
numerically determined effective mass $m_\text{eff}\approx15 \,\mathrm{ng} $%
, in very good agreement to the measured value of $260$. The extrapolated
thermal effect, in contrast, drops more than four orders of magnitude below
the observed mechanical displacement at $\Omega_\mathrm{m} $ (cf. figure~\ref%
{f:response}). At mechanically non-resonant frequencies, this modulation
would be largely dominated by a thermorefractive effect as compared to
thermoelastic effects, as $(d n/d T) /n \alpha>10$. Even if enhanced by the
mechanical resonance, thermoelastic contributions may therefore be estimated
at or below the $10^{-2}$-level.

We finally note that thermally induced forces that may be related to the
identified thermal effects would be out of phase (by nearly $\pi/2$) with
the driving pump laser modulation, since the relevant radio frequencies are
well beyond the thermal cutoff frequencies. As a direct consequence, the
observed interference effects between mechanical displacement and the
Kerr-induced modulation would form a single symmetric resonant peak,
fundamentally different to the observed dispersive shape. This again
confirms the dominance of radiation pressure in this system.

\clearpage

\section{Resolved-sideband cooling}

\label{s:rsc}

The successful demonstration of cooling by dynamical backaction immediately
raises the question of how strongly the temperature of the mechanical
oscillator can be reduced. A series of theoretical and experimental
investigations revealed the fundamental \cite{Wilson-Rae2007, Marquardt2007}
and technical \cite{Schliesser2008, Diosi2008, Rabl2009} limitations of
radiation-pressure cooling of mechanical oscillators. In essence, if all
technical sources of heating are avoided, the quantum fluctuations of the
cooling light field provide a fluctuating force, driving the oscillator to
random motion and therefore compete with the laser cooling effect. An
advantageous ratio of these two effects can however be achieved in the
so-called \emph{resolved-sideband} (RSB) regime, in which the mechanical
oscillation frequency $\Omega_\mathrm{m} $ exceeds the cavity linewidth $%
\kappa$. In this case, cooling of the mechanical oscillator to its quantum
mechanical ground state is possible. In the following, a brief outline of
the required theoretical considerations will be given after a motivation by
well-known results from atomic physics. In the main part of this section,
experimental results---including the first demonstration of resolved
sideband laser cooling of an optomechanical device---will be presented.

\subsection{Ground state cooling: the atomic physics case}

The quantum mechanical expectation value of the energy of a harmonic
mechanical oscillator of $\Omega_\mathrm{m} $ and mass $m_\text{eff}$ is
given by 
\begin{align}
\langle H_\mathrm{mech} \rangle &=\frac{1}{2} \frac{ \left\langle\hat
p^2\right\rangle }{m_\text{eff}}+\frac{1}{2} {m_\text{eff} \Omega_\mathrm{m}
^2} \left\langle\hat x^2\right\rangle= \\
&=\hbar \Omega_\mathrm{m}  \left(\langle \hat n \rangle+\frac{1}{2}\right)
\end{align}
where the phonon number operator $\hat n=\hat b^\dagger  \hat b $ is given
by the creation and annihilation operators 
\begin{align}  \label{e:xzpfdef}
\hat b^\dagger  &=\frac{1}{2 x_\mathrm{ZPF} }\left( {\hat x }-i \frac{\hat p 
}{m_\text{eff}\Omega_\mathrm{m} }\right) \\
\intertext{with the so-called zero-point fluctuations}
x_\mathrm{ZPF} &=\sqrt{\frac{\hbar}{2 m_\text{eff} \Omega_\mathrm{m} }},
\end{align}
which are of the order of $100\,\mathrm{am} $ for typical silica
microtoroidal resonators ($m_\text{eff}=10\,\mathrm{ng} $, $\Omega_\mathrm{m}
/2\pi=40\,\mathrm{MHz} $).

The question raised by the cooling results presented in the previous section
is whether it is possible to reduce $\langle H_\mathrm{mech}\rangle$ to
levels comparable with the ground-state energy $\hbar \Omega_\mathrm{m} /2$.
In other words, is it possible to reach the quantum ground state, in which
the occupation number (the number of excitation quanta, phonons) $\langle
n\rangle$ reaches zero? In this case, deviations from the classical cooling
behavior described in the previous section are clearly expected in order to
prevent cooling to reach arbitrarily low energy states.

To answer this question, it is instructive to consider the results obtained
in the context of laser cooling of ions (or atoms) \cite{Hansch1975,
Wineland1975, Wineland1979, Stenholm1986}. Trapped in a harmonic potential,
these elementary particles constitute mechanical oscillators as well, their
eigenfrequency $\Omega_\mathrm{m} $ being given by the tightness of the
trap. Lasers can be used to drive electronic transitions of energy $\hbar
\omega _0$ and lifetime $\kappa^{-1}$, and optical and mechanical degrees of
freedom are coupled by the Doppler shift of the optical resonance if the ion
is moving and by the momentum transfer of absorbed and emitted photons.

\begin{figure}[bt]
\centering
\includegraphics[width=.8 \linewidth]{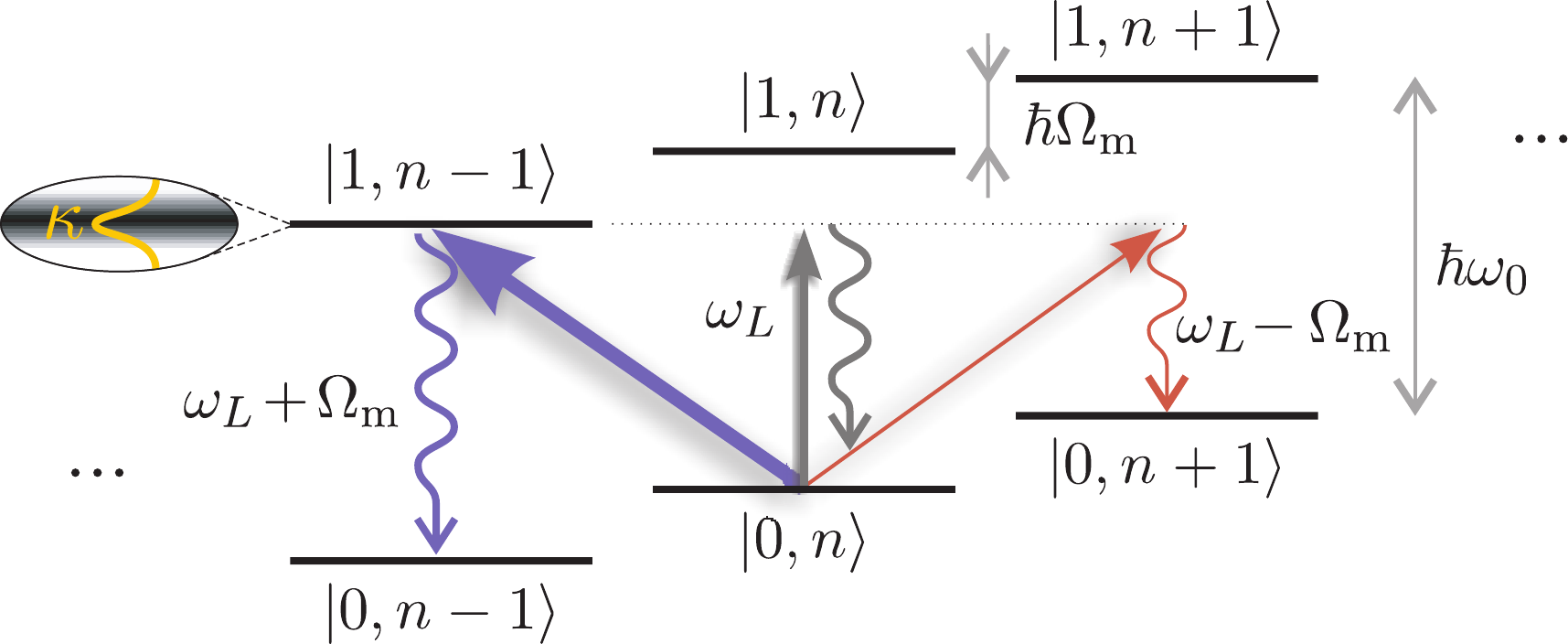}
\caption{Optical sideband cooling. Laser photons of energy $\hbar \protect%
\omega_\mathrm{l} =\hbar(\protect\omega _0-\Omega_\mathrm{m} )$
preferentially induce ``red-sideband'' transitions $|0,n\rangle%
\rightarrow|1,n-1\rangle$ (blue arrow) if they are detuned from the carrier $%
|0,n\rangle\rightarrow|1,n\rangle$ (gray arrow) and blue sideband $%
|0,n\rangle\rightarrow|1,n+1\rangle$ (red arrow) transitions. As a
consequence, the phonon occupation $n$ of the mechanical oscillator is
reduced when the photon gets absorbed by the atom, or, in the optomechanical
case, by the cavity. Re-emission of the photon (wavy lines), on average,
does not change the phonon occupation (neglecting recoil), so that a detuned
laser provides cooling. Figure from ref.\ \protect\cite{Schliesser2008}.}
\label{f:levelScheme}
\end{figure}

If the laser is detuned from the optical resonance, cooling can occur by
favoring the absorption of a photon only in conjunction with the
annihilation of a mechanical excitation quantum. Essentially, it is the
phonon energy $\hbar \Omega_\mathrm{m} $ which makes up for the energy mismatch $%
\hbar(\omega _0-\omega_\mathrm{l} )$ of the incoming photon to drive the
electronic transition. If absorption takes place, the subsequently
re-emitted photon has an average energy of $\hbar \omega _0$ (neglecting
recoil). It therefore carries away the additional energy of the phonon, and
leaves the mechanical oscillator in a state of lower excitation $n$ (figure~%
\ref{f:levelScheme}).

This method is usually referred to as optical sideband cooling: an ion
oscillating in its trap exhibits absorption sidebands at frequencies $\omega
_0\pm j \Omega_\mathrm{m} $, $j\in \mathbb{N}$, very similar to the case of
a cavity discussed in section \ref{ss:classical}. If the laser is tuned to
the red sideband at $\omega _0-\Omega_\mathrm{m} $, cooling transitions are
resonantly enhanced as shown in figure~\ref{f:levelScheme}. The quantum
theory of laser cooling of trapped atoms or ions reveals that this method
allows ground state cooling $\langle n\rangle\rightarrow 0$ provided that $%
\Omega_\mathrm{m} \gg\kappa$ (neglecting recoil) \cite{Wineland1979,
Stenholm1986}. In this case, the lowest average occupation that can be
achieved is given by 
\begin{equation}
\langle n\rangle =\frac{\kappa^2}{16 \Omega_\mathrm{m} ^2}.
\end{equation}
Interestingly, this result can also be viewed as being due to the
competition of the cooling effect of the detuned laser, due to an effective
viscous force, and a ``heating'' effect due to quantum fluctuations of the
light beam, giving rise to a fluctuating radiation-pressure force \cite%
{Itano1992}.

\subsection{Limits of radiation-pressure cooling using dynamical backaction}

A similar analysis can be applied to cooling by dynamical backaction. To
assess the fundamental cooling limits, we consider the radiation pressure
force as it is obtained from the quantum Langevin equations: 
\begin{align}
\delta\! \hat F_\mathrm{rp} (\Omega )&= -\hbar g_0  \bar a  \left(\frac{%
\delta \hat s_\mathrm{in} (\Omega )\tau_\mathrm{ex} ^{-1/2}+\delta \hat s_%
\mathrm{vac} (\Omega )\tau_0 ^{-1/2}}{-i(\bar \Delta +\Omega )+\kappa/2} +%
\frac{\delta \hat s_\mathrm{in}^{\dagger} (\Omega )\tau_\mathrm{ex}
^{-1/2}+\delta \hat s_\mathrm{vac}^{\dagger} (\Omega )\tau_0 ^{-1/2}}{%
+i(\bar \Delta -\Omega )+\kappa/2}\right)  \notag \\
&\qquad+i \hbar g_0 ^2 \bar a ^2\, \delta \hat x (\Omega ) \left(\frac{1}{%
-i(\bar \Delta +\Omega )+\kappa/2} -\frac{1}{+i(\bar \Delta -\Omega
)+\kappa/2}\right).
\end{align}
In essence, the expression in the first line correspond to quantum
backaction due to quantum fluctuations of the intracavity photon number,
while the expression on the second line is due to dynamical backaction
(proportional to the mechanical displacement). The latter force
contribution, is usually absorbed into the effective susceptibility $\chi_%
\text{eff}$ from equation (\ref{e:chieff}), while the spectrum of the force
quantum fluctuations is calculated as 
\begin{align}
\bar S_{FF}^{\mathrm{ba,qn}}(\Omega )&=\hbar^2 g_0 ^2 \bar a ^2 \frac{\kappa%
}{2}\left( \frac{1}{(\bar \Delta +\Omega )^2+(\kappa/2)^2}+\frac{1}{(\bar
\Delta -\Omega )^2+(\kappa/2)^2}\right) \\
&= \frac{\hbar^2}{x_\mathrm{ZPF} ^2}\frac{1}{2} \underbrace{ \left( \frac{%
g_0 ^2 \bar a ^2 x_\mathrm{ZPF} ^2 \kappa}{(\bar \Delta +\Omega
)^2+(\kappa/2)^2}\right.}_{A_-} + \underbrace{ \left. \frac{g_0 ^2 \bar a ^2
x_\mathrm{ZPF} ^2 \kappa}{(\bar \Delta -\Omega )^2+(\kappa/2)^2} \right)}%
_{A_+}
\end{align}
using the known correlation functions for the input quantum noise. We have introduced
here the rates $A_-$ and $A_+$, which can be shown to correspond to rates of
anti-Stokes and Stokes scattering events in which phonons are annihilated or
created \cite{Wilson-Rae2007, Marquardt2007, Genes2008, Clerk2008}.

The remaining energy stored in the mode after cavity cooling can be
evaluated by integrating the mechanical displacement spectrum. This is
possible analytically if a hot mechanical reservoir $\hbar \Omega_\mathrm{m}
\ll k_\mathrm{B} T$ is assumed and the effective susceptibility still
corresponds to a high-quality Lorentzian. Without further proof we note here
that this can be safely assumed as long as the magnitude of the optomechanical
coupling parameter 
\begin{equation}
G =2 \bar a  g_0  x_\mathrm{ZPF}   \label{e:lcr}
\end{equation}
is much less than $\kappa$. For higher values of $G$, hybridization of
optical and mechanical modes set in, a regime treated in detail in \cite%
{Dobrindt2008, Groblacher2009a}. In the limits $\Omega_\mathrm{m} 
,\kappa\gg G ,\Gamma_\mathrm{m} $ relevant to the work presented in the
following, one eventually obtains \cite{Genes2008} 
\begin{align}
\frac{1}{2}m_\text{eff} \Omega_\mathrm{m} ^2 {\langle \delta \hat x ^2
\rangle} &=\int |\chi_\text{eff}(\Omega )|^2\left(\bar S_{FF}^\mathrm{th}%
(\Omega ) +\bar S_{FF}^\mathrm{ba,qn}(\Omega )\right) \frac{d\Omega }{2\pi}=
\\
&=\frac{1}{\Gamma_\mathrm{eff} }\left(\Gamma_\mathrm{m}  {k_\mathrm{B} T}+%
\frac{A_-+A_+}{2}\hbar \Omega_\mathrm{m} \right) \\
&=\frac{\Gamma_\mathrm{m} }{\Gamma_\mathrm{eff} } {k_\mathrm{B} T}+\frac{A_-%
}{\Gamma_\mathrm{eff} }\hbar \Omega_\mathrm{m} +\frac{\hbar \Omega_\mathrm{m}
}{2}.
\end{align}
For significant cooling, one has  $A_-\gg A_+\gg \Gamma_\mathrm{m} $, and finally 
\begin{align}
\langle n \rangle\approx\frac{\Gamma_\mathrm{m} }{\Gamma_\mathrm{eff} }%
\underbrace{\frac{k_\mathrm{B}T}{\hbar \Omega_\mathrm{m} }}_{n_\mathrm{bath}%
} +\underbrace{\frac{A_+}{A_- -A_+}}_{ n_\mathrm{min}}.
\end{align}

As a consequence, even for very strong cooling $\Gamma_\mathrm{eff}
\gg\Gamma_\mathrm{m} $, the phonon occupation cannot be reduced to
arbitrarily low numbers; instead it is bound by $n_\mathrm{min}$. Two simple
limits are derived depending on the ratio of mechanical resonance frequency
and optical cavity linewidth: 
\begin{align}
n_\mathrm{min} &\approx\frac{\kappa}{4 \Omega_\mathrm{m} } & & \kappa\gg
\Omega_\mathrm{m}  & \text{unresolved sidebands} \\
n_\mathrm{min} &\approx \frac{\kappa^2}{16 \Omega_\mathrm{m} ^2} & &
\kappa\ll \Omega_\mathrm{m}  & \text{resolved sidebands}.
\end{align}
Evidently, the ground state $\langle n \rangle\rightarrow 0$ can only be
reached in the resolved-sideband regime, where $\kappa\ll\Omega_\mathrm{m} $%
. Due to the non-zero occupation $n_\mathrm{bath}\approx k_\mathrm{B}
T/\hbar \Omega_\mathrm{m} $ when in equilibrium with the thermal bath, more
stringent requirements arise in laboratory experiments. In particular, both $%
\kappa > \Gamma_\mathrm{m}  n_\mathrm{bath}$ \cite{Dobrindt2008} and $\Omega_%
\mathrm{m}  >\Gamma_\mathrm{m}  n_\mathrm{bath}$ \cite{Marquardt2007} are
necessary to enable sufficiently high cooling rates without being limited by
the cavity decay rate or an effectively overdamped mechanical resonator,
respectively. Apart from these limits, it is desirable to work with optical
resonances as narrow as possible to enable ground state cooling.

\begin{figure}[tb]
\centering
\includegraphics[width=\linewidth]{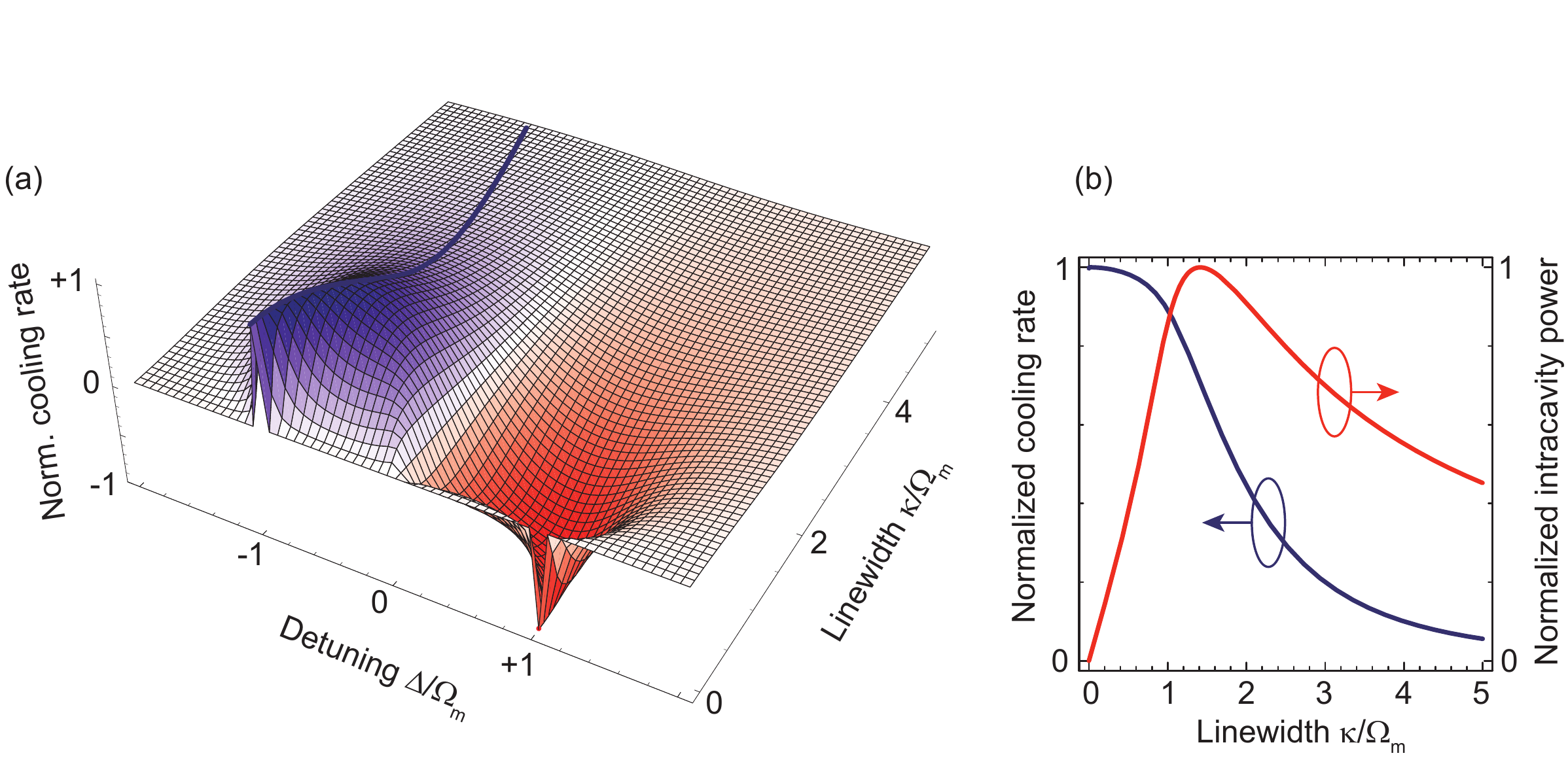} 
\caption{ (a) Normalized cooling rate as a function of detuning and
linewidth, for a fixed mechanical resonance frequency $\Omega_\mathrm{m} $
assuming $G <\protect\kappa$. Blue line indicates the optimum detuning for a
given cavity linewidth. (b) Normalized cooling rate (blue line) and
intracavity power (red line) as a function of linewidth for fixed $\Omega_%
\mathrm{m} $, if the detuning is kept at the optimum value.}
\label{f:cr3d-all}
\end{figure}

Another advantage of working in the resolved-sideband regime is that the
cooling rate increases monotonically for decreasing $\kappa$ if all other
parameters are fixed. As shown in figure~\ref{f:cr3d-all}, if the detuning
is chosen for an optimum cooling rate, the intracavity power reduces as the
laser is effectively detuned further from the WGM cavity resonance. This is
advantageous to keep undesired nonlinearities at a low level, but also to
prevent heating effects due to absorbed light. This effect is particularly
relevant in cryogenic environments, where small absorbed powers may already
significantly alter the temperature of the system, see section \ref%
{ss:measba}.

Finally, another technical imperfection in this cooling scheme may lead to a
limit in the attainable occupation number. If the cooling laser frequency
exhibits technical frequency noise, it is translated by the cavity into
radiation pressure force noise. Laser frequency fluctuations $\bar S_{\omega
\omega }(\Omega )$, with $ \bar S_{qq}^\mathrm{in}(\Omega )=1+{4 |\bar s_%
\mathrm{in} |^2} \bar S_{\omega \omega }(\Omega )/{\Omega ^2} $ lead to
additional radiation-pressure force fluctuations with the spectrum 
\begin{equation}  \label{e:sffbafn}
\bar S_{FF}^\mathrm{ba,fn}(\Omega )=\frac{\hbar^2 \bar a ^4 g_0 ^2}{\Omega }
\left(\frac{\bar \Delta }{(\bar \Delta -\Omega )^2+(\kappa/2)^2}-\frac{\bar
\Delta }{(\bar \Delta +\Omega )^2+(\kappa/2)^2}\right)\bar S_{\omega \omega
}(\Omega ),
\end{equation}
which can be derived from the intracavity fluctuations of the $p$%
-quadrature. A lower limit for the occupation can then be estimated by
integration of the resulting displacement noise spectrum, taking into
account the modified susceptibility of the mechanical oscillator (in
particular its damping $\Gamma_\mathrm{dba}\approx \Gamma_\mathrm{eff} $,
cf.\ equation (\ref{e:Gdba}). In the linear cooling regime \cite%
{Dobrindt2008}, the integrand is simply a Lorentzian, and we assume that the
force noise is approximately constant within the frequency band in which it interacts
with the
mechanical oscillator. Then one obtains 
\begin{equation}
m_\text{eff} \Omega_\mathrm{m} ^2 \langle \delta x ^2 \rangle\approx\frac{%
\bar S_{FF}^\mathrm{ba,fn}(\Omega_\mathrm{m} )}{2 m_\text{eff} \Gamma_%
\mathrm{dba}}= \underbrace{\bar S_{\omega \omega }(\Omega_\mathrm{m} )\, 
\frac{\bar a ^2 |\bar \Delta |}{\kappa \Omega_\mathrm{m} }}_{n_\mathrm{fn}}
\hbar \Omega_\mathrm{m} .
\end{equation}
As will be discussed later, this result gets modified by the additional
presence of thermal noise \cite{Schliesser2008}, not taken into account in
other discussions of frequency-noise induced cooling limits which \cite%
{Diosi2008, Rabl2009}.

\subsection{Resolved-sideband cooling of a silica microtoroid}

\label{ss:rsc}

Implementing an optomechanical system which operates in the
resolved-sideband regime is advantageous to suppress the effects of
nonlinearities and absorption-induced heating present in most laboratory
experiments. Furthermore, the analysis at the quantum level has shown that
this regime is mandatory for cooling to the quantum ground state. It is
interesting to note that this finding is quite generally valid, and applies
in an analogous manner to a large class of systems in which a mechanical
oscillator is cooled by parametric coupling to physical systems as diverse
as an ion \cite{Tian2004}, a quantum dot \cite{Wilson-Rae2004}, a Josephson
qubit \cite{Martin2004a}, a superconducting quantum interference device
(SQUID) \cite{Blencowe2007}, a superconducting single-electron transistor
(sSET) \cite{Blencowe2005, Naik2006}, or a resonant electric circuit \cite%
{Brown2007}. The excitations of all these systems need to have a lifetime $%
\kappa^{-1}$ exceeding the oscillation period $\sim\Omega_\mathrm{m} ^{-1}$
of the mechanical oscillator to enable ground-state cooling.

Experimental implementation of the resolved-sideband regime however proves
non-trivial: in a classical optomechanical system involving a Fabry-Perot\
resonator, increasing the cavity length $L$ in principle reduces $\kappa$,
but at the same time also the coupling strength $g_0 =-\omega_\mathrm{c} /L$%
. Also, it typically increases the waist size of the beam, which may induce
diffraction losses (and therefore increase $\kappa$ again) if, as it is
typically the case, the micromechanical oscillator constitutes one of the
cavity mirrors. In an integrated system such as silica microtoroids,
increasing the cavity size (within the relevant range of 60 to 100 $\,%
\mathrm{\mu m} $\ diameter) typically leads only to modest improvements of $%
\kappa$, as the losses are dominated by absorption along the light
propagation path. In addition to the reduced coupling $g_0 =-\omega_\mathrm{c%
} /R$, the mechanical resonance frequency of the RBM is reduced for larger
cavity size. The first experiment demonstrating efficient resolved-sideband
laser cooling was eventually performed in 2008 with silica microresonators 
\cite{Schliesser2008}, while other systems, such as superconducting
microwave cavities \cite{Teufel2008} or Fabry-Perot\ resonators \cite%
{Groblacher2009} followed soon thereafter.

\begin{figure}[hbtp]
\centering
\includegraphics[width= .5\linewidth]{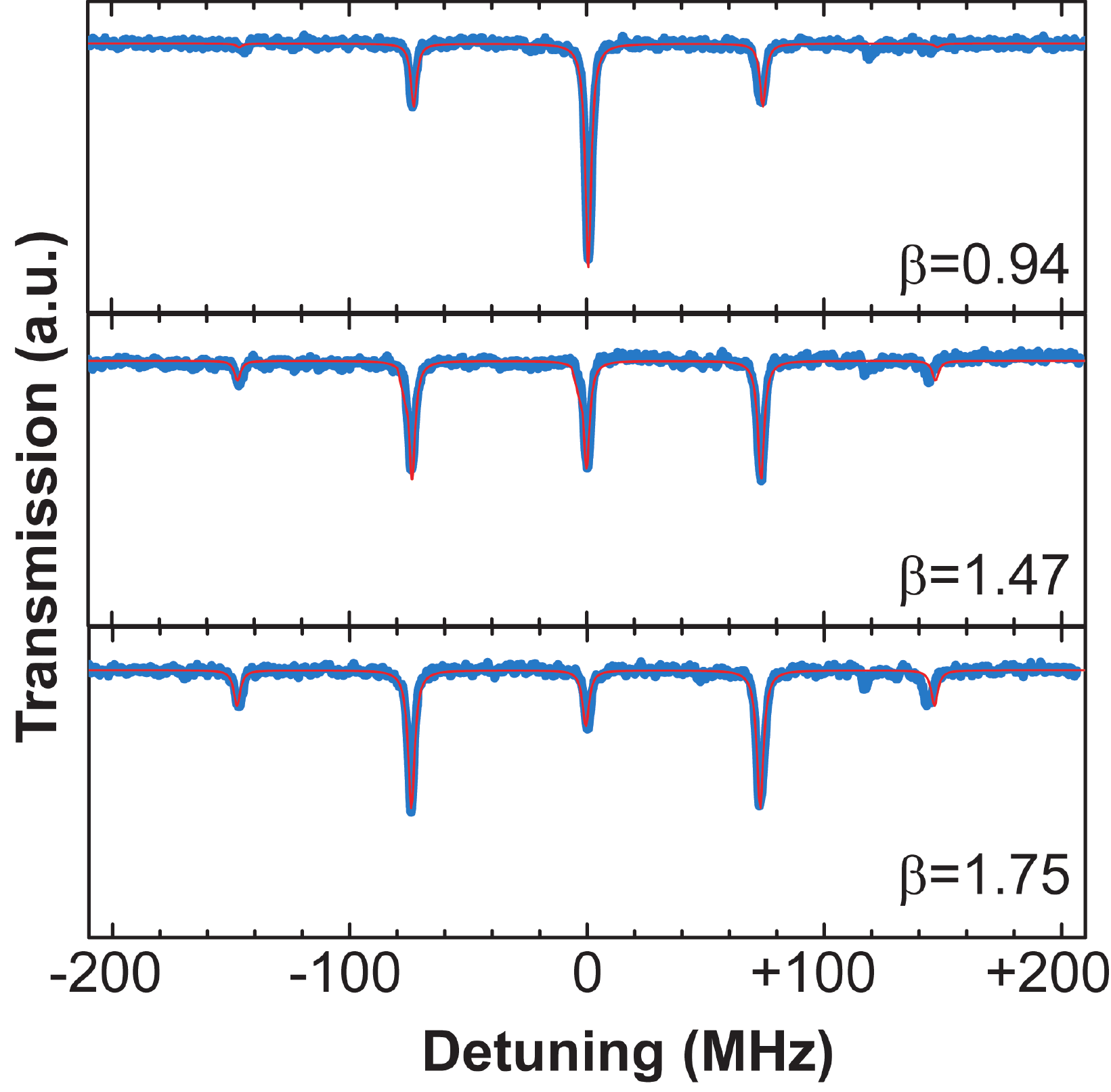}
\caption{Steady-state transmission spectrum of a microtoroid when the
mechanical degree of freedom is excited to picometer-scale amplitude
oscillations at its resonance frequency of $\Omega_\mathrm{m} /2\protect\pi%
=73{.}5 \,\mathrm{MHz} $ using an auxiliary laser. The linewidth of the
resonances corresponds to the optical decay rate of $\protect\kappa/2\protect%
\pi=3{.}2\,\mathrm{MHz} $, placing this device deeply into the resolved
sideband regime. Blue points are experimental data, and the red lines are
fits using the Bessel function expansion (equation (\protect\ref%
{e:oscillatingTranmission})) with $\protect\beta$-parameters given in each
panel. Figure from ref.\ \protect\cite{Schliesser2008}. }
\label{f:resolvedSidebands}
\end{figure}

Figure~\ref{f:resolvedSidebands} shows data obtained with a $47\,\mathrm{\mu
m}  $-diameter toroid hosting a RBM at $\Omega_\mathrm{m} /2\pi=73.5\,%
\mathrm{MHz} $. A first laser drives the RBM to regenerative
oscillations, by pumping a WGM in a blue-detuned manner, leading to the
well-known oscillatory instability \cite{Kippenberg2005, Carmon2005,
Rokhsari2005, Rokhsari2006a,Hossein-Zadeh2006}. A second laser, launched
into the same tapered fibre, is used to probe a different, \mbox{high-Q} WGM
resonance of the oscillating toroid. Using optical filters to extract only
the transmission signal of the probe laser we obtained  the traces shown in figure~\ref%
{f:resolvedSidebands} as this laser is swept over the resonance.

Instead of a single dip, several dips are observed if the electronic signal
is filtered with a low-pass filter with a cutoff far below $\Omega_\mathrm{m}
$. From the intracavity mode amplitude of the oscillating cavity (equation (%
\ref{e:sidebands})) the low-frequency component of the transmitted power 
\begin{equation}
|\bar s_\mathrm{out}^{\mathrm{DC}}|^2=\left(1-\eta_\mathrm{c} (1-\eta_%
\mathrm{c} ) \sum_n \frac{\kappa^2 J_n(\beta)^2}{(\Delta+n \Omega_\mathrm{m}
)^2+(\kappa/2)^2}\right)|\bar s_\mathrm{in} ^2|
\label{e:oscillatingTranmission}
\end{equation}
is calculated \cite{Schliesser2008}, where $J_n$ are the Bessel functions
and $\beta=g_0 x_0/\Omega_\mathrm{m} $. From fits using this model, it is
possible to derive the amplitude $x_0$ of the mechanical oscillation as
about $5.3$, $8.3$ and $9.9\,\mathrm{pm} $ for the three traces,
respectively. At the same time, these traces clearly show that the
mechanical resonance frequency $\Omega_\mathrm{m} $---by which the
Lorentzian dips are spaced---largely exceeds the optical cavity linewidth $%
\kappa$. Indeed, the fits yield a resolved-sideband factor $\Omega_\mathrm{m}
/\kappa\approx 23$, due to the very high cavity finesse of $440{,}000$.

\begin{figure}[btp]
\centering
\includegraphics[width= .8\linewidth]{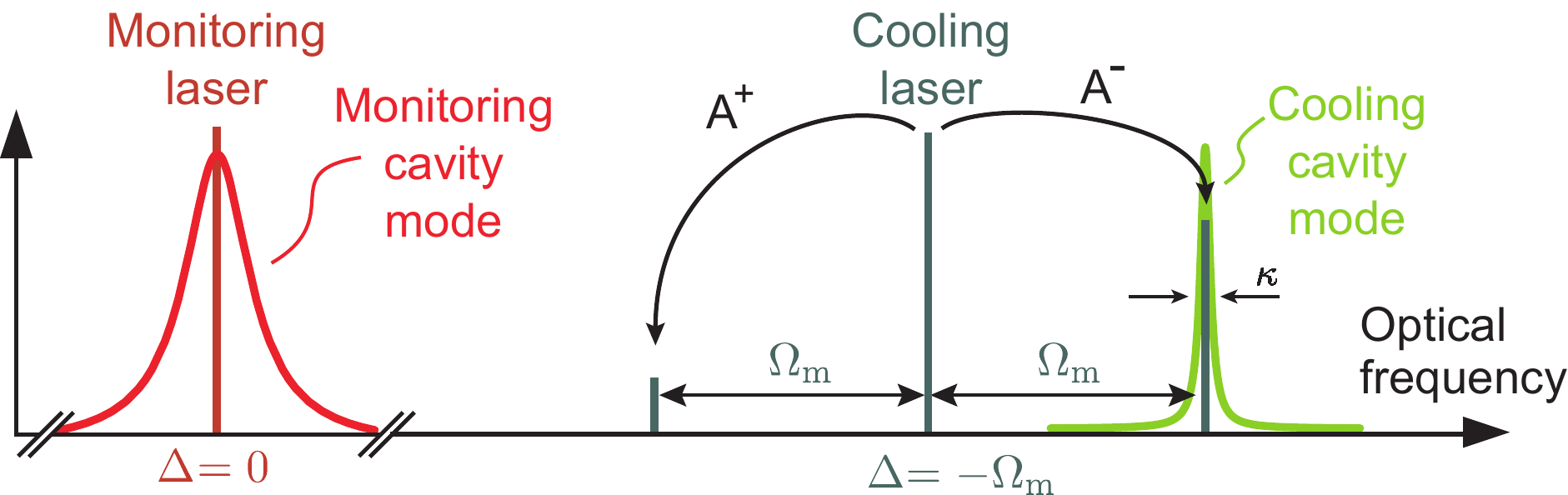}
\caption{Scheme used for resolved-sideband cooling. The cooling laser at $%
\protect\lambda\approx 980\,\mathrm{nm} $ (green line) is tuned to the lower
sideband ($\bar \Delta =-\Omega_\mathrm{m} $) of a high-Q optical resonance.
Resonant anti-Stokes scattering into the cavity mode, at rate $A_-$
dominates over Stokes scattering at rate $A_+$. A second laser is tuned in
resonance with a different WGM at $\protect\lambda=1064\,\mathrm{nm} $, and
used to monitor the mechanical displacements. Figure from ref.\ \protect\cite%
{Schliesser2008}.}
\label{f:rscScheme}
\end{figure}

For resolved-sideband cooling, the two lasers are used in a different way.
The cooling laser is tuned below the narrow WGM, to the lower sideband at $%
\bar \Delta =-\Omega_\mathrm{m} $. The second laser is used to sensitively
monitor mechanical displacements and therefore is tuned in resonance with a
different WGM (figure~\ref{f:rscScheme}). Using two completely independent
laser sources helps to rule out artifacts in the displacement measurements,
due to potentially induced noise correlations between the cooling beam and
induced motion of the mechanical oscillator, referred to as ``squashing'' in
feedback cooling \cite{Poggio2007}.

\begin{figure}[btp]
\centering
\includegraphics[width= .75\linewidth]{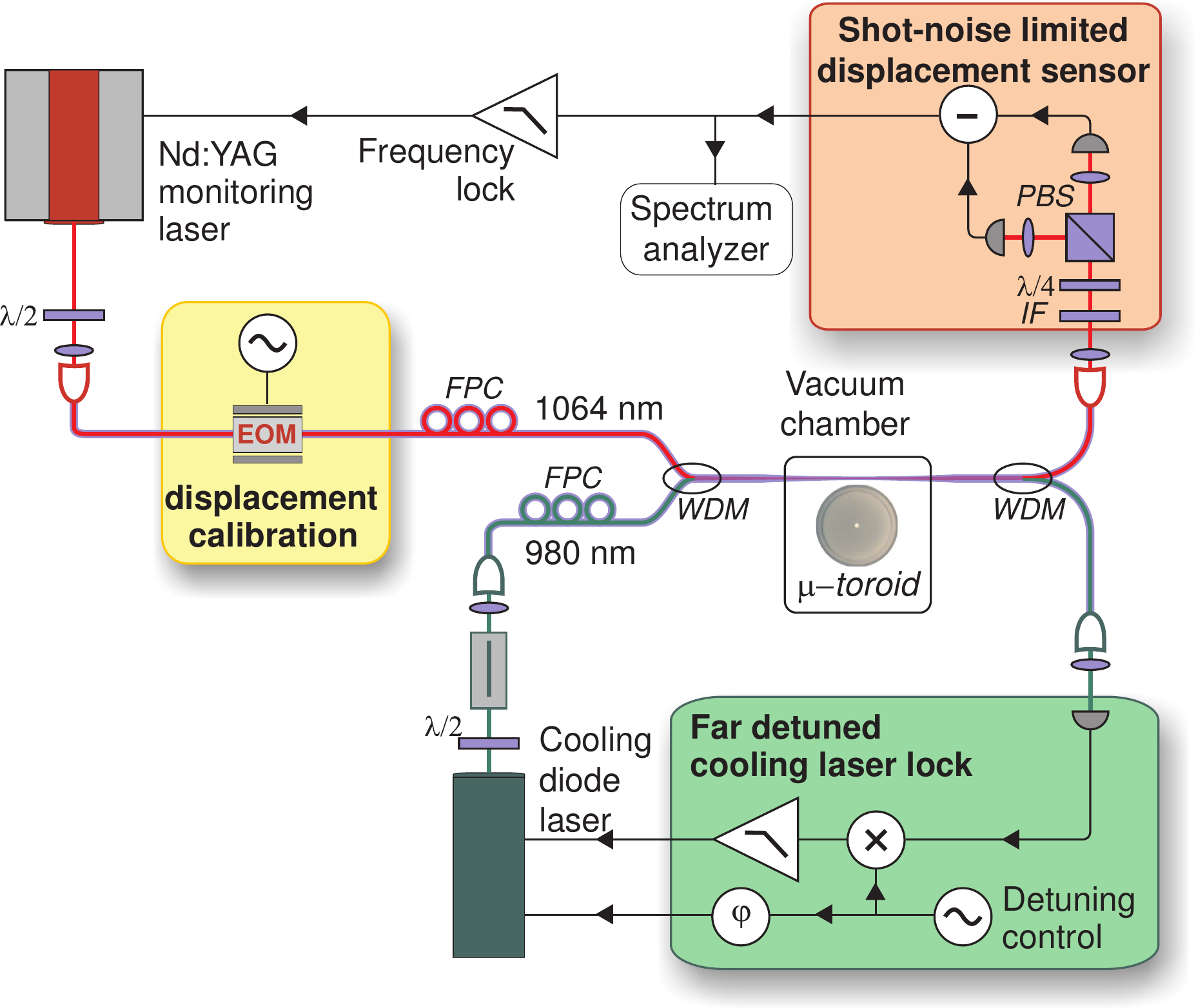}
\caption{Setup used to implement resolved-sideband cooling and
high-sensitivity monitoring of the RBM of silica toroidal microresonators.
Two lasers are used, a diode laser at $980\,\mathrm{nm} $ for cooling by
tuning and locking it to the lower sideband of a high-Q WGM, and a Nd:YAG
laser at $1064\,\mathrm{nm} $ monitoring mechanical displacements using the
H\"ansch-Couillaud technique. More details are given in the text. Figure
from ref.\ \protect\cite{Schliesser2008}.}
\label{f:rscSetup}
\end{figure}

In this experiment, sensitive monitoring of mechanical motion is
accomplished using the H\"ansch-Couillaud technique described in section \ref%
{ss:HC}. As shown in figure~\ref{f:rscSetup}, a low-noise Nd:YAG laser is
used for this purpose, which is locked to the line-center using feedback to
a piezo in contact with the YAG crystal. In order to lock the cooling laser to a
detuning much larger than the resonance linewidth $\kappa$, we use an
experimental technique, in which the signal obtained by frequency modulation
spectroscopy \cite{Bjorklund1983} is used as an error signal. 

\begin{figure}[btp]
\centering
\includegraphics[width= .8\linewidth]{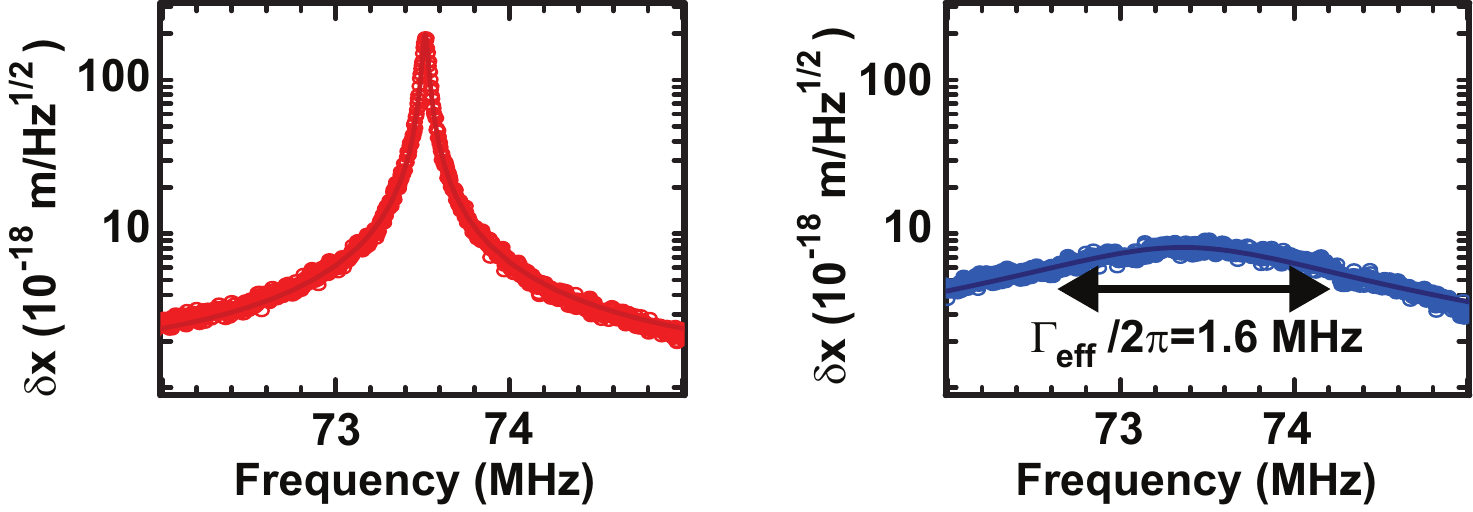}
\caption{Displacement noise spectra of the RBM as recorded by the monitoring
laser with the cooling laser off (left panel), and running at a power of $2{.%
}7\,\mathrm{mW} $, when detuned to the lower sideband (right panel). Cooling
rates up to $1.6\,\mathrm{MHz} $ can be achieved. Circles are data points
and lines Lorentzian fits. Figure from ref.\ \protect\cite{Schliesser2008}.}
\label{f:rscStrongCooling}
\end{figure}

Figure~\ref{f:rscStrongCooling} shows cooling results obtained with the $%
73.5\,\mathrm{MHz} $-sample. Note that due to the strongly detuned
operation, only a fraction of about $\sim(4(\Omega_\mathrm{m}
/\kappa)^2+1)^{-1}\approx 5 \times 10^{-4}$ of the launched power of $3\,%
\mathrm{mW} $, i.\ e.\ circa $1.5\,\mathrm{\mu W} $, is coupled into the
cavity. Nonetheless, very high cooling rates up to $\Gamma_\mathrm{dba}%
/2\pi=1{.}56\,\mathrm{MHz} $ can be achieved. The mode temperatures achieved
with this sample, however, remained above $20\,\mathrm{K} $, due to the
modest mechanical quality factor of about $Q_\mathrm{m}=2{,}100$, and a
heating mechanism described below.

\begin{figure}[btp]
\centering
\includegraphics[width= .4\linewidth]{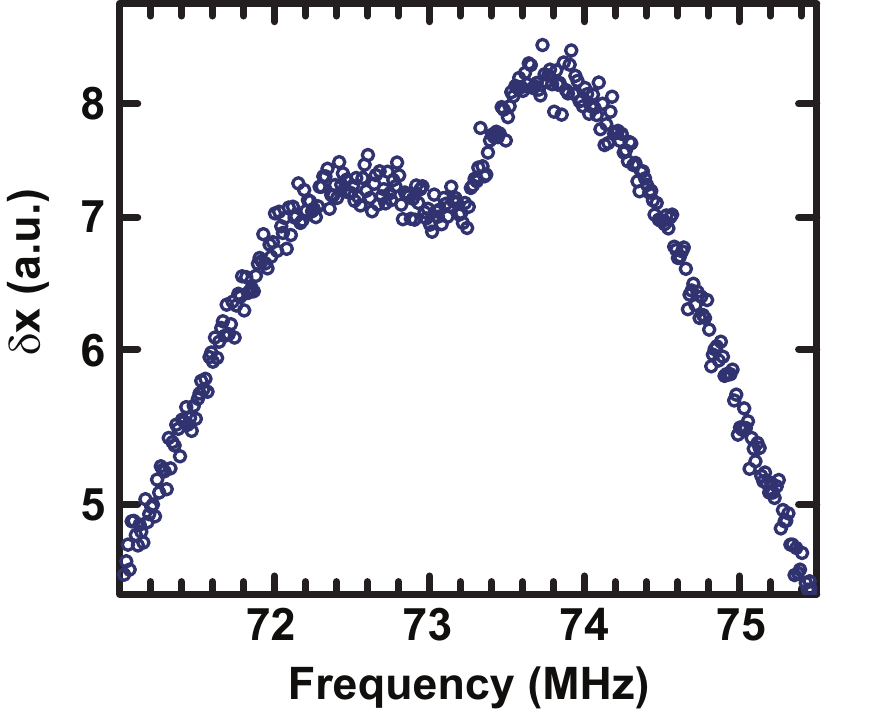}
\caption{Resonance doublet observed in the power spectrum of the cooling
laser for strong cooling powers. See text for more information. }
\label{f:nms}
\end{figure}

We have also recorded noise spectra of the mechanical oscillator by directly
analyzing the transmitted power of the cooling laser (as explained in
section \ref{s:dynba}). Qualitatively the same behavior of the mechanical
spectra is observed during cooling, however with significantly worse
signal-to-noise ratio. This renders, for example, the determination of the
mode temperature extremely difficult for strong cooling. Furthermore, for
cooling laser powers on the order of $3\,\mathrm{mW} $, significant
deformations of the mechanical spectra from their originally Lorentzian
shape were observed. The particular shapes were strongly dependent on the
detuning and coupling conditions. Figure \ref{f:nms} shows one example in
which a resonance doublet appeared. We note that such an doublet would in
principle be expected when entering the regime of strong optomechanical
coupling, in which optical and mechanical modes hybridize \cite%
{Dobrindt2008, Groblacher2009a}.

\begin{figure}[btp]
\centering
\includegraphics[width= .8\linewidth]{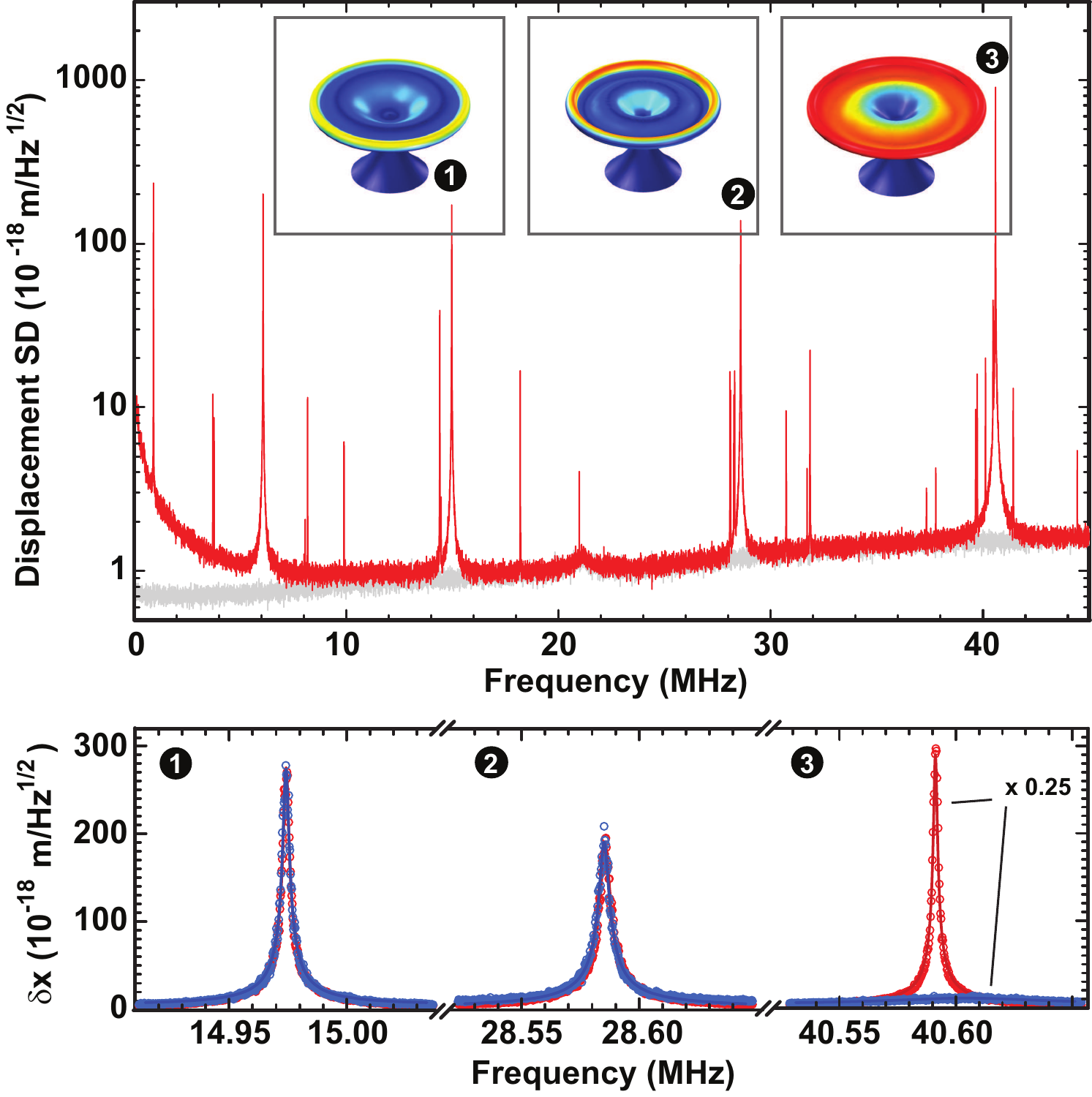}
\caption{Resolved-sideband cooling of the RBM of a microtoroidal oscillator.
Top panel shows a broadband displacement noise spectrum recorded using the
H\"ansch-Couillaud technique (red trace). Gray trace is background trace
with the fiber taper retracted from the WGM near field. It is dominated by
shot noise, its dependence on Fourier frequency is due to the weaker
transduction of displacement fluctuations to phase fluctuations at Fourier
frequencies beyond the cavity cutoff $\protect\kappa/2$. The sensitivity is
at the $10^{-18} \,\mathrm{m/\protect\sqrt{Hz}} $-level. Insets show the
displacement patterns of three radially symmetric modes, with the
corresponding strain indicated in the color code. The bottom panel shows
zooms on the displacement noise of these three modes, when the cooling laser
is off (red traces) and on (blue traces), at a detuning of $\bar \Delta
=-\Omega_\mathrm{m} $. Clearly, only the RBM mode (number 3) is affected.
The lowest achieved occupation number in these measurements was $\langle
n\rangle\approx 5900$. Figure from ref.\ \protect\cite{Schliesser2008}.}
\label{f:rsc}
\end{figure}

To achieve lower mode temperatures, a second cooling run was initiated. We
used a larger toroid (radius $R=28\,\mathrm{\mu m}  $) with a RBM at $\Omega_%
\mathrm{m} /2\pi=40{.}6 \,\mathrm{MHz} $ and $\Gamma_\mathrm{m} /2\pi =1{.}%
3\,\mathrm{kHz} $, corresponding to a very high mechanical quality factor of 
$30{,}000$. The broadband displacement spectrum recorded with this sample
using the H\"ansch-Couillaud technique (cooling laser is off) is shown in
figure~\ref{f:rsc}. A displacement sensitivity at the $10^{-18}\,\mathrm{m/%
\sqrt{Hz}} $-level is achieved. At low Fourier frequencies, the noise
spectrum again reveals a thermorefractive background already discussed in
section \ref{s:UHS}.

The strong peaks observed at $14.96$, $28{.}58$ and $40.59\,\mathrm{MHz} $
can be assigned to different radially symmetric modes in this sample. The
latter peak belongs to the RBM, which has an effective mass of about $10\,%
\mathrm{ng} $. If the cooling laser is tuned to the lower sideband of a $%
\kappa/2\pi=5{.}8 \,\mathrm{MHz} $-wide WGM, cooling of the RBM is evident
by the reduction of the RBM's thermal noise (figure~\ref{f:rsc}). It is
interesting to note that the thermal noise in the other radially symmetric
modes is not affected. In the resolved sideband regime, this is possible due
to the fact that the absorption sidebands of the individual mechanical modes
do not overlap, as they are as narrow as the optical resonance. Furthermore,
we note that effects analogous to ``sympathetic cooling''---due to, for
example a preferential coupling of the RBM to other radially symmetric
modes---is not observed.

The mode temperature $T_\mathrm{m}$ of the RBM is subsequently evaluated by
integrating the displacement noise spectrum $\bar S_{xx}(\Omega )$ (cf.\
section \ref{ss:coolingByDBA}). The lowest mode temperature obtained with
this sample is $T_\mathrm{m}\approx 11\,\mathrm{K} $, corresponding to a
residual occupation of $\langle n\rangle\approx 5900$. As in the experiments
with the first sample, this is significantly above the value expected from
the high cooling rates, reaching up to $\Gamma_\mathrm{dba}/2\pi=119\,%
\mathrm{kHz} $ with this sample. This discrepancy is attributed to excess
heating by the cooling laser's frequency noise.

Indeed, an independent measurement reveals frequency fluctuations of the
cooling laser at the level of $\sqrt{\bar S_{\omega  \omega }(\Omega_\mathrm{%
m} )}=2 \pi\cdot 200 \,\mathrm{Hz/\sqrt{Hz}} $, a value consistent with
earlier measurements on the frequency noise at radio frequencies of a
grating stabilized diode laser \cite{Zhang1995}. For a finite bath
temperature $T$, the lowest occupation that can be attained can be estimated
by 
\begin{equation}
\langle n \rangle\approx\frac{\bar S_{FF}^{\mathrm{ba,fn}}(\Omega_\mathrm{m}
)+\bar S_{FF}^\mathrm{th}(\Omega_\mathrm{m} )} {2 m_\text{eff} \Gamma_%
\mathrm{dba}}.
\end{equation}
In the resolved-sideband limit $\kappa\ll\Omega_\mathrm{m} $, the force
noise due to frequency fluctuations from equation (\ref{e:sffbafn})
simplifies to 
\begin{equation}
\bar S_{FF}^{\mathrm{ba,fn}}(\Omega_\mathrm{m} )\approx \frac{4\eta_\mathrm{c%
} ^2 \bar S_{\omega \omega }(\Omega_\mathrm{m} ) P_\mathrm{in}^2}{R^2 \Omega_%
\mathrm{m} ^4},
\end{equation}
and the cooling rate is approximately given by 
\begin{equation}
\Gamma_\mathrm{dba}\approx \frac{4 g_0 ^2 \bar a ^2 x_\mathrm{ZPF} ^2}{\kappa%
}=\frac{2 g_0 ^2 \eta_\mathrm{c}  P_\mathrm{in}}{m_\text{eff} \omega_\mathrm{%
l}  \Omega_\mathrm{m} ^3}.
\end{equation}
As a consequence, the lowest temperature is obtained at a finite input
power, and one finds $n_\mathrm{min}\approx\sqrt{2 k_\mathrm{B} T m_\text{eff%
} \Gamma_\mathrm{m}  \bar S_{\omega \omega }(\Omega_\mathrm{m} )}/\hbar |g_0
|$ \cite{Schliesser2008}. For the parameters of this experimental run, $%
T=300\,\mathrm{K} $, $m_\text{eff}=10\,\mathrm{ng} $, $\Gamma_\mathrm{m}
/2\pi=1.3\,\mathrm{kHz} $, $\sqrt{\bar S_{\omega \omega }(\Omega_\mathrm{m}
)/\Omega_\mathrm{m} ^2}\approx 4\,\mathrm{\mu rad/\sqrt{Hz}} $, $R=38\,%
\mathrm{\mu m}  $, $\Omega_\mathrm{m} /2\pi=40.6\,\mathrm{MHz} $, $\omega
/2\pi\approx300\,\mathrm{THz} $, one obtains $n_\mathrm{min}\approx 5200$,
close to the lowest observed occupation number.

\subsection{Direct sideband spectroscopy}

If the ground state of the mechanical oscillator is approached, it becomes
increasingly difficult to reliably measure its occupation number. For
example, if the method demonstrated here---using an independent motion
transducer based on an optical phase measurement---the requirements on its
performance are quite challenging. Even for a perfect implementation of such
a measurement, in which the standard quantum limit can be reached (cf.\
section \ref{s:UHS}), the signal-to-background ratio at the mechanical
resonance frequency would be only equal to unity at the mechanical resonance
frequency when the resonator is in its ground state: for $\langle
n\rangle\rightarrow 0$, the spectrum of displacement fluctuations is expected
to peak at a value of 
\begin{equation}
\bar S_{xx}^0(\Omega_\mathrm{m} )=\frac{\hbar}{m_\text{eff} \Gamma_\mathrm{m}
\Omega_\mathrm{m} },
\end{equation}
which exactly equals $\bar S_{xx}^\mathrm{SQL}(\Omega_\mathrm{m} )$.

We note that our experiment reaches an imprecision level close to the SQL:
For example, for the sample described above, we calculate $\sqrt{\bar
S_{xx}^0(\Omega_\mathrm{m} )}\approx 2.2 \,\mathrm{am/\sqrt{Hz}} $, and
routinely achieve imprecision noise at the same level (cf.\ figure~\ref%
{f:rsc}). However, as in most other experiments reporting similarly low
imprecision levels \cite{Regal2008, Teufel2009}, backaction of these
measurements is difficult to quantify and may impede cooling to the quantum
ground state.

As an alternative approach to measure the residual occupation number for
small $\langle n \rangle $, it has been suggested to directly monitor the
individual motional sidebands of the cooling laser \cite{Wilson-Rae2007}. In
theory, the optical spectrum\footnote{%
Note that here the high-$Q$ approximation for the mechanical susceptibility
was used, $((\Omega_\mathrm{m} ^2-\Omega ^2)^2-\Omega ^2 \Gamma_\mathrm{eff}
^2)^{-1}\approx (4 \Omega_\mathrm{m} ^2 ( (\Omega -\Omega_\mathrm{m}
)^2+(\Gamma_\mathrm{eff} /2)^2))^{-1}$. We also emphasize that this spectrum
lies in the optical domain, instead of the RF-domain of the spectra
discussed previously.} of the cooling laser light coupled back into the
taper is described by the expression 
\begin{align}
S_{\Phi\Phi}(\omega )&=\eta_\mathrm{c}  \left(\delta(\omega -\omega_\mathrm{l%
} )\left(\frac{1}{\eta_\mathrm{c} }-\frac{(1-\eta_\mathrm{c} ) \kappa^2}{%
\bar \Delta ^2+(\kappa/2)^2}\right)\bar s_\mathrm{in} ^2 \right.  \notag \\
&\quad\quad\quad+\frac{ \Gamma_\mathrm{eff} /(2\pi)}{(\omega -(\omega_%
\mathrm{l} +\Omega_\mathrm{m} ))^2+(\Gamma_\mathrm{eff} /2)^2}\,A_- \langle
n\rangle  \notag \\
&\quad\quad\quad\left.+ \frac{\Gamma_\mathrm{eff} /(2\pi)}{(\omega -(\omega_%
\mathrm{l} -\Omega_\mathrm{m} ))^2+(\Gamma_\mathrm{eff} /2)^2}\,A_+
\left(\langle n\rangle+1\right)\right),
\end{align}
where $\Phi$ is the photon flux emerging from the cavity \cite%
{Wilson-Rae2007}. Next to the central line at the carrier frequency $\omega_\mathrm{l} $,
two sidebands
are expected, their lineshape being determined by the effective susceptibility of
the mechanical oscillator. Most notably, the spectrally integrated photon
flux of the individual sidebands is given by $A_- \,\langle n\rangle$ and $%
A_+ \left(\langle n\rangle+1\right)$ for the upper and lower sidebands,
respectively. As a consequence, the sideband asymmetry, given initially by
the asymmetry in $A_-$ and $A_+$, becomes balanced for sufficiently low $%
\langle n \rangle$. The change in the ratio of the sideband amplitudes could
therefore serve as a gauge of the occupation \cite{Wilson-Rae2007}.

Accessing the individual sidebands in a measurement necessitates the ability
to individually resolve them against the much stronger carrier signal at the
laser frequency. For typical parameters of an optomechanical experiment,
this is difficult, as in the RSB regime the power even in the resonantly
enhanced upper sideband is weaker by $\sim \langle n \rangle g_0 ^2 4 \eta_%
\mathrm{c} ^2 x_\mathrm{ZPF} ^2/\Omega_\mathrm{m} ^2\sim\langle n\rangle 
\mathcal{O}(10^{-9})$ than the carrier, and the lower sideband is again
weaker by a factor of $\kappa^2/16\Omega_\mathrm{m} ^2$. As a consequence,
sufficient suppression of the carrier using a (single) filtering cavity is
hardly possible, as it would require sub-kHz cavity linewidth, which could
not even simultaneously collect all light in the sidebands (typically $%
\Gamma_\mathrm{eff} /2\pi>1 \,\mathrm{kHz} )$.

An alternative way of individually resolving the sidebands is to use a
heterodyne technique. A similar technique has been demonstrated to enable
resolving motional sidebands\footnote{%
Note however that only micro-motion sidebands could be detected for a cold
ion. To observe the secular motion, the ion was driven to large-amplitude
oscillations in the trap.} of a laser-cooled ion \cite{Raab2000}. In such an
experiment, the cooling light which couples back to the fiber taper is mixed
with a strong local oscillator beam at a different frequency $\omega_\mathrm{%
l} -\Omega _\mathrm{AOM}$ with $\Omega _\mathrm{AOM}>\Omega_\mathrm{m} $.
Then the upper sideband, carrier, and lower sideband signals are detected at
the radio frequencies $\Omega _\mathrm{AOM}+\Omega_\mathrm{m} $, $\Omega _%
\mathrm{AOM}$ and $\Omega _\mathrm{AOM}-\Omega_\mathrm{m} $, respectively.

\begin{figure}[bt]
\centering
\includegraphics[width= .7 \linewidth]{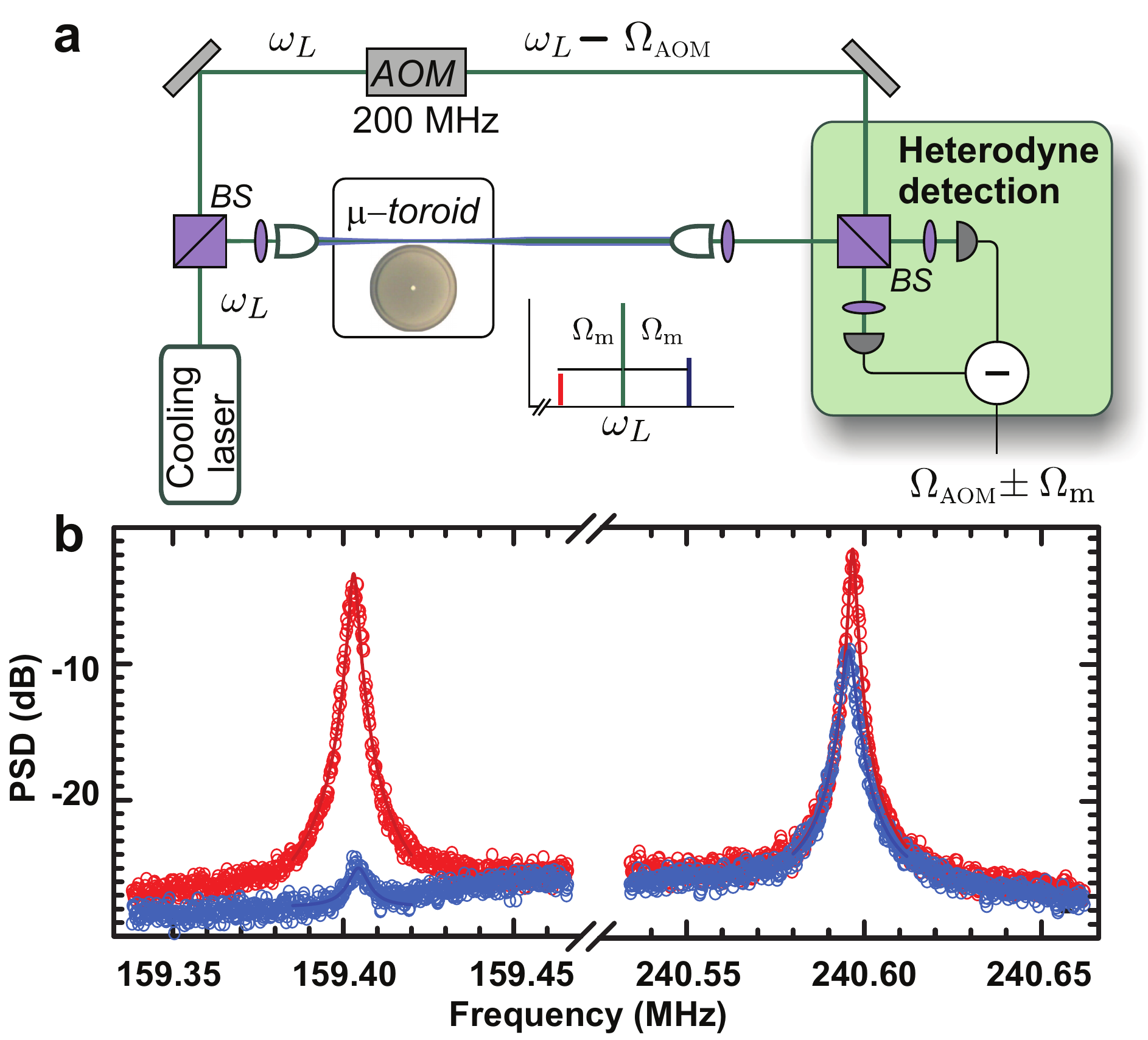}
\caption{Heterodyne spectroscopy of the motional sidebands during cooling.
(a) Experimental setup, in which a frequency-shifted laser beam is used as a
heterodyne local oscillator. (b) Experimental results, showing the power
spectral density of the differential photocurrent recorded in the heterodyne
receiver. Red points are recorded with the cooling laser close to resonance,
and blue points are recorded with a detuned cooling laser. Figure from ref.\ 
\protect\cite{Schliesser2008}.}
\label{f:sidebandSpectroscopy}
\end{figure}

Figure~\ref{f:sidebandSpectroscopy} shows the implementation and results of
a proof-of-principle experiment using this technique. When the laser is
tuned close to resonance, the motional sidebands have roughly the same
amplitude, which are again independently calibrated using a
frequency-modulation technique. As $\langle n\rangle \gg 1$, the residual
asymmetry in this case is attributed to a finite detuning from the WGM
resonance. When the laser is further detuned, the asymmetry becomes more
pronounced, as $A_-\gg A_+$. In this experiment, an asymmetry of more than $%
15\,\mathrm{dB} $ is reached. Higher values could not be observed due to the
limited SBR. This limitation is due to the classical frequency noise of the
cooling laser, which could be shown to induce the relatively high background
noise level in this measurement.

We note however that, in contrast to the experiment performed on trapped
ions \cite{Raab2000}, this experiment was performed without actively driving
the motion of the mechanical oscillator. In our case, the oscillator is
exclusively driven by thermal noise. In that sense, this measurement
technique beautifully reveals in a very direct manner the physical process
underlying cooling by dynamical backaction: enhanced anti-Stokes scattering
and suppressed Stokes-scattering.

\clearpage

\section{Approaching the quantum ground state}

\label{s:cryo}

The work presented in the previous sections has clearly demonstrated the
potential of resolved-sideband cooling. In the following, we present
experimental results in which this technique is successfully implemented to
cool a mechanical oscillator close to its quantum ground state, rivalling
the results achieved with nanoelectromechanical systems engineered for this
purpose for nearly a decade \cite{Knobel2003, LaHaye2004, Schwab2005,
Naik2006, Rocheleau2009}. In our laboratory these results \cite%
{Schliesser2009a} were achieved by combining laser cooling with efficient
cryogenic precooling, a technique traditionally applied to
nanoelectromechanical systems, but novel in the context of optomechanics.

As a second crucial feature, the vastly superior displacement sensitivity of
our optomechanical system warrants the ability to monitor such an ultracold
oscillator at levels close to the amplitude of its zero-point fluctuations $%
x_\mathrm{ZPF} =\sqrt{\hbar/2m_\text{eff}\Omega_\mathrm{m} }$---in spite of
the fact that its mass is more than 1000-times higher than in typical
nanomechanical systems.

Finally, considered from the perspective of quantum measurement, we are able
to assess the backaction of the measurement performed on the ultracold
oscillator. We will show that considering both imprecision and backaction of
our measurement, one can conclude that the optical displacement transduction
performs in a near-ideal manner.

\subsection{Implementation}

From the theoretical considerations of the previous sections, we can
summarize the limits in cooling by dynamical backaction in the simple
expression 
\begin{equation}
\langle n \rangle \geq \frac{\Gamma_\mathrm{m} }{\Gamma_\mathrm{m} +\Gamma_%
\mathrm{dba}} \frac{k_\mathrm{B} T}{\hbar \Omega_\mathrm{m} } +n_\mathrm{fn}
+\frac{A_+}{A_--A_+}.
\end{equation}
Reaching $\langle n \rangle\rightarrow 0$ therefore requires low mechanical
dissipation $\Gamma_\mathrm{m} $, high cooling rates $\Gamma_\mathrm{dba}$,
a low environment temperature $T$, high mechanical resonance frequency $%
\Omega_\mathrm{m} $, low frequency noise $S_{\omega \omega }(\Omega_\mathrm{m%
} )\propto n_\mathrm{fn}$ of the cooling laser (or, more general, no
technical noise of any kind), and operation in the resolved-sideband regime
to achieve low $A_+/(A_--A_+)\approx \kappa^2/16\Omega_\mathrm{m} ^2$.

After the optimization of the silica microtoroids for low dissipation, high
frequency and operation deeply in the resolved-sideband regime, and the
demonstration of very high cooling rates, the most significant performance
improvements are expected for operation in a cryogenic experiment, and using
a cooling laser devoid of any technical noise at the relevant radio
frequencies. In the following, we briefly describe the implementation
of these two advances.

\subsubsection{Cryogenic optomechanical experiment}

For the successful cryogenic operation of an optomechanical cooling
experiment, it is important that the concrete technical implementation
provides a proper thermalization of the sample to the cryogen, even in
presence of a microwatt-scale heat load through residual light absorption in
the sample. This may be a non-trivial task, as strong thermal anchoring of
the mechanically compliant device would in many cases imply designs opposite
to the requirement of low clamping losses of the mechanical oscillator. For
example, the highest mechanical quality factors in silica microresonators
were achieved with toroids held only on a sub-micrometric pillar \cite%
{Schliesser2008}, or suspended from silica spokes \cite{Anetsberger2008}.
Heat transport from the region of the optical mode in the torus to a heat
sink below the silicon chip can be expected to be prohibitively low.

\begin{figure}[btp]
\centering
\includegraphics[width=  \linewidth]{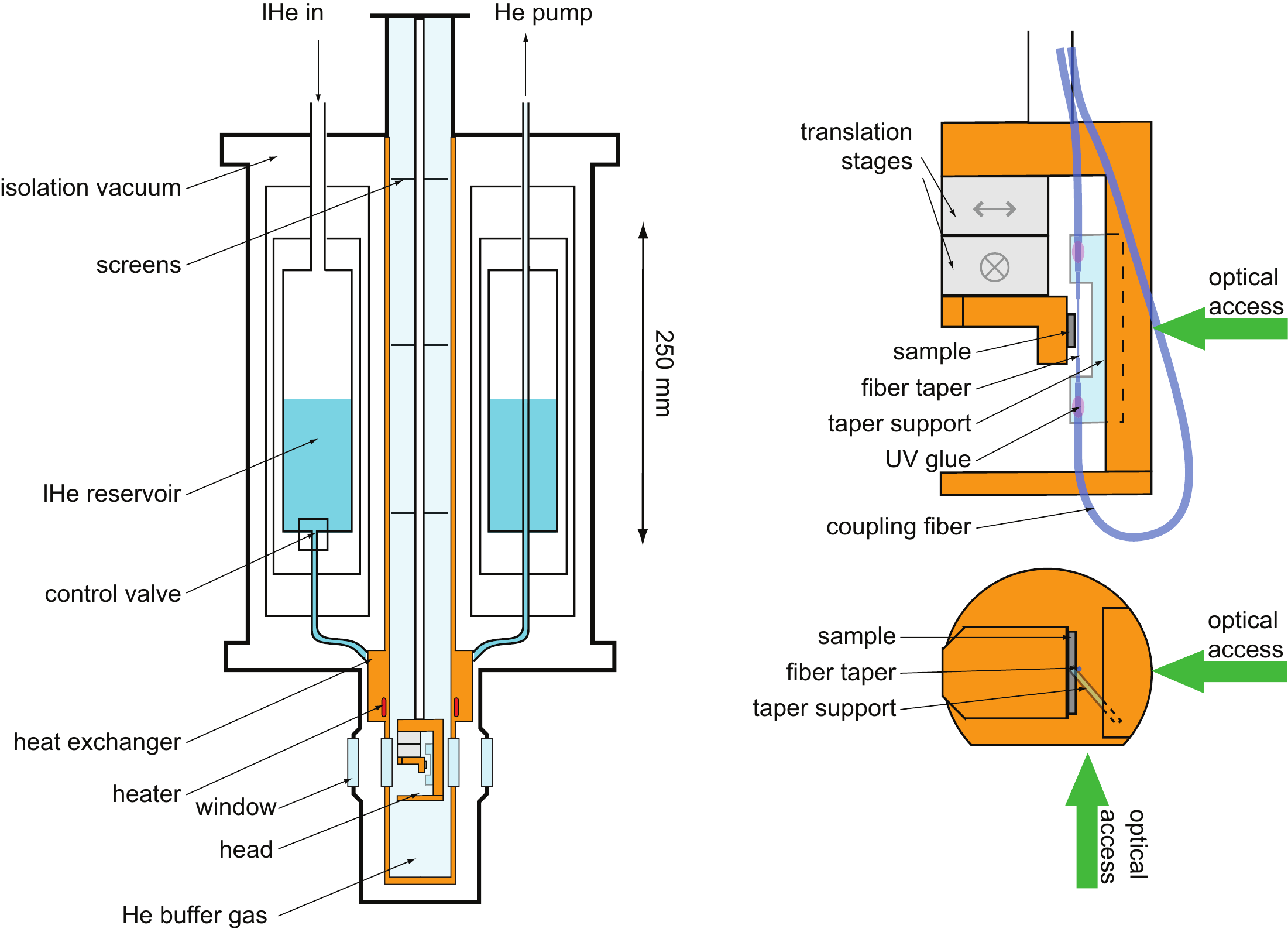}
\caption{Implementation of cryogenic cooling. Left panel: schematic drawing
of the employed buffer gas cryostat. Liquid ${}^4\mathrm{He}$, kept in an
isolated reservoir, evaporates when pumped through a capillary (``heat
exchanger'') and thereby cools low-pressure ($0{.}1\,\mathrm{mbar} $-$50\,%
\mathrm{mbar} $) ${}^4\mathrm{He}$-gas in the sample space to temperatures
down to $1{.}6 \,\mathrm{K} $. However, a heater can also be used to
stabilize the temperature of the buffer gas to a different value in the full
range up to $300\,\mathrm{K} $. The experimental assembly (``head'') is
directly immersed into the buffer gas. Right panel: side (top) and bottom
(bottom) view of the experimental head. The coupling taper, glued to a glass
taper mount, is fixed, while the position of the chip containing the silica
microresonators can be adjusted using stick-slip piezoelectric actuators.
Optical access from two orthogonal directions allows convenient addressing
the toroids on a chip, and precision positioning for coupling to the tapered
fiber.}
\label{f:cryostat}
\end{figure}

For this reason, we chose to directly immerse the sample into the cryogen, a
dilute helium gas, similar to an early experiment with silica microspheres 
\cite{Treussart1998} immersed in superfluid helium (figure~\ref{f:cryostat}%
). In contrast to other cryogenic experiments with optical microcavities
employing a cold finger cryostat \cite{Srinivasan2007,Groblacher2008}, this
allows us to reliably thermalize the sample to the base temperature of the
cryostat ($1{.}6\,\mathrm{K} $). At the same time, due its low pressure
(typically held in the range $0{.}1\,\mathrm{mbar} $-$50\,\mathrm{mbar} $),
and the increased intrinsic damping of the mechanical mode due to two-level
fluctuators (see section \ref{sss:dissipation}), the helium gas present in
the sample chamber does not induce significant damping of the mechanical
mode.

The employed commercial Optistat SXM cryostat (Oxford instruments)
implements such a cooling scheme by providing two completely independent
helium containers (cf.\ figure~\ref{f:cryostat}): A larger (volume $4{.}3$ liters),
 thermally insulated reservoir contains liquid helium. From
this reservoir, a rotary pump continuously draws helium through a capillary
contained in a copper block. During this process, the helium evaporates and
cools to temperatures down to $1{.}6\,\mathrm{K} $. The copper block has a
large central bore ($49\,\mathrm{mm} $ diameter), and constitutes part of
the wall of the second helium container---an approximately $50\,\mathrm{cm} $-long
tube filled with low-pressure helium gas. It therefore serves as a heat
exchanger between the cold He gas from the reservoir and the buffer gas in
the central chamber in which the sample is held. The temperature of the heat
exchanger can be continuously tuned by controlling the evaporation rate of
the He using a control valve. In addition, an electric heater in the copper
block allows electronic temperature stabilization, and makes the whole
temperature range up to $300\,\mathrm{K} $ accessible.

Successful coupling to the WGM in the toroids requires a stable,
micro\-meter-scale gap between the fiber taper and the edge of the toroid.
Low vibrations in the experimental assembly are therefore an important
selection criterion. The chosen system has successfully been operated in
many labs for the purpose of low-temperature scanning-probe
microscopy---obviously very vibration-sensitive applications. Both the
suspension of the experimental head from a $\sim 0.5\,\mathrm{m} $-long
metallic rod into the buffer gas, and our compact head design (figure~\ref%
{f:cryostat}) indeed renders the cryogenic coupling setup extremely stable 
\cite{Riviere2009}. Standard techniques to reduce the coupling of vibrations
from the running pumps (heavy masses on the transfer pipes between cryostat
and pump) are sufficient to eliminate vibrations to a degree that they
cannot be observed in any experimental signal. At the same time, this
cryostat allows convenient optical access from two orthogonal directions,
providing both a side and top view of the silicon chip. Standard machine
vision products (Navitar $12\times$) with large working distance ($\sim9\,%
\mathrm{cm} $) allow observation of the coupling region with up to $\times 14
$-magnification, sufficient for convenient addressing and alignment of the
toroids versus the fixed fiber taper.

Preparation of a cryogenic cooling experiment includes two critical tasks:
selection of an adequate cooling sample and preparation of a fiber taper
mounted into the experiment. In particular, the resonators on the chip are
tested for high-Q WGM resonances allowing to enter the RSB regime ($%
\kappa<\Omega_\mathrm{m} $) and for low-mass, high-quality RBM modes. These
measurements are done in standard room-temperature coupling setups. If an
adequate sample is found, it is mounted into the cryogenic head.

The compactness of the cryogenic experiment renders the use of the bulky
metallic taper holders used during taper fabrication impossible. Instead,
tapers have to be transferred to a compact glass support made of a
microscopy glass slide which, importantly, features the same coefficient of
thermal expansion as the taper. In that manner, the taper tension remains
constant during cooldown of the cryostat.

\begin{figure}[tb]
\centering
\includegraphics[width= .9 \linewidth]{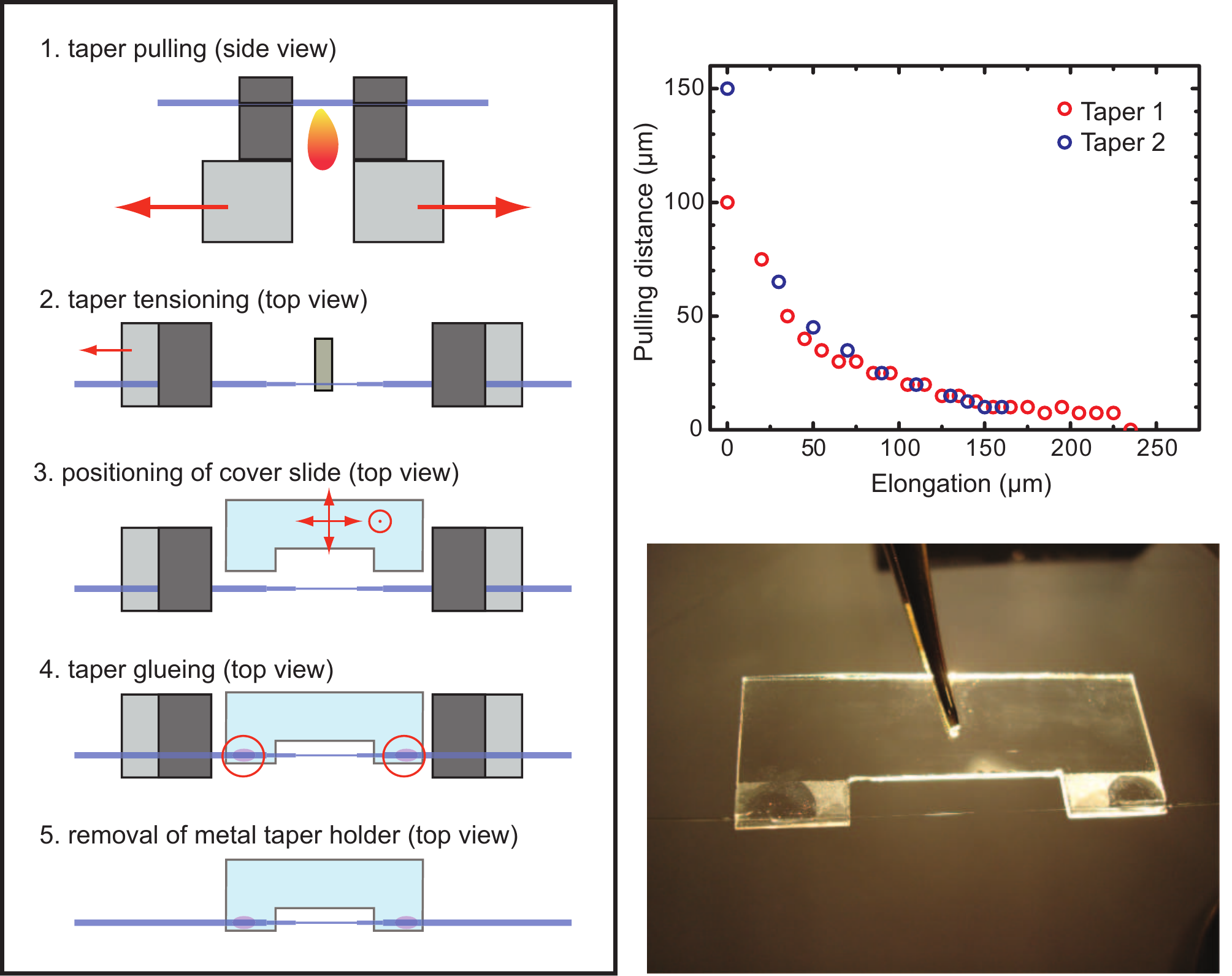}
\caption{Five steps in the fabrication and transfer of a fiber taper to the
compact glass support used in the cryogenic experiments (left panel). After
pulling the taper over the flame of a hydrogen torch, its strain is tested
using an auxiliary chip with microtoroids (upper right graph). If properly
strained, the glass support (prepared from a cover slide) is positioned
underneath the taper, and the fiber is glued to the support with an epoxy.
After hardening the epoxy with UV light, the metal clamps of the fabrication
holder are opened to release the taper now only held by the glass support
(photograph bottom right). It can then be mounted into the experimental
head. See text for more details.}
\label{f:taperTransfer}
\end{figure}

The fabrication and transfer of the fiber taper proceeds in several steps
which are illustrated in figure~\ref{f:taperTransfer}. After the standard
fabrication procedure in a hydrogen flame, the taper, still in the
fabrication holder, is placed in an auxiliary testing setup. Here, a
microtoroid is approached until it touches the taper at its central
position. If retracted again, the toroid pulls the taper with it for a
certain distance due to proximity forces. Using a micrometer drive, the
taper is strained by increasing the separation of the clamps (``elongation''
in figure~\ref{f:taperTransfer}), and the tests are repeated until the
distance over which the taper can be displaced by the toroid is in the range
of $10\,\mathrm{\mu m}  $. While not particularly quantitative, this simple
test was found very helpful in the delicate task of mounting a well-strained
taper into the cryostat: floppy tapers make coupling at low temperatures
impossible, while too tightly strained tapers are prone to rupture during
cooldown. The glass support, fabricated from a simple microscopy slide is
then positioned underneath the taper with a 3d-translation stage. The taper
is glued to the support with UV-hardening epoxy applied first on one side,
hardened, and then on the other side. Then the fiber is released from the
metal holder and transferred to the cryostat. We finally note that due to
its reduced diameter and potentially lower quality of the available
single-mode fiber material, creating a cryogenic fiber taper at shorter
wavelength (e.\ g. $780\,\mathrm{nm} $ as in this work) is significantly
more difficult than at near-IR wavelengths ($1{.}5\,\mathrm{\mu m}  $).

The slide carrying the taper is then mounted into the experimental head, and
the fiber ends are guided to the top of the sample insert and leave the
cryostat via a helium-tight feedthrough. We have achieved total optical
transmission through the cryostat up to $80\%$ from fiber end to fiber end.
A toroid from the mounted sample can then be approached using the
piezoelectric stick-slip actuators (Attocube systems) that carry the sample
chip. For testing purposes, and to check the correct alignment of the taper
mount (position, tilt), these experiments can also be done with the sample
insert (experimental head, suspension rod, and top vacuum flange) held
outside the cryostat.

To verify the effectiveness of the buffer gas, a sample with microtoroids,
and a coupling taper suited for $1550\,\mathrm{nm} $-light was mounted into
the cryostat. The cryostat was cooled down, and the mode temperature of this
sample's RBM was determined using a displacement measurement based on the
PDH technique (see section \ref{sss:PDH}). To enhance the sensitivity, the
weak probing light ($<2\,\mathrm{\mu W} $) from the employed low-noise fiber
laser (BASIK, Koheras) was amplified using an erbium-doped amplifier after
leaving the cryostat, allowing us to measure displacement spectra down to the
base temperature of the cryostat \cite{Arcizet2009a}. In figure~\ref%
{f:thermalization}, the mode temperature $T_\mathrm{m}$ extracted from the
calibrated spectra is shown to closely follow the temperature of the helium
buffer gas down to about $1{.}8\,\mathrm{K} $, or $\langle n \rangle\approx
600$. In spite of being exclusively in contact with the cold helium gas,
this $62\,\mathrm{MHz} $-RBM could therefore be thermalized to occupation numbers
which are, for lower frequency oscillators, often only attainable in
dilution refrigerator systems.

\begin{figure}[hbt]
\centering
\includegraphics[width= .6 \linewidth]{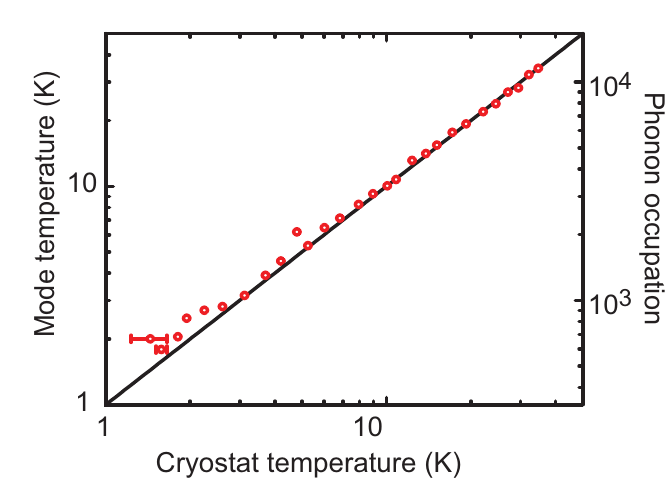}
\caption{Thermalization of the $62\,\mathrm{MHz} $-RBM of a silica
microtoroid. The mode temperature $T_\mathrm{m}$ follows the temperature of
the helium buffer gas down to $1{.}8 \,\mathrm{K} $, or $\langle n
\rangle\approx 600$, proving the effectiveness of the buffer gas cooling.
Figure from ref.\ \protect\cite{Schliesser2009a}. }
\label{f:thermalization}
\end{figure}

Cryogenic cooling of the silica microresonators implies severe modifications
of the properties of these devices. As discussed already in section \ref%
{sss:dissipation}, the mechanical quality factor critically depends on the
temperature of the structure, due to coupling of the mechanical modes to
two-level systems in the glass. In contrast, the change in the sound
velocity, and therefore the mechanical resonance frequencies, is a
relatively weak effect at the $\mathcal{O}(10^{-2})$-level \cite%
{Arcizet2009a}. At the same time, the thermorefractive ($dn/dT$) and
thermoelastic ($\alpha$) coefficients are strongly modified. While $\alpha$
tends to zero at low temperatures ($T<2 \cdot 10^{-9} \,\mathrm{K^{-1}} $ at 
$1{.}6\,\mathrm{K} $), the thermorefractive coefficient continuously drops
and changes sign at a temperature of about $8\,\mathrm{K} $.\footnote{%
Note these observations \cite{Arcizet2009a} differ from a previous report 
\cite{Treussart1998}.} The resulting non-trivial temperature dependence of
the optical resonance frequency gives rise to interesting thermal
multistability effects when an optical resonance is probed using a laser
powerful enough to heat the cavity above this temperature \cite{Arcizet2009a}%
. In the context of the cooling experiments, it is important to note that
negative $dn/dT$ implies that the red wing of an optical resonance is
dynamically self-stable, relaxing dramatically the requirements on the laser
frequency stability and locking speed when working with red detunings.

\subsubsection{Optical system}

For a cooling experiment with the aim of reaching very low occupation numbers, it is
mandatory to avoid excess backaction noise in the cooling laser beam, as
this leads to a limit \cite{Schliesser2008, Diosi2008,Rabl2009} in the
cooling performance. The grating-stabilized diode lasers employed in our
previous cooling work was observed to exhibit excess frequency noise (cf.\
section \ref{ss:rsc}). One possible strategy to eliminate this noise is to
measure the noise using an independent, quiet cavity, and fast feedback on
the frequency of the laser using e.\ g.\ the diode laser current as an
actuator. Feedback with the required bandwidth exceeding $50\,\mathrm{MHz} $
however is technically difficult, as a few meters of cable alone would already
induce phase lags in the loop sufficient to render it ineffective.

As an alternative approach, a quiet cavity could be used in transmission to
filter out the noise. In this manner, reduction of phase noise by $85\,%
\mathrm{dB} $ has been demonstrated at $1\,\mathrm{MHz} $ Fourier frequency 
\cite{Hald2005}. However, the filtering reduces the available power by more
than a factor of 5, so that the remaining available power of $\sim 2\,%
\mathrm{mW} $ is problematically low for cooling experiments. At the same
time, the complex filtering setup derogates one important asset of a
grating-stabilized diode laser: its convenient, wide-range, and fast
tunability.

\begin{figure}[tb]
\centering
\includegraphics[width= \linewidth]{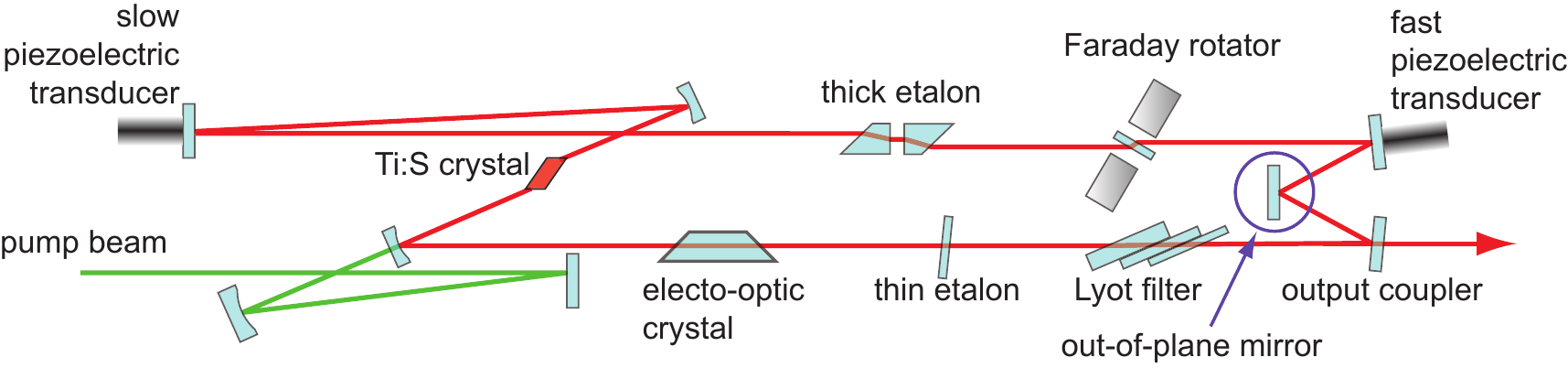}
\caption{Layout of the titanium-sapphire laser used for the cryogenic
cooling experiments. The Ti:S crystal in the ring cavity is pumped optically
at $532\,\mathrm{nm} $ (green beam). In order to select a single
longitudinal mode within the tuning range of $750$ to $870\,\mathrm{nm} $
(determined by the cavity mirrors), a Lyot filter, thick etalon and thin
etalon are used. A Faraday rotator, one reflection point lying outside the
plane of the rest of the cavity mode, and a number of Brewster-angled
surfaces ensure unidirectional lasing action. The precise frequency of the
laser can be actuated by displacing cavity mirrors using a fast, a slow
piezo, as well as an intracavity electro-optic modulator.}
\label{f:TiS}
\end{figure}

For these reasons, we chose to use a solid-state pumped continuous wave
titanium-sapphire (Ti:S) laser (Matisse TX, Sirah Lasertechnik), which
combines very wide tunability ($750$ to $870\,\mathrm{nm} $) with high
output power ($>0.5 \,\mathrm{W} $) and quantum-limited noise in both
amplitude and phase at the relevant Fourier frequencies. Figure~\ref{f:TiS}
shows the schematic setup of this device. The laser is pumped with up to $6\,%
\mathrm{W} $ at $532\,\mathrm{nm} $ from a frequency-doubled diode-pumped
solid-state laser (Millenia, Spectra Physics). The frequency of the laser is
adjusted by a combination of intracavity filters: a Lyot birefringent filter
narrows down the frequency range in which gain exceeds loss to a few hundred
GHz, and a thin and a thick etalon, with free-spectral ranges of about $250\,%
\mathrm{GHz} $ and $20\,\mathrm{GHz} $, respectively, single out one
longitudinal mode of the laser to oscillate. Once the desired laser mode
oscillates, electronic feedback loops are used to lock the etalons to keep
their transmission maximum resonant with the laser mode. These loops are
implemented by the controller of the laser and can be adjusted on a computer
using an USB-interface.

Two mirrors mounted on piezo-electric transducers can then be used to
fine-adjust the frequency of the laser. In particular, the ``slow''
transducer allows frequency scans up to about $60\,\mathrm{GHz} $. The
``fast'' transducer, together with the intracavity electro-optic modulator,
can be used for fast feedback to the laser frequency. In normal operation,
these actuators are employed in a feedback-loop to lock the laser to an
external, temperature-stabilized, $\sim 30\,\mathrm{cm} $-long high-finesse (%
$\mathcal{F}\sim 300$) cavity via the PDH technique, in order to eliminate
low-frequency frequency noise of the laser, reducing the laser linewidth to $%
<30\,\mathrm{kHz} $. In this case, the laser frequency is controlled by
tuning the resonance frequency of the reference cavity, again by displacing
a cavity mirror mounted on a piezoelectric transducer. The custom version of
the laser set up in our laboratory allows alternatively to directly lock the
laser to a WGM in a silica microtoroid, by feeding an error signal (derived
using one of the above-described methods) to the laser controller.

\begin{figure}[btp]
\centering
\includegraphics[width= .9 \linewidth]{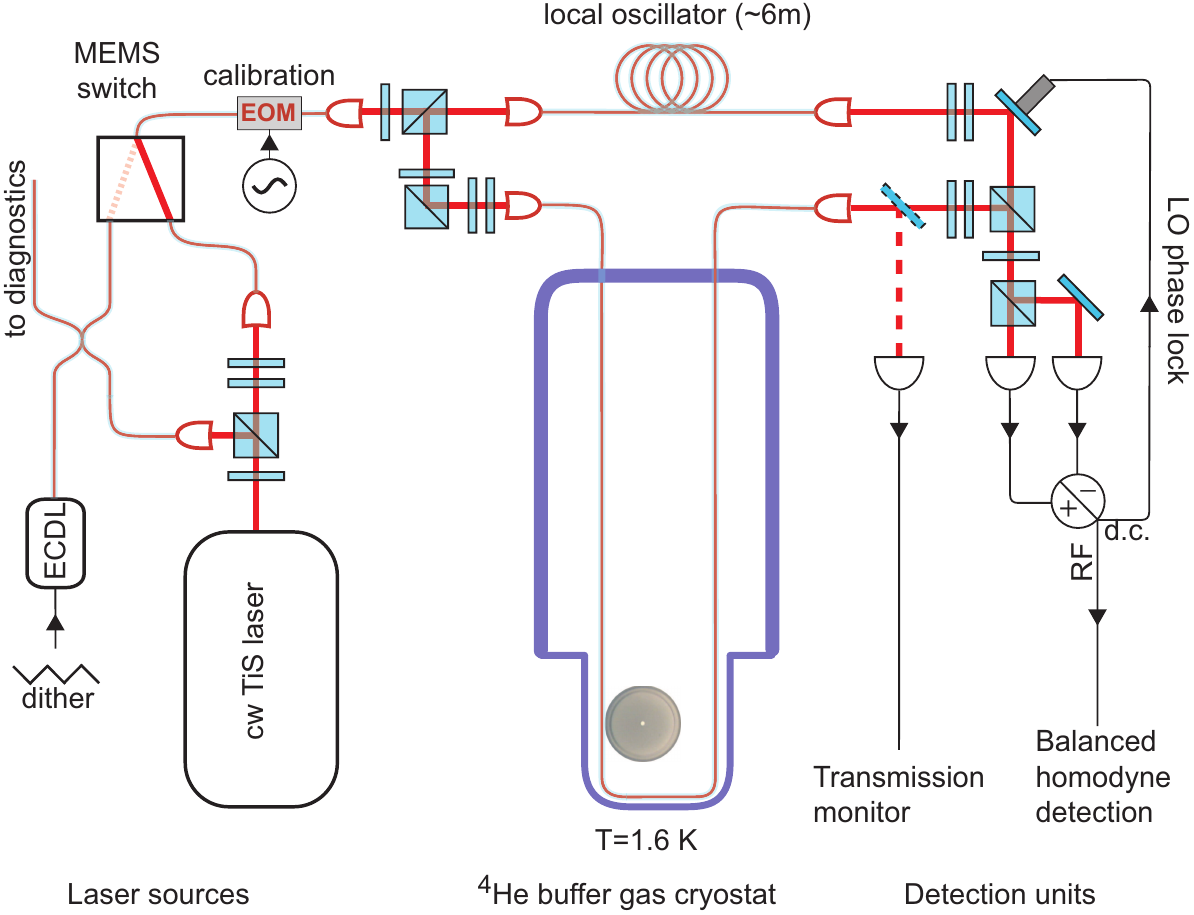}
\caption{Schematic of the setup used for resolved-sideband laser cooling in
a cryogenic environment. An external cavity diode laser (ECDL) is used for
precharacterization of the optical modes. When a suited WGM resonance is
found, the Ti:S laser is tuned to the frequency of the ECDL using the
``diagnostics'' optical output (see text). Then the input to the experiment
is switched to the Ti:S using a MEMS switch. The subsequent optical setup is
essentially a Mach-Zehnder interferometer, one arm of which contains the
cryostat with the taper coupled to the WGM in a microtoroid. For testing
purposes, the output of the taper can also directly be monitored by guiding
the light to an auxiliary photoreceiver (dashed beam path). In normal
operation, the transmitted light is spatially overlapped with the local
oscillator in a polarizing beam splitter cube, albeit in orthogonal
polarization modes. Projected into a different polarization basis using a
waveplate and another polarizing cube, the modes interfere, and the
interference signal is recorded with a balanced receiver. The phase of the
local oscillator---and therefore also the detected signal quadrature---is
locked using the DC-signal from the balanced receiver. The
radio-frequency-part is sent to a spectrum analyzer to obtain the
displacement noise spectra, calibrated again by the frequency modulation of
probing laser using an elecro-optic modulator (EOM). Waveplates and
polarizing beam splitter cubes are used at various positions in the optical
setup to adjust powers and polarizations of the beams.}
\label{f:setup}
\end{figure}

The Ti:S laser can be tuned to arbitrary frequencies within its tuning
range; however, continuous scanning over the whole range is not possible, as
the various frequency selectors (Lyot and etalon filters) have to be
readjusted if the laser has to be tuned by more than about $30\,\mathrm{GHz} 
$. We therefore resort to a continuously tunable grating-stabilized diode
laser (DL) in Littman-Metcalf configuration (TLB-6300, New Focus) for
pre-characterization of the toroids (cf.\ figure~\ref{f:setup}). Once a
well-suited WGM resonance is found with the DL, the Ti:S laser is tuned
close to the frequency of this resonance.

For this purpose, the Ti:S frequency is coarsely adjusted first, using a
grating spectrograph. For the subsequent fine-adjustment, it was found very
useful to monitor the interference signal of the DL---scanning over a
broader frequency range including the WGM resonance---and the slowly tuning
Ti:S laser. When the DL scans through the Ti:S frequency, a transient
interference signal is observed as long as the difference frequency is
within the bandwidth of the receiver ($125\,\mathrm{MHz} $). This transient
beat serves as a convenient marker of the Ti:S frequency, while the standard
transmission signal of the DL indicates the WGM resonance frequency.

After the Ti:S has been tuned close to the WGM resonance, an optical switch
based on MEMS\footnote{%
MEMS: microelectromechanical system}-technology is used to switch to the
Ti:S laser as input to the experiment. The optical setup of the experiment
corresponds essentially to a homodyne measurement (figure~\ref{f:setup}). To
achieve the highest possible SBR, good interference contrast between the
local oscillator and signal beams is crucial. For this reason, identical
collimators in both beams are employed. Careful alignment then allowed to
achieve interference contrast up to $90\%$ (figure~\ref%
{f:interferenceContrast}).

\begin{figure}[htb]
\centering
\includegraphics[width= \linewidth]{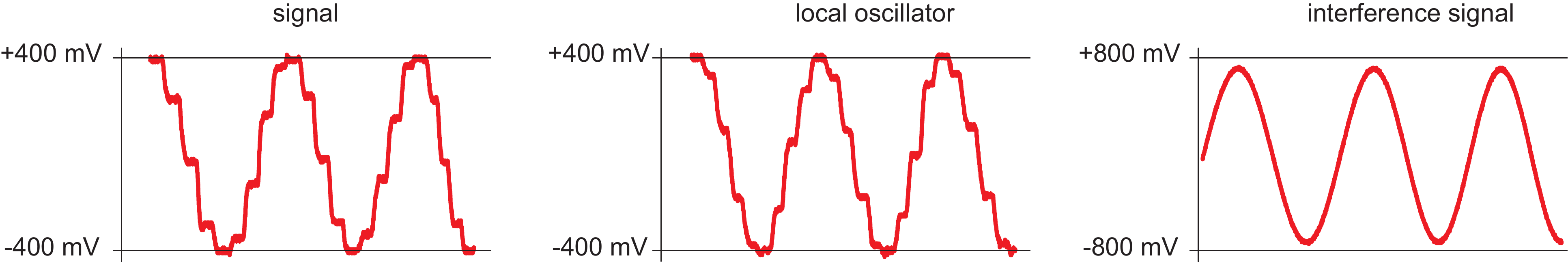}
\caption{Interference contrast of the homodyne receiver. ``Signal'' and
``local oscillator'' traces are recorded by manually rotating the last $%
\protect\lambda/2$-waveplate in front of the last polarizing beam splitter
cube in steps of $10^\circ$, with the other input beam blocked. For this
test, the powers in both beams are adjusted to yield a maximum signal of $%
400 \,\mathrm{mV} $. If both beams are opened, and the detection unit is
properly aligned, a total signal amplitude of $1440\,\mathrm{mV} $,
corresponding to $90\%$ interference contrast, is recorded.}
\label{f:interferenceContrast}
\end{figure}

The phase of the local oscillator is locked by using the DC-part of the
homodyne signal from the balanced receiver, forcing the average detected
differential photocurrent to zero for a phase measurement. Note that this is
slightly different to the technique described in \ref{ss:homodyne}, where an
orthogonal polarization component (in both signal and LO beams) was used to
derive the LO phase error signal. In this experiment, the length of the
employed fibers renders this approach problematic due to apparent drifts in
the polarization rotation in the fibers. Note also that it is necessary to
match the total length of the fibers in both arms of the interferometer
(excluding the potential propagation length in the toroid) in order to
measure the correct signal quadrature at all Fourier frequencies.

\begin{figure}[tb]
\centering
\includegraphics[width= .5 \linewidth]{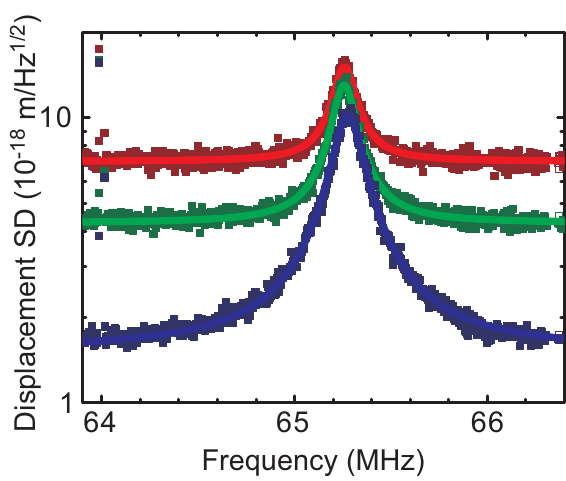}
\caption{Displacement noise spectra of the RBM of a silica microtoroid held
at a temperature of $2{.}4\,\mathrm{K} $. The power used to probe the RBM
was about $3\,\mathrm{\protect\mu W} $ (red trace), $10\,\mathrm{\protect\mu %
W} $ (green trace) and $100\,\mathrm{\protect\mu W} $ (blue trace). Points
are measured data, lines are Lorentzian fits. Figure from ref.\ \protect\cite%
{Schliesser2009a}.}
\label{f:lowTprobing}
\end{figure}

Figure~\ref{f:lowTprobing} shows typical displacement spectra of the $65.3\,%
\mathrm{MHz} $-RBM of a silica microtoroid probed by homodyne spectroscopy
with the Ti:S laser. The sample is held in the cryostat, at a temperature of 
$2{.}4\,\mathrm{K} $, corresponding to $\langle n\rangle \approx 770$.
Sensitivity at the $10^{-18}\,\mathrm{m/\sqrt{Hz}} $-level, as well as a SBR
of nearly $20\,\mathrm{dB} $ (in noise power) is achieved, in spite of a
mechanical quality factor which was as low as 540 in this case. Note that
displacement noise measured at resonance is the sum of the thermal noise in
the mechanical mode \emph{plus} the imprecision background noise due to
quantum noise in the detection, which is reduced at higher probing powers.

\subsection{Cooling towards the quantum ground state}

\label{ss:gsc}

To demonstrate the performance of combined optical and cryogenic cooling, we
used a $52\,\mathrm{\mu m}  $-diameter sample with a WGM of $5{.}5\,\mathrm{%
MHz} $ intrinsic decay rate and $9\,\mathrm{MHz} $ mode splitting, loaded
down to $\kappa/2\pi\approx 19\,\mathrm{MHz} $ using the fiber taper,
corresponding to a finesse of about $70{,}000$. A room-temperature reference
measurement (figure~\ref{f:lowTprobing}) of its RBM at $65{.}2\,\mathrm{MHz} 
$ reveals an effective mass of $5{.}6\,\mathrm{ng} $, in good agreement with
the simulated value of $4{.}9\,\mathrm{ng} $. Note that the experimentally
determined mass can only be as accurate as is the displacement calibration and
the knowledge of the actual temperature of the RBM. We generally estimate
the resulting systematic error to a level of $30\%$, arising from
imperfections in the modulation scheme and temperature changes induced by
dynamical backaction or absorption-induced heating.

\begin{figure}[bt]
\centering
\includegraphics[width= \linewidth]{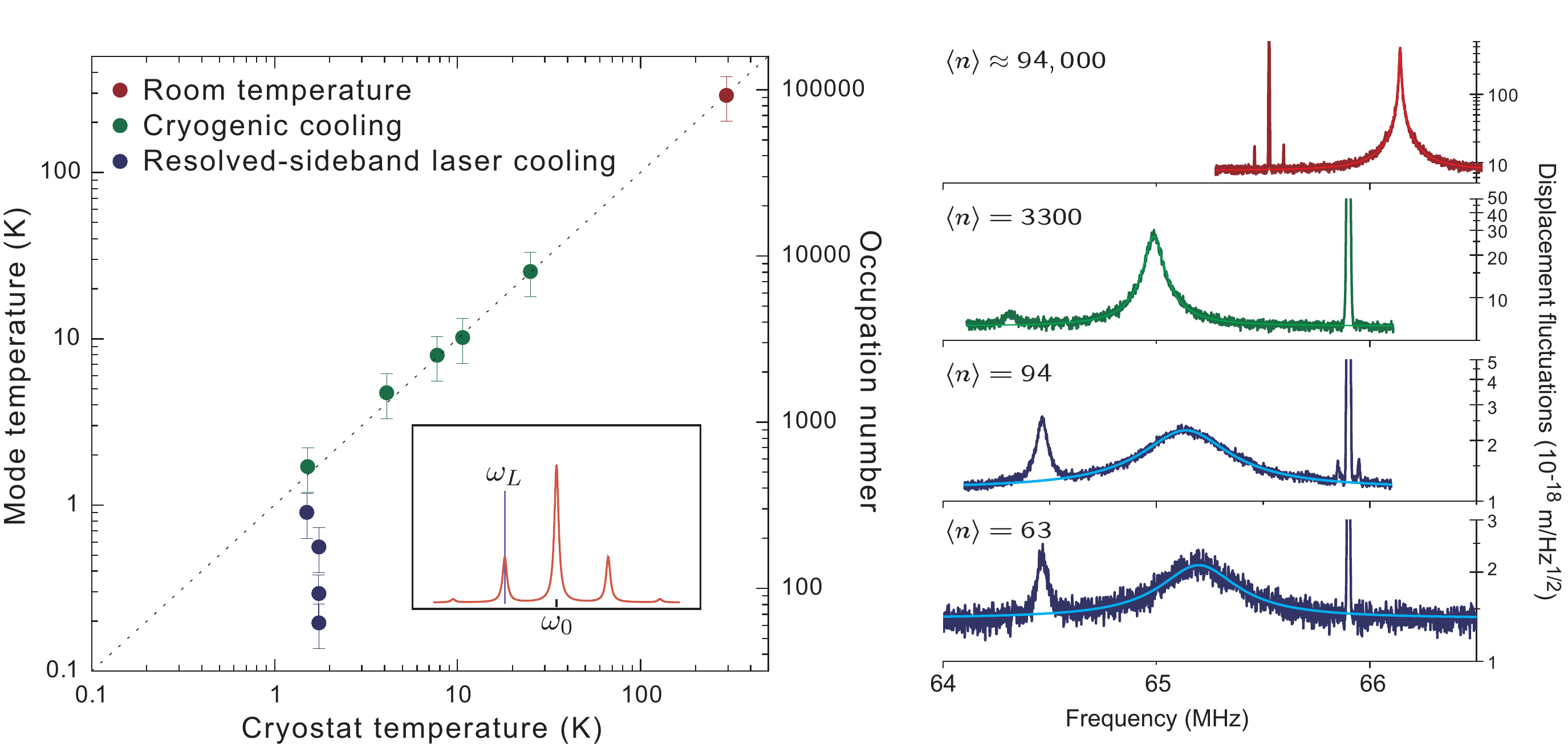}
\caption{Left panel: Mode temperature and occupation number of the RBM,
measured using noise thermometry, versus the temperature of the buffer gas
in the cryostat, including a reference measurement at room temperature (red
point). Green data points are taken with the laser tuned to the optical
resonance, so no optical cooling takes place. When tuned to the lower
mechanical sideband (inset), additional laser cooling is observed (blue
points). Panels on the right show a selection of displacement noise spectra,
from which the mode temperatures of the measurements were derived. Apart
from the calibration peak these spectra also reveal a nearby second
mechanical mode. Nonetheless, the spectrum of the RBM can be well-fit with a
Lorentzian (thin lies). Figure from ref.\ \protect\cite{Schliesser2009a}.}
\label{f:GFAllNew}
\end{figure}

Using this effective mass, it is possible to derive the mode temperature
from subsequent displacement measurements during cooldown of the cryostat.
The results again confirm proper thermalization of the sample (figure~\ref%
{f:lowTprobing}), down to an occupation of $\langle n \rangle \approx 550$.
For further cooling, the Ti:S laser is tuned to the lower sideband at $\bar
\Delta =-\Omega_\mathrm{m} $. The signal from the homodyne receiver can
still be used to measure mechanical displacements, however, due to the
detuned operation, the spectrum of the phase quadrature now displays a more
complicated dependence from the mechanical noise spectrum. In contrast to
the simple expression (\ref{e:sqqresonant}) for $\bar \Delta =0$, we now
obtain 
\begin{align}  \label{e:sqqdetuned}
\bar S_{qq}^\mathrm{out}(\Omega )&=1+ \frac {4 \bar a ^2 g_0 ^2 \eta_\mathrm{%
c}  \kappa}{\bar \Delta ^2+(2\eta_\mathrm{c} -1)^2(\kappa/2)^2}\times  \notag
\\
&\qquad\frac{\bar \Delta ^4+2 \bar \Delta ^2 (2 \eta_\mathrm{c}
-1)(\kappa/2)^2+(2\eta_\mathrm{c} -1)^2(\kappa/2)^2 ((\kappa/2)^2+\Omega ^2)%
}{ \bar \Delta ^4+2\bar \Delta ^2((\kappa/2)^2-\Omega
^2)+((\kappa/2)^2+\Omega ^2)^2}\,\bar S_{xx}(\Omega ),
\end{align}
where the first term, equal to unity, again is due to quantum noise. Importantly, the transduction of a
frequency modulation into $\bar S_{qq}$ has the same dependence on the
relevant experimental parameters as the transduction of displacements $\bar
S_{xx}$, so that the calibration scheme remains valid also in the detuned
case.

Rewriting equation (\ref{e:sqqdetuned}) as a quantum-noise induced
imprecision in the displacement measurement one obtains 
\begin{align}  \label{e:sxximdetuned}
\bar S_{xx}^\mathrm{im,qn}(\Omega )&= \frac {\bar \Delta ^2+(2\eta_\mathrm{c}
-1)^2(\kappa/2)^2} {4 \bar a ^2 g_0 ^2 \eta_\mathrm{c}  \kappa}\times  \notag
\\
&\qquad \frac {\bar \Delta ^4+2\bar \Delta ^2((\kappa/2)^2-\Omega
^2)+((\kappa/2)^2+\Omega ^2)^2} {\bar \Delta ^4+2 \bar \Delta ^2 (2\eta_%
\mathrm{c} -1)(\kappa/2)^2+(2\eta_\mathrm{c} -1)^2(\kappa/2)^2
((\kappa/2)^2+\Omega ^2)}.
\end{align}
For $\kappa\gg\Gamma_\mathrm{eff} $, which is typically in the case in the
experiments presented here, the spectral shape of this function can be
assumed to be flat over the frequency range of interest, and we find as a
useful approximation 
\begin{align}
\bar S_{xx}^\mathrm{im,qn}(\Omega_\mathrm{m} )&\approx \frac{\Omega_\mathrm{m%
} ^2}{4 \eta_\mathrm{c} ^2 g_0 ^2}\frac{\hbar \omega }{P_\mathrm{in}}
\end{align}
in the resolved sideband regime ($\bar \Delta =-\Omega_\mathrm{m} \gg\kappa$%
). It is noteworthy that this value is only a factor of 4 higher than
expected for resonant probing $\bar \Delta =0$ in this regime. However, in
the resolved sideband regime, only a fraction of $\sim(\kappa/4\Omega_%
\mathrm{m} )^2$ of the launched power is coupled to the cavity.

Figure~\ref{f:GFAllNew} shows displacement spectra and corresponding mode
temperatures recorded during such a cooling run. As expected, the damping of
the mechanical modes is optically increased, when the laser is detuned to $%
\bar \Delta =-\Omega_\mathrm{m} $. Note that for these experiments, active
stabilization of the laser frequency to the optical sideband of the WGM has
not been necessary. Instead, a second electronic spectrum analyzer was used
to demodulate the homodyne signal at the frequency $\Omega _\mathrm{mod}$ of
the calibration modulation. As expected from equation (\ref{e:sxximdetuned}%
), this signal displays local maxima at $\bar \Delta =0,\pm \Omega _\mathrm{%
mod}$. After the laser is tuned to $\bar \Delta =-\Omega_\mathrm{m} $, the
system is sufficiently stable during the averaging of typically several tens
of seconds (small drifts can be manually compensated).

For the highest launched powers of $\sim 0{.}2 \,\mathrm{mW} $, the total
damping rate was increased to $\Gamma_\mathrm{eff} /2\pi=370\,\mathrm{kHz} $%
. At the same time, the mode temperature was reduced to $T_\mathrm{m}=200\pm
60\,\mathrm{mK} $, corresponding to an occupation number of $\langle n \rangle =
63\pm 20$ \cite{Schliesser2009a}. The error interval is due to potential
calibration errors in the displacement measurement and the determination of
the effective mass. This value is comparable to the lowest occupation number
achieved in nanoelectromechanical systems, yet those experiments have to
rely more heavily on advanced cryogenic machinery \cite{Naik2006}, and
suffer from the insufficient signal-to-noise ratio.

\subsection{Assessing measurement backaction}

\label{ss:measba}

The strong suppression of thermal noise in the measurements presented here
allows an interesting analysis of the data from the perspective of quantum
measurement \cite{Braginsky1992}. As discussed in section \ref%
{ss:sensitivetheory}, fundamentally, the achieved total uncertainty in the
measurement of the displacement of the mechanical oscillator is limited by
two effects: imprecision and backaction. The imprecision in our experiment
is given by detection shot noise, in the detuned case discussed here it is
described by equation (\ref{e:sxximdetuned}).

On the other hand, the light used to measure the mechanical oscillator
exerts backaction on it. Inevitably, intracavity photon number fluctuations
give rise to a fluctuating force, the spectrum of which can be derived as
 \cite{Schliesser2009a} 
\begin{align}
\bar S_{FF}^\mathrm{ba, qn}(\Omega )&=\hbar^2 g_0 ^2 \bar a ^2 \kappa \frac{%
\bar \Delta ^2+(\kappa/2)^2+\Omega ^2}{\bar \Delta ^4+2\bar \Delta
^2((\kappa/2)^2-\Omega ^2)+((\kappa/2)^2+\Omega ^2)^2}.
\end{align}
These force fluctuations are usually referred to as quantum backaction,
simplifying to 
\begin{align}
\bar S_{FF}^\mathrm{ba, qn}(\Omega_\mathrm{m} )&\approx \frac{2 g_0 ^2 P_%
\mathrm{in} \eta_\mathrm{c}  \hbar}{\omega  \Omega_\mathrm{m} ^2}
\end{align}
in the resolved-sideband regime ($|\bar \Delta |=\Omega_\mathrm{m} \gg \kappa
$). Note that the spectra of imprecision and backaction noise, as well as
their possible correlation $S_{xF}(\Omega )$ reflect properties of the
measurement device (the cavity pumped by a laser field)---independent of the
mechanical oscillator.

Other sources of measurement backaction include excess noise in the
intracavity photon number, which may arise from laser frequency fluctuations 
\cite{Schliesser2008, Diosi2008, Rabl2009}. This effect can be ruled out in
this work as the employed Ti:S is known to exhibit only quantum fluctuations
at the Fourier frequencies of interest. Another possible source of
measurement backaction is heating of the torus due to light absorption. This
increases the temperature of the structure, and raises the level of Langevin
force fluctuations driving the mechanical oscillator.

\begin{figure}[bt]
\centering
\includegraphics[width= \linewidth]{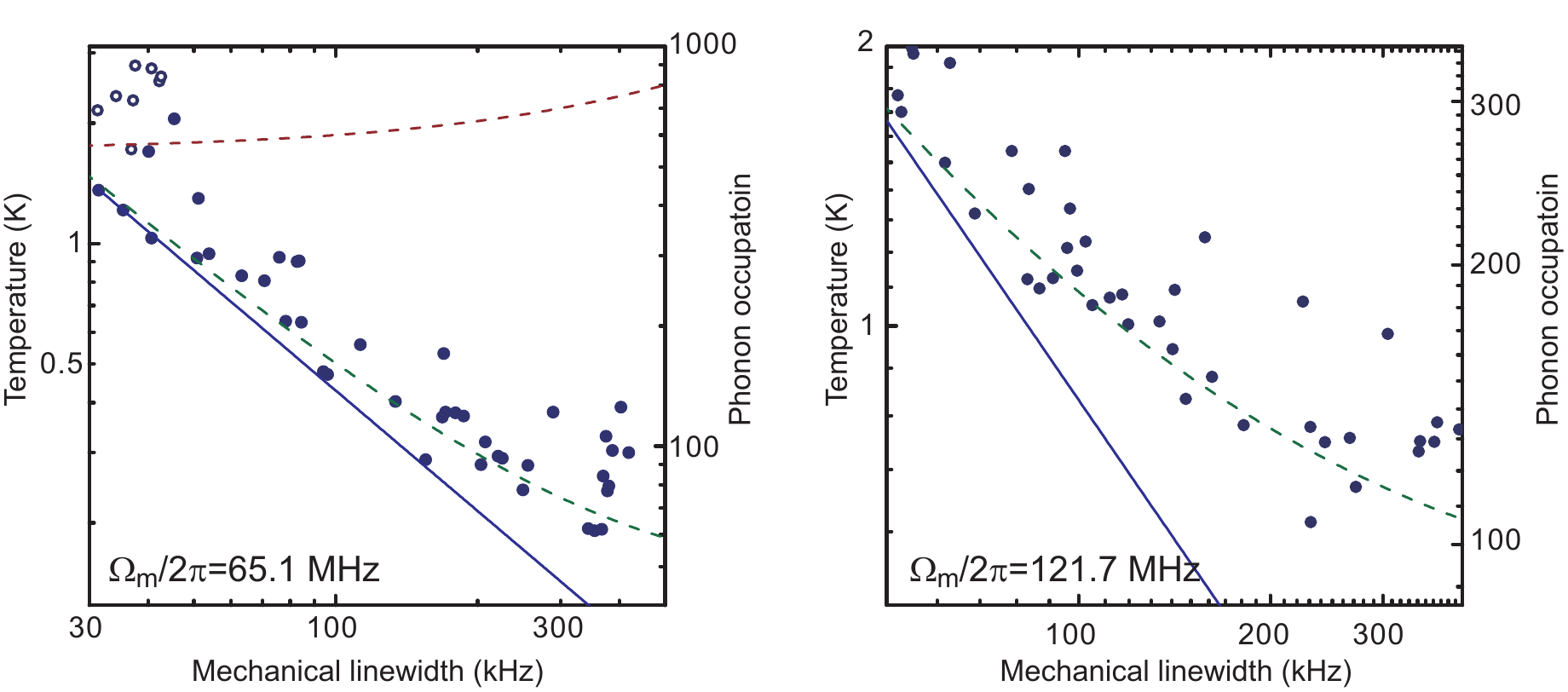}
\caption{Resolved-sideband cooling of two samples with frequencies of $%
\Omega_\mathrm{m} /2\protect\pi=65.1\,\mathrm{MHz} $ (left panel) and $%
\Omega_\mathrm{m} /2\protect\pi=121{.}7\,\mathrm{MHz} $ (right panel). The
graphs show the mechanical linewidth $\Gamma_\mathrm{eff} $ (abscissa)
versus the derived mode temperature $T_\mathrm{m}$ (ordinate) during a
cooling run. Open points correspond to measurements with the laser tuned
close to the optical resonance (no optical cooling), and filled points to
measurements with the laser tuned close to the lower sideband. When varying
the power of the cooling laser, both linewidth and mode temperature are
changed. Deviation from the simple cooling behavior (blue line) is
attributed to an increase of the structures' temperature (red dashed line),
taken into account in a more elaborate model (green dashed line). Scatter is
due to varying operation conditions, uncertainty in phonon occupation for
each point is $<30\%$. Figure from ref.\ \protect\cite{Schliesser2009a}.}
\label{f:resultsNew}
\end{figure}

A series of cooling measurements (figure~\ref{f:resultsNew}) indeed reveals
a deviation from the simple relation (\ref{e:simplecooling}), which can
however be reproduced by introducing a heating term 
\begin{equation}
T_\mathrm{m}=\frac{\Gamma_\mathrm{m} (T^{\prime })}{\Gamma_\mathrm{m}
(T^{\prime })+\Gamma_\mathrm{dba}} T^{\prime },
\end{equation}
with $T^{\prime }=T+ \Delta T_\mathrm{abs}$, where $\Delta T_\mathrm{abs}$
is proportional to the power circulating in the cavity. In addition, we have
taken also the temperature dependence of the mechanical damping $\Gamma_%
\mathrm{m} (T^{\prime })$ into account. For the $65\,\mathrm{MHz} $%
-oscillator below $2\,\mathrm{K} $, with \cite{Arcizet2009a} $d\Gamma_%
\mathrm{m} /dT\approx 2 \pi 16\,\mathrm{kHz/K} $, we find a heating of about 
$5\,\mathrm{K} $ per Watt of circulating power. Similar values were
extracted from studies of optical multi-stability at low temperature at a
wavelength of $1.5\,\mathrm{\mu m}  $, corroborating the attribution of the
observed backaction effect to laser-induced heating.

A comparison with a second cooling run with a different sample with $\Omega_%
\mathrm{m}  /2\pi=121{.}7\,\mathrm{MHz} $ and $Q_\mathrm{m} =2{,}200$, but a
broader WGM resonance ($\kappa/2\pi=155\,\mathrm{MHz} $) emphasizes the
importance of the resolved-sideband regime for the efficiency of cooling in
the presence of laser absorption: A significantly more pronounced heating
effect prevents reaching occupation numbers below $\langle n \rangle=100$, in spite
of the higher mechanical frequency (figure~\ref{f:resultsNew}).

From the data of the $\langle n \rangle \approx 63$-cooling run we can now
extract quantitative values of the corresponding backaction force
fluctuations. As a very conservative upper limit, we may use the \emph{total}
thermal force fluctuations $\bar S_{FF}^\mathrm{the}(\Omega )=2 m_\text{eff}
\Gamma_\mathrm{m}  k_\mathrm{B} T^{\prime }$, and find a value of $\sqrt{%
\bar S_{FF}^\mathrm{the}(\Omega_\mathrm{m} )}=8 \,\mathrm{fN/\sqrt{Hz}} $.
If we consider only the temperature \emph{rise} $\Delta T_\mathrm{abs}$
caused by the laser absorption as the backaction of the measurement, a lower
value of $\sqrt{\bar S_{FF}^\mathrm{ba}(\Omega_\mathrm{m} )}=4 \,\mathrm{fN/%
\sqrt{Hz}} $ is found. In these assessments, we benefit from the low
occupation which allows us to extract the effect of measurement backaction,
as it is large enough to be observed on top of the background of the thermal
noise.

It is interesting to compare these findings with fundamental limits. 
Quantum mechanics imposes an inequality on imprecision and backaction noise,
which, for the particular case of an optical measurement of a mechanical
oscillator's displacement, can be written as \cite{Braginsky1992} 
\begin{equation}  \label{e:sxsf}
\bar S_{xx}^\mathrm{im}(\Omega )\cdot \bar S_{FF}^\mathrm{ba}(\Omega )\geq 
\frac{\hbar^2}{4}.
\end{equation}
This relation can be considered a manifestation of the Heisenberg
uncertainty principle in the context of continuous position measurement \cite%
{Braginsky1992}.

Taking the force noise extracted from our data, and the experimental
imprecision of $\sqrt{\bar S_{xx}^\mathrm{im}(\Omega_\mathrm{m} )}\approx 1{.%
}4\,\mathrm{am/\sqrt{Hz}} $ achieved in the same measurement, an upper limit
from the backaction-imprecision product of $\sqrt{\bar S_{xx}^\mathrm{im}%
(\Omega_\mathrm{m} )\cdot \bar S_{FF}^\mathrm{the}(\Omega_\mathrm{m} )}%
\approx 220 \,\hbar / 2$ is found. Considering only the absorption-induced
heating as a backaction mechanism, an even lower value of $\sqrt{\bar S_{xx}^%
\mathrm{im}(\Omega_\mathrm{m} )\cdot \bar S_{FF}^\mathrm{ba}(\Omega_\mathrm{m%
} )}\approx 100\, \hbar / 2$ is found from our experiments. This is an order
of magnitude lower than the values achieved with nanomechanical oscillators
cooled in dilution refrigerators: Readout with an atomic point contact \cite%
{Flowers-Jacobs2007} achieved a backaction-imprecision product of $%
1700\pm400 \,\hbar /2$, while measurements using a superconducting single-electron
transistor \cite{Naik2006} have
achieved a value\footnote{%
Note that the number quoted in this manuscript, $15\,\hbar/2$, is a
theoretical estimate if shot-noise limited detection was possible. As
discussed in the supplementary information, the actual imprecision noise in
the experiment was $50$-times higher \cite{Naik2006, Clerk2008}.} of $\sim
800\,\hbar/2$.

\clearpage

\section{Conclusion}

\label{s:conclusion}

In this chapter, we have reported on optomechanical interactions in
high-finesse optical whispering gallery mode resonators. Careful analysis
and understanding of these devices' properties has allowed us to optimize
their performance for the purpose of cavity optomechanics. Among the various
systems now designed and studied in this context, they offer a unique
combination of high-frequency ($30$--$120\,\mathrm{MHz} $), high-quality ($Q_%
\mathrm{m} $ up to $80{,}000$) mechanical modes and ultra-high finesse
(intrinsic finesse $\mathcal{F}_0\approx 0{.}9\cdot 10^6$) optical
resonances. These key figures rival even the best optical cavities developed
in the context of cavity quantum electrodynamics \cite{Rempe1992} in terms
of finesse, and the mechanical quality factors of state-of-the art nano- and
micromechanical oscillators \cite{Naik2006, Verbridge2007} in the same
frequency range.

Together with an intrinsically strong parametric coupling of optical and
mechanical degrees of freedom---with optical resonance frequency shifts of
typically more than $10\,\mathrm{kHz} $ per femtometer displacement---this
has enabled ultrasensitive optical monitoring of the mechanical mode.
Adapting powerful tools from quantum optics as optical displacement meters
limited only by optical quantum noise, sensitivities at the level of $%
10^{-18}\,\mathrm{m/\sqrt{Hz}} $ were achieved \cite{Schliesser2008,
Schliesser2008b}. This experimentally demonstrated imprecision is below the
expected noise level associated with the zero-point fluctuations of the
mechanical mode, which so far has been achieved only with much lighter
nanomechanical oscillators \cite{Naik2006, Clerk2008}. Reaching such a
sensitivity is a crucial precondition for the experimental verification of
the concepts of quantum measurement in the context of displacement
measurements, such as quantum backaction and the emergence of the standard
quantum limit \cite{Braginsky1992, Tittonen1999, Schwab2005}. Furthermore,
this exquisite sensitivity \cite{Schliesser2008b} may also be exploited to
monitor nanomechanical oscillators brought into the near-field of the cavity
mode, such as silicon nitride nanobeams or -membranes \cite{Anetsberger2009}%
, graphene sheets or diamond nanowires. Beyond mechanical effects, this
sensitivity has allowed us to study fundamental thermal noise mechanisms, such
as thermorefractive noise, which are of interest for the application of
silica microresonators as frequency references \cite{Vassiliev1998,
Matsko2007}, as biophysical sensors \cite{Armani2007, Schroter2008,
Vollmer2009} or for the proposed demonstration of Kerr squeezing \cite%
{Kippenberg2004a}.

\begin{figure}[btp]
\caption{Cooling experiments performed in different laboratories around the
globe, including both experiments based on dynamical backaction and active
feedback cooling. The cooled oscillators span about 8 orders of magnitude in
frequency. Open symbols indicate the reservoir temperature of the
experiments, distinguishing cryogenic from room-temperature experiments.
Full symbols indicate the lowest mode temperature achieved when
optomechnical cooling is applied. The dashed line indicates the temperature
for which $\hbar \Omega_\mathrm{m} \approx k_\mathrm{B} T$. The individual
results are described in references: AURIGA \protect\cite{Vinante2008},
Cornell \protect\cite{Hertzberg2009}, IBM \protect\cite{Poggio2007}, IQOQI 
\protect\cite{Gigan2006, Groblacher2008,Groblacher2009}, JILA \protect\cite%
{Teufel2008}, LKB \protect\cite{Cohadon1999,Arcizet2006,Arcizet2006a}, LMU 
\protect\cite{Hohberger2004, Favero2007}, MIT \protect\cite%
{Corbitt2007,Corbitt2007a}, MPQ \protect\cite%
{Schliesser2006,Schliesser2008,Schliesser2009a}, NIST \protect\cite%
{Brown2007}, Oregon \protect\cite{Park2009}, Stanford \protect\cite{Weld2006}%
, UCSB \protect\cite{Kleckner2006}, UMD \protect\cite{Naik2006}, UWA 
\protect\cite{Mow-Lowry2008}, Yale \protect\cite{Thomson2007}.}
\label{f:groundState}\centering
\includegraphics[width= \linewidth]{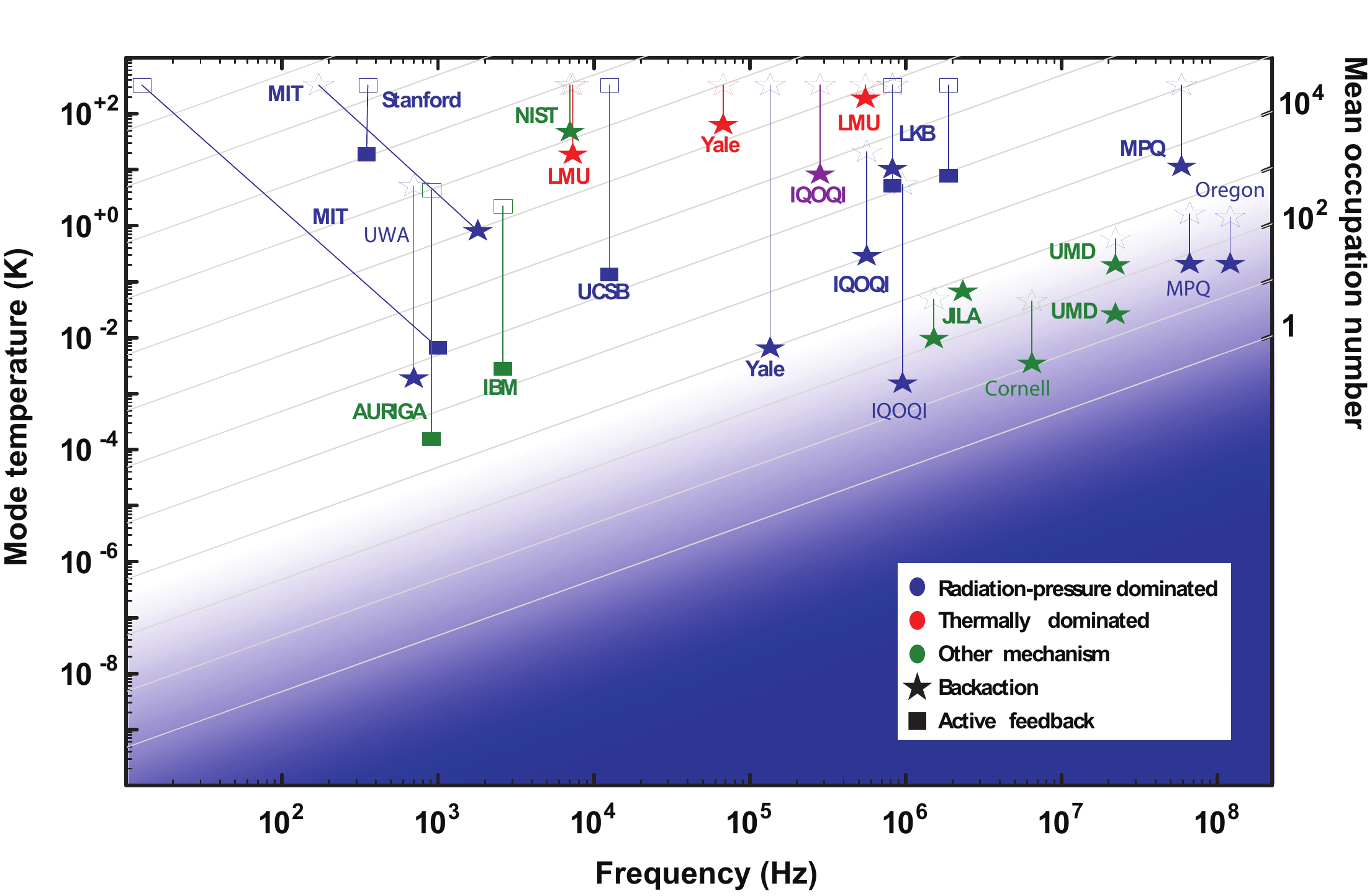}
\end{figure}

Due to the high-finesse, the dramatically enhanced intracavity radiation
pressure exerts a readily detectable force on the mechanical mode. This
effect was directly measured in a ``pump-probe''-type measurement, in which
the displacements induced by modulated radiation-pressure is probed with a
second laser. It was shown that the nonlinear cross-coupling of the two
light fields is strongly dominated by radiation pressure induced mechanical
displacement, and more than two orders of magnitude stronger than the
well-known Kerr effect \cite{Schliesser2006}. Furthermore, radiation
pressure has also been shown to induce a modification of the dynamics of the
mechanical mode, changing both its effective spring constant (optical
spring) and its damping (amplification and cooling). Predicted as early as
1967 by Braginsky \cite{Braginskii1967}, this dynamical backaction was
systematically studied over a wide regime of experimental parameters
(detuning, photon storage time, mechanical oscillation period). Our
experiments have demonstrated, for the first time, efficient optical cooling
of a mechanical mode induced by dynamical backaction, both in the
``Doppler'' \cite{Schliesser2006} and the resolved-sideband regime \cite%
{Schliesser2008}. These techniques are now widely employed in experiments
which aim to demonstrate ground-state cooling of a mechanical device (figure %
\ref{f:groundState}).

In our experiments, we have identified several important barriers on the way
towards this ultimate goal. Frequency noise of the driving electromagnetic
field, practically relevant in many systems \cite{Schliesser2008, Diosi2008,
Rabl2009}, is shown here to be eliminated by resorting to a quantum-noise
limited laser system. Furthermore, much in contrast to trapped atoms or
ions, even very high-Q macroscopic oscillators are not very well isolated
from their finite-temperature environment. Fluctuating thermal forces thus
compete with laser cooling, limiting the occupancies achieved in
room-temperature experiments to a few thousand quanta (figure \ref%
{f:groundState}).

To overcome this limit, we have implemented resolved-sideband laser cooling
in a cryogenic environment \cite{Schliesser2009a}. The mechanical oscillator
is cooled to an occupation number of $\langle n \rangle\approx 63\pm 20$. The
optical detection scheme provides small enough displacement imprecision to
enable monitoring even such an ultracold oscillator with appreciable
signal-to-noise ratio. Additionally, due to the low occupation number, we are able to
extract an upper limit on the \emph{backaction} of the displacement
measurement. The product of backaction and imprecision noise lies only a
factor of $100$ above the fundamental quantum limit \cite{Braginsky1992},
and constitutes the lowest value reported in the literature \cite{Clerk2008}.

In conclusion, we have introduced an optomechanical system suited for the
exploration of quantum effects in mesoscopic mechanical oscillators, by
experimentally demonstrating \emph{(i)} the possibility to monitor its
displacements with an imprecision below the standard quantum limit, \emph{%
(ii)} a combination of cooling techniques based on radiation-pressure and
standard cryogenics capable of cooling the oscillator close to its quantum
ground state, and \emph{(iii)} a near-ideal operation of the displacement
transducer in the sense of quantum measurement (imprecision-backaction
product). Leveraging quantum optical techniques, we have therefore
established a route into the quantum realm of mesoscopic oscillators, which
previously was thought to be uniquely accessible with nano-electromechanical
systems \cite{Schwab2005}.

\clearpage

\section{Outlook}

\label{s:outlook}

Notwithstanding such progress, true quantum effects of radiation pressure and the 
quantum states of a mechanical oscillator have not been observed in any experiment today.%
\footnote{%
There are, however, experiments that could be considered ``quantum
simulations'' of mesoscopic mechanical oscillators---based on ultracold
atoms \cite{Murch2008, Brennecke2008}.}%
In our experiments involving WGM micoresonators, it appears that cooling to the 
quantum ground state, and the
observation of quantum backaction are closest to fruition.
In the following, we briefly outline the strategies pursued
in our laboratory to achieve these goals, and, concluding this chapter,
point to other interesting optomechanical effects.

The most severe antagonist to ground state cooling is the intrinsic damping $%
\Gamma_\mathrm{m} $ of the mechanical mode, coupling it to the environment
at a finite temperature $T$. This coupling tries to maintain the mode and
the environment in thermal equilibrium, essentially feeding the power $%
\Gamma_\mathrm{m}  k_\mathrm{B}(T-T_\mathrm{m})$ into the mechanical mode,
which has to be continuously removed by laser cooling. There are two obvious
ways to reduce this load: reducing the environment temperature $T$, and
reducing the mechanical damping $\Gamma_\mathrm{m} $. While the former is a
technical task, the latter exhibits complex dependence on material, geometry
and operating conditions.

In particular we have observed (sections \ref{sss:dissipation} and \ref%
{ss:gsc}) a strong increase of the mechanical damping at cryogenic
temperatures due to relaxation of two-level systems (TLS) present in
silica material \cite{Pohl2002,Arcizet2009a}. This flaw is known, however,
to ameliorate at sufficiently low temperatures $\lesssim 1\,\mathrm{K} $, at
which the damping due to TLS falls off with $\Gamma_\mathrm{m} \propto
\Omega_\mathrm{m} /T^3$. Thus, lowering the operating temperature of the
experiment may be expected to enhance the cooling performance with a scaling
up to $T^{-4}$. For this reason, a ${}^3\mathrm{He}$-cryogenic system is
presently being tested in our laboratory. A base temperature of $600\,\mathrm{mK} $%
and significantly higher mechanical quality factors ($>10^4$) than in the $%
{}^4\mathrm{He}$-cryostat were already measured. Resolved-sideband cooling,
with the cooling rates of $1.5\,\mathrm{MHz} $ already demonstrated, would
place a $70\,\mathrm{MHz} $ RBM directly in the quantum ground state. A very
crucial issue in this context, however, is the suppression of heating by
laser absorption, requiring very high-quality samples and operation deeply
in the resolved sideband regime.

Optical absorption could be reduced by using ultrapure crystalline materials
for the WGM resonators. Indeed, pioneering work at the Jet Propulsion
Laboratory has resulted in WGM resonators made of quartz \cite%
{Ilchenko2008a}, $\mathrm{MgF}_{2}$ and $\mathrm{CaF}_{2}$ \cite%
{Grudinin2006a} having optical quality factors up to $6\cdot 10^{10}$.
Machining and polishing WGM resonators down to a diameter of $80 \,\mathrm{%
\mu m}  $ has also been achieved. At the same time, the pristine crystalline
structure avoids mechanical losses due to two-level systems. At $\sim 100\,%
\mathrm{kHz} $-frequencies, mechanical quality factors $Q_\mathrm{m} \gtrsim
10^8$ were measured in bulk $\mathrm{CaF}_{2}$ samples, both at room and
cryogenic temperatures \cite{Nawrodt2007}.

\begin{figure}[tb]
\centering
\includegraphics[width=\linewidth]{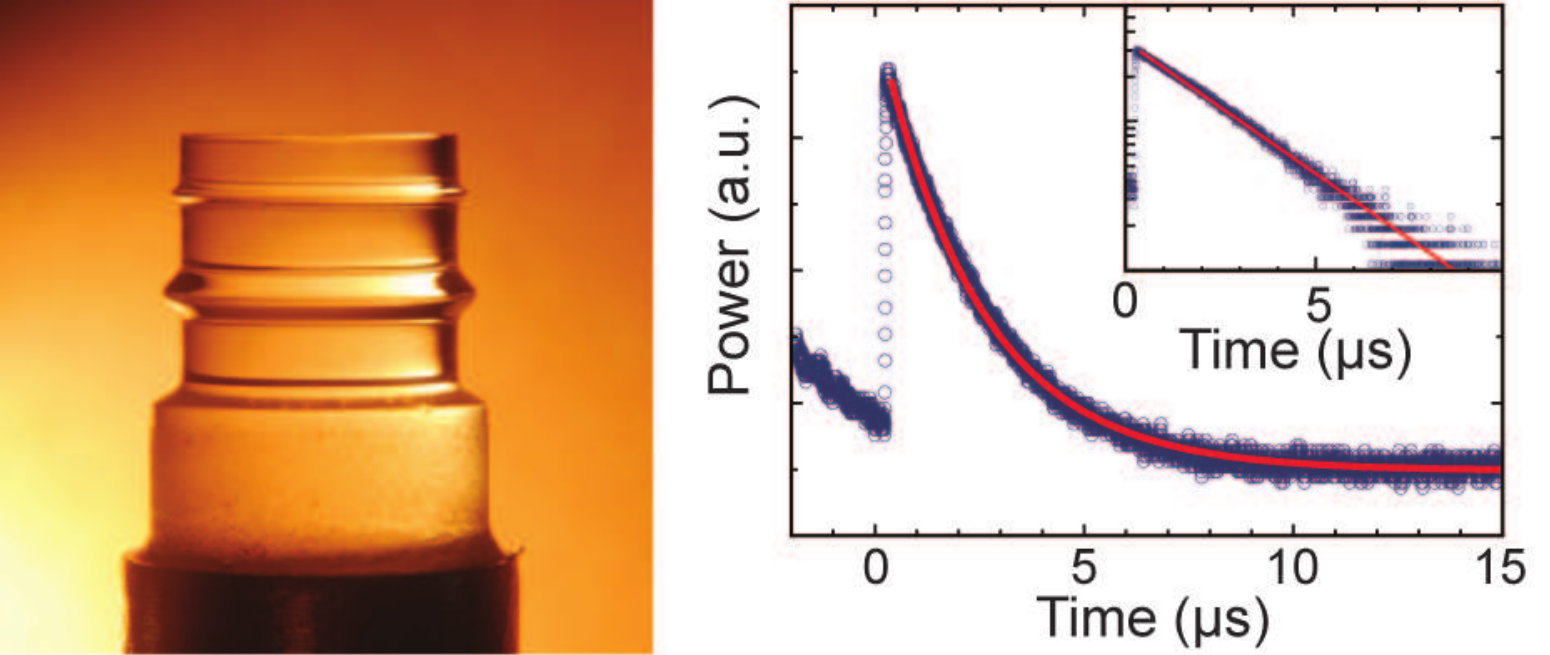}  
\caption{Whispering gallery mode resonator made of calcium fluoride. The
rims of the protrusions support WGM with extreme photon storage times. Left
panel is a photograph of a $6\,\mathrm{mm} $-diameter prototype resonator
produced at MPQ. Right panel shows a ringdown trace for a critically coupled
resonator. The ringdown time of $2.5\,\mathrm{\protect\mu s} $ corresponds
to an intrinsic quality factor of $1{.}2\cdot 10^{10}$ \protect\cite%
{Hofer2009}.}
\label{f:10tt10}
\end{figure}

We have therefore started to explore the optomechanical properties of
crystalline WGM resonators in our laboratory. Several different geometries
were fabricated using a precision lathe and polished using diamond slurry.
An example of such a resonator is shown in figure \ref{f:10tt10}. We have
achieved optical quality factors up to $Q_0=1{.}2\cdot 10^{10}$,
corresponding to a linewidth of $\kappa/2\pi=24\,\mathrm{kHz} $ in a $R=1{.}%
8 \,\mathrm{mm} $ resonator (intrinsic Finesse $\mathcal{F}_0=760{,}000$).
We have also produced an $800\,\mathrm{\mu m}  $-diameter, $100\,\mathrm{\mu
m}  $ thick disk and achieved $Q_0\approx 10^9$ ($\mathcal{F}_0=400{,}000$).
Optical transduction techniques described in section \ref{s:UHS} are used to
measure mechanical modes in these structures, which are typically found in
the range between $0{.}5$ and $5\,\mathrm{MHz} $. The highest measured
mechanical quality factor was $Q_\mathrm{m} =136{,}000$. Already an
encouraging value, there is strong evidence for this value to be still
limited by clamping losses which can be mitigated by a more suitable design
and suspension of the resonators. Furthermore, the rather high effective
masses (around $600\,\mathrm{\mu g} $) and the slightly weaker
optomechanical coupling ($|g_0 |/2\pi\sim 1.5 \,\mathrm{kHz/fm} )$ call for
further miniaturization of the structures. For example, a $80\,\mathrm{\mu m}
$-diameter, $10\,\mathrm{\mu m}  $ thick disk would possess $m_\text{eff}%
=90\,\mathrm{ng} $ and $\Omega_\mathrm{m} /2\pi=63 \,\mathrm{MHz} $. If $Q_%
\mathrm{m} \sim 10^8$ can be reached, the power required to cool such a
device from $T=T_\mathrm{m}=1{.}6\,\mathrm{K} $ to $T_\mathrm{m} \approx
\hbar \Omega_\mathrm{m} /k_\mathrm{B}$ is as low as $10\,\mathrm{\mu W} $ in
the resolved-sideband regime. Heating due to absorbtion is likely to be
totally negligible considering the optical quality of the crystals.

We finally point out that many other crystalline materials may be amenable
to this approach, combining optomechanical coupling with yet other
functionality. As an example, we have tested various polished diamond
spheres, and have observed WGMs with quality factors up to $3\cdot 10^6$ in
a 3~mm-diameter sphere.

A somewhat opposite approach consists in decoupling optical and mechanical
degrees of freedom. Keeping the silica WGM resonators as optical cavities,
but placing an external mechanical oscillator in the near-field of the WGM
allows for independently engineered mechanical properties. For example,
light-weight ($m_\text{eff}\approx 1 \,\mathrm{pg} $) nanomechanical
oscillators such as strained SiN strings, have been shown to result in  $%
Q_\mathrm{m} \approx 10^6$ at $\sim1\,\mathrm{MHz} $-resonance frequencies 
\cite{Verbridge2008}. Placed in the near-field of a WGM, movement of the
polarizable oscillator in the optical field gradient induces frequency
shifts on the order of $|g_0 |\approx 10\,\mathrm{MHz/nm} $, without
inducing detectable optical loss of the WGM. A measurement imprecision below
the noise level associated with zero-point fluctuations and dynamical
backaction induced by the optical dipole force have been demonstrated with
such a system in our laboratory \cite{Anetsberger2009}.

These results are very encouraging for studies pertaining to the observation
of quantum backaction. Usually masked by the much stronger thermal noise,
this effect may become detectable if the ratio $\bar S_{FF}^\mathrm{ba,qn}%
(\Omega_\mathrm{m} )/\bar S_{FF}^\mathrm{the}(\Omega_\mathrm{m} )\approx
\bar a ^2 g_0 ^2 \kappa \hbar^2 \Omega_\mathrm{m} ^{-2}/2 m_\text{eff}
\Gamma_\mathrm{m}  k T$ (assuming the limit $\Omega_\mathrm{m}  \gtrsim
\kappa/2$) approaches unity. Leveraging the low mass and high quality factor
available with nanomechanical oscillators, it appears feasible to approach
this regime, with moderate optical probing powers of $\sim 100\,\mathrm{\mu W%
} $ even \emph{at room temperature}. To differentiate the added noise
induced by quantum backaction from potentially present absorption induced
heating, correlation measurements \cite{Verlot2008} or backaction
cancellation \cite{Caniard2007} between two oscillators with slightly
different frequencies may be employed.

A final example of near-future research projects enabled by the progress
presented in this thesis is an investigation of the regime of strong
optomechanical coupling \cite{Dobrindt2008} in the yet unexplored optical
domain. As our analysis shows, the strong---but tunable---coupling of the
optical mode to the mechanical mode via the mean photon field $\bar a $
gives rise to features very similar to the effect of electromagnetically
induced transparency (EIT) in atomic physics \cite%
{Zhang2003,Fleischhauer2005}.

\begin{figure}[tb]
\centering
\includegraphics[width=.8\linewidth]{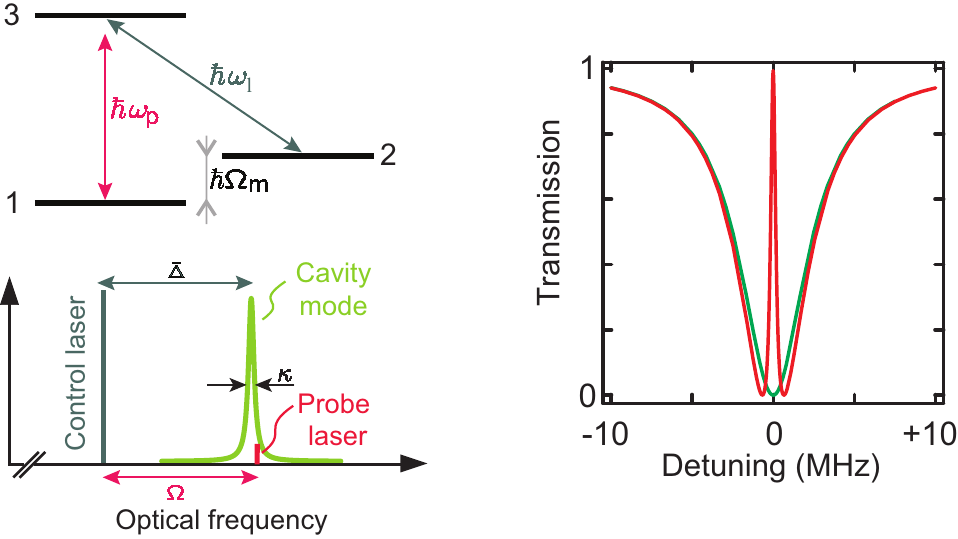}  
\caption{Optomechanically induced transparency. Left: A coupling laser of
frequency $\protect\omega_\mathrm{l} $ drives red sideband transitions
between (sets of) levels 2 and 3, which differ both in the number of optical
and mechanical excitation quanta (each time a photon is added to the
coupling field, a phonon in the mechanical degree of freedom is removed). A
probing laser tuned close to the $1\rightarrow 3$ transition of the
unperturbed systems actually interacts with a superposition of levels 2 and
3, leading to the opening of a transmission window in the case of an
optomechanical system. Right panel shows the resulting power transmission of
the probe laser through a taper-toroid system for $\bar \Delta =-\Omega_%
\mathrm{m} $ versus the detuning $(\Omega -\Omega_\mathrm{m} )/2\protect\pi$
of the probe laser. Numerical parameters are typical for silica
microtoroids, with $\Omega_\mathrm{m} /2\protect\pi=40\,\mathrm{MHz} $, $%
\Gamma_\mathrm{m} /2\protect\pi=1.3\,\mathrm{kHz} $, $\protect\kappa/2%
\protect\pi= 5\,\mathrm{MHz} $, $R=$, $m_\text{eff}=10\,\mathrm{ng} $, $%
R=40\,\mathrm{\protect\mu m}  $ and $\bar s_\mathrm{in} ^2 \hbar \protect%
\omega_\mathrm{l} =300\,\mathrm{\protect\mu W} $ (red line) or $\bar s_%
\mathrm{in} ^2 \hbar \protect\omega_\mathrm{l} =0\,\mathrm{\protect\mu W} $
(green line).}
\label{f:omit}
\end{figure}

The basic idea of such an experiment is illustrated in figure \ref{f:omit}.
The strong field $\bar a $ of the coupling (formerly ``cooling'') laser
oscillates at frequency $\omega_\mathrm{l} =\omega_\mathrm{c} -\Omega_%
\mathrm{m} $, and thereby couples the levels $2\leftrightarrow 3$ by
processes in which a phonon is removed upon the addition of a photon to the
intracavity field (red-sideband transitions). A second, very weak laser
oscillating at $\omega_\mathrm{p}=\omega_\mathrm{l} +\Omega $, probes
``carrier'' transitions $1\leftrightarrow 2$, in which originally no phonons
are added or removed. Due to the strong optomechanical coupling induced by
the coupling field, it effectively interacts with a hybrid optomechanical
resonance. Similar to EIT, this hybridization opens up a \emph{tunable}
transmission window for the probing laser at the center of the optical
resonance. We therefore refer to this effect as ``optomechanically induced
transparency''.

A simple model for this scheme can be set up directly from the Langevin
equations (\ref{e:eomx})--(\ref{e:aux00}). Neglecting thermal and quantum noise for an
elementary analysis, one obtains for the (power) transmission of the probing
laser
\begin{align}
  T_\mathrm{p}&=\left|
1-\frac{1+i f(\Og)}
        {-i(\bD+\Og)+\kappa/2+2\bD f(\Og)} \etac \kappa
\right|^2 \intertext{with}
  f(\Og)&=\hbar \dwdx^2 \ba^2\,\frac{\chi(\Og)}{i(\bD-\Og)+\kappa/2}.%
\end{align} 
At a basic level, this effect can be understood from the fact that for $%
\omega_\mathrm{p}-\omega_\mathrm{l} =\Omega \approx \Omega_\mathrm{m} $, the
beat of coupling and probing laser drives the mechanical oscillator. The
anti-Stokes field generated, in turn, from the coupling field interferes
with the incoming probing light.

The formal analogy with atomic EIT is even more obvious in the limiting case
of: \textit{(i)} a high-$Q$ oscillator $\Omega_\mathrm{m} \gg\Gamma_\mathrm{m%
} $ 
\textit{(ii)} the resolved-sideband regime $\Omega_\mathrm{m} \gg\kappa$ and 
\textit{(iii)} the detuning $\bar \Delta =-\Omega_\mathrm{m} $. Using the
abbreviation $\Delta^{\prime }\equiv\Omega -\Omega_\mathrm{m} $ one can then
simplify the equation system to 
\begin{align}
\left(-i \Delta^{\prime }+\kappa/2\right)\delta \hat a (\Omega ) &=+i g_0 
\bar a  \delta \hat x (\Omega )+\sqrt{\eta_\mathrm{c}  \kappa}\,\delta \hat
s_\mathrm{in} (\Omega ) \\
(-i \Delta^{\prime }+ \Gamma_\mathrm{m} ) \delta \hat x (\Omega )&= i \frac{%
\hbar g_0  \bar a }{m_\text{eff} \Omega_\mathrm{m} } \delta \hat a (\Omega )
\end{align}
directly analogous to the EIT case \cite{Milonni2005}, with the well-known
solution 
\begin{equation}
\delta \hat a (\Omega )=\frac{\eta_\mathrm{c}  \kappa}{(- i \Delta^{\prime
}+\kappa/2)+\frac{2\bar a ^2 g_0 ^2 x_\mathrm{ZPF} ^2}{\-i \Delta^{\prime
}+\Gamma_\mathrm{m} }}.
\end{equation}
Mechanical and optical oscillator play the role of two dipole transitions,
where a pump laser couples two of the involved three levels. As in EIT, the
modification of the transmission for the probe laser may be used for tuning
optical group velocities to generate slow or fast light \cite{Kash1999,
Hau1999, Milonni2005}, or, in designs hosting large arrays of optomechanical
systems, even storage of light in mechanical states \cite{Fleischhauer2000,
Liu2001, Phillips2001} may be envisioned.

\renewcommand{\mdagger}{\dagger} \renewcommand{\mhat}{\hat}

\section*{Acknowledgements}

\addcontentsline{toc}{section}{Acknowledgements}

We are grateful to many people for their contributions to the work presented
in this chapter. In particular, we would like to thank our collaborators
Olivier Arcizet, R\'emi Rivi\`ere, Georg Anetsberger, Johannes Hofer and
Stefan Weis for their experimental and theoretical contributions, and Thomas
Becker and J\"org Kotthaus for infrastructure support. We are indebted to
Nima Nooshi, Ignacio Wilson-Rae and Wilhelm Zwerger for their input on
theoretical aspects of this work. We also benefited from stimulating
discussions with J\"org Kotthaus and Theodor H\"ansch. This work was funded
by the Max-Planck Society, the Deutsche Forschungsemeinschaft, the European
Union MINOS Project and a Marie Curie Excellence Grant.

\addcontentsline{toc}{section}{References} 
\bibliographystyle{elsart-harv}
\bibliography{/Users/aschlies/Documents/Literature/microCavities}

\begin{thebibliography}{221}
\expandafter\ifx\csname natexlab\endcsname\relax\def\natexlab#1{#1}\fi
\expandafter\ifx\csname url\endcsname\relax
  \def\url#1{\texttt{#1}}\fi
\expandafter\ifx\csname urlprefix\endcsname\relax\def\urlprefix{URL }\fi

\bibitem[{Anetsberger et~al.(2009)Anetsberger, Arcizet, Unterreithmeier,
  Rivi{\`e}re, Schliesser, Weig, Kotthaus, and Kippenberg}]{Anetsberger2009}
Anetsberger, G., Arcizet, O., Unterreithmeier, Q.~P., Rivi{\`e}re, R.,
  Schliesser, A., Weig, E.~M., Kotthaus, J.~P., Kippenberg, T.~J., 2009.
  Near-field cavity optomechanics with nanomechanical oscillators. Nature
  Physics 5, 909--914.

\bibitem[{Anetsberger et~al.(2008)Anetsberger, Rivi{\`e}re, Schliesser,
  Arcizet, and Kippenberg}]{Anetsberger2008}
Anetsberger, G., Rivi{\`e}re, R., Schliesser, A., Arcizet, O., Kippenberg,
  T.~J., 2008. Ultralow-dissipation optomechanical resonators on a chip. Nature
  Photonics 2, 627--633.

\bibitem[{Aoki et~al.(2006)Aoki, Bayan, Wilcut, Bowen, Parkins, Kippenberg,
  Vahala, and Kimble}]{Aoki2006}
Aoki, T., Bayan, D., Wilcut, E., Bowen, W.~P., Parkins, A.~S., Kippenberg,
  T.~J., Vahala, K.~J., Kimble, H.~J., 2006. Observation of strong coupling
  between one atom and a monolithic microresonator. Nature 443, 671--674.

\bibitem[{Arcizet(2007)}]{Arcizet2007}
Arcizet, O., 2007. Mesure optique ultrasensible et refroidissement par pression
  de radiation d{'}un micro-r{\'e}sonateur m{\'e}canique. Ph.D. thesis,
  Universit{\'e} Paris VI.

\bibitem[{Arcizet et~al.(2006{\natexlab{a}})Arcizet, Cohadon, Briant, Pinard,
  and Heidmann}]{Arcizet2006a}
Arcizet, O., Cohadon, P.-F., Briant, T., Pinard, M., Heidmann, A.,
  2006{\natexlab{a}}. Radiation-pressure cooling and optomechanical instability
  of a micromirror. Nature 444, 71--74.

\bibitem[{Arcizet et~al.(2006{\natexlab{b}})Arcizet, Cohadon, Briant, Pinard,
  Heidmann, Mackowski, Michel, Pinard, Fran{c}ais, and Rousseau}]{Arcizet2006}
Arcizet, O., Cohadon, P.-F., Briant, T., Pinard, M., Heidmann, A., Mackowski,
  J.-M., Michel, C., Pinard, L., Fran{c}ais, O., Rousseau, L.,
  2006{\natexlab{b}}. High-sensitivity optical monitoring of a micromechanical
  resonator with a quantum-limited optomechanical sensor. Physical Review
  Letters 97, 133601.

\bibitem[{Arcizet et~al.(2008)Arcizet, Molinelli, Briant, Heidmann, Mackowksi,
  Michel, Pinard, Francais, and Rousseau}]{Arcizet2008a}
Arcizet, O., Molinelli, C., Briant, T. anf~Cohadon, P.-F., Heidmann, A.,
  Mackowksi, J.-M., Michel, C., Pinard, L., Francais, O., Rousseau, L., 2008.
  Experimental optomechanics with silicon micromirrors. New Journal of Physics
  10, 125021.

\bibitem[{Arcizet et~al.(2009)Arcizet, Rivi{\`e}re, Schliesser, Anetsberger,
  and Kippenberg}]{Arcizet2009a}
Arcizet, O., Rivi{\`e}re, R., Schliesser, A., Anetsberger, G., Kippenberg,
  T.~J., 2009. Cryogenic properties of optomechanical silica microcavities.
  Physical Review A 80, 021803(R).

\bibitem[{Armani et~al.(2007)Armani, Kulkarni, Fraser, Flagan, and
  Vahala}]{Armani2007}
Armani, A.~M., Kulkarni, R.~P., Fraser, S.~E., Flagan, R.~C., Vahala, K.~J.,
  2007. Label-free, single-molecule detection with optical microcavities.
  Science 317, 783--787.

\bibitem[{Arnold et~al.(2003)Arnold, Khoshima, Teraoka, Holler, and
  Vollmer}]{Arnold2003}
Arnold, S., Khoshima, M., Teraoka, I., Holler, S., Vollmer, F., 2003. Shift of
  whispering-gallery modes in microspheres by protein adsorption. Optics
  Letters 28, 272--274.

\bibitem[{Ashkin(1970)}]{Ashkin1970}
Ashkin, A., 1970. Acceleration and trapping of particles by radiation pressure.
  Physical Review Letters 24, 156--159.

\bibitem[{Ashkin(1978)}]{Ashkin1978}
Ashkin, A., 1978. Trapping of atoms by resonance radiation pressure. Physical
  Review Letters 40, 729--732.

\bibitem[{Bartell and Hunklinger(1982)}]{Bartell1982}
Bartell, U., Hunklinger, S., 1982. Pressure dependence of the low-temperature
  acoustic anomalies in vitreous silica. Journal de Physique Colloques 43, 498.

\bibitem[{Bhattacharya et~al.(2008{\natexlab{a}})Bhattacharya, Giscard, and
  Meystre}]{Bhattacharya2008a}
Bhattacharya, M., Giscard, P.~L., Meystre, P., {JAN} 2008{\natexlab{a}}.
  Entanglement of a {L}aguerre-{G}aussian cavity mode with a rotating mirror.
  Physical Review A 77~({1}), 013827.

\bibitem[{Bhattacharya et~al.(2008{\natexlab{b}})Bhattacharya, Giscard, and
  Meystre}]{Bhattacharya2008}
Bhattacharya, M., Giscard, P.~L., Meystre, P., {MAR} 2008{\natexlab{b}}.
  Entangling the rovibrational modes of a macroscopic mirror using radiation
  pressure. Physical Review A 77~({3}), 030303.

\bibitem[{Bhattacharya and Meystre(2007)}]{Bhattacharya2007a}
Bhattacharya, M., Meystre, P., {AUG 17} 2007. Trapping and cooling a mirror to
  its quantum mechanical ground state. Physical Review Letters 99~({7}),
  073601.

\bibitem[{Bjorklund et~al.(1983)Bjorklund, Levenson, Lenth, and
  Ortiz}]{Bjorklund1983}
Bjorklund, D.~C., Levenson, M.~D., Lenth, W., Ortiz, C., 1983. Frequency
  {M}odulation ({FM}) {S}pectroscopy. Applied Physics B 32, 145--152.

\bibitem[{Black(2001)}]{Black2001}
Black, E.~D., 2001. An introduction to {P}ound-{D}rever-{H}all laser frequency
  stabilization. American Journal of Physics 69, 79--87.

\bibitem[{Blair et~al.(1995)Blair, Ivanov, Tobar, Turner, van Kann, and
  Heng}]{Blair1995}
Blair, D.~G., Ivanov, E.~N., Tobar, M.~E., Turner, P.~J., van Kann, F., Heng,
  I.~S., 1995. High sensitivity gravitational wave antenna with parametric
  transducer readout. Physical Review Letters 74, 1908--1911.

\bibitem[{Blencowe and Buks(2007)}]{Blencowe2007}
Blencowe, M.~P., Buks, E., 2007. Quantum analysis of a linear dc squid
  mechanical displacement detector. Physical Review B 76, 014511.

\bibitem[{Blencowe et~al.(2005)Blencowe, Imbers, and Armour}]{Blencowe2005}
Blencowe, M.~P., Imbers, J., Armour, A.~D., 2005. Dynamics of a nanomechanical
  resonator coupled to a superconducting single-electron transistor. New
  Journal of Physics 7, 236.

\bibitem[{Bocko and Onofrio(1996)}]{Bocko1996}
Bocko, M.~F., Onofrio, R., 1996. On the measurement of a weak classical force
  coupled to a harmonic oscillator: experimental progress. Review of Modern
  Physics 68, 755--799.

\bibitem[{B{\"o}hm et~al.(2006)B{\"o}hm, Gigan, Blaser, Zeilinger, Aspelmeyer,
  Langer, B{\"a}uerle, Hertzberg, and Schwab}]{Bohm2006}
B{\"o}hm, H.~R., Gigan, S., Blaser, F., Zeilinger, A., Aspelmeyer, M., Langer,
  G., B{\"a}uerle, D., Hertzberg, J.~B., Schwab, K.~C., 2006. High reflectivity
  high-{Q} micromechanical {B}ragg mirror. Applied Physics Letters 89, 223101.

\bibitem[{Bose et~al.(1997)Bose, Jacobs, and Knight}]{Bose1997}
Bose, S., Jacobs, K., Knight, P.~L., 1997. Preparation of nonclassical states
  in cavities with a moving mirror. Physical Review A 56, 4175--4186.

\bibitem[{Braginskii and Manukin(1967)}]{Braginskii1967}
Braginskii, V.~B., Manukin, A.~B., 1967. Ponderomotive effects of
  electromagnetic radiation. Soviet Physics JETP Letters 25~(4), 653--655.

\bibitem[{Braginskii et~al.(1970)Braginskii, Manukin, and
  Tikhonov}]{Braginskii1970}
Braginskii, V.~B., Manukin, A.~B., Tikhonov, M.~Y., 1970. Investigation of
  dissipative ponderomotive effects of electromagnetic radiation. Soviet
  Physics JETP 31, 829--830.

\bibitem[{Braginsky et~al.(2000)Braginsky, Gorodetsky, and
  Vyatchanin}]{Braginsky2000}
Braginsky, V.~B., Gorodetsky, M., Vyatchanin, S., 2000. Thermo-refractive noise
  in gravitational wave antennae. Physics Letters A 271, 303--307.

\bibitem[{Braginsky et~al.(1989)Braginsky, Gorodetsky, and
  Ilchenko}]{Braginsky1989}
Braginsky, V.~B., Gorodetsky, M.~L., Ilchenko, V.~S., 1989. Quality-factor and
  nonlinear properties of optical whispering-gallery modes. Physics Letters A
  137~(7-8), 393--397.

\bibitem[{Braginsky et~al.(2003)Braginsky, Gorodetsky, Khalili, Matsko, Thorne,
  and Vyatchanin}]{Braginsky2003}
Braginsky, V.~B., Gorodetsky, M.~L., Khalili, F.~Y., Matsko, A.~B., Thorne,
  K.~S., Vyatchanin, S.~P., 2003. Noise in gravitational-wave detectors and
  other classical-force measurements is not influenced by test-mass
  quantization. Physical Review D 67, 082001.

\bibitem[{Braginsky et~al.(1999)Braginsky, Gorodetsky, and
  Vyatchanin}]{Braginsky1999}
Braginsky, V.~B., Gorodetsky, M.~L., Vyatchanin, S.~P., 1999. Thermodynamical
  fluctuations and photo-thermal shot noise in gravitational wave antennae.
  Physics Letters A 264, 1--10.

\bibitem[{Braginsky and Khalili(1992)}]{Braginsky1992}
Braginsky, V.~B., Khalili, F.~Y., 1992. Quantum {M}easurement. Cambridge
  University Press.

\bibitem[{Braginsky and Khalili(1996)}]{Braginsky1996}
Braginsky, V.~B., Khalili, F.~Y., 1996. Quantum nondemolition measurements: the
  route from toys to tools. Reviews of Modern Physics 68, 1--11.

\bibitem[{Braginsky and Manukin(1977)}]{Braginsky1977}
Braginsky, V.~B., Manukin, A.~B., 1977. Measurement of {W}eak {F}orces in
  {P}hysics {E}xperiments. University of Chicago Press.

\bibitem[{Braginsky et~al.(2001)Braginsky, Strigin, and
  Vyatchanin}]{Braginsky2001}
Braginsky, V.~B., Strigin, S.~E., Vyatchanin, V.~P., 2001. Parametric
  oscillatory instability in {F}abry-{P}erot interferometer. Physics Letters A
  287~(5-6), 331--338.

\bibitem[{Braginsky et~al.(1977)Braginsky, Vorontsov, and
  Khalili}]{Braginsky1977a}
Braginsky, V.~B., Vorontsov, Y.~I., Khalili, F.~Y., 1977. Soviet Physics JETP
  46, 705.

\bibitem[{Braginsky et~al.(1980)Braginsky, Vorontsov, and
  Thorne}]{Braginsky1980}
Braginsky, V.~B., Vorontsov, Y.~I., Thorne, K., 1980. Quantum {N}ondemolition
  {M}easurements. Science 209, 547--557.

\bibitem[{Braginsky and Vyatchanin(2002)}]{Braginsky2002}
Braginsky, V.~B., Vyatchanin, S.~P., 2002. Low quantum noise tranquilizer for
  {F}abry-{P}erot interferometer. Physics Letters A 293, 228--234.

\bibitem[{Brennecke et~al.(2008)Brennecke, Ritter, Donner, and
  Esslinger}]{Brennecke2008}
Brennecke, F., Ritter, S., Donner, T., Esslinger, T., 2008. Cavity
  optomechanics with a {B}ose-{E}instein condensate. Science 322, 235--238.

\bibitem[{Briant et~al.(2003)Briant, Cohadon, Heidmann, and
  Pinard}]{Briant2003b}
Briant, T., Cohadon, P.-F., Heidmann, A., Pinard, M., 2003. Optomechanical
  characterization of acoustic modes in a mirror. Physical Review A 68, 033823.

\bibitem[{Brown et~al.(2007)Brown, Britton, Epstein, Chiaverini, Leibfried, and
  Wineland}]{Brown2007}
Brown, K.~R., Britton, J., Epstein, R.~J., Chiaverini, J., Leibfried, D.,
  Wineland, D.~J., 2007. Passive cooling of a micromechanical oscillator with a
  resonant electric circuit. Physical Review Letters 99~(13), 137205.
\newline\urlprefix\url{http://link.aps.org/abstract/PRL/v99/e137205}

\bibitem[{Caniard et~al.(2007{\natexlab{a}})Caniard, Briant, Cohadon, Pinard,
  and Heidmann}]{Caniard2007a}
Caniard, T., Briant, T., Cohadon, P.-F., Pinard, M., Heidmann, A.,
  2007{\natexlab{a}}. Ultrasensitive optical measurement of thermal and quantum
  noises. Optics and Spectroscopy 103, 225--230.

\bibitem[{Caniard et~al.(2007{\natexlab{b}})Caniard, Verlot, Briant, Cohadon,
  and Heidmann}]{Caniard2007}
Caniard, T., Verlot, P., Briant, T., Cohadon, P.-F., Heidmann, A.,
  2007{\natexlab{b}}. Observation of back-action noise cancellation in
  interferometric and weak force measurements. Physical Review Letters 99,
  110801.

\bibitem[{Carmon et~al.(2007)Carmon, Cross, and Vahala}]{Carmon2007a}
Carmon, T., Cross, M.~C., Vahala, K.~J., 2007. Chaotic quivering of
  micron-scaled on-chip resonators excited by centrifugal optical pressure.
  Physical Review Letters 98, 167203.

\bibitem[{Carmon et~al.(2005)Carmon, Rokhsari, Yang, Kippenberg, and
  Vahala}]{Carmon2005}
Carmon, T., Rokhsari, H., Yang, L., Kippenberg, T.~J., Vahala, K.~J., 2005.
  Temporal behavior of radiation-pressure-induced vibrations of an optical
  microcavity phonon mode. Physical Review Letters 94~(22), 223902.

\bibitem[{Carmon and Vahala(2007)}]{Carmon2007}
Carmon, T., Vahala, K.~J., 2007. Modal spectroscopy of optoexcited vibrations
  of a micron-scale on-chip resonator at greater than 1 {G}{H}z frequency.
  Physical Review Letters 98, 123901.

\bibitem[{Carmon et~al.(2004)Carmon, Yang, and Vahala}]{Carmon2004a}
Carmon, T., Yang, L., Vahala, K.~J., 2004. Dynamical thermal behavior and
  thermal selfstability of microcavities. Optics Express 12, 4742--4750.

\bibitem[{Caves(1980)}]{Caves1980}
Caves, C.~M., Jul 1980. Quantum-mechanical radiation-pressure fluctuations in
  an interferometer. Physical Review Letters 45~(2), 75--79.

\bibitem[{Caves(1981)}]{Caves1981}
Caves, C.~M., 1981. Quantum-mechanical noise in an interferometer. Physical
  Review D 23, 1693.

\bibitem[{Chang and Campillo(1996)}]{Chang1996}
Chang, R.~K., Campillo, A.~J., 1996. Optical processes in microcavities. World
  Scientific.

\bibitem[{Chu et~al.(1985)Chu, Hollberg, Bjorkholm, Cable, and
  Ashkin}]{Chu1985}
Chu, S., Hollberg, L., Bjorkholm, J.~E., Cable, A., Ashkin, A., Jul 1985.
  Three-dimensional viscous confinement and cooling of atoms by resonance
  radiation pressure. Physical Review Letters 55~(1), 48--51.

\bibitem[{Clerk et~al.(2008)Clerk, Devoret, Girvin, Marquardt, and
  Schoelkopf}]{Clerk2008}
Clerk, A.~A., Devoret, M.~H., Girvin, S.~M., Marquardt, F., Schoelkopf, R.~J.,
  2008. Introduction to {Q}uantum {N}oise, {M}easurement and {A}mplification.
  arXiv:0810.4729.

\bibitem[{Cohadon et~al.(1999)Cohadon, Heidmann, and Pinard}]{Cohadon1999}
Cohadon, P.-F., Heidmann, A., Pinard, M., 1999. Cooling of a mirror by
  radiation pressure. Physical Review Letters 83, 3174--3177.

\bibitem[{Cole et~al.(2008)Cole, Gr{\"o}blacher, Gugler, and
  Aspelmeyer}]{Cole2008}
Cole, G.~D., Gr{\"o}blacher, S., Gugler, K.~Gigan, S., Aspelmeyer, M., 2008.
  Monocrystalline {A}l$_{x}${G}a$_{1-x}${A}s heterostructures for
  high-reflectivity high-{Q} micromechanical resonators in the megahertz
  regime. Applied Physics Letters 92, 261108.

\bibitem[{Collot et~al.(1993)Collot, Lef{\`e}vre-Seguin, Brune, Raimond, and
  Haroche}]{Collot1993}
Collot, L., Lef{\`e}vre-Seguin, V., Brune, M., Raimond, J.~M., Haroche, S.,
  1993. Very high-{Q} whispering-gallery mode resonances observed on fused
  silica microshperes. Europhysics Letters 23~(5), 327--334.

\bibitem[{Corbitt et~al.(2007{\natexlab{a}})Corbitt, Chen, Innerhofer,
  Muller-Ebhardt, Ottaway, Rehbein, Sigg, Whitcomb, Wipf, and
  Mavalvala}]{Corbitt2007}
Corbitt, T., Chen, Y., Innerhofer, E., Muller-Ebhardt, H., Ottaway, D.,
  Rehbein, H., Sigg, D., Whitcomb, S., Wipf, C., Mavalvala, N.,
  2007{\natexlab{a}}. An all-optical trap for a gram-scale mirror. Physical
  Review Letters 98~(15), 150802.

\bibitem[{Corbitt et~al.(2007{\natexlab{b}})Corbitt, Wipf, Bodiya, Ottaway,
  Sigg, Smith, Whitcomb, and Mavalvala}]{Corbitt2007a}
Corbitt, T., Wipf, C., Bodiya, T., Ottaway, D., Sigg, D., Smith, N., Whitcomb,
  S., Mavalvala, N., 2007{\natexlab{b}}. Optical dilution and feedback cooling
  of a gram-scale oscillator to 6.9 m{K}. Physical Review Letters 99~(16),
  160801.
\newline\urlprefix\url{http://link.aps.org/abstract/PRL/v99/e160801}

\bibitem[{Cuthbertson et~al.(1996)Cuthbertson, Tobar, Ivanov, and
  Blair}]{Cuthbertson1996}
Cuthbertson, B.~D., Tobar, M.~E., Ivanov, E, N., Blair, D.~G., 1996. Parametric
  back-action effects in a high-{Q} cyrogenic sapphire transducer. Review of
  Scientific Instruments 67, 2435--2442.

\bibitem[{Dantan et~al.(2008)Dantan, Genes, Vitali, and Pinard}]{Dantan2008}
Dantan, A., Genes, C., Vitali, D., Pinard, M., 2008. Self-cooling of a movable
  mirror to the ground state using radiation pressure. Physical Review A 77,
  011804(R).

\bibitem[{Diedrich et~al.(1989)Diedrich, Bergquist, Itano, and
  Wineland}]{Diedrich1989}
Diedrich, F., Bergquist, J.~C., Itano, W.~M., Wineland, D.~J., Jan 1989. Laser
  cooling to the zero-point energy of motion. Physical Review Letters 62~(4),
  403--406.

\bibitem[{Di{\'o}si(2008)}]{Diosi2008}
Di{\'o}si, L., 2008. Laser linewidth hazard in optomechanical cooling. Physical
  Review A 78, 021801.

\bibitem[{Dixon(1967)}]{Dixon1967}
Dixon, R.~W., 1967. Photoelastic properties of selected materials and their
  relevance for applications to acoustic light modulators and scanners. Journal
  of Applied Physics 38, 5149.

\bibitem[{Dobrindt and Kippenberg(2009)}]{Dobrindt2009}
Dobrindt, J., Kippenberg, T.~J., 2009. Theory of quantum backaction enhancement
  and displacement measurement using a multiple cavity mode transducer. arxiv.

\bibitem[{Dobrindt et~al.(2008)Dobrindt, Wilson-Rae, and
  Kippenberg}]{Dobrindt2008}
Dobrindt, J.~M., Wilson-Rae, I., Kippenberg, T.~J., 2008. Parametric
  normal-mode splitting in cavity optomechanics. Physical Review Letters 101,
  263602.

\bibitem[{Dorsel et~al.(1983)Dorsel, McCullen, Meystre, Vignes, and
  Walther}]{Dorsel1983}
Dorsel, A., McCullen, J.~D., Meystre, P., Vignes, E., Walther, H., 1983.
  Optical bistability and mirror confinement induced by radiation pressure.
  Physical Review Letters 51~(17), 1550--1553.

\bibitem[{Drever et~al.(1983)Drever, Hall, Kowalski, Hough, Ford, Munley, and
  Ward}]{Drever1983}
Drever, R. W.~P., Hall, J.~L., Kowalski, F.~V., Hough, J., Ford, G.~M., Munley,
  A.~J., Ward, H., 1983. Laser {P}hase and {F}requency {S}tabilization {U}sing
  an {O}ptical {R}esonantor. Applied Physics B 31, 97--105.

\bibitem[{Dykman(1978)}]{Dykman1978}
Dykman, M.~I., 1978. Heating and cooling of local and quasilocal vibrations by
  a nonresonance field. Soviet Physics - Solid State 20, 1306--1311.

\bibitem[{Eichenfield et~al.(2009{\natexlab{a}})Eichenfield, Camacho, Chan,
  Vahala, and Painter}]{Eichenfield2009a}
Eichenfield, M., Camacho, R., Chan, J., Vahala, K., Painter, O.,
  2009{\natexlab{a}}. A picogram and nanometer scale photonic crystal
  opto-mechanical cavity. Nature 459, 550--555.

\bibitem[{Eichenfield et~al.(2009{\natexlab{b}})Eichenfield, Chan, Camacho,
  Vahala, and Painter}]{Eichenfield2009}
Eichenfield, M., Chan, J., Camacho, R.~M., Vahala, K.~J., Painter, O.,
  2009{\natexlab{b}}. Optomechanical crystals. Nature 462, 78--82.

\bibitem[{Eichenfield et~al.(2007)Eichenfield, Michael, Perahia, and
  Painter}]{Eichenfield2007}
Eichenfield, M., Michael, C.~P., Perahia, R., Painter, O., 2007. Actuation of
  micro-optomechanical systems via cavity-enhanced optical dipole forces.
  Nature Photonics 1, 416--422.

\bibitem[{Ekinci and Roukes(2005)}]{Ekinci2005}
Ekinci, K.~L., Roukes, M.~L., 2005. Nanoelectromechanical systems. Review of
  Scientific Instruments 76, 061101.

\bibitem[{Fabre et~al.(1994)Fabre, Pinard, Bourzeix, Heidmann, Giacobino, and
  Reynaud}]{Fabre1994}
Fabre, C., Pinard, M., Bourzeix, S., Heidmann, A., Giacobino, E., Reynaud, S.,
  1994. Quantum-noise reduction using a cavity with a movable mirror. Physical
  Review A 49, 1337--1343.

\bibitem[{Favero et~al.(2007)Favero, Metzger, Camerer, K{\"o}nig, Lorenz,
  Kotthaus, and Karrai}]{Favero2007}
Favero, I., Metzger, C., Camerer, S., K{\"o}nig, D., Lorenz, H., Kotthaus,
  J.~P., Karrai, K., 2007. Optical cooling of a micromirror of wavelength size.
  Applied Physics Letters 90, 104101.

\bibitem[{Favero et~al.(2009)Favero, Stapfner, Hunger, Paulitschke, Reichel,
  Lorenz, Weig, and Karrai}]{Favero2009}
Favero, I., Stapfner, S., Hunger, D., Paulitschke, P., Reichel, J., Lorenz, H.,
  Weig, E.~M., Karrai, K., 2009. Fluctuating nanomechanical systems in a high
  finesse optical microcavity. Optics Express 17, 12813--12820.

\bibitem[{Fleischhauer et~al.(2005)Fleischhauer, Imamoglu, and
  Marangos}]{Fleischhauer2005}
Fleischhauer, M., Imamoglu, A., Marangos, J.~P., 2005. Electromagnetically
  induced transparency: {O}ptics in coherent media. Review of Modern Physics
  77, 633--673.

\bibitem[{Fleischhauer and Lukin(2000)}]{Fleischhauer2000}
Fleischhauer, M., Lukin, M.~D., 2000. Dark-state polaritons in
  electromagnetically induced transparency. Physical Review Letters 84,
  5094--5097.

\bibitem[{Flowers-Jacobs et~al.(2007)Flowers-Jacobs, Schmidt, and
  Lehnert}]{Flowers-Jacobs2007}
Flowers-Jacobs, N.~E., Schmidt, D.~R., Lehnert, K.~W., 2007. Intrinsic noise
  properties of atomic point contact displacement detectors. Physical Review
  Letters 98, 096804.

\bibitem[{Gardiner and Zoller(2004)}]{Gardiner2004}
Gardiner, C.~W., Zoller, P., 2004. Quantum Noise. Springer.

\bibitem[{Genes et~al.(2008)Genes, Vitali, Tombesi, Gigan, and
  Aspelmeyer}]{Genes2008}
Genes, C., Vitali, D., Tombesi, P., Gigan, S., Aspelmeyer, M., 2008.
  Ground-state cooling of a micromechanical oscillator: Comparing cold-damping
  and cavity-assisted cooling schemes. Physical Review A 77, 033804.

\bibitem[{Gigan et~al.(2006)Gigan, B{\"o}hm, Paternosto, Blaser, Langer,
  Hertzberg, Schwab, B{\"a}uerle, Aspelmeyer, and Zeilinger}]{Gigan2006}
Gigan, S., B{\"o}hm, H.~R., Paternosto, M., Blaser, F., Langer, G., Hertzberg,
  J.~B., Schwab, K.~C., B{\"a}uerle, D., Aspelmeyer, M., Zeilinger, A., 2006.
  Self-cooling of a micromirror by radiation pressure. Nature 444, 67--70.

\bibitem[{Gillespie and Raab(1995)}]{Gillespie1995}
Gillespie, A., Raab, F., 1995. Thermally excited vibrations of the mirrors of
  laser interferometer gravitational wave detectors. Physical Review D 52,
  577--585.

\bibitem[{Giovannetti and Vitali(2001)}]{Giovannetti2001}
Giovannetti, V., Vitali, D., 2001. Phase-noise measurement in a cavity with a
  movable mirror undergoing quantum {B}rownian motion. Physical Review A 63,
  023812.

\bibitem[{Gorodetsky and Grudinin(2004)}]{Gorodetsky2004}
Gorodetsky, M.~L., Grudinin, I.~S., 2004. Fundamental thermal fluctuations in
  microspheres. Journal of the Optical Society of America B 21, 697--705.

\bibitem[{Gozzini et~al.(1985)Gozzini, Maccarrone, Mango, Longo, and
  Barbarino}]{Gozzini1985}
Gozzini, A., Maccarrone, F., Mango, F., Longo, I., Barbarino, S., 1985.
  Light-pressure bistability at microwave frequencies. Journal of the Optical
  Society of America B 2, 1841.

\bibitem[{Gr{\"o}blacher et~al.(2008)Gr{\"o}blacher, Gigan, B{\"o}hm,
  Zeilinger, and Aspelmeyer}]{Groblacher2008}
Gr{\"o}blacher, S., Gigan, S., B{\"o}hm, H.~R., Zeilinger, A., Aspelmeyer, M.,
  2008. Radiation pressure self-cooling of a micromirror in a cryogenic
  environment. Europhysics Letters 81, 54003.

\bibitem[{Gr{\"o}blacher et~al.(2009{\natexlab{a}})Gr{\"o}blacher, Hammerer,
  Vanner, and Aspelmeyer}]{Groblacher2009a}
Gr{\"o}blacher, S., Hammerer, K., Vanner, M.~R., Aspelmeyer, M.,
  2009{\natexlab{a}}. Observation of strong coupling between a micromechanical
  resonator and an optical cavity field. Nature 460, 724--727.

\bibitem[{Gr{\"o}blacher et~al.(2009{\natexlab{b}})Gr{\"o}blacher, Hertzberg,
  Vanner, Gigan, Schwab, and Aspelmeyer}]{Groblacher2009}
Gr{\"o}blacher, S., Hertzberg, J.~B., Vanner, M.~R., Gigan, S., Schwab, K.~C.,
  Aspelmeyer, M., 2009{\natexlab{b}}. Demonstration of an ultracold
  micro-optomechanical oscillator in a cryogenic cavity. Nature Physics 5,
  485--488.

\bibitem[{Grudinin et~al.(2006)Grudinin, Ilchenko, and Maleki}]{Grudinin2006a}
Grudinin, I., Ilchenko, V.~S., Maleki, L., 2006. Ultrahigh optical {Q} factors
  of crystalline resonators in the linear regime. Physical Review A 74, 063806.

\bibitem[{Hadjar et~al.(1999)Hadjar, Cohadon, Aminoff, Pinard, and
  Heidmann}]{Hadjar1999}
Hadjar, Y., Cohadon, P.~F., Aminoff, C.~G., Pinard, M., Heidmann, A., 1999.
  High-sensitivity optical measurement of mechanical {B}rownian motion.
  Europhysics Letters 47~(5), 545--551.

\bibitem[{Hahtela et~al.(2004)Hahtela, Nera, and Tittonen}]{Hahtela2004}
Hahtela, O., Nera, K., Tittonen, I., 2004. Position measurement of a cavity
  mirror using polarization spectroscopy. Journal of Optics A 6, S115--S120.

\bibitem[{Hald and Ruseva(2005)}]{Hald2005}
Hald, J., Ruseva, V., 2005. Efficient suppression of diode-laser phase noise by
  optical filtering. Journal of the Optical Society of America B 22~(11),
  2338--2344.

\bibitem[{Hamann et~al.(1998)Hamann, Haycock, Klose, Pax, Deutsch, and
  Jessen}]{Hamann1998}
Hamann, S.~E., Haycock, D.~L., Klose, G., Pax, P.~H., Deutsch, I.~H., Jessen,
  P.~S., May 1998. Resolved-sideband {R}aman cooling to the ground state of an
  optical lattice. Physical Review Letters 80~(19), 4149--4152.

\bibitem[{H{\"a}nsch and Couillaud(1980)}]{Hansch1980}
H{\"a}nsch, T.~W., Couillaud, B., 1980. Laser frequency stabilization by
  polarization spectroscopy of a reflecting reference cavity. Optics
  Communications 35~(3), 441--444.

\bibitem[{H{\"a}nsch and Schawlow(1975)}]{Hansch1975}
H{\"a}nsch, T.~W., Schawlow, A.~L., 1975. Cooling of gases by laser radiation.
  Optics Communications 13, 68--69.

\bibitem[{Harris et~al.(2007)Harris, Zwickl, and Jayich}]{Harris2007}
Harris, J. G.~E., Zwickl, B.~M., Jayich, A.~M., 2007. Stable, mode-matched,
  medium-finesse optical cavity incorporating a micromechanical cantilever.
  Review of Scientific Instruments 78, 013107.

\bibitem[{Hau et~al.(1999)Hau, Harris, Dutton, and Behroozi}]{Hau1999}
Hau, L.~V., Harris, S.~E., Dutton, Z., Behroozi, C.~H., 1999. Light speed
  reduction to 17 metres per second in an ultracold atomic gas. Nature 397,
  594--598.

\bibitem[{Haus(1984)}]{Haus1984}
Haus, H.~A., 1984. Waves and fields in optoelectronics. Prentice-Hall.

\bibitem[{Heidmann et~al.(1997)Heidmann, Hadjar, and Pinard}]{Heidmann1997}
Heidmann, A., Hadjar, Y., Pinard, M., 1997. Quantum non-demolition measurement
  by optomechanical coupling. Applied Physics B 64, 173--180.

\bibitem[{Hertzberg et~al.(2009)Hertzberg, Rocheleau, Ndukum, Savva, Clerk, and
  Schwab}]{Hertzberg2009}
Hertzberg, J.~B., Rocheleau, T., Ndukum, T., Savva, M., Clerk, A.~A., Schwab,
  K.~C., 2009. Back-action evading measurements of nanomechanical motion.
  arxiv:0906.0967.

\bibitem[{Hofer et~al.(2009)Hofer, Schliesser, and Kippenberg}]{Hofer2009}
Hofer, J., Schliesser, A., Kippenberg, T.~J., 2009. Cavity optomechanics with
  ultra-high {Q} crystalline micro-resonators. arXiv:0911.1178.

\bibitem[{H{\"o}hberger~Metzger and Karrai(2004)}]{Hohberger2004}
H{\"o}hberger~Metzger, C., Karrai, K., 2004. Cavity cooling of a microlever.
  Nature 432, 1002--1005.

\bibitem[{Hossein-Zadeh et~al.(2006)Hossein-Zadeh, Rokhsari, Hajimiri, and
  Vahala}]{Hossein-Zadeh2006}
Hossein-Zadeh, M., Rokhsari, H., Hajimiri, A., Vahala, K.~J., 2006.
  Characterization of a radiation-pressure-driven micromechanical oscillator.
  Physical Review A 74, 023813.

\bibitem[{Hossein-Zadeh and Vahala(2007)}]{Hossein-Zadeh2007}
Hossein-Zadeh, M., Vahala, K., 2007. Observation of optical spring effect in a
  microtoroidal optomechanical resonator. Optics Letters 32~(12), 1611--1613.

\bibitem[{Hunklinger et~al.(1973)Hunklinger, Arnold, and
  Stein}]{Hunklinger1973}
Hunklinger, S., Arnold, W., Stein, S., 1973. Anomalous ultrasonic attenuation
  in vitreous silica at low temperatures. Physics Letters 45A, 311--312.

\bibitem[{Hutchinson(1979)}]{Hutchinson1979}
Hutchinson, J.~R., 1979. Axisymmetric flexural vibrations of a thick free
  circular plate. Journal of Applied Mechanics 46, 139--144.

\bibitem[{Hutchinson(1980)}]{Hutchinson1980}
Hutchinson, J.~R., 1980. Vibrations of solid cylinders. Journal of Applied
  Mechanics 47, 901--907.

\bibitem[{Ilchenko et~al.(1998)Ilchenko, Volikov, Velichanski, Treussart,
  Lef{\`e}vre-Seguin, Raimond, and Haroche}]{Ilchenko1998}
Ilchenko, V., Volikov, P.~S., Velichanski, V.~L., Treussart, V.~L.,
  Lef{\`e}vre-Seguin, V., Raimond, J.-M., Haroche, S., 1998. Strain-tunable
  high-{Q} optical microresonator. Optics Communications 145, 86--90.

\bibitem[{Ilchenko et~al.(1994)Ilchenko, Gorodetsky, and
  Vyatchanin}]{Ilchenko1994}
Ilchenko, V.~S., Gorodetsky, M.~L., Vyatchanin, S.~P., 1994. Coupling and
  tunability of optical whispering-gallery modes: a basis for coordinate meter.
  Optics Communications 107, 41--48.

\bibitem[{Itano et~al.(1992)Itano, Bergquist, Bollinger, and
  Wineland}]{Itano1992}
Itano, W.~M., Bergquist, J.~C., Bollinger, J.~J., Wineland, D.~J., 1992. Laser
  cooling of trapped ions. Laser Manipulation of Atoms and Ions. North-Holland,
  Amsterdam, pp. 519--537S.

\bibitem[{J{\"a}ckle(1972)}]{Jackle1972}
J{\"a}ckle, J., 1972. On the ultrasonic attenuation in glasses at low
  temperature. Zeitschrift f{\"u}r Physik 257, 212--223.

\bibitem[{Kash et~al.(1999)Kash, Sautenkov, Zibrov, Hollberg, Welch, Lukin,
  Rostovtsev, Fry, and Scully}]{Kash1999}
Kash, M.~M., Sautenkov, V.~A., Zibrov, A.~S., Hollberg, L., Welch, G.~R.,
  Lukin, M.~D., Rostovtsev, Y., Fry, E.~S., Scully, M.~O., 1999. Ultraslow
  group velocity and enhanced nonlinear optical effects in a coherently driven
  hot atomic gas. Physical Review Letters 82, 5229--5232.

\bibitem[{Kippenberg et~al.(2005)Kippenberg, Rokhsari, Carmon, Scherer, and
  Vahala}]{Kippenberg2005}
Kippenberg, T.~J., Rokhsari, H., Carmon, T., Scherer, A., Vahala, K.~J., 2005.
  Analysis of {R}adiation-{P}ressure {I}nduced {M}echanical {O}scillation of an
  {O}ptical {M}icrocavity. Physical Review Letters 95, 033901.

\bibitem[{Kippenberg et~al.(2006)Kippenberg, Rokhsari, and
  Vahala}]{Kippenberg2006}
Kippenberg, T.~J., Rokhsari, H., Vahala, K., 2006. Scanning probe microscopy of
  thermally excited mechanical modes of an optical microcavity. arxiv:0602234.

\bibitem[{Kippenberg et~al.(2004)Kippenberg, Spillane, and
  Vahala}]{Kippenberg2004a}
Kippenberg, T.~J., Spillane, S.~M., Vahala, K.~J., 2004. Kerr-nonlinearity
  optical parametric oscillation in an ultrahigh-{Q} toroid microcavity.
  Physical Review Letters 93~(8), 083904.

\bibitem[{Kippenberg and Vahala(2007)}]{Kippenberg2007}
Kippenberg, T.~J., Vahala, K., 2007. Cavity {O}pto-{M}echanics. Optics Express
  15, 17172--17205.

\bibitem[{Kippenberg and Vahala(2008)}]{Kippenberg2008}
Kippenberg, T.~J., Vahala, K.~J., 2008. Cavity {O}ptomechanics: {B}ack-{A}ction
  at the {M}esoscale. Science 321, 1172--1176.

\bibitem[{Kleckner and Bouwmeester(2006)}]{Kleckner2006}
Kleckner, D., Bouwmeester, D., 2006. Sub-kelvin optical cooling of a
  micromechanical resonator. Nature 444, 75--78.

\bibitem[{Kleckner et~al.(2006)Kleckner, Marshall, De~Dood, Khodadad, Pors,
  Irvine, and Bouwmeester}]{Kleckner2006a}
Kleckner, D., Marshall, W., De~Dood, J.~M.~A., Khodadad, N.~D., Pors, B.-J.,
  Irvine, W.~T.~M., Bouwmeester, D., 2006. High finesse opto-mechanical cavity
  with a movable thirty-micron-size mirror. Physical Review Letters 96, 173901.

\bibitem[{Knobel and Cleland(2003)}]{Knobel2003}
Knobel, R.~G., Cleland, A.~N., 2003. Nanometre-scale displacement sensing using
  a single-electron transistor. Nature 424, 291--293.

\bibitem[{LaHaye et~al.(2004)LaHaye, Buu, Camarota, and Schwab}]{LaHaye2004}
LaHaye, M.~D., Buu, O., Camarota, B., Schwab, K.~C., 2004. Approaching the
  quantum limit of a nanomechanical resonator. Science 304, 74--77.

\bibitem[{Landau and Lifshitz(1970)}]{Landau1970}
Landau, L.~D., Lifshitz, E.~M., 1970. Theory of elasticity, 2nd Edition. Vol.~7
  of Course of Theoretical Physics. Pergamon Press.

\bibitem[{Landau and Lifshitz(1980)}]{Landau1980}
Landau, L.~D., Lifshitz, E.~M., 1980. Statistical Physics, 3rd Edition. Vol.~5
  of Course of Theoretical Physics. Pergamon Press.

\bibitem[{Landau and Lifshitz(1984)}]{Landau1984}
Landau, L.~D., Lifshitz, E.~M., 1984. Electrodynamics of continuous media, 2nd
  Edition. Vol.~8 of Course of Theoretical Physics. Pergamon Press.

\bibitem[{Landau and Lifshitz(1987)}]{Landau1987}
Landau, L.~D., Lifshitz, E.~M., 1987. Fluid Mechanics, 2nd Edition. Vol.~6 of
  Course of Theoretical Physics. Pergamon Press.

\bibitem[{Law(1995)}]{Law1995}
Law, C.~K., 1995. Interaction between a moving mirror and radiation pressure: A
  {H}amiltonian formulation. Physical Review A 51, 2537--2541.

\bibitem[{Lebedew(1901)}]{Lebedew1901}
Lebedew, P., 1901. Untersuchungen {\"u}ber die {D}ruckkr{\"a}fte des {L}ichtes.
  Annalen der Physik 311, 433--458.

\bibitem[{Leibfried et~al.(2003)Leibfried, Blatt, Monroe, and
  Wineland}]{Leibfried2003}
Leibfried, D., Blatt, R., Monroe, C., Wineland, D., Mar 2003. Quantum dynamics
  of single trapped ions. Review of Modern Physics 75~(1), 281--324.

\bibitem[{Li et~al.(2008)Li, Pernice, Xiong, Baehr-Jones, Hochberg, and
  Tang}]{Li2008}
Li, M., Pernice, W. H.~P., Xiong, C., Baehr-Jones, T., Hochberg, M., Tang,
  H.~X., 2008. Harnessing optical forces in integrated photonic circuits.
  Nature 456, 480--484.

\bibitem[{{LIGO~Scientific~Collaboration}(2009)}]{Abbott2009b}
{LIGO~Scientific~Collaboration}, 2009. Observation of a kilogram-scale
  oscillator near its quantum ground state. New Journal of Physics 11~(7),
  073032.

\bibitem[{Lin et~al.(2009)Lin, Rosenberg, Jiang, Vahala, and Painter}]{Lin2009}
Lin, Q., Rosenberg, J., Jiang, X., Vahala, K.~J., Painter, O., 2009. Mechanical
  oscillation and cooling actuated by the optical gradient force.
  arxiv:0905.2716.

\bibitem[{Liu et~al.(2001)Liu, Dutton, Behroozi, and Hau}]{Liu2001}
Liu, C., Dutton, Z., Behroozi, C., Hau, L., 2001. Observation of coherent
  optical information storage in an atomic medium using halted light pulses.
  Nature 409, 490--493.

\bibitem[{Locke and Tobar(2004)}]{Locke2004}
Locke, C.~R., Tobar, M.~E., 2004. Measurement of the strain-induced coefficient
  of permittivity of sapphire using whispering gallery modes excited in a
  high-{Q} acoustic sapphire oscillator. Measurement Science and Technology 15,
  2145--2149.

\bibitem[{Locke et~al.(1998)Locke, Tobar, Ivanov, and Blair}]{Locke1998}
Locke, C.~R., Tobar, M.~E., Ivanov, E.~N., Blair, D.~G., 1998. Parametric
  interaction of the electric and acoustic fields in a sapphire monocrystal
  transducer with a microwave readout. Journal of Applied Physics 84,
  6523--6527.

\bibitem[{Love(1906)}]{Love1906}
Love, A. E.~H., 1906. A treatise on the mathematical theory of elasticity.
  Cambridge University Press.

\bibitem[{Ludwig et~al.(2008{\natexlab{a}})Ludwig, Kubala, and
  Marquardt}]{Ludwig2008a}
Ludwig, M., Kubala, B., Marquardt, F., 2008{\natexlab{a}}. The optomechanical
  instability in the quantum regime. New Journal of Physics 10, 095013.

\bibitem[{Ludwig et~al.(2008{\natexlab{b}})Ludwig, Neuenhanh, Metzger, Ortlieb,
  Favero, Karrai, and Marquardt}]{Ludwig2008}
Ludwig, M., Neuenhanh, C., Metzger, C., Ortlieb, A., Favero, I., Karrai, K.,
  Marquardt, F., 2008{\natexlab{b}}. Self-induced oscillations in an
  optomechanical system driven by bolometric backaction. Physical review
  Letters 101, 133903.

\bibitem[{Ma et~al.(2007)Ma, Schliesser, Del'Haye, Dabirian, Anetsberger, and
  Kippenberg}]{Ma2007}
Ma, R., Schliesser, A., Del'Haye, P., Dabirian, A., Anetsberger, G.,
  Kippenberg, T., 2007. Radiation-pressure-driven vibrational modes in
  ultrahigh-{Q} silica microspheres. Optics Letters 32, 2200--2202.

\bibitem[{Mancini et~al.(2002)Mancini, Giovannetti, Vitali, and
  Tombesi}]{Mancini2002}
Mancini, S., Giovannetti, V., Vitali, D., Tombesi, P., 2002. Entangling
  {M}acroscopic {O}scillators {E}xploiting {R}adiation {P}ressure. Physical
  Review Letters 88, 120401.

\bibitem[{Mancini and Tombesi(1994)}]{Mancini1994}
Mancini, S., Tombesi, P., 1994. Quantum noise reduction by radiation pressure.
  Physical Review A 49, 4055--4065.

\bibitem[{Marquardt et~al.(2007)Marquardt, Chen, Clerk, and
  Girvin}]{Marquardt2007}
Marquardt, F., Chen, J.~P., Clerk, A.~A., Girvin, S.~M., 2007. Quantum theory
  of cavity-assisted sideband cooling of mechanical motion. Physical Review
  Letters 99, 093902.

\bibitem[{Marquardt et~al.(2006)Marquardt, Harris, and Girvin}]{Marquardt2006}
Marquardt, F., Harris, J. G.~E., Girvin, S.~M., 2006. Dynamical multistability
  induced by radiation pressure in high-finesse micromechanical optical
  cavities. Physical Review Letters 96~(10), 103901.
\newline\urlprefix\url{http://link.aps.org/abstract/PRL/v96/e103901}

\bibitem[{Marshall et~al.(2003)Marshall, Simon, Penrose, and
  Bouwmeester}]{Marshall2003}
Marshall, W., Simon, C., Penrose, R., Bouwmeester, D., 2003. Towards {Q}uantum
  {S}uperpositions of a {M}irror. Physical Review Letters 91, 130401.

\bibitem[{Martin et~al.(2004)Martin, Shnirman, Tian, and Zoller}]{Martin2004a}
Martin, I., Shnirman, A., Tian, L., Zoller, P., 2004. Ground-state cooling of
  mechanical oscillators. Physical Review B 69, 125339.

\bibitem[{Matsko et~al.(2007)Matsko, Savchenko, Yu, and Maleki}]{Matsko2007}
Matsko, A.~B., Savchenko, A.~A., Yu, N., Maleki, L., 2007.
  Whispering-gallery-mode resonators as frequency references. {I}. fundamental
  limitations. Journal of the Optical Society of America B 24, 1324--1335.

\bibitem[{Milatz and van Zolingen(1953)}]{Milatz1953a}
Milatz, J. M.~W., van Zolingen, J.~J., 1953. The {B}rownian motion of
  electrometers. Physica 19, 181--194.

\bibitem[{Milatz et~al.(1953)Milatz, van Zolingen, and van Iperen}]{Milatz1953}
Milatz, J. M.~W., van Zolingen, J.~J., van Iperen, B.~B., 1953. The reduciton
  in the {B}rownian motion in electrometers. Physica 19, 195--207.

\bibitem[{Milonni(2005)}]{Milonni2005}
Milonni, P.~W., 2005. Fast light, slow light and left-handed light. Taylor and
  Francis.

\bibitem[{M{\"o}hl et~al.(1980)M{\"o}hl, Petrucci, Thorndahl, and van~der
  Meer}]{Mohl1980}
M{\"o}hl, D., Petrucci, G., Thorndahl, L., van~der Meer, S., 1980. Physics and
  technique of stochastic cooling. Physics Reports 58, 73--102.

\bibitem[{Monroe et~al.(1995)Monroe, Meekhof, King, Jefferts, Itano, Wineland,
  and Gould}]{Monroe1995}
Monroe, C., Meekhof, D.~M., King, B.~E., Jefferts, S.~R., Itano, W.~M.,
  Wineland, D.~J., Gould, P., 1995. Resolved-sideband raman cooling of a bound
  atom to the 3d zero point energy. Physical Review Letters 75~(22),
  4011--4014.

\bibitem[{Mow-Lowry et~al.(2008)Mow-Lowry, Mullavey, Goßler, Gray, and
  McClelland}]{Mow-Lowry2008}
Mow-Lowry, C.~M., Mullavey, A.~J., Goßler, S., Gray, M.~B., McClelland, D.~E.,
  2008. Cooling of a gram-scale cantilever flexure to 70 m{K} with a
  servo-modified optical spring. Physical Review Letters 100, 010801.

\bibitem[{Murch et~al.(2008)Murch, Moore, Gupta, and Stamper-Kurn}]{Murch2008}
Murch, K.~W., Moore, K.~L., Gupta, S., Stamper-Kurn, D.~M., 2008. Observation
  of quantum-measurement backaction with an ultracold atomic gas. Nature
  Physics 4, 561--564.

\bibitem[{Naik et~al.(2006)Naik, Buu, LaHaye, Armour, Clerk, Blencowe, and
  Schwab}]{Naik2006}
Naik, A., Buu, O., LaHaye, M.~D., Armour, A.~D., Clerk, A.~A., Blencowe, M.~P.,
  Schwab, K.~C., 2006. Cooling a nanomechanical resonator with quantum
  back-action. Nature 443, 193--196.

\bibitem[{Nawrodt et~al.(2007)Nawrodt, Zimmer, Koettig, Nietzsche, Th{\"u}rk,
  Vodel, and Seidel}]{Nawrodt2007}
Nawrodt, R., Zimmer, A., Koettig, T., Nietzsche, S., Th{\"u}rk, M., Vodel, M.,
  Seidel, P., 2007. High mechanical {Q}-factor measurements on calcium fluoride
  at cryogenic temperatures. European Physical Journal - Applied Physics 38,
  53--59.

\bibitem[{Nichols and Hull(1901)}]{Nichols1901}
Nichols, E.~F., Hull, G.~F., 1901. A preliminary communication on the pressure
  of heat and light radiation. Physical Review 13, 307--320.

\bibitem[{Nichols and Hull(1903{\natexlab{a}})}]{Nichols1903a}
Nichols, E.~F., Hull, G.~F., 1903{\natexlab{a}}. The pressure due to radiation.
  Physical Review 17, 26--50.

\bibitem[{Nichols and Hull(1903{\natexlab{b}})}]{Nichols1903b}
Nichols, E.~F., Hull, G.~F., 1903{\natexlab{b}}. The pressure due to radiation.
  Physical Review 17, 91--104.

\bibitem[{Nowacki(1975)}]{Nowacki1975}
Nowacki, W., 1975. Dynamic problems of thermoelasticity. Springer.

\bibitem[{Park and Wang(2009)}]{Park2009}
Park, Y.-S., Wang, H., 2009. Resolved-sideband and cryogenic cooling of an
  optomechanical resonator. Nature Physics 5, 489--493.

\bibitem[{Park and Wang(2007)}]{Park2007a}
Park, Y.~S., Wang, H.~L., 2007. Radiation pressure driven mechanical
  oscillation in deformed silica microspheres via free-space evanescent
  excitation. Optics Express 15, 16471--16477.

\bibitem[{Pfeifer et~al.(2007)Pfeifer, Nieminen, Heckenberg, and
  Rubinsztein-Dunlop}]{Pfeifer2007}
Pfeifer, R. N.~C., Nieminen, T.~A., Heckenberg, N.~R., Rubinsztein-Dunlop, H.,
  2007. Colloquium: {M}omentum of an electromagnetic wave in dielectric media.
  Reviews of Modern Physics 79, 1197--1216.

\bibitem[{Phillips et~al.(2001)Phillips, Fleischhauer, Mair, Walsworth, and
  Lukin}]{Phillips2001}
Phillips, D.~F., Fleischhauer, A., Mair, A., Walsworth, R.~L., Lukin, M.~D.,
  2001. Storage of light in atomic vapor. Physical Review Letters 86, 783--786.

\bibitem[{Pinard et~al.(2000)Pinard, Cohadon, Briant, and
  Heidmann}]{Pinard2000}
Pinard, M., Cohadon, P.~F., Briant, T., Heidmann, A., 2000. Full mechanical
  characterization of a cold damped mirror. Physical Review A 63, 013808.

\bibitem[{Pinard et~al.(2005)Pinard, Dantan, Vitali, Arcizet, Briant, and
  Heidmann}]{Pinard2005}
Pinard, M., Dantan, A., Vitali, D., Arcizet, O., Briant, T., Heidmann, A.,
  2005. Entangling movable mirrors in a double-cavity system. Europhysics
  Letters 72, 747--753.

\bibitem[{Pinard et~al.(1999)Pinard, Hadjar, and Heidmann}]{Pinard1999}
Pinard, M., Hadjar, Y., Heidmann, A., 1999. Effective mass in quantum effects
  of radiation pressure. European Physics Journal D 7, 107--116.

\bibitem[{Poggio et~al.(2007)Poggio, Degen, Mamin, and Rugar}]{Poggio2007}
Poggio, M., Degen, L., Mamin, J.~J., Rugar, D., 2007. Feedback cooling of a
  cantilever's fundamental mode below 5 m{K}. Physical Review Letters 99,
  017201.

\bibitem[{Pohl et~al.(2002)Pohl, Liu, and Thompson}]{Pohl2002}
Pohl, R.~O., Liu, X., Thompson, E., 2002. Low-temperature thermal conductivity
  and acoustic attenuation in amorphous solids. Review of Modern Physics 74,
  991--1013.

\bibitem[{Raab et~al.(2000)Raab, Eschner, Bolle, Oberst, Schmidt-Kaler, and
  Blatt}]{Raab2000}
Raab, C., Eschner, J., Bolle, J., Oberst, H., Schmidt-Kaler, F., Blatt, R.,
  2000. Motional sidebands and direct measurement of the cooling rate in the
  resonance fluorescence of a single trapped ion. Physical Review Letters
  85~(3), 538.

\bibitem[{Rabl et~al.(2009)Rabl, Genes, Hammerer, and Aspelmeyer}]{Rabl2009}
Rabl, P., Genes, C., Hammerer, K., Aspelmeyer, M., 2009. Phase-noise induced
  limitations in resolved-sideband cavity cooling of mechanical resonators.
  arxiv:0903.1637.

\bibitem[{Regal et~al.(2008)Regal, Teufel, and Lehnert}]{Regal2008}
Regal, C.~A., Teufel, J.~D., Lehnert, K.~W., 2008. Measuring nanomechanical
  motion with a microwave cavity interferometer. Nature Physics 4, 555--560.

\bibitem[{Rehbein et~al.(2005)Rehbein, Harms, Schnabel, and
  Danzmann}]{Rehbein2005}
Rehbein, H., Harms, J., Schnabel, R., Danzmann, K., 2005. Optical transfer
  functions of {K}err nonlinear cavities and interferometers. Physical Review
  Letters 95, 193001.

\bibitem[{Rempe et~al.(1992)Rempe, Thomson, Kimble, and Lalezari}]{Rempe1992}
Rempe, G., Thomson, R.~J., Kimble, H.~J., Lalezari, R., 1992. Measurement of
  ultralow losses in an optical interferometer. Optics Letters 17, 363--365.

\bibitem[{Rivi{\`e}re et~al.(2010)Rivi{\`e}re, Arcizet, Schliesser,
  Anetsberger, and Kippenberg}]{Riviere2009}
Rivi{\`e}re, R., Arcizet, O., Schliesser, A., Anetsberger, G., Kippenberg, T.,
  2010. Design and operation of a cryogenic taper-coupling setup for cavity
  optomechanical experiments(in preparation).

\bibitem[{Rokhsari et~al.(2005)Rokhsari, Kippenberg, Carmon, and
  Vahala}]{Rokhsari2005}
Rokhsari, H., Kippenberg, T.~J., Carmon, T., Vahala, K.~J., 2005.
  Radiation-pressure-driven micro-mechanical oscillator. Optics Express 13,
  5293--5301.

\bibitem[{Rokhsari et~al.(2006)Rokhsari, Kippenberg, Carmon, and
  Vahala}]{Rokhsari2006a}
Rokhsari, H., Kippenberg, T.~J., Carmon, T., Vahala, K.~J., 2006. Theoretical
  and experimental study of radiation pressure-induced mechanical oscillations
  (parametric instability) in optical microcavities. IEEE Journal of Selected
  Topics in Quantum Electronics 12~(1), 96--107.

\bibitem[{Sandoghdar et~al.(1996)Sandoghdar, Treussart, Hare,
  Lef{\`e}vre-Seguin, Raimond, and Haroche}]{Sandoghdar1996}
Sandoghdar, V., Treussart, F., Hare, J., Lef{\`e}vre-Seguin, V., Raimond,
  J.-M., Haroche, S., 1996. Very low threshold whispering-gallery-mode
  microsphere laser. Physical Review A 54~(3), R1777--R1780.

\bibitem[{Schliesser et~al.(2008{\natexlab{a}})Schliesser, Anetsberger,
  Rivi{\`e}re, Arcizet, and Kippenberg}]{Schliesser2008b}
Schliesser, A., Anetsberger, G., Rivi{\`e}re, R., Arcizet, O., Kippenberg,
  T.~J., 2008{\natexlab{a}}. High-sensitivity monitoring of micromechanical
  vibration using optical whispering gallery mode resonators. New Journal of
  Physics 10, 095015.

\bibitem[{Schliesser et~al.(2009)Schliesser, Arcizet, Rivi{\`e}re, Anetsberger,
  and Kippenberg}]{Schliesser2009a}
Schliesser, A., Arcizet, O., Rivi{\`e}re, R., Anetsberger, G., Kippenberg, T.,
  2009. Resolved-sideband cooling and position measurement of a micromechanical
  oscillator close to the {H}eisenberg uncertainty limit. Nature Physics 5,
  509--514.

\bibitem[{Schliesser et~al.(2006)Schliesser, Del'Haye, Nooshi, Vahala, and
  Kippenberg}]{Schliesser2006}
Schliesser, A., Del'Haye, P., Nooshi, N., Vahala, K., Kippenberg, T., 2006.
  Radiation pressure cooling of a micromechanical oscillator using dynamical
  backaction. Physical Review Letters 97, 243905.

\bibitem[{Schliesser et~al.(2008{\natexlab{b}})Schliesser, Rivi{\`e}re,
  Anetsberger, Arcizet, and Kippenberg}]{Schliesser2008}
Schliesser, A., Rivi{\`e}re, R., Anetsberger, G., Arcizet, O., Kippenberg, T.,
  2008{\natexlab{b}}. Resolved-sideband cooling of a micromechanical
  oscillator. Nature Physics 4, 415--419.

\bibitem[{Schr{\"o}ter et~al.(2008)Schr{\"o}ter, Reich, Arcizet, R{\"a}dler,
  Nickel, and Kippenberg}]{Schroter2008}
Schr{\"o}ter, B., Reich, C., Arcizet, O., R{\"a}dler, J.~O., Nickel, B.,
  Kippenberg, T.~J., 2008. Chip based, lipid bilayer functionalized
  microresonators for label-free, ultra sensitive and time-resolved molecular
  detection. submitted.

\bibitem[{Schwab and Roukes(2005)}]{Schwab2005}
Schwab, K.~C., Roukes, M.~L., 2005. Putting mechanics into quantum mechanics.
  Physics Today 58~(7), 36--42.

\bibitem[{Sheard et~al.(2004)Sheard, Gray, Mow-Lowry, McClelland, and
  Whitcomb}]{Sheard2004}
Sheard, B.~S., Gray, M.~B., Mow-Lowry, C.~M., McClelland, D.~E., Whitcomb,
  S.~E., 2004. Observation and characterization of an optical spring. Physical
  Review A 69, 051801.

\bibitem[{Spillane et~al.(2003)Spillane, Kippenberg, Painter, and
  Vahala}]{Spillane2003}
Spillane, S.~M., Kippenberg, T.~J., Painter, O.~J., Vahala, K.~J., 2003.
  Ideality in a fiber-taper-coupled microresonator system for application to
  cavity quantum electrodynamics. Physical Review Letters 91~(4), 043902.

\bibitem[{Spillane et~al.(2002)Spillane, Kippenberg, and Vahala}]{Spillane2002}
Spillane, S.~M., Kippenberg, T.~J., Vahala, K.~J., 2002. Ultralow-threshold
  {R}aman laser using a spherical dielectric microcavity. Nature 415~(6872),
  621--623.

\bibitem[{Srinivasan and Painter(2007)}]{Srinivasan2007}
Srinivasan, K., Painter, O., 2007. Optical fiber taper coupling and
  high-resolution wavelength tuning of microdisk resonators at cryogenic
  temperatures. Applied Physics Letters 90, 031114.

\bibitem[{Stenholm(1986)}]{Stenholm1986}
Stenholm, S., 1986. The semiclassical theory of laser cooling. Reviews of
  Modern Physics 58, 699--739.

\bibitem[{Tamura(2009)}]{Tamura2009}
Tamura, S.-I., 2009. Vibrational cavity modes in a free cylindrical disk.
  Physical Review B 79, 054302.

\bibitem[{Teufel et~al.(2009)Teufel, Donner, Castellanos-Beltran, Harlow, and
  Lehnert}]{Teufel2009}
Teufel, J.~D., Donner, R., Castellanos-Beltran, M.~A., Harlow, J.~W., Lehnert,
  K.~W., 2009. Nanomechanical motion measured with precision beyond the
  standard quantum limit. arxiv.

\bibitem[{Teufel et~al.(2008)Teufel, Harlow, Regal, and Lehnert}]{Teufel2008}
Teufel, J.~D., Harlow, J.~D., Regal, C.~A., Lehnert, K.~W., 2008. Dynamical
  backaction of microwave fields on a nanomechanical oscillator. Physical
  Review Letters 101, 197203.

\bibitem[{Thomson et~al.(2008)Thomson, Zwickl, Jayich, Marquardt, Girvin, and
  Harris}]{Thomson2007}
Thomson, J.~D., Zwickl, B.~M., Jayich, A.~M., Marquardt, F., Girvin, S.~M.,
  Harris, J. G.~E., 2008. Strong dispersive coupling of a high finesse cavity
  to a micromechanical membrane. Nature 452, 72--75.

\bibitem[{Tian and Zoller(2004)}]{Tian2004}
Tian, L., Zoller, P., 2004. Coupled ion-nanomechanical systems. Physical Review
  Letters 93~(26), 266403.

\bibitem[{Tielb{\"u}rger et~al.(1992)Tielb{\"u}rger, Merz, Ehrenfels, and
  Hunklinger}]{Tielburger1992}
Tielb{\"u}rger, D., Merz, R., Ehrenfels, R., Hunklinger, S., 1992. Thermally
  activated relaxation processes in vitreous silica: {A}n investigation by
  {B}rillouin scattering at high pressures. Physical Review B 45, 2750--2760.

\bibitem[{Tittonen et~al.(1999)Tittonen, Breitenbach, Kalkbrenner, M{\"u}ller,
  Conradt, Schiller, Steinsland, Blanc, and de~Rooij}]{Tittonen1999}
Tittonen, I., Breitenbach, G., Kalkbrenner, T., M{\"u}ller, T., Conradt, R.,
  Schiller, S., Steinsland, E., Blanc, N., de~Rooij, N.~F., 1999.
  Interferometric measurements of the position of a macroscopic body: Towards
  observations of quantum limits. Physical Review A 59, 1038--1044.

\bibitem[{Treussart et~al.(1998)Treussart, Ilchenko, Roch, Hare,
  Lef{\`e}vre-Seguin, Raimond, and Haroche}]{Treussart1998}
Treussart, F., Ilchenko, V.~S., Roch, J.-F., Hare, J., Lef{\`e}vre-Seguin, V.,
  Raimond, J.-M., Haroche, S., 1998. Evidence for intrinsic {K}err bistability
  of high-{Q} resonators in suprafluid helium. European Physical Journal D 1,
  235--238.

\bibitem[{Vacher et~al.(2005)Vacher, Courtens, and Foret}]{Vacher2005}
Vacher, R., Courtens, E., Foret, M., 2005. Anharmonic versus relaxational sound
  damping in glasses. {I}{I}. {V}itreous silica. Physical Review B 72, 214205.

\bibitem[{Vahala et~al.(2009)Vahala, Herrmann, Kn{\"u}nz, Batteiger, Saathoff,
  H{\"a}nsch, and Udem}]{Vahala2009}
Vahala, K., Herrmann, M., Kn{\"u}nz, S., Batteiger, V., Saathoff, G.,
  H{\"a}nsch, T.~W., Udem, T., 2009. A phonon laser. Nature Physics Advance
  online publication.

\bibitem[{Vahala(2008)}]{Vahala2008}
Vahala, K.~J., 2008. Back-action limit of linewidth in an optomechanical
  oscillator. Physical Review A 78, 023832.

\bibitem[{Vassiliev et~al.(1998)Vassiliev, Velichansky, Ilchenko, Gorodetsky,
  Hollberg, and Yarovitsky}]{Vassiliev1998}
Vassiliev, V.~V., Velichansky, V.~L., Ilchenko, V.~S., Gorodetsky, M.~L.,
  Hollberg, L., Yarovitsky, A.~V., 1998. Narrow-line-width diode laser with a
  high-{Q} microsphere resonator. Optics Communications 158~(1-6), 305--312.

\bibitem[{Verbridge et~al.(2008)Verbridge, Craighead, and
  Parpia}]{Verbridge2008}
Verbridge, S.~S., Craighead, H.~G., Parpia, J.~M., 2008. A megahertz
  nanomechanical resonator with room temperature quality factor over a million.
  Applied Physics Letters 92, 013112.

\bibitem[{Verbridge et~al.(2007)Verbridge, Shapiro, Craighead, Parpia, and
  Jeevak}]{Verbridge2007}
Verbridge, S.~S., Shapiro, D.~F., Craighead, H.~G., Parpia, J.~M., Jeevak, M.,
  2007. Macroscopic tuning of nanomechanics: {S}ubstrate bending for reversible
  control of frequency and quality factor of nanostring resonators. Nano
  Letters 7, 1728--1735.

\bibitem[{Verlot et~al.(2008)Verlot, Tavernarakis, Briant, Cohadon, and
  Heidmann}]{Verlot2008}
Verlot, P., Tavernarakis, A., Briant, T., Cohadon, P.-F., Heidmann, A., 2008.
  Scheme to probe optomechanical correlations between two optical beams down to
  the quantum level. Physical Review Letters 102, 103601.

\bibitem[{Vernooy et~al.(1998{\natexlab{a}})Vernooy, Furusawa, Georgiades,
  Ilchenko, and Kimble}]{Vernooy1998a}
Vernooy, D.~W., Furusawa, A., Georgiades, A.~P., Ilchenko, V.~S., Kimble,
  H.~J., 1998{\natexlab{a}}. Cavity {Q}{E}{D} with high-{Q} whispering gallery
  modes. Physical Review A 57, R2293--R2296.

\bibitem[{Vernooy et~al.(1998{\natexlab{b}})Vernooy, Ilchenko, Mabuchi, Sreed,
  and Kimble}]{Vernooy1998}
Vernooy, D.~W., Ilchenko, V.~S., Mabuchi, H., Sreed, W.~W., Kimble, H.~J.,
  1998{\natexlab{b}}. High-{Q} measurements of fused-silica microspheres in the
  near infrared. Optics Letters 23~(4), 247--249.

\bibitem[{Vinante et~al.(2008)Vinante, Bignotto, Bonaldi, Cerdonio, Conti,
  Falferi, Liguori, Longo, Mezzena, Ortolan, Prodi, Salemi, Taffarello,
  Vedovato, Vitale, and Zendri}]{Vinante2008}
Vinante, A., Bignotto, M., Bonaldi, M., Cerdonio, M., Conti, L., Falferi, P.,
  Liguori, N., Longo, R., Mezzena, R., Ortolan, A., Prodi, G.~A., Salemi, F.,
  Taffarello, L., Vedovato, G., Vitale, S., Zendri, J.-P., 2008. Feedback
  cooling of the normal modes of a massive electromechanical system to
  submillikelvin temperature. Physical Review Letters 101, 033601.

\bibitem[{Vitali et~al.(2007)Vitali, Gigan, Ferreira, Bohm, Tombesi, Guerreiro,
  Vedral, Zeilinger, and Aspelmeyer}]{Vitali2007}
Vitali, D., Gigan, S., Ferreira, A., Bohm, H.~R., Tombesi, P., Guerreiro, A.,
  Vedral, V., Zeilinger, A., Aspelmeyer, M., 2007. Optomechanical entanglement
  between a movable mirror and a cavity field. Physical Review Letters 98,
  030405.

\bibitem[{Vollmer et~al.(2009)Vollmer, Arnold, and Keng}]{Vollmer2009}
Vollmer, F., Arnold, S., Keng, D., 2009. Single virus detection from the
  reactive shift of a whispering-gallery mode. PNAS 105, 20701--20704.

\bibitem[{Vollmer et~al.(2002)Vollmer, Braun, Libchaber, Khoshsima, Teraoka,
  and Arnold}]{Vollmer2002}
Vollmer, F., Braun, D., Libchaber, A., Khoshsima, M., Teraoka, I., Arnold, S.,
  2002. Protein detection by optical shift of a resonant microcavity. Applied
  Physics Letters 80~(21), 4057--4059.

\bibitem[{Weld and Kapitulnik(2006)}]{Weld2006}
Weld, D.~M., Kapitulnik, A., 2006. Feedback control and characterization of a
  microlever using optial radiation pressure. Applied Physics Letters 89,
  164102.

\bibitem[{White(1975)}]{White1975}
White, G.~K., 1975. Thermal expansion of vitreous silica at low temperatures.
  Physical Review Letters 34, 204--205.

\bibitem[{Wilson-Rae(2008)}]{Wilson-Rae2008}
Wilson-Rae, I., 2008. Intrinsic dissipation in nanomechanical resonators due to
  phonon tunneling. Physical Review B 77, 245418.

\bibitem[{Wilson-Rae et~al.(2007)Wilson-Rae, Nooshi, Zwerger, and
  Kippenberg}]{Wilson-Rae2007}
Wilson-Rae, I., Nooshi, N., Zwerger, W., Kippenberg, T.~J., 2007. Theory of
  ground state cooling of a mechanical oscillator using dynamical backaction.
  Physical Review Letters 99~(9), 093901.

\bibitem[{Wilson-Rae et~al.(2004)Wilson-Rae, Zoller, and
  Imamoglu}]{Wilson-Rae2004}
Wilson-Rae, I., Zoller, P., Imamoglu, A., 2004. Laser cooling of a
  nanomechanical resonator mode to its quantum ground state. Physical Review
  Letters 92~(7), 075507.

\bibitem[{Wineland and Dehmelt(1975)}]{Wineland1975}
Wineland, D.~J., Dehmelt, H., 1975. Proposed $10^{14} \delta\nu<\nu$ {L}aser
  {F}luorescence {S}pectroscopy on {T}l$^+$ {I}on {M}ono-{O}scillator. Bulletin
  of the American Physical Society 20, 637.

\bibitem[{Wineland and Itano(1979)}]{Wineland1979}
Wineland, D.~J., Itano, W.~M., 1979. Laser cooling of atoms. Physical Review A
  20, 1521--1540.

\bibitem[{Woodruff(1968)}]{Woodruff1968}
Woodruff, A.~E., 1968. The radiometer and how it does not work. The Physics
  Teacher 6, 358--363.

\bibitem[{Yuen and Chan(1983)}]{Yuen1983}
Yuen, H.~P., Chan, V. W.~S., 1983. Noise in homodyne and heterodyne detection.
  Optics Letters 8, 177--179.

\bibitem[{Zener(1937)}]{Zener1937}
Zener, C., 1937. Internal friction in solids i. theory of internal friction in
  reeds. Physical Review 52, 230--235.

\bibitem[{Zener(1938)}]{Zener1938}
Zener, C., 1938. Internal friction in solids {I}{I}. {G}eneral theory of
  thermoelastic internal friction. Physical Review 53, 90--99.

\bibitem[{Zhang et~al.(2003)Zhang, Peng, and Braunstein}]{Zhang2003}
Zhang, J., Peng, K., Braunstein, S.~L., 2003. Quantum-state transfer from light
  to macroscopic oscillators. Physical Review A 68, 013808.

\bibitem[{Zhang et~al.(1995)Zhang, Poizat, et~al.}]{Zhang1995}
Zhang, T.~C., Poizat, J.~P., et~al., 1995. Quantum noise of free-running and
  externally-stabilized laser diodes. Quantum and semiclassical optics 7,
  601--613.

\bibitem[{Zhou et~al.(2009)Zhou, Arcizet, Schliesser, Rivi{\`e}re, and
  Kippenberg}]{Zhou2009}
Zhou, X., Arcizet, O., Schliesser, A., Rivi{\`e}re, R., Kippenberg, T., 2009.
  in preparation.

\bibitem[{Zwickl et~al.(2008)Zwickl, Shanks, Jayich, Yang, Bleszynski~Jayich,
  Thomson, and Harris}]{Zwickl2008}
Zwickl, B.~M., Shanks, W.~E., Jayich, A.~M., Yang, C., Bleszynski~Jayich, C.,
  Thomson, J.~D., Harris, J. G.~E., 2008. High quality mechanical and optical
  properties of commercial silicon nitride membranes. Applied Physics Letters
  92, 103125.

\end{thebibliography}

\end{document}